\documentclass[12pt,twoside,openright]{book}
\usepackage{simplewick}
\usepackage{dsfont}
\usepackage{subcaption, csquotes, setspace}
\usepackage{amsmath, amssymb, mathrsfs, bbm}
\usepackage{graphicx}
\usepackage[usenames,dvipsnames]{color}
\definecolor{nicegray}{cmyk}{0,0,0,0.5}

\newcommand{\tcb}{\textcolor{blue}}

\usepackage[lmargin=3.5cm,rmargin=2.5cm,tmargin=3.5cm,bmargin=2.5cm]{geometry}
\usepackage{framed}

\usepackage{titlesec}
\titleformat{\chapter}[display]
    {\flushright \LARGE \bfseries\sffamily \color{nicegray}}{\chaptertitlename\ \thechapter}{6pt}{ \flushleft \Huge}
   \titleformat{\section}
  {\sffamily\Large\bfseries}
  {\thesection}{1em}{}
\usepackage{fancyhdr}
\usepackage{cite}
\usepackage{tocloft}
\usepackage[bookmarks]{hyperref}

\newcommand{\bi}{\begin{itemize}}
\newcommand{\ei}{\end{itemize}}
\newcommand{\bea}{\begin{align}}
\newcommand{\eea}{\end{align}}
\newcommand{\be}{\begin{equation}}
\newcommand{\ee}{\end{equation}}

\newcommand{\pl}{{\partial}}

\fancypagestyle{plain}{ %
  \fancyhf{} 
  
}

\renewcommand{\chaptermark}[1]%
         {\markboth{\thechapter.\ #1}{}}
\renewcommand{\sectionmark}[1]%
         {\markright{\thesection\ #1}}
\newcommand\lrpar{\raise .8ex\hbox{$^\leftrightarrow$} \hspace{-9pt}
\partial}

\newcommand*\longhookrightarrow{\ensuremath{\lhook\joinrel\relbar\joinrel\rightarrow}}

\titleformat*{\subsection}{\large \bfseries\sffamily}
\titleformat*{\subsubsection}{\bfseries\sffamily}

\newcommand{\MYTitle}[9]{
  \thispagestyle{empty}
  \vspace*{\stretch{1}}
  {\parindent0cm
   \rule{\linewidth}{.0ex}}
  \begin{flushright}

    \vspace*{\stretch{1}}
    \sffamily\bfseries\Huge
    #1\\
     \vspace*{0.25cm}
    \rule{7.9cm}{0.05ex}\\
    \vspace*{0.35cm} 
    \sffamily\bfseries\large
     Dissertation by #2
    \vspace*{\stretch{1}}
  \end{flushright}
  \rule{\linewidth}{.0ex}
  \vspace*{\stretch{5}}
  \begin{flushright}
    \includegraphics[width=2in]{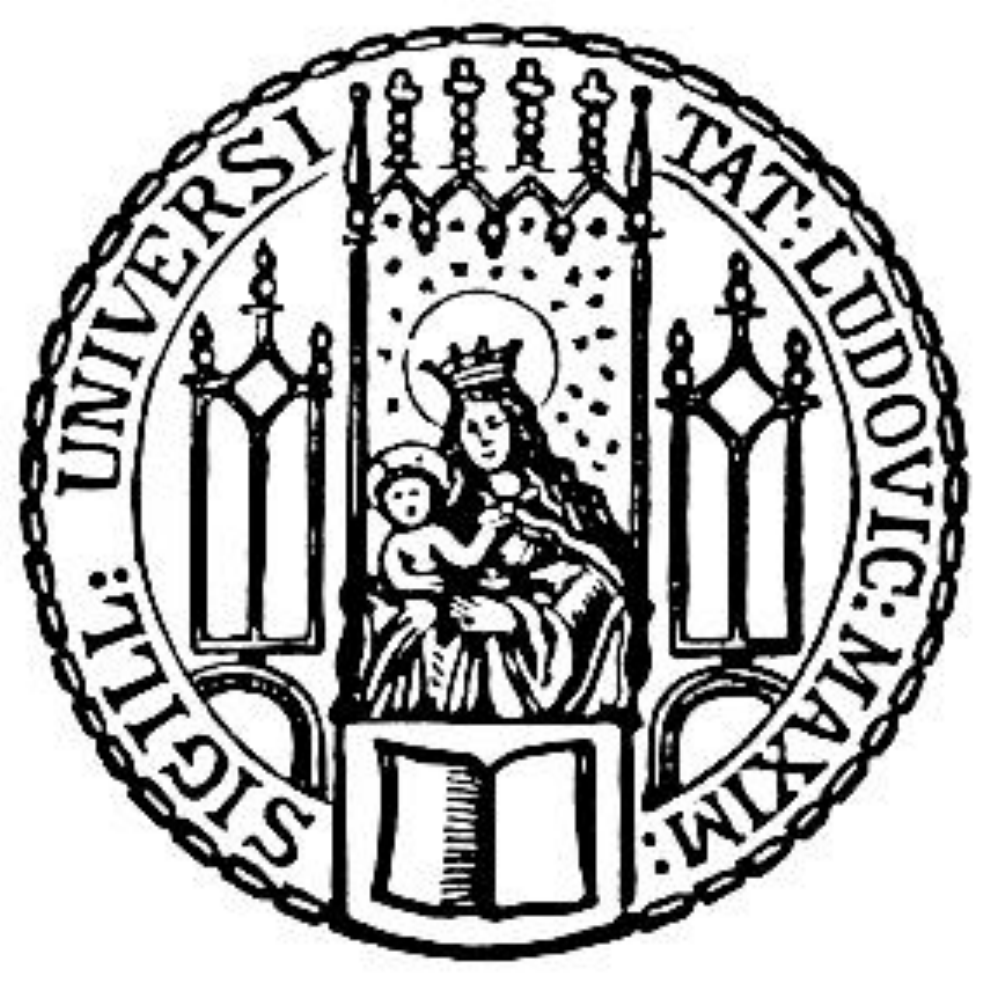}
      \end{flushright}
       \vspace*{\stretch{1}}
      
  \newpage  
  \thispagestyle{empty}

  {\parindent0cm}
      \vspace*{\stretch{0.25}}
  \begin{flushright}
    \sffamily\bfseries\Huge
    #1\\
    \vspace*{\stretch{0.3}}
  \end{flushright}
  \begin{center}  \sffamily
    \Large Dissertation\\
    \Large an der #4\\
    \Large der Ludwig--Maximilians--Universit\"at\\
    \Large M\"unchen\\
    \vspace*{\stretch{0.25}}
    \Large vorgelegt von\\
    \Large #2\\
    \Large aus #3\\
    \vspace*{\stretch{0.25}}
    \Large M\"unchen, den #6
        \vspace*{\stretch{.5}}
  \end{center}
    \newpage
  \thispagestyle{empty}

  \vspace*{\stretch{1}}

  \begin{flushleft}
  \includegraphics[width=2.1cm]{siegel.pdf}\\
  \vspace*{0.6cm}
   \sffamily
 {\large Dissertation} \\[7mm]
 Submitted to the faculty of physics of the\\
Ludwig--Maximilians--Universit\"at M\"unchen \\[6mm]
by Charlotte Sleight \\[6mm]
Supervised by Prof. Dr. Johanna Karen Erdmenger\\
Max-Planck-Institut f\"ur Physik, M\"unchen \\[6mm]

    1st Referee:  #7 \\[1mm]
    2nd Referee: #8 \\[6mm]
    Date of submission: May 2nd 2016\\[1mm]
    Date of oral examination: July 26th 2016 \\
  \end{flushleft}

}

\begin{document}

  \frontmatter

  \MYTitle
{\flushleft Interactions in Higher-Spin Gravity: \\ \flushright A Holographic Perspective}
{Charlotte Sleight}                      
{Hull, Gro\ss britannien}
{Fakult\"at f\"ur Physik}
{M\"unchen 2016}
{2. Mai 2016}                            
{Prof. Dr. Johanna Karen Erdmenger}                          
{Prof. Dr. Dieter L\"ust}                         
 {??}                         

\newpage 
  \thispagestyle{empty}
  
  {
      \vspace*{\stretch{5}}
  \begin{flushright}
   
    \emph{To my dad, sister, and the wonderful memory of my mum.}
   
  \end{flushright}}
\newpage

\thispagestyle{empty}
  
  {
      \vspace*{7.5cm}
  \begin{flushleft}
   
   \hspace*{1.5cm} \emph{``Die Probleme werden gel\"ost, nicht durch Beibringen neuer Erfahrungen, sondern durch Zusammenstellung des l\"angst Bekannten.''}
        \end{flushleft}
      \begin{flushright}
      --- Ludwig Wittgenstein
       \end{flushright}
        
        }

 \addcontentsline{toc}{chapter}{\protect Thesis publications}

\thispagestyle{empty}

\noindent
\textcolor{nicegray}{{\bf \sffamily \huge Thesis publications}}\\

\noindent
This thesis is based on some of the author's work  that was carried out during the period of May 2013 - March 2016 as a Ph.D. student at the Max-Planck-Institut f\"ur Physik (Werner-Heisenberg-Institut) M\"unchen. The corresponding publications are:\\

\noindent
\cite{Bekaert:2014cea} {\bf Towards holographic higher-spin interactions: Four-point functions and \hspace*{0.7cm} higher-spin exchange.}\\
     \hspace*{0.7cm} Xavier Bekaert, Johanna Erdmenger, Dmitry Ponomarev and Charlotte Sleight. \\
     \hspace*{0.7cm} Published in JHEP: \href{http://link.springer.com/article/10.1007\%2FJHEP03\%282015\%29170}{\color{blue} JHEP {\bf 1503} (2015) 170 }\\

\noindent
\cite{Bekaert:2015tva} {\bf Quartic AdS Interactions in Higher-Spin Gravity from CFT}\\
     \hspace*{0.7cm} Xavier Bekaert, Johanna Erdmenger, Dmitry Ponomarev and Charlotte Sleight. \\
     \hspace*{0.7cm} Published in JHEP: \href{http://link.springer.com/article/10.1007\%2FJHEP11\%282015\%29149}{\color{blue} JHEP {\bf 1511} (2015) 149}\\
     
\noindent
\cite{Bekaert:2016ezc} {\bf Bulk quartic vertices from boundary four-point correlators}\\
     \hspace*{0.7cm} Xavier Bekaert, Johanna Erdmenger, Dmitry Ponomarev and Charlotte Sleight. \\ \hspace*{0.7cm} \href{https://arxiv.org/abs/1602.08570}{\color{blue}arXiv:1602.08570 [hep-th]}, published in World Scientific.
     \\

\noindent
\cite{Sleight:2016dba} {\bf Higher-spin Interactions from CFT: The Complete Cubic Couplings}\\
     \hspace*{0.7cm} Charlotte Sleight and Massimo Taronna. \\
     \hspace*{0.7cm} Published in Physical Review Letters: \href{https://arxiv.org/abs/1603.00022}{\color{blue}Phys.Rev.Lett. 116 (2016)} no.18, 181602\\
     \vspace*{-0.25cm} 
    
\hspace*{-.45cm}Subsequent developments, which appeared after thesis submission:

\vspace*{0.1cm} 
\noindent
\cite{Sleight:2016xqq} {\bf Higher-Spin Algebras, Holography and Flat Space}\\
     \hspace*{0.7cm} Charlotte Sleight and Massimo Taronna\\
     \hspace*{0.7cm} Published in JHEP:  \href{https://link.springer.com/article/10.1007\%2FJHEP02\%282017\%29095}{\color{blue}JHEP {\bf 1702} (2017) 095}\\
     
  \newpage 
    \tableofcontents
 \markboth{Contents}{Contents}

\newpage
  \cleardoublepage

   \mainmatter\setcounter{page}{1}
   \chapter{Introduction}

\vspace*{0.5cm}

\section{A tale of two theories}

Existing physical theories provide a powerful framework to describe all phenomena observed so far in Nature. To date, the range of length scales accessed by physical experiment is colossal: From the deep subatomic distances of $\sim 10^{-19}$ m probed by collisions of highly accelerated particles, to $\sim 10^{25}$ m from deep sky surveys.

The two monumental pillars behind our theoretical understanding of Nature thus far are Quantum Field Theory (QFT) and General Relativity:

 QFT describes the dynamics of elementary particles. All known elementary particles and their interactions are encoded in the Lagrangian of the Standard Model of particles physics \cite{Glashow:1961tr,Weinberg:1967tq,salam}, and consists of three fundamental ingredients: Matter fields, the Brout-Englert-Higgs boson responsible for particle masses \cite{PhysRevLett.13.321,HIGGS1964132,PhysRevLett.13.508,PhysRevLett.13.585} and force carriers (gauge bosons). The gauge bosons account for three of the four known fundamental interactions amongst the constituents of visible matter: The electromagnetic, strong nuclear and weak nuclear forces.

 The remaining fundamental force is gravity. Einstein's theory of General Relativity \cite{Einstein:1916vd} provides a beautiful theory of gravitation at the classical level, which elegantly geometrises dynamics. Gravity is a dominant force at large, cosmic, scales and indeed, General Relativity is the backbone of the cosmological standard model (or $\Lambda$CDM model). The latter provides a coherent framework for the history of the Universe from Big Bang nucleosynthesis up to the present day.

In spite of these successes, the theoretical tools of QFT --- that have been so effective in describing interactions at the subatomic scale --- are unfortunately not directly applicable to the gravitational force. The characteristic gravitational scale is the Planck scale $\ell_{pl} \sim 10^{-35}$ m, with the gravitational coupling being $\sim \ell_p^2 / L^2$ in a process at length scales $L$. Unlike for the Standard Model, General Relativity therefore suffers a lack of predictive power at Planckian scales, where it becomes strongly coupled (in other words, it is non-renormalisable \cite{Goroff:1985th,Goroff:1985sz}). An understanding of how to reconcile QFT and General Relativity is imperative, since gravity couples to matter and consequently we cannot forgo the need for a  quantum description of it.

One approach, which has proven to be particularly successful in improving UV behaviour, is to consider coupling gravity to an enlarged spectrum of gauge and matter fields. A notable example is ${\cal N} = 8$ supergravity, which contains one graviton, 8 gravitinos (which have spin $3/2$), 28 vector bosons, 56 fermions and 70 scalar fields. Current results indicate its finiteness at seven-loop order \cite{Bern:2007hh,Green:2006yu,Bern:2009kd,Beisert:2010jx,Kallosh:2011dp}, with speculations that it is all-loop finite. These promising results illustrate that the idea of increasing the particle content, and symmetry, is one worth pursuing.

\section{Higher-spins and strings}

From a Wilsonsian perspective, both the Standard Model and General Relativity are effective field theories describing the relevant degrees of freedom up to a certain energy scale. The ultimate physical theory is viewed as having an onion-like structure, with different layers corresponding to different energies, with different degrees of freedom being excited and governed by different Lagrangians. In this regard, the break down of General relativity at the Planck scale indicates that it should be substituted by a new, more fundamental theory.

The problem of formulating a consistent quantum theory that includes as its low-energy limit both the Standard Model and General Relativity is known as the quantum gravity problem. Because energies at which quantum-gravitational effects become significant are not directly accessible via existing experiments, the development of the subject has been largely driven by Gedankenexperiments and symmetry principles.

For example, considering General Relativity as an effective quantum theory with cut-off $M_{pl}$, counter-terms to the Einstein-Hilbert action appear in the form of higher curvature invariants (due to diffeomorphism invariance) and therefore additional derivatives. At one loop for example, the only correction that is generated is the Gauss-Bonnet term,
\begin{equation}
S_{\text{1-loop}} = S_{EH} + S_{GB} = \frac{1}{\ell^{d-2}_{p}} \int \sqrt{-g}\left[ R + \alpha \left(R^2 - 4 R^{\mu \nu} R_{\mu \nu}+ R_{\mu \nu \rho \sigma}R^{\mu \nu \rho \sigma}\right)\right],
\end{equation}
with $\alpha >> \ell^2_p$. Such corrections (${\cal O}\left(R^2, R^3, ...\right)$) lead to short distance violations of causality at the classical level \cite{Camanho:2014apa}, inducing (Shapiro) time \emph{advances}. Causality cannot be restored by adding local higher curvature terms, but instead by introducing an infinite number of new massive particles of spins $s > 2$ and mass $\sim 1/\sqrt{\alpha}$.

We see from above example that basic consistency requirements seem to imply that a theory of quantum gravity  should admit a completion to a theory with an enlarged spectrum, which includes an infinite tower of higher-spin excitations.

A promising research program which addresses the problem of quantum gravity is string theory, which indeed has a spectrum containing infinitely many massive excitations of increasing mass and spin. These arise from the different vibrational modes of the fundamental string, in an enticing unified manner. The string states dispose themselves on Regge trajectories in the (mass)$^2$ vs. spin plane, where for the first Regge trajectory we have, e.g. for the open bosonic spin in flat space,
\begin{equation}
\alpha^\prime m^2_s = s - 1, \label{lregge}
\end{equation}
with $\alpha^\prime = l_s^2$ the square of the string length. In fact, the higher-spin resonances are responsible for the improved UV behaviour \cite{Veneziano:1968yb,Amati:1987wq,Gross:1987kza,Gross:1987ar,Amati:1987uf,Amati:191788,PhysRevD.37.359,Sundborg:1988tb,Moeller:2005ez} compared to General Relativity above, essentially because the string length $l_s$ acts as a natural UV-cutoff.

An important observation from our experience with QFTs is that improved UV behaviour is typically the signal of an underlying symmetry. In addition to the example of supersymmetry mentioned previously, another well-known example of this phenomenon is given by 
the weak interaction: Fermi's four-fermion theory is non-renormalisable, but describes the weak interaction remarkably well up to $\sim 100$ GeV. Together with QED, it can be considered as the low-energy effective theory for the Standard Model of the electroweak interactions. In the latter, as opposed to the effective four-point interaction of Fermi's theory, weak interactions are mediated by massive $W^{\pm}$ and $Z$ bosons. These arise from the spontaneous breaking of an $ SU\left(2\right) \times U\left(1\right)$ gauge symmetry, which emerges at very high energies.

One may then ask whether a similar phenomenon occurs in string theory. This question was indeed posed by Gross in 1988 \cite{Gross:1988ue}, and subsequently explored in a number of works (for instance: \cite{Gross:1989ge,Moore:1993ns,Moeller:2005ez,Taronna:2010qq,Sagnotti:2010at,Taronna:2012gb,Gaberdiel:2015mra,Gaberdiel:2015wpo}). Just like the broken $SU\left(2\right) \times U\left(1\right)$ symmetry of the electroweak interactions can be seen by examining weak scattering amplitudes at energies high enough so that the masses of $W^{\pm}$ and $Z$ can be neglected, by studying the high energy ($\alpha^\prime \rightarrow \infty$) behaviour of string scattering, Gross argued for the existence of an infinite dimensional symmetry group that gets restored at high energies. Should this be the case, the symmetry would be generated by gauge symmetries associated to the infinite family of  higher-spin gauge fields that emerge in this limit.\footnote{For example, recall that on the first Regge trajectory in flat space we have $m^2_s \propto 1/\alpha^{\prime} \rightarrow 0$.}

With the presence of an infinite dimensional symmetry comes a lot of control. Uncovering such an underlying symmetry principle behind string theory could therefore lead to a greater understanding of quantum gravity at high energies. For example, from the space-time viewpoint string theory is still formulated in a background-dependent manner, without an analogue of the powerful geometric background independent presentation of Einstein gravity.
Further more, at present both the action principle and equations of motion for the second quantised form of string theory (string field theory) require the addition of infinitely many auxiliary fields.\footnote{At the free level, the second quantised formulation of all string models was completed in the `80s \cite{Kato:1982im,Siegel:1985xj,Siegel:1985tw,Banks:1985ff,Ohta:1985af,Ohta:1985zw,NEVEU1986573,NEVEU1986125}. While at the interacting level, see for example: \cite{Witten:1985cc,Zwiebach:1992ie,Erler:2016ybs}.} In this sense there is much to be unravelled about the properties of high energy interactions in quantum gravity.

\section{This thesis}
In this thesis we aim to shed some light on the nature of interactions in the presence of higher-spin symmetry. To this end we employ a particular consequence of string theory known as \emph{holography}, focusing on its most celebrated incarnation: The AdS/CFT correspondence \cite{Maldacena:1997re,Gubser:1998bc,Witten:1998qj}.

\begin{figure}[h]
  \centering
  \includegraphics[scale=0.5]{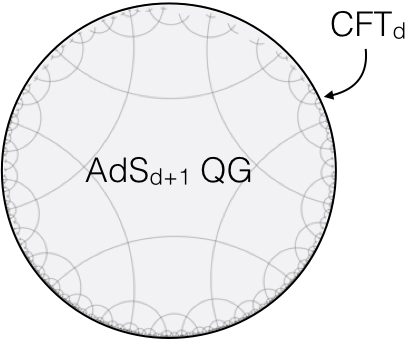}
  \caption{For a given dual pair \eqref{adscft0}, we can think of the CFT$_{d}$ as living on the boundary (solid black boundary of disc) of the gravity theory in asymptotically AdS$_{d+1}$ (entire disc). This picture can be obtained by considering Euclidean AdS and compactifying the $\mathbb{R}^d$ boundary to $S^d$ by adding a point at infinity.} \label{hsduality}
\end{figure}

 In its most general form, the AdS/CFT correspondence is the statement that two systems are equivalent:
\begin{equation}
    \text{AdS$_{d+1}$ QG = CFT$_{d}$.}\label{adscft0}
\end{equation}
Quantum gravity in asymptotically anti-de Sitter spacetime AdS$_{d+1}$ and non-gravitational conformal field theory\footnote{A conformal field theory is a quantum field theory without any intrinsic scale.} (CFT) in a flat $d$-dimensional spacetime. This relationship is called a duality. It is holographic since the gravitational theory lives in (at least) one extra dimension. Since AdS space is conformally flat, we can in fact think of the dual CFT as living on the boundary of the gravity theory in AdS. This is often  represented as in figure \ref{hsduality}.

The duality \eqref{adscft0} is remarkable as it opens up the possibility to study quantum gravity through its dual CFT, and vice versa. In particular, the dimensionless coupling $\lambda$ in the CFT$_d$ is typically related to the string length of the gravity theory in AdS via 
\begin{equation}
    \lambda \quad \sim \quad \left(\frac{R^2}{\alpha^\prime}\right)^{d/2},\label{lastr}
\end{equation}
where $R$ is the AdS radius. This relationship between the CFT coupling and the string length illustrates the duality's strong-weak nature: For example, in taking the point-particle limit $\alpha^\prime / R^2 \rightarrow 0$, the CFT coupling $\lambda$ grows large. The latter limit therefore opens up the possibility to study strongly coupled systems from the perspective of the comparably well-understood General Relativity (at least when quantum corrections are suppressed), and has been extensively explored since the birth of the AdS/CFT correspondence.\footnote{See, for example, \cite{Ammon:2015wua,Nastase:2015wjb} for pedagogical introductions to applying the AdS/CFT correspondence to stongly coupled systems.}

On the other hand, taking the opposite limit: $\alpha^\prime/R^2 \rightarrow \infty$, equation \eqref{lastr} tells us that the high-energy limit of string theory may be probed by taking the free theory limit $\lambda \rightarrow 0$ in the dual CFT \cite{HaggiMani:2000ru,Sundborg:2000wp,witten60thschw,Tseytlin:2002gz,Karch:2002vn,Gopakumar:2003ns,Bonelli:2003zu,Sagnotti:2003qa,PhysRevD.70.025010,Gopakumar:2004ys,PhysRevD.72.066008,Aharony:2006th,Yaakov:2006ce,PhysRevD.75.106006}. As will become apparent later, free CFTs are constrained by higher-spin symmetry owing to the emergence of an infinite tower of conserved currents of increasing spin in the free limit. Since the dual quantum gravity theory should be governed by the same symmetry as its CFT counterpart, this makes the existence of a highly symmetric phase of string theory even more plausible and more tractable to study.

The most famous manifestation of the holographic duality \eqref{adscft0}, which was crystallised by Maldacena in 1997 \cite{Maldacena:1997re}, is the conjecture that ${\cal N} = 4$ Super Yang-Mills theory in $d=3+1$ dimensions is equivalent to type IIB superstring theory on AdS$_5 \times S^5$. In the high energy limit $\alpha^\prime/R^2 \rightarrow \infty$, the string spectrum on the AdS$_5 \times S^5$ background has been extrapolated, and could be identified with the operator spectrum of planar\footnote{I.e. when the number of colours goes to infinity. As shall be explained later, this limit corresponds to the weak coupling regime of the dual string theory.} ${\cal N} = 4$ Super Yang-Mills theory at weak coupling ($\lambda \rightarrow 0$)  \cite{Bianchi:2003wx,Beisert:2003te,Bianchi:2005yh,Beisert:2004di,Bianchi:2004xi,Bianchi:2005ze,Bianchi:2006gk}. With control of the spectrum, studies towards understanding the spontaneous breaking of the local higher-spin symmetries in AdS were also undertaken from a holographic vantage point. From the CFT perspective, this corresponds to switching on the Yang-Mills coupling ($\lambda \ne 0$), leading to the anomalous violation of the higher-spin ($s>2$) conserved currents. These studies were made possible by a remarkable feature of ${\cal N} = 4$ Super Yang-Mills, which is that the spectrum of local operators is calculable for any coupling $\lambda$ in the planar limit \cite{Beisert:2010jr}. See \cite{Gaberdiel:2015mra,Gaberdiel:2015uca,Gaberdiel:2015wpo} for the more recent parallel storyline unfolding in AdS$_3$, where further relevant progress has been made.

While the above example provides a concrete framework to study string theory in the $\alpha^\prime \rightarrow \infty$ limit, in practice this is complicated by the exponentially growing spectrum, composed of infinitely many Regge trajectories. A streamlined version of this set-up is provided by instead considering a minimal interacting higher-spin-symmetric spectrum on AdS,\footnote{See \cite{Vasiliev:2004qz,Sorokin:2004ie,Bekaert:2005vh,Campoleoni:2009je,Bekaert:2010hw,Gaberdiel:2012uj,Giombi:2012ms,Sagnotti:2013bha,Rahman:2013sta,Gomez:2014dwa,Didenko:2014dwa,Rahman:2015pzl} for a selection of reviews on various aspects of higher-spin theories.} which consists of a single infinite tower of higher-spin gauge fields. This is believed to provide an effective description of the first Regge trajectory of string theory in the tensionless limit. Such theories have been conjectured to be dual to the free fixed points of vector models\cite{witten60thschw,Sezgin:2002rt,Klebanov:2002ja,Leigh:2003gk,Sezgin:2003pt,Giombi:2011kc,Anninos:2011ui,Chang:2012kt},\footnote{See also \cite{Konstein:2000bi,Shaynkman:2001ip,Sezgin:2001zs,Vasiliev:2001zy,Mikhailov:2002bp} for earlier closely related work.} which indeed have linear spectra as opposed to the exponentially growing spectra of Matrix models, like that of ${\cal N} = 4$ Super Yang-Mills theory. The duality between minimal higher-spin gauge field theories and vector models therefore provides a toy model for investigating the nature of string interactions in its putative maximally symmetric phase. Moreover, there have been recent results on embedding such higher-spin gauge theories into string theory \cite{Gaberdiel:2014cha,Baggio:2015jxa}.

The construction of consistent interactions among higher-spin gauge fields has a long history, with some of their properties remaining quite elusive. While on a flat background the existence of consistent higher-spin gauge field theories remains to be clarified,\footnote{In particular, in the view of the various no-go theorems \cite{PhysRev.135.B1049,Coleman:1967ad,PhysRev.186.1337,Velo:1970ur,Velo:1972rt,Aragone:1979hx,Deser:1990bk,Weinberg:1980kq,Porrati:2008rm,Bekaert:2010hp,Porrati:2010hm,Taronna:2011kt,Joung:2013nma}, with \cite{Bekaert:2010hw} being a nice review. However see \cite{Metsaev:1991mt,Metsaev:1991nb,Ponomarev:2016jqk,Conde:2016izb,Bengtsson:2016hss,Sleight:2016xqq,Ponomarev:2016lrm} for further developments in four-dimensions using non-covariant formalisms.} significant progress has been made on an AdS background: Thus far all possible structures for cubic interactions of bosonic gauge fields have been constructed for any triplet of spin \cite{Fradkin:1986qy,Fradkin:1987ks,Manvelyan:2004mb,Fotopoulos:2007yq,Boulanger:2008tg,Zinoviev:2008ck,Manvelyan:2009tf,Fotopoulos:2010nj,Bekaert:2010hk,Alkalaev:2010af,Boulanger:2011qt,Boulanger:2011se,Vasilev:2011xf,Joung:2011ww,Joung:2012rv,Joung:2012fv,Taronna:2012gb,Joung:2012hz,Boulanger:2012dx,Lopez:2012pr,Joung:2013doa} and classified \cite{Vasilev:2011xf,Joung:2012hz,Boulanger:2012dx,Joung:2013nma}, with consistent propagation at the non-linear level requiring the spectrum to be unbounded in spin \cite{Berends:1984rq,Metsaev:1991mt,Metsaev:1991nb}.\footnote{This comes from the structure of the higher-spin algebra, which
is to large extent unique \cite{Fradkin:1986ka,Konstein:1989ij,Boulanger:2013zza}.} Little is known about the properties of quartic and higher-order interactions in (A)dS - in spite of the existence of background-independent equations of motion \cite{Vasiliev:1988xc,Vasiliev:1990en,Vasiliev:1990vu,Vasiliev:1992av,Vasiliev:1992ix,Vasiliev:1995dn,Prokushkin:1998bq,Vasiliev:1999ba,Engquist:2002vr,Vasiliev:2003ev,Sezgin:2003pt,Chang:2012kt}, which employ an infinite set of auxiliary fields.\footnote{See also \cite{Boulanger:2011dd,Boulanger:2012bj,Didenko:2015pjo,Bonezzi:2015igv,Arias:2016ajh} for proposals of an action, using the machinery of auxiliary fields.}

A complete understanding of higher-spin gauge theories beyond cubic order is a problem which has remained open since the ball began rolling for studies of higher-spin interactions in the `70s. In particular, like the tail of $\alpha^\prime$ corrections in string theory, one expects the emergence of non-local interactions unbounded in their number of derivatives, whose locality properties have not yet been fully understood \cite{Barnich:1993vg,Polyakov:2009pk,Taronna:2011kt,Dempster:2012vw,Vasiliev:2015wma,Bekaert:2015tva,Skvortsov:2015lja,Taronna:2016ats,Bekaert:2016ezc,Vasiliev:2016xui,Taronna:2016xrm,Bengtsson:2016hss,Sleight:2016xqq,Ponomarev:2016lrm}. The latter behaviour requires careful consideration for the consistency of non-trivially interacting higher-spin gauge theories.

\subsection{Summary of main results}
It is the goal of this work to employ the holographic duality \eqref{adscft0} to shed some light on the above questions regarding the nature of interactions in higher-spin gauge theories.

For simplicity, we consider the most streamlined set-up of the so-called type A minimal bosonic higher-spin theory on AdS$_{d+1}$, which is a theory of even spin gauge fields represented by totally symmetric double-traceless\footnote{This is the algebraic constraint: $\varphi_{ \mu_1 ... \mu_{s-4} \rho \sigma}{}^{\rho \sigma} = 0$, whose role will be explained later.} tensors $\varphi_{ \mu_1 ... \mu_s}$. These have linearised gauge transformations
\begin{equation}
  \delta \varphi_{ \mu_1 ... \mu_s} = \nabla_{\left(\right.\mu_1}\xi_{\mu_2 ... \mu_s \left. \right)}, \qquad s = 2, 4, 6, ...\, .
\end{equation}
In addition to the gauge fields above, the type A spectrum includes a parity even scalar, which we denote by $\varphi_0$. This theory has been conjectured to be dual to the free scalar $O\left(N\right)$ vector model in $d$-dimensions \cite{Sezgin:2002rt,Klebanov:2002ja}, which is a CFT of an $N$-component real scalar field transforming in the fundamental representation of $O\left(N\right)$.

\begin{figure}[h]
  \centering
  \includegraphics[scale=0.55]{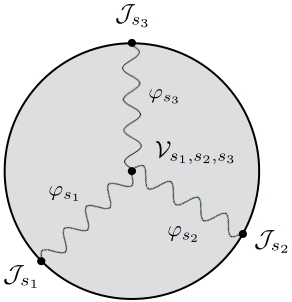}
  \caption{Holographic interpretation of the three-point function of operators ${\cal J}_{s_i}$ in the free $O\left(N\right)$ model at large $N$. The operators ${\cal J}_{s_i}$ are inserted on the boundary of AdS$_{d+1}$,  sourced by the boundary values of their dual fields ${\varphi_{s_i}}$ in AdS. The wavy lines are the propagators of the ${\varphi_{s_i}}$. With the knowledge of $\langle {\cal J}_{s_1}{\cal J}_{s_2}{\cal J}_{s_3} \rangle$ the on-shell form of the cubic interaction ${\cal V}_{s_1,s_2,s_3}$ can then in principle be determined. } \label{intro1233pt}
\end{figure}

The leitmotiv of this thesis is that correlation functions of the free scalar $O\left(N\right)$ vector model in the large $N$ limit are considered equivalent to a perturbative Feynman diagram expansion in the type A minimal bosonic higher-spin theory on AdS with the appropriate boundary conditions. Assuming this holographic duality holds, with the knowledge of correlators in the free $O\left(N\right)$ model (which, in a free theory, are straightforward to compute by Wick's theorem) one can then in principle infer the on-shell form of the interactions in the higher-spin theory. The basic idea  - which is made more precise in the following chapter - is illustrated for the extraction of cubic interactions in AdS in figure \ref{intro1233pt}.

Taking the above approach, in this work we determine the complete set of cubic couplings
and the quartic self interaction of the parity even scalar $\varphi_0$ in the action of the type A minimal bosonic higher-spin theory on AdS. In other words, using holography we fix the on-shell action
\begin{align}\nonumber
    S\left[\varphi\right] & = \sum\limits_{s} \int_{\text{AdS}}  \frac{1}{2} \varphi_{s} \left(\Box - m^2_{s} + ... \right) \varphi_{s} + \sum\limits_{s_i}\int_{\text{AdS}}  {\cal V}_{s_1,s_2,s_3}\left(\varphi\right) + \int_{\text{AdS}}  {\cal V}_{0,0,0,0}\left(\varphi_0\right) + ...,
\end{align}
where the $\ldots$ in the free quadratic action denote terms which depend on the gauge fixing \cite{Fronsdal:1974ew}. The cubic interaction for each triplet of spins takes the schematic form
\begin{equation} \label{arxcub}
  {\cal V}_{s_1,s_2,s_3}\left(\varphi\right) = g_{s_1, s_2, s_3} \left[ \nabla^{\mu_1} ... \nabla^{\mu_{s_3}} \varphi_{\nu_1 ... \nu_{s_1}} \nabla^{\nu_1} ... \nabla^{\nu_{s_1}} \varphi_{\rho_1 ... \rho_{s_2}} \nabla^{\rho_1} ... \nabla^{\rho_{s_2}} \varphi_{\mu_1 ... \mu_{s_1}} + {\cal O}(\Lambda) \right],
\end{equation}
where the terms of ${\cal O}(\Lambda)$ and higher in the bracket are descending in the number of derivatives, and are fixed uniquely. The overall coupling is given by
\begin{align}\label{g0123int}
    g_{s_1, s_2, s_3} = \frac{1}{\sqrt{N}}\frac{\pi ^{\frac{d-3}{4}}2^{\tfrac{3 d-1}{2}}\left(-2\Lambda\right)^{\left(s_1+s_2+s_3\right)/2}}{ \Gamma (d+s_1+s_2+s_3-3)}\sqrt{\frac{\Gamma(s_1+\tfrac{d-1}{2})}{\Gamma\left(s_1+1\right)}\frac{\Gamma(s_2+\tfrac{d-1}{2})}{\Gamma\left(s_2+1\right)}\frac{\Gamma(s_3+\tfrac{d-1}{2})}{\Gamma\left(s_3+1\right)}}.
\end{align}
Put in the context of existing results for cubic vertices of higher-spin gauge fields, the above result is new in the sense that the coefficients \eqref{g0123int} together with the precise combination of structures in \eqref{arxcub} were previously unknown in any higher-spin gauge theory on AdS. Notably, the couplings \eqref{g0123int} are finite and each vertex \eqref{arxcub} is local with a finite number of derivatives for fixed spins.

Furthermore, in \cite{Sleight:2016dba} it was verified that the rigid structure constants implied by the holographically reconstructed cubic action \eqref{arxcub} coincide with the
the known expressions \cite{Vasiliev:2003ev,Joung:2014qya} for higher-spin algebra structure constants, which are unique
in general dimensions \cite{Boulanger:2013zza}. This extends to general dimensions (and to the metric-like formalism) the tree-level three-point function test \cite{Giombi:2009wh} of Giombi and Yin in AdS$_4$. In particular, the test confirms
that the holographically reconstructed cubic couplings uniquely solve the Noether procedure at the
quartic order.

The quartic self interaction of the scalar takes the schematic form
\begin{align}
    {\cal V}_{0,0,0,0}\left(\varphi_0\right) = \sum\limits_{r,\,n}\, a_{r, n} \left( \varphi_0 \nabla_{\mu_1} ... \nabla_{\mu_r} \varphi_0 + ... \right) \Box^n \left( \varphi_0 \nabla^{\mu_r} ... \nabla^{\mu_1} \varphi_0 + ... \right), \quad r \in 2 \mathbb{N}, \label{qvintro}
\end{align}
where the $\dots$ denote a finite number of terms with no more than $r$ derivatives. As opposed to the cubic vertices in AdS$_{d+1}$ above, to extract the quartic interaction we work in AdS$_4$, in which the conformally coupled scalar affords certain simplifications.  The coefficients in the derivative expansion can be expressed in terms of a generating function $a_r\left(z\right) = \sum_n a_{r,n}\,z^n$, whose explicit form is quite involved:
 
\begin{align} \nonumber
 a_{r}\left(z\right) = & \frac{\Lambda^{r}}{ N} \frac{\left(\pi e^{\gamma\left(2r+1\right)}\Gamma\left(r\right)^2\left(\left(2r+1\right)^2+\left(\Lambda z+r+\tfrac{9}{4}\right)^2\right)^2 P\left(r+\tfrac{1}{2};z\right)^2   - 4^2\right)}{2^{r-4}\Gamma\left(r\right)^2\left(\left(\Lambda z+r+\tfrac{9}{4}\right)^2+\left(r-\tfrac{1}{2}\right)^2\right)}
\\  \nonumber
&-\frac{ i^r \pi ^{2} }{2\sqrt{2}N} \frac{ \Gamma\left(r+\frac{3}{2}\right)}{\Gamma\left(\frac{r}{2}+1\right)^2} e^{\gamma\left(2r+\frac{5}{2}\right)}P\left(r+\tfrac{1}{2};z\right)P\left(\tfrac{3}{4}; z\right)\\ \nonumber & \hspace*{4cm}\times \left(\left(2r+1\right)^2+\left(2\Lambda z+2r+\tfrac{9}{2}\right)^2\right) \left(9+\left(2\Lambda z+2r+\tfrac{9}{2}\right)^2\right),
\end{align}

\noindent
where we introduced
\begin{align}
P\left(a|z\right) = \prod^{\infty}_{k=0} \left[\left(1+\frac{a}{k}\right)^2+\left(\frac{4\Lambda z+4r+9}{4k}\right)^2\right]e^{-2a/k},
\end{align}
and $\gamma$ is the Euler-Mascheroni constant.

Consistent with standard expectations, the quartic interaction \eqref{qvintro} is non-local in the sense that it is unbounded in the number of derivatives. 
Drawing on the interpretation of CFT correlation functions as scattering amplitudes in AdS, the non-locality of a vertex can be quantified  by studying the contributions in the conformal partial wave expansion of its four-point amplitude. Our findings suggest that the holographic duality rules out the appearance of certain pathological non-localities, however there are some subtleties which remain to be clarified. This is discussed in the final chapter of this work.

\subsubsection{Intermediate results}

In obtaining the above results, we took a number of intermediate steps which may be of interest / applicable in a broader context. For example:

\begin{itemize}
    \item Techniques to evaluate tree-level three-point amplitudes in AdS, involving external fields of arbitrary integer spin.
    \item Propagators in AdS for gauge fields of arbitrary integer spin, in a basis of harmonic functions.
    \item Methods to decompose a general tree-level four-point amplitude (exchange or contact) in AdS into conformal partial waves.
\end{itemize}

\subsubsection{Ambient space}

We emphasise that obtaining many of the results in this work was facilitated by the application of the ``ambient space formalism'' \cite{Dirac:1936fq}. This is a powerful approach, with the basic idea being that fields in Euclidean AdS$_{d+1}$, or their dual CFT operators, can be expressed in terms of fields in an ambient
Minkowski space $\mathbb{M}^{d+2}$. The action of the AdS$_{d+1}$ isometry group, or of the conformal group
SO$\left(d + 1, 1\right)$, can then be realised as the group of linear Lorentz transformations. For calculations in AdS, a key feature of this approach is the ability to represent expressions that are intrinsic to the AdS manifold (e.g. involving non-commuting covariant derivatives) in terms of simpler-flat space ones of the ambient space (e.g. commuting partial derivatives). Because of these simplifying features, the ambient framework has enjoyed a wide variety of applications in the AdS,  higher-spin and conformal field theory literature. See for example: \cite{Fronsdal:1978vb,Metsaev:1995re,Metsaev:1997nj,Bars:1997bz,Biswas:2002nk,Bekaert:2003uc,Hallowell:2005np,Barnich:2006pc,Fotopoulos:2006ci,Alkalaev:2009vm,Cornalba:2009ax,Weinberg:2010fx,Penedones:2010ue,Paulos:2011ie,Costa:2011mg,Costa:2011dw,Fitzpatrick:2011ia,SimmonsDuffin:2012uy,Costa:2014kfa}.

\subsection{Outline}

The outline of this thesis is as follows:\\

\noindent
In chapter \tcb{\ref{chap::adscft}} we explain our approach to holographically reconstruct higher-spin interactions in detail. We review the precise dictionary between CFT correlation functions and amplitudes in the dual gravity theory on AdS, as well as the identification of CFT operators with fields in AdS. We further elaborate on the relevant aspects of the duality between higher-spin theories on AdS and free CFTs.
\\

\noindent
In chapter \tcb{\ref{chapt::CFT}} we give a self contained review of the techniques in CFT which are relevant for this work. We recall how the conformal symmetry constrains two-, three- and four-point functions of operators, paying particular attention to operators of arbitrary integer spin. We further review the conformal partial wave expansion of four-point correlation functions, and introduce useful techniques for later application.
\\

\noindent
In chapter \tcb{\ref{chapt::vec}} we introduce the free scalar $O\left(N\right)$ vector model. We apply the review of CFT correlators in chapter \tcb{\ref{chapt::CFT}}, together with useful methods to apply Wick's theorem, to establish the two- and three-point functions of all single-trace conserved currents of arbitrary spin in the theory. We also compute the four-point function of the scalar single-trace operator, and determine all coefficients in its operator product expansion. With the latter result, we determine the conformal partial wave expansion and express it in an integral form for later application. This chapter is based on aspects of the original works: \cite{Bekaert:2015tva,Sleight:2016dba}. \\

\noindent
Chapter \tcb{\ref{chapt::witten}} is dedicated to the computation of Witten diagrams. We review existing results for three-point Witten diagrams involving external scalars, and develop an approach to extend them to external fields of arbitrary integer spin. We then turn to four-point Witten diagrams, deriving bulk-to-bulk propagators for gauge fields of arbitrary spin in a basis of AdS harmonic functions. We then demonstrate how to expand a general four-point Witten diagram into conformal partial waves. We apply the latter techniques to Witten diagrams with identical external scalars, including four-point exchanges of gauge fields with arbitrary spin and a general contact interaction dressed with derivatives. This chapter is based on aspects of the original works: \cite{Bekaert:2014cea,Bekaert:2015tva,Sleight:2016dba}.
\\

\noindent
Chapter \tcb{\ref{chapt::hr}} is a culmination of the preceding chapters, in which the intermediate results established throughout are put together to extract all cubic interactions in the minimal bosonic higher-spin theory on AdS$_{d+1}$, and the quartic self-interaction of the scalar for the theory on AdS$_4$. This chapter is based on aspects of the original works: \cite{Bekaert:2015tva,Sleight:2016dba}.
\\

\noindent
In chapter \tcb{\ref{chapt::summ}} we summarise the main results of this work, and discuss the non-locality of higher-spin interactions from a holographic perspective. This chapter is based on on-going work with M. Taronna (see \cite{Sleight:2017pcz}), and aspects of the original works: \cite{Bekaert:2015tva,Bekaert:2016ezc}.\\

\noindent
Various technical details are relegated to the appendices (\tcb{\ref{Aambient}}-\tcb{\ref{appendix::quarticbasis}}). In particular, a detailed review of the ambient formalism is given in appendix \tcb{\ref{Aambient}}.

   \chapter{The AdS/CFT correspondence}
\label{chap::adscft}
As stated in the introduction, the AdS/CFT correspondence is the conjecture that quantum gravity in asymptotically anti-de Sitter spacetime AdS$_{d+1}$ is equivalent to a non-gravitational conformal field theory (CFT) in a flat $d$-dimensional spacetime,
\begin{equation}
    \text{AdS$_{d+1}$ QG = CFT$_{d}$.}\label{adscft}
\end{equation}
The aim of this chapter is to review the most pertinent aspects of this duality,  relevant for the applications this thesis.

\section{Practicalities}
\label{sec::gen}

For practical applications, the holographic correspondence is more conveniently formulated in terms of the generating functions of the two dual theories, in Euclidean signature.

In a CFT, the generating function $F_{\text{CFT}}\left[{\bar \varphi}\right]$ of connected correlators admits the path-integral representation , 
\begin{align}
    \exp\left(-F_{\text{CFT}}\left[{\bar \varphi}\right]\right) = \int D\phi\, \exp\left(-S_{\text{CFT}}\left[\phi\right] + \int d^dy\, {\bar \varphi}\left(y\right){\cal O}\left(y\right)\right). \label{qftgen}
\end{align}
 We use $\phi$ to denote the fundamental field(s) in the theory, governed by the conformally invariant action $S_{\text{CFT}}\left[\phi\right]$. The operator ${\cal O}$ is built from the fields $\phi$, and is sourced by ${\bar \varphi}\left(y\right)$. The source is not dynamical, but a fixed function which is under our control.\footnote{For example, it might be a background electric
or magnetic field, or a background pressure density.}

The basic idea behind the holographic duality \eqref{adscft} is to breathe life into the source ${\bar \varphi}$. We can think of the CFT$_d$ as living on the boundary of the higher-dimensional AdS$_{d+1}$, in which the source is promoted to a dynamical field $\varphi\left(y,z\right)$ governed by the bulk action $S_{\text{AdS}}\left[ \varphi \right]$ at the classical level. The only control we have over the field is the boundary value ${\bar \varphi}\left(y\right)$ at $z=0$. 
The duality then states that the physical quantity $F_{\text{CFT}}\left[{\bar \varphi}\right]$ in the CFT coincides with the AdS one $\Gamma_{\text{AdS}}\left[{\bar \varphi}\right]$, \cite{Gubser:1998bc,Witten:1998qj}. I.e. $
 F_{\text{CFT}}\left[{\bar \varphi}\right] = \Gamma_{\text{AdS}}\left[{\bar \varphi}\right]$, with 
\begin{align}
     \exp\left(-\Gamma_{\text{AdS}}\left[{\bar \varphi}\right]\right) & = \int_{\varphi|_{\partial \text{AdS}} =  {\bar \varphi}} {\cal D} \varphi \exp\left(-\frac{1}{G}S_{\text{AdS}}\left[\varphi\right]\right),
\end{align}
and $G$ the gravitational constant. We make the above identification more precise in the following section.

The crucial point, central to the work behind this thesis, is that connected correlators in the CFT can be computed holographically by instead functionally differentiating the bulk quantity $\Gamma_{\text{AdS}}\left[{\bar \varphi}\right]$
\begin{equation}
    \langle {\cal O}\left(y_1\right) ... {\cal O}\left(y_n\right) \rangle_{\text{conn.}} = (-1)^n \frac{\delta}{\delta {\bar \varphi}\left(y_1\right)} ... \frac{\delta}{\delta {\bar \varphi}\left(y_n\right)} \Gamma_{\text{AdS}}\left[{\bar \varphi}\right]\Big|_{{\bar \varphi}=0}, \label{corgen0}
\end{equation}
with each differentiation sending a $\varphi\left(y,z\right)$ particle into AdS.

Quantum effects in the gravity theory on AdS are measured in the dual CFT by the number of degrees of freedom
\begin{equation}
 \left(\frac{R}{\ell_p}\right)^{d-1} \quad \sim \quad N_{\text{dof.}}. \label{dofeq0}
\end{equation}
This can be understood by the fact that we can roughly measure the degrees of freedom in a CFT by the overall coefficient ${\sf C}_{T}$ of the energy momentum tensor two-point function.\footnote{Consider for example free theory: The stress tensors add, so increasing the number of fields (degrees of freedom) increases the two-point function coefficient.} From the preceding discussion, under the holographic duality we identify
\begin{equation}
\langle T T \rangle_{\text{CFT}} \quad \sim \quad {\sf C}_{T} \quad \sim \quad N_{\text{dof.}}  \qquad \leftrightarrow \qquad   \langle g g \rangle_{\text{AdS}} \quad \sim \quad \frac{R^{d-1}}{G} \quad \sim \quad \left(\frac{R}{\ell_p}\right)^{d-1},
\end{equation}
since the CFT energy momentum tensor is dual to the graviton in AdS (\S \tcb{\ref{subsec::fom}}). This gives the relation \eqref{dofeq0}.

The weak coupling expansion $G << R$ in the bulk, with dominant contribution\footnote{I.e. neglecting loops.}
\begin{align}
     \exp\left(-\Gamma_{\text{AdS}}\left[{\bar \varphi}\right]\right) & \quad \approx \quad \exp\left(-\frac{1}{G}S_{\text{AdS}}\left[\varphi\right]\right)\Big|_{\varphi|_{\partial \text{AdS}} =  {\bar \varphi}},
\end{align}
therefore corresponds to a large $N_{\text{dof}}$ expansion in the dual CFT, controlled by the dimensionless coupling
\begin{equation}
   g := R^{1-d}G \: \sim \: 1/N_{\text{dof}}+ {\cal O}\left(1/N^2_{\text{dof}}\right). 
\end{equation}
From the corresponding $1/N_{\text{dof}}$ expansion admitted by $F_{\text{CFT}}$,
\begin{equation}
  F_{\text{CFT}} = N_{\text{dof}}\, F^{\left(0\right)}_{\text{CFT}} + F^{\left(1\right)}_{\text{CFT}} + \frac{1}{N_{\text{dof}}}\, F^{\left(2\right)}_{\text{CFT}} + ...,
\end{equation}
we can identify
\begin{equation}
  \exp\left(-N_{\text{dof}}\, F^{\left(0\right)}_{\text{CFT}}\left[{\bar \varphi}\right]\right) \quad = \quad \exp\left(-\frac{1}{G}S_{\text{AdS}}\left[\varphi\right]\right)\Big|_{\varphi|_{\partial \text{AdS}} =  {\bar \varphi}}.\label{treelevholo}
\end{equation}
Correlators in the CFT may thus be computed in the large $N_{\text{dof}}$ limit via a tree-level  Feynman diagram expansion in AdS space with boundary conditions that are introduced in the following section.

Equation \eqref{treelevholo} is key to the results established in this work. It is from this equation that we extract interactions in the action of the minimal bosonic higher-spin theory from correlation functions in its dual CFT, as will be explained in more detail in the following sections. To this end, we first make more precise the identification between operators in CFT and fields on AdS in the large $N_{\text{dof}}$ limit.

\subsection{Field-operator map}
\label{subsec::fom}

What does it take for an operator ${\cal O}$ of scaling dimension $\Delta$ and spin-$s$ to be described by some field $\varphi$ in an asymptotically AdS$_{d+1}$ space? First of all, the CFT operator must be gauge invariant. This restricts the spectrum of local operators to be composed of traces. In the large $N_{\text{dof}}$ limit, single-trace operators are identified with single particle states in the gravity theory, while multi-trace operators with multi-particle states.

An important feature of the AdS/CFT duality is that both theories are governed by the same symmetry group: The conformal group is isomorphic to $SO\left(d,2\right)$, which is the isometry group of AdS. A second natural requirement for CFT operators and bulk fields to be identified, is then that they sit in the same representation of $SO\left(d,2\right)$.
This has the implication that $\varphi$ is also spin-$s$. Furthermore, it must transform in the same way as ${\cal O}$ under dilatations, which correspond to boundary limit of the energy generators $E$ in AdS:

The action of the dilatation generator $D$ on ${\cal O}$ is derived in \S \tcb{\ref{sec::repo}}, and is given by
\begin{align}
    \left[D,{\cal O}\left(y\right)\right] = - i \left(\Delta + y \cdot \partial \right){\cal O}\left(y\right). \label{indila}
\end{align}
It acts on the bulk field $\varphi$ through a Lie derivative with respect to the AdS Killing field $E$ (which can be extracted from \eqref{a7}, $E = J_{\left(d+1\right)\, d}$):\footnote{For concreteness we work in Poincar\'e co-ordinates for AdS$_{d+1}$  
\begin{align}
   ds^2 = \frac{R^2}{z^2}\left(dz^2+dy_idy^i\right),
\end{align}
with $R$ the AdS radius. The boundary of AdS is located at $z=0$, with boundary directions $y^i$, $i = 1, ..., d$.}
\begin{align}
    {\cal L}_{E}\;\varphi_{\mu_1 ... \mu_s} = - i\left( z \cdot \partial_z + y \cdot \partial_y +s \right)\varphi_{\mu_1 ... \mu_s}.
\end{align}
Comparing the above with \eqref{indila}, to preserve conformal invariance as $z \rightarrow 0$, we require for boundary directions $i$
\begin{align}
    \left(z \cdot \partial_z+s\right) \varphi_{i_1 ... i_s} \quad \sim \quad \Delta\, \varphi_{i_1 ... i_s} \quad \implies \quad \varphi_{i_1 ... i_s}\quad \sim \quad z^{\Delta-s}. \label{as1}
\end{align}
The coefficient of $z^{\Delta-s}$ is identified with the expectation value $\langle {\cal O}\left(y\right) \rangle $, since  both have the same transformation properties under the
conformal group.

The source ${\bar \varphi}$ of ${\cal O}$ has scaling dimension $d-\Delta$, which can be verified by demanding that \eqref{qftgen} is invariant under dilatations. For $\varphi$ to describe the source, we thus require\footnote{Note that here we do not have $-s$ (unlike in \eqref{as1}) since the $\mu$ indices are \emph{raised} and receive a minus sign in the Lie derivative.}
\begin{align}
    \varphi^{i_1 ... i_s}\quad \sim \quad z^{d+s-\Delta} \quad \text{as} \quad z \rightarrow 0, \label{sb}
\end{align}
with the coefficient of $z^{d+s-\Delta}$ given by the source ${\bar \varphi}$.

The above near-boundary behaviour places restrictions on the mass of the field $\varphi_s$, which can be seen as follows. At the linearised level, we have 
\begin{align}
    \left(-\Box + m^2_s\right)\varphi_{\mu_1 ...\mu_s} = 0.
\end{align}
To extract the boundary behaviour, we need only consider the asymptotic form of the equation of motion as $z \rightarrow 0$, which in the boundary directions reads
\begin{align}
  &  -\left(\frac{z}{R}\right)^{d+1} \partial_{z}\left(\left(\frac{R}{z}\right)^{d-1}\partial^{z}\varphi_{i_1 ... i_s}\right) - \frac{2sz}{R^2} \partial_ z\varphi_{i_1 ... i_s} \\ \nonumber
    & \hspace*{7cm} + \frac{sd}{R^2}\varphi_{i_1 ... i_s} + \left(m^2_s-\frac{1}{R^2}s\left(s-1\right)\right)\varphi_{i_1 ... i_s}  = 0.
\end{align}
The solution is given in terms of two integration constants $A_{i_1 ...i_s}\left(y\right)$ and $B_{i_1 ... i_s}\left(y\right)$,\footnote{This can be obtained by Fourier transforming with respect to the boundary directions, and making an ansatz of the type ${\hat \varphi}_s\left(p,z\right) \quad \sim \quad {\hat A}\left(p\right) z^{\Delta+s}$.}
\begin{align}
    \varphi_{i_1 ... i_s} \quad \sim \quad A_{i_1 ... i_s}\left(y\right)z^{\Delta-s}\: +\: B_{i_1 ... i_s}\left(y\right)z^{d-\Delta-s},
\end{align}
where $\Delta$ satisfies
\begin{equation}
    \Delta\left(\Delta-d\right) - s = m^2_sR^2. \label{md}
\end{equation}
Should this hold, from the above discussion we can thus interpret $A_{i_1 ... i_s}\left(y\right)$ as 
the expectation value $\langle {\cal O}_{i_1 ... i_s}\left(y\right)\rangle $ in the presence of the source $B_{i_1 ... i_s}\left(y\right) = {\bar \varphi}_{i_1 ... i_s}\left(y\right)$.

To summarise, small excitations around a given CFT state are dual to perturbations of the gravitational background, where the latter satisfy equations of motion derived from the gravitational action.

\subsection{Gauge fields and conserved currents}
\label{subsec::gfcc}

A special entry in the field-operator dictionary is reserved for operators/fields with spin-$s$ and dimension/energy $\Delta = s+d-2$. Such representations of $SO\left(d,2\right)$ are \emph{short} representations due to the appearance of zero norm states in the Fock space/ The latter form an invariant submodule, which can be factored out. This leaves us with a ``shorter'' unitary representation. For the bulk fields, this multiplet shortening can be seen as the emergence of a gauge symmetry
\begin{align}
   \delta \varphi_{\mu_1 ... \mu_s} = \nabla_{\left(\mu_1\right.}\xi_{\mu_2 ... \mu_s\left.\right)}. \label{ghs}
\end{align}
In the dual CFT, this shortening manifests itself in the fact that the corresponding operators are \emph{conserved}. This is explained further in  \S \tcb{\ref{subsec:unitarity}}, however for now let us note that invariance under the transformations induced by the bulk gauge symmetry \eqref{ghs} ensures conservation of the dual CFT operator:
\begin{align}\nonumber
 0 & =   \delta \int_{\partial \text{AdS}} d^dy \,{\bar \varphi}^{i_1 ...i_s} {\cal O}_{i_1 ...i_s} = - \int_{\partial \text{AdS}} d^dy\, \xi^{i_2 ...i_s} \partial^{i_1} {\cal O}_{i_1 ...i_s} \quad \implies \quad \partial^{i_1} {\cal O}_{i_1 ...i_s} = 0.
\end{align}
Such representations saturate the unitarity bound for spin-$s >$ 0 unitary irreducible representations of $SO\left(d,2\right)$. States with lower dimension / energies will have negative norm, and should be excluded from the physical spectrum. This is justified from a CFT perspective in \S \tcb{\ref{subsec:unitarity}}.

As a final comment, note that the above discussion applies only to representations of non-zero spin. The celebrated  result for the unitarity bound of scalar fields in AdS was derived earlier by Breitenlohner and Freedman \cite{Breitenlohner:1982bm}. It reads
\begin{equation}
    \Delta \ge  \frac{1}{2}\left(d-2\right),
\end{equation}
or more familiarly in terms of the mass of the bulk scalar field,
\begin{equation}
    m^2 > - \left(\frac{d}{2R}\right)^2.
\end{equation}
This is known as
the Breitenlohner-Freedman (or BF) bound, and informed us that fields can be a little bit tachyonic in AdS and still be stable.

\section{Higher-spin / vector model holography}
\label{sec::ce}

 With the dictionary between CFT operators and bulk fields in place, we are ready to explore the holographic duality in more detail. As outlined in the introduction, in this work we are interested in studying interactions in the high energy limit $\alpha^\prime / R^2 \rightarrow \infty$ through the holographic looking-glass. The logic is clear: In this regime the dual CFT with coupling $\lambda$ is free (recall \eqref{lastr})
 \begin{equation}
   \lambda \quad \sim \quad \left(\frac{R^2}{\alpha^\prime}\right)^{d/2} \quad \rightarrow \quad 0,
 \end{equation}
 and one may then hope that its simplicity can be used to help us gain some insight in the bulk.

Let us first consider this limit in the context of well-known conjectured equivalence between ${\cal N}=4$ Super-Yang Mills in $d = 3+1$ dimensions with $SU\left(N\right)$ gauge group, and the type IIB super-string on AdS$_5 \times S^5$. Amongst the single-trace operators in the former, in the free limit there emerges an infinite tower of conserved currents of increasing spin. For example, bi-linears of the schematic form\footnote{Note that we use the partial derivative $\partial$ as opposed to the gauge-covariant derivative $\nabla$, since we are working in the free theory.}\:\footnote{Since the scalar $\boldsymbol{\phi}^a$ has scaling dimension $\frac{d}{2}-1$, the quantum numbers of \eqref{bl} coincide with that for a spin-$s$ conserved current \S \tcb{\ref{subsec:unitarity}}.}
\begin{equation}
    {\cal J}_{i_1 ... i_s} = \frac{1}{N} \sum\limits^6_{a=1} \text{Tr}\left[ \boldsymbol{\phi}^a \partial_{\left(i_1\right.} ...\partial_{\left.i_s\right)} \boldsymbol{\phi}^a\right] + ...\,,
\quad s=2, 4, 6, ...\,,\label{bl}
\end{equation}
where the $\dots$ denote terms which make the operator a primary one (\S \tcb{\ref{subsec::pridesc}}), and the $\boldsymbol{\phi}^a$ are the six scalar fields\footnote{Analogous gauge-invariant bi-linears also exist for the $N^2-1$ free gauge fields and $4\left(N^2-1\right)$ free complex fermions. At the free level different fundamental fields propagate independently so it is consistent to restrict attention to a subset of them.} in the adjoint representation of $SU\left(N\right)$. Recalling the field-operator map for conserved currents (\S \tcb{\ref{subsec::gfcc}}), each conserved current \eqref{bl} is dual to a gauge field of the same spin in the bulk. In this way the correspondence confirms that string theory in the regime $\alpha^\prime / R^2 \rightarrow \infty$ possesses an infinite tower of gauge fields: one for each spin $s$. As these states saturate the unitarity bound, they have the lowest mass for given spin and therefore lie on the first Regge trajectory.

Since the fields in ${\cal N}=4$ SYM are matrix valued, the single-trace operators in the theory are not restricted to bi-linears in the elementary fields. In general, any single-trace operator built from $\boldsymbol{\phi}^a$ can be expressed as a linear combination of the operators
\begin{align}
    \frac{1}{N^{n/2}} \text{Tr}\left[ \left( \partial^{i_{1}} ... \partial^{i_{m_1}} \boldsymbol{\phi}^{a_1}\right)\, \left( \partial^{j_{1}} ... \partial^{j_{m_2}} \boldsymbol{\phi}^{a_2}\right) ...\, \left( \partial^{k_{1}} ... \partial^{k_{m_n}} \boldsymbol{\phi}^{a_n}\right)\right], \label{ol}
\end{align}
which are multi-linear in the elementary fields. These correspond to the exponentially growing number of single-particle states on sub-leading (for $n > 2$) Regge trajectories in the bulk AdS theory, forming matter multiplets of the higher-spin algebra.

While the above set up gives a complete holographic description of the high-energy limit of string theory, the study of interactions is complicated by the shear magnitude of the spectrum. A crucial simplifying observation is that the bi-linears \eqref{bl} form a closed subsector in the free theory at $\lambda \equiv 0$.\footnote{By this we mean that the OPEs of the bi-linears \eqref{bl} contain no contributions from single-trace operators with $k>2$. This can be confirmed by direct computation of the OPE coefficients via Wick's theorem, or see \cite{Mikhailov:2002bp}.} This suggests that it is possible to consistently restrict attention to the first Regge trajectory in the tensionless limit $\alpha^\prime / R^2 \rightarrow \infty$,\footnote{In fact, this was verified recently \cite{Gaberdiel:2014cha} for the case of string theory on AdS$_3 \times S^3 \times \mathbb{T}^4$.} which contains a single gauge field for each integer spin. The dynamics of the first Regge trajectory in this regime is thus captured by non-linear theories of higher-spin gauge fields on AdS, and their study may shed some light on string theory in the tensionless limit (or equivalently, its maximially symmetric phase).

From a CFT perspective, the proliferation of single-trace operators \eqref{ol} is a consequence of the elementary fields sitting in the adjoint of $SU\left(N\right)$. A toy model of the above scenario was proposed by Szegin and Sundell \cite{Sezgin:2002rt}, and Klebanov and Polyakov \cite{Klebanov:2002ja}, who contemplated the holographic duals of CFTs in which the elementary fields are instead in the \emph{fundamental} representation. In this case the only possible class of single-trace operators are bi-linears in the elementary fields, and one could therefore conceive that they give a holographic description of higher-spin gauge theories on AdS. More concretely, they proposed:
\begin{framed}
\begin{quote}
\centering The free scalar $O\left(N\right)$ vector model in $d$-dimensions \\
\emph{is equivalent to} \\
The type A minimal bosonic higher-spin theory on AdS$_{d+1}$.
\end{quote}
\end{framed}

\noindent
The theory on AdS is one of even spin gauge fields, represented by symmetric tensors $\varphi_s$, of spins $s = 2, 4, 6, ...$ . It also contains a parity even scalar, which we denote by $\varphi_0$.

The free scalar $O\left(N\right)$ vector model is a theory of an $N$ component free real scalar field $\phi^a$, transforming in the fundamental representation of $O\left (N\right)$. The single-trace spectrum consists of a scalar single-trace operator ${\cal O} = \phi^a \phi^a$ of scaling dimension $\Delta = d-2$, as well as an infinite tower of even-spin conserved currents
\begin{equation} \label{adssec::js}
    {\cal J}_{i_1 ... i_s} = \frac{1}{\sqrt{N}} \phi^a \partial_{i_1} ...\, \partial_{i_s} \phi^a + ..., \quad \partial^{i_1} {\cal J}_{i_1 ... i_s} = 0, \quad s = 2, 4, 6, ...\,,
\end{equation}
of scaling dimension $\Delta_s = s +\Delta = s+d-2$.

While the crucial requirement that the CFT single-trace and bulk single-particle states are in one-to-one correspondence is satisfied, the bulk fields must be prescribed the near-boundary behaviour (\S \tcb{\ref{subsec::fom}}):
\begin{align}
   \varphi_{i_1 ... i_s} \quad \sim \quad z^{\Delta_s-s},
\end{align}
in order to be able to identify
\begin{align} \nonumber
   \varphi_s \quad & \leftrightarrow \quad {\cal J}_{s}, \\ \nonumber
   \varphi_0 \quad & \leftrightarrow \quad {\cal O}.
\end{align}

\subsubsection{Generalisations}

The above duality is the simplest of its kind in the context of AdS higher-spin / vector model dualities. In particular, the spectrum of gauge fields in the bulk is truncated to those involving only even spin. The full non-minimal theory contains a higher-spin gauge field for \emph{each} integer spin, and its dual CFT description is given by the free $U\left(N\right)$ invariant theory with a \emph{complex} $N$ component scalar field $\phi^a$. The latter also contains conserved currents of odd spin, which accommodates for the extension of the bulk spectrum to include gauge fields of odd spin. Further generalisations of this set up have been considered, including for instance: Having a parity odd scalar in the bulk (the so-called type-B theory), Chan-Paton factors and supersymmetry \cite{Leigh:2003gk,Sezgin:2003pt,Giombi:2011kc,Chang:2012kt}.\footnote{Some approaches towards deriving the higher spin/vector model duality
from first principles were investigated in \cite{Das:2003vw,Douglas:2010rc,Koch:2010cy,Jevicki:2011ss,Jevicki:2011aa,deMelloKoch:2012vc,Zayas:2013qda,Sachs:2013pca,Leigh:2014tza,Leigh:2014qca,Koch:2014aqa,Mintun:2014gua,Jin:2015aba}.} However in this work we focus on the simplest, most streamlined, case presented above, free from the latter extra decorations. See: \cite{Bekaert:2012ux,Giombi:2012ms,Giombi:2016ejx} for comprehensive reviews of the higher-spin / vector model dualities.

\subsubsection{Tests of the duality}

Before we proceed to applications of the higher-spin / free vector model duality, let us note that although the duality is still a conjecture, its validity has withstood a number of non-trivial checks. The first was the tree-level three-point function test by Giombi and Yin \cite{Giombi:2009wh,Giombi:2010vg}, who verified using Vasiliev's system that certain three-point amplitudes in the bulk higher-spin theory on AdS$_4$ correctly reproduce the corresponding three-point functions in the free scalar $O\left(N\right)$ vector model in $3d$.\footnote{See also \cite{Chang:2011mz,Ammon:2011ua} for subsequent tree-level three point function tests in the context of the higher-spin holographic dualities in AdS$_3$.} Using the result for the cubic order action in this thesis, the latter test of the duality was later extended to general dimensions in the metric-like formalism in \cite{Sleight:2016xqq}. The duality has also been verified at the level of one-loop vacuum energy \cite{Giombi:2013fka,Giombi:2014iua,Giombi:2014yra,Beccaria:2014xda,Giombi:2016pvg,Pang:2016ofv,Gunaydin:2016amv} (see also \cite{Bae:2016rgm,Bae:2016hfy} in the context of adjoint-valued elementary fields).

\subsection{AdS higher-spin interactions from CFT}
\label{adscft::hsicft}

As explained in the introduction, there is much still to be understood about interactions of higher-spin gauge fields. This makes the higher-spin / vector model duality even more profound, as it opens up the opportunity to study the possible higher-spin interactions on AdS space through the consideration of a drastically simpler (free) CFT. The role of this subsection is to put this idea on a more concrete footing.

Assuming the existence of a standard action principle $S_{\text{HS AdS}}\left[\varphi\right]$ for the typa A minimal bosonic higher-spin theory, we perform a weak field expansion around an AdS background in the power of the fields,\footnote{To do so, we re-define the fields $\varphi \rightarrow \sqrt{G}\varphi$. This also ensures a canonically normalised kinetic term.}
\begin{equation}
  S_{\text{HS AdS}}\left[\varphi\right] = G S^{\left(2\right)}_{\text{HS AdS}}\left[\varphi\right] + G^{3/2} S^{\left(3\right)}_{\text{HS AdS}}\left[\varphi\right] + G^{2} S^{\left(4\right)}_{\text{HS AdS}}\left[\varphi\right] + ...\,.
\end{equation}
Recalling the holographic equality \eqref{treelevholo}, with the knowledge of the leading contribution to the generating function, $F^{\left(0\right)}_{\text{free}\;O\left(N\right)}$, of connected correlators in the large $N$ free scalar $O\left(N\right)$ vector model, one may in principle iteratively extract the on-shell interactions in the conjectured classical bulk higher-spin theory: 
\begin{align} \label{adscft::ke}
    \exp\left(-\frac{1}{\sqrt{N}}\, F^{\left(0\right)}_{\text{free}\;O\left(N\right)}\right) = \prod^\infty_{n=2} \exp\left(-\sqrt{G}^{\;n-2}S^{\left(n\right)}_{\text{HS AdS}}\left[\varphi\right]\right)\Big|_{\varphi|_{\partial \text{AdS}} =  {\bar \varphi}}.
\end{align}
With the quadratic kinetic term $S^{\left(2\right)}_{\text{HS AdS}}\left[\varphi\right]$ known, we begin this procedure at cubic order ($n=3$). The cubic action of the minimal bosonic higher-spin theory takes the form
\begin{align}
    S^{\left(3\right)}_{\text{HS AdS}}\left[\varphi\right] = \sum^{\infty}_{s_i=0}\int_{\text{AdS}} {\cal V}_{s_1,s_2,s_3}\left[\varphi_{s_i}\right],
\end{align}
where by applying the Noether procedure the possible structures that can enter the cubic vertex ${\cal V}_{s_1,s_2,s_3}$ between gauge fields of spins $s_1$, $s_2$ and $s_3$ have already been determined \cite{Vasilev:2011xf,Joung:2011ww,Boulanger:2012dx}. In this work, building on the latter results we completely fix the coupling constants between all vertex structures by solving
\begin{equation}
\langle {\cal J}_{s_1}{\cal J}_{s_2}{\cal J}_{s_3}  \rangle =  \sqrt{G} \frac{\delta}{\delta {\bar \varphi}_{s_1}}\frac{\delta}{\delta {\bar \varphi}_{s_2}}\frac{\delta}{\delta {\bar \varphi}_{s_3}} S^{\left(3\right)}_{\text{HS AdS}}\left[\varphi\right]\Big|_{\varphi_{s_i}|_{\partial \text{AdS}} =  {\bar \varphi}_{s_i}},
\end{equation}
which equates the three-point correlator of the spin-$s_i$ conserved currents \eqref{adssec::js} in the free scalar $O\left(N\right)$ vector model, to the tree-level three-point amplitude of the dual spin-$s_i$ gauge fields $\varphi_{s_i}$, generated by the bulk interaction ${\cal V}_{s_1,s_2,s_3}$. 
In particular, to this end we determine the three-point correlators of all single-trace conserved currents ${\cal J}_{s_i}$, and develop techniques to systematically compute three-point bulk amplitudes involving external fields of arbitrary integer spin.

With the above result for the cubic action, we may then proceed to quartic order. We restrict to the simplest case of solving for the quartic self-interaction of the scalar $\varphi_0$, which is dual to the scalar single-trace operator ${\cal O}$. Moreover, we focus on the set-up in AdS$_4$, where the conformally coupled scalar provides further simplifications (which will become clear in \S \tcb{\ref{chapt::witten}}.). For clarity, the equation we solve is given diagrammatically below
\begin{align} \label{adscft::4ptextr}
   & \langle {\cal O}\left(y_1\right){\cal O}\left(y_2\right){\cal O}\left(y_3\right){\cal O}\left(y_4\right) \rangle_{\text{conn.}}   \\ \nonumber \\ \nonumber
    & \hspace*{1.25cm}  \includegraphics[scale=0.425]{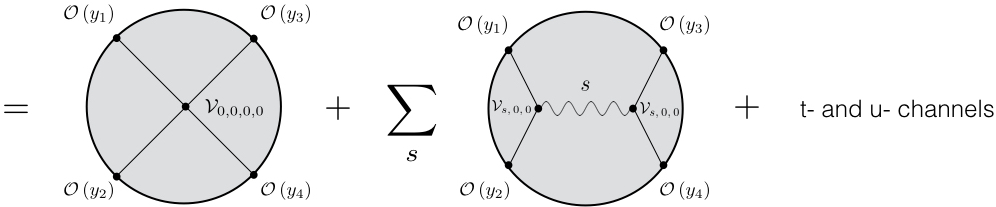}.
\end{align}
The diagrams contributing to the right hand side of the equality consist of: the contact amplitude generated by the quartic interaction ${\cal V}_{0,0,0,0}$ we seek to establish, and the four-point exchanges of each spin-$s$ gauge field between two pairs of the bulk scalar. The iterative nature of this reconstruction procedure is thus apparent, for in order to compute the exchange diagrams we require all cubic interactions between two real scalars and a spin-$s$ gauge field in the theory.

There is a gap in technical difficulty in applying this holographic approach to extracting the cubic vertices ${\cal V}_{s_1,s_2,s_3}$, and extending it to the quartic vertex above. In particular, in order to match the four-point correlator of the operator ${\cal O}$ with the dual four-point bulk amplitudes as shown in equation \eqref{adscft::4ptextr}, we put both sides of the equation on an equal footing by decomposing them into conformal partial waves. While it is well understood how to decompose CFT correlators into conformal partial waves (especially in free theories), as an intermediate step we tailor methods to establish conformal partial wave expansions of the bulk amplitudes.

We therefore spend the following chapters gathering the necessary ingredients and sharpening technical tools. This culminates in \S \tcb{\ref{chapt::hr}}, in which we put together the latter intermediate results, to fix completely the cubic action and the scalar quartic self-interaction in the type A minimal bosonic higher-spin theory as described above.

  \chapter{Correlation functions in CFT}
\label{chapt::CFT}
The work of this thesis is to a large part underpinned by techniques and concepts in conformal field theory (CFT). This chapter is intended to give a self-contained overview of the relevant ideas, in particular of the constraints imposed by conformal symmetry on two-, three- and four-point functions of local operators. The first few sections cover the basics of CFT, which can be safely skipped by the knowledgeable reader. Later sections move on to discuss conformal block expansions, and more recent developments relevant for applications in this work.

\section{Conformal algebra and transformations}

A \emph{conformal transformation} is a change of co-ordinates that locally rescales the metric tensor,
\begin{equation}
    g_{\mu \nu}\left(x\right) \rightarrow \Omega^2\left(x\right)g_{\mu \nu}\left(x\right). \label{weyltransf}
\end{equation}
A conformal field theory is a field theory that is invariant under these transformations, and whose physics looks the same at all length scales as a consequence. In this thesis we are concerned with CFTs defined on a fixed flat background, and for the remainder of this discussion we therefore restrict to the case $g_{\mu \nu}\left(x\right) = \eta_{\mu \nu}$.

 The transformations \eqref{weyltransf} form a group, which for a fixed Minkowski background includes the Poincar\'e group as a subgroup, since the metric is invariant under Lorentz transformations and translations ($\Omega\left(x\right)=1$). In addition to the latter, the conformal group is generated by scale transformations
 \begin{equation}
     x^{\mu} \rightarrow \lambda x^{\mu}, \qquad \Omega\left(x\right) = \lambda, \qquad \lambda \in \mathbb{R}, 
 \end{equation}
 and special conformal transformations
 \begin{equation}
     x^{\mu} \rightarrow  \frac{x^{\mu}+b^\mu x^2}{1+2 b\cdot x + b^2x^2}, \qquad \Omega\left(x\right) = \frac{1}{x^2}.
 \end{equation}
 The conformal algebra can be determined by considering the infinitesimal form of the above transformations. For example, the generator $D$ associated to scale transformations can be obtained by considering its action on a test function
 \begin{align}\label{Dact}
  &  \delta f\left(x\right) =  e^{i\lambda D}f\left(x\right)e^{-i\lambda D} = \lambda \left[iD,f\left(x\right)\right] + {\cal O}\left(\lambda^2\right) = \lambda \left(x \cdot \partial \right) f\left(x\right) + {\cal O}\left(\lambda^2\right),
 \end{align}
 which implies $i D =  \left(x \cdot \partial \right)$. Similarly, the remaining generators can be determined
 \begin{align}
     i P_{\mu} = \partial_{\mu},\qquad iM_{\mu \nu} = x_{\mu} \partial_{\nu} - x_{\nu} \partial_{\mu}, \qquad i K_{\mu} = 2 x_{\mu}\left( x \cdot \partial\right) - x^2 \partial_{\mu}, \label{PMKact}
 \end{align}
 which are associated to translations, special conformal transformations and Lorentz transformations, respectively. These annihilate the conformal vacuum. With the explicit form of the generators, the conformal algebra can be worked out to be
 \begin{align}\label{confalg}
 &\left[D, K_{\mu}\right] = iK_{\mu}, \quad  \left[D, P_{\mu}\right] = -iP_{\mu}, \quad  \left[P_{\mu}, K_{\nu}\right] = 2iM_{\mu \nu}-2i\eta_{\mu \nu}D\\ \nonumber
    & \left[M_{\mu \nu}, P_{\sigma}\right] = -i\left(\eta_{\mu \sigma}P_{\nu} - \eta_{\nu \sigma}P_{\mu}\right), \quad \hspace*{0.2cm} \left[M_{\mu \nu}, K_{\sigma}\right] = -i\left(\eta_{\mu \sigma}K_{\nu} - \eta_{\nu \sigma}K_{\mu}\right),\\ \nonumber
     & \left[M_{\mu \nu}, M_{\rho \sigma}\right] = i\left(\eta_{\nu \rho}M_{\mu \sigma} +\eta_{\mu \sigma}M_{\nu \rho} - \eta_{\mu \rho}M_{\nu \sigma} - \eta_{\nu \sigma}M_{\mu \rho}\right),
 \end{align}
 with all other commutators vanishing.

From an AdS/CFT point of view, perhaps the most important property of the
conformal group on $\mathbb{R}^{d-1,1}$ is that it is isomorphic to $SO\left(d,2\right)$, which (as shown in \S \tcb{\ref{Aambient}}) is the isometry group of AdS$_{d+1}$. This can be seen at the level of the generators by defining $J_{AB}$, $A,B = 0, ..., d+1$
 \begin{align}
     J_{\mu \nu} = M_{\mu \nu}, \qquad J_{\mu d} = \frac{1}{2}\left(K_{\mu}-P_{\mu}\right), \qquad  J_{\mu \left(d+1\right)} = \frac{1}{2}\left(K_{\mu}+P_{\mu}\right), \qquad J_{ \left(d+1\right)d} = D,
 \end{align}
 which satisfy the $\mathfrak{so}\left(d,2\right)$ commutation relations,
 \begin{equation}
     \left[J_{AB},J_{CD}\right] = i\left(\eta_{B C}J_{A D} +\eta_{A D}J_{B C} - \eta_{A C}J_{B D} - \eta_{B D}J_{A C}\right),\label{sod2alg}
 \end{equation}
 where $\eta_{AB} = \left(-++...+-\right)$.

\section{Representations on operators}
\label{sec::repo}
With the knowledge of the conformal algebra and its generators, we can classify operators according to their representations. These can be derived as induced representations of the subgroup that leave a point in space-time invariant, the \emph{stability subgroup}.

The stability subgroup is generated by Lorentz generators $M_{\mu \nu}$, dilatations $D$ and special conformal transformations $K_{\mu}$. This can be verified using their explicit forms \eqref{Dact} and \eqref{PMKact}. Operators at a fixed space-time point form a representation of this subgroup. On the other hand, the translation generator $P_\mu$ acts exclusively on the co-ordinates $x^\mu$, so that
\begin{equation}
    \left[P_\mu,{\cal O}^a\left(x\right)\right] = -i \partial_{\mu}{\cal O}^a\left(x\right). \label{pa}
\end{equation}
This implies that an operator ${\cal O}^a$ at a generic point in space-time can be obtained from its value at, for example, the origin via
\begin{equation}
   {\cal O}^a\left(x\right) = e^{i x\cdot P} {\cal O}^a\left(0\right) e^{-i x\cdot P}.\label{Prep}\end{equation}
Therefore, the representation of an operator under the full conformal algebra can be determined (or \emph{induced}) from its representation under the stability subalgebra through the action \eqref{Prep} of the translation generator. To wit, for a conformal generator $G$
\begin{equation}
    \left[G,{\cal O}^a\left(x\right)\right] = e^{i x\cdot P}[{\widetilde G},{\cal O}^a\left(0\right)]e^{-i x\cdot P},
\end{equation}
with 
\begin{align}
  {\widetilde G} =  e^{-i x\cdot P}\,G\,e^{i x\cdot P} = \sum\limits^{\infty}_{n=0} \frac{\left(-i\right)^n}{n!} \underbrace{\left[x \cdot P\right.,\left[x \cdot P\right., ... ,\left[x \cdot P\right.,}_{\text{$n$ times}} G\left.\right] ...\left.\right]\left.\right], \label{hausd}
\end{align}
using the Baker-Campbell-Hausdorff formula. In practice, for all generators of the conformal algebra this series truncates at either the first or second term.

\subsubsection*{Warm up: Poincar\'e  Representations}

In a Poincar\'e invariant QFT, local operators at the origin transform in irreducible representations of the Lorentz group,
\begin{equation}
    \left[M_{\mu \nu}, {\cal O}^{a}\left(0\right)\right] = \left(\Sigma_{\mu \nu}\right){}^{a}{}_{b}{\cal O}^b\left(0\right).
\end{equation}
This defines the spin of the operator ${\cal O}^a$, given matrices $\Sigma_{\mu \nu}$ forming a finite dimensional representation of the Lorentz algebra. Away from the origin, we then have
\begin{align} \nonumber
     \left[M_{\mu \nu}, {\cal O}^{a}\left(x\right)\right] & = e^{i x \cdot P} \left[e^{-i x \cdot P}M_{\mu \nu}e^{i x \cdot P},{\cal O}^a\left(0\right)\right]e^{-i x \cdot P} \\ \nonumber
     & = e^{i x \cdot P} \left[M_{\mu \nu}+ x_{\mu}P_{\nu} - x_{\nu}P_{\mu},{\cal O}^a\left(0\right)\right]e^{-i x \cdot P}\\ \label{ma}
     & = \left(\Sigma_{\mu\nu}{}^a{}_b - ix_{\mu}\partial_{\nu}+ix_{\nu}\partial_{\mu}\right){\cal O}\left(x\right),
\end{align}
where in the second equality we employed \eqref{hausd}.

\subsubsection*{Dilatations and special conformal transformations}
In a scale invariant theory, it is natural to diagonalise the dilatation operator acting on operators at the origin,
\begin{equation}
    \left[D,{\cal O}\left(0\right)\right] = - i \Delta {\cal O}\left(0\right). \label{do}
\end{equation}
$\Delta$ is referred to as the \emph{scaling dimension} of ${\cal O}$.

Applying again the Baker-Campbell-Hausdorff formula, we find
\begin{align}
    \left[D,{\cal O}\left(x\right)\right] = - i \left(\Delta + x \cdot \partial \right){\cal O}\left(x\right). \label{dila}
\end{align}
Likewise, for the special conformal transformation,
\begin{align} \label{ka}
    \left[K_\mu,{\cal O}\left(x\right)\right] = \left( 2 x^\nu \Sigma_{\nu \mu}+2i\Delta x_\mu -i x^2 \partial_\mu + i x_\mu x \cdot \partial + {\cal K}_{\mu} \right) {\cal O}\left(x\right),
\end{align}
where we define
\begin{equation}
   \left[K_\mu,{\cal O}\left(0\right)\right] =  {\cal K}_{\mu} {\cal O}\left(0\right).
\end{equation}
\subsection{Primaries and descendants}
\label{subsec::pridesc}
In fact, $K_{\mu}$ is a lowering operator for scaling dimension,
\begin{align}
    \left[D,\left[K_\mu,{\cal O}\left(0\right)\right]\right] & = \left[\left[D,K_\mu\right]+K_{\mu}D,{\cal O}\left(0\right)\right] \\ \nonumber
    & = -i \left(\Delta-1\right) \left[K_{\mu},{\cal O}\left(0\right)\right].
\end{align}
As we shall see in \S \tcb{\ref{subsec:unitarity}}, for unitary CFTs scaling dimensions are bounded from below. This implies that each representation of the conformal algebra must have an operator of lowest dimension, which is then annihilated by $K_\mu$ at $x=0$,
\begin{equation}
    \left[K_\mu,{\cal O}\left(0\right)\right] = {\cal K}_{\mu}{\cal O}\left(0\right) = 0. \label{pri}
\end{equation}
This is the definition of a \emph{primary operator}.

Translation generators on the other hand are raising operators for scaling dimension,
\begin{align}
    \left[D,\left[P_\mu,{\cal O}\left(0\right)\right]\right] & =  -i \left(\Delta+1\right) \left[P_{\mu},{\cal O}\left(0\right)\right].
\end{align}
Given a primary operator ${\cal O}$ of dimension $\Delta$, we can construct operators of higher dimension by acting with translation generators,
\begin{align}
{\cal O}\left(0\right) \rightarrow (-i)^n \partial_{\mu_1} ... \partial_{\mu_n}{\cal O}\left(0\right), \quad \Delta \rightarrow \Delta+n.\label{desc}
\end{align}
These operators are called \emph{descendants} of the operator ${\cal O}$. A representation of the conformal algebra is thus constructed from ${\cal O}\left(0\right)$ and its descendants.

\subsubsection{Finite conformal transformations}
In the preceding section, we derived the action of infinitesimal conformal transformations on operators, given by \eqref{pa}, \eqref{ma}, \eqref{dila} and \eqref{ka}. For a primary operator of scaling dimension $\Delta$ and rotation matrices $\Sigma_{\mu\nu}$, these can be summarised as
\begin{equation}
    \left[Q_{v},{\cal O}\left(x\right)\right] = -i\left( v \cdot \partial + \frac{\Delta}{d} \left(\partial \cdot v \right) - \frac{i}{2} \partial^{\mu} v^{\nu} \Sigma_{\mu \nu}  \right){\cal O}\left(x\right),
\end{equation}
where $v_{\mu}$ is a general conformal Killing vector
\begin{align}
    v_{\mu} = a_{\mu} + \omega_{\mu \sigma}x^{\sigma} + \lambda x_{\mu} + \left(b_{\mu}x^2 - 2x_{\mu} b \cdot x\right).
\end{align}
This corresponds to a combination of infinitesimal translations $a_{\mu}$, rotations $\omega_{\mu \nu}$, scalings $\lambda$ and special conformal transformations parametrised by $b_{\mu}$.

The corresponding finite transformations are obtained by exponentiating the charge, \begin{align}
   {\cal O}^{\prime}\left(x^{\prime}\right) =  U {\cal O}\left(x\right) U^{-1} = \Omega\left(x^{\prime}\right)^{\Delta} D\left(R\left(x^{\prime}\right)\right) {\cal O}\left(x^{\prime}\right), \quad U = e^{Q_v},\label{finite}
\end{align} 
where 
\begin{align}
    \frac{\partial x^{\prime \mu}}{\partial x^{\nu}} = \Omega\left(x^\prime\right) R^{\mu}{}_{\nu}\left(x^\prime\right), \qquad R^{\mu}{}_{\nu} \in \text{SO}\left(d\right).
\end{align}
In \eqref{finite}, $D\left(R\right)$ is a matrix implementing the action of $R$ in the $\text{SO}\left(d\right)$ representation of ${\cal O}$. For example,
\begin{align} \nonumber
    &\text{scalar representation:} \hspace*{3.5cm} D\left(R\right) = 1 \\ \nonumber
    & \text{vector representation:}  \hspace*{3cm} D\left(R\right)_{\mu}{}^{\nu} = R_{\mu}{}^{\nu}
\\ \nonumber
& \hspace*{1.75cm} \vdots \hspace*{6.8cm} \vdots \\ \nonumber
& \text{spin-$s$ representation:}  \hspace*{1.7cm} D\left(R\right)_{\mu_1 ... \mu_s}{}^{\nu_1 ... \nu_s} = R_{\left(\right.\mu_1}{}^{\nu_1} ...\, R_{\left.\mu_s\right)}{}^{\nu_s} - \text{traces.}
\end{align}

\section{Radial quantisation} 
In this section we introduce a particular way of quantising CFTs, which proves useful in establishing various properties of conformally invariant theories. Examples of which, that we rely upon later, include: The state-operator correspondence (\S \tcb{\ref{subsec::soc}}), the subsequent convergence of the operator product expansion (\S \tcb{\ref{sec::cftOPE}}) and unitarity bounds (\S \tcb{\ref{subsec:unitarity}}).

The commutation relations considered in the previous section hold in any quantisation. The standard way to quantise in QFT is to choose a specific space-time foliation, with each leaf endowed with its own Hilbert space. Hilbert spaces on different leaves are connected by a unitary evolution operator.

In practice, it is convenient to choose quantisations that respect the symmetries of a theory: if all surfaces are related by a symmetry transformation, the Hilbert space is the same on each surface. For example, in a theory with Poincar\'e symmetry typically we divide space-time into surfaces of equal time. The evolution operator is then the Hamiltonian, which moves us from one fixed time slice to another.

In a theory with conformal symmetry in $\mathbb{R}^d$, it is natural to foliate space-time with concentric spheres. States are then evolved from smaller spheres to larger spheres using the dilatation operator. This is naturally parameterised by co-ordinates $\left(r,\text{\bf n}\right)$, where $r \in \mathbb{R}$ is a radial co-ordinate and $\text{\bf n}^{\mu}$ is a unit vector on $S^{d-1}$. The space-time metric is then
\begin{equation}
    ds^2 = dr^2 + r^2 d{\bf n}^2.
\end{equation}
To gain a more intuitive grasp of why the dilatation generator defines an evolution operator, we introduce a time co-ordinate $t = \ln r $. The metric then becomes
\begin{equation}
    ds^2 = e^{2t}\left[dt^2+d{\bf n}^2\right],
\end{equation}
which is conformally equivalent to a cylinder. This maps $\mathbb{R}^d$ to $\mathbb{R} \times S^{d-1}$, with the $S^{d-1}$ spheres precisely those which define the foliation above. In the Schr\"odinger picture, states on the cylinder evolve as
\begin{equation}
    \frac{\partial}{\partial t}| \psi\left(t\right) \rangle = -H |\psi\left(t\right)\rangle.
\end{equation}
The crucial observation is that a time translation $t \rightarrow t + \tau$ on the cylinder is generated by a rescaling $r \rightarrow e^{\tau}r$. In other words, the dilatation operator displaces points along the time direction of the cylinder,
\begin{equation}
    D = -i x \cdot \frac{\partial}{\partial x} \: \rightarrow \: -i \frac{\partial}{\partial t}.
\end{equation}
We can hence identify the Hamiltonian $H$ with the dilatation operator 
\begin{equation}
    H = i D.
\end{equation}
We can use this observation to quantise a CFT on the cylinder, and in the next section we construct the Hilbert space. Quantising a CFT is this manner is known as \emph{radial quantisation}, which dates back to Fubini, Hanson and Jackiw  \cite{PhysRevD.7.1732}.

\subsection{State-operator correspondence}
\label{subsec::soc}

Having laid down the basic framework of radial quantisation, we now construct a complete basis of states on a sphere of fixed radius. To begin, we assume the existence of a vacuum state $|0\rangle$ that is invariant under all global conformal transformations
\begin{equation}
    P_{\mu}|0\rangle = K_\mu |0\rangle = M_{\mu \nu}|0\rangle = D|0\rangle=0.
\end{equation}
The $SO\left(d\right)$ angular momentum operators $M_{\mu\nu}$ are the only generators that commute with $D$. States in the Hilbert space ${\cal H}$ can therefore be classified according to two quantum numbers: Their scaling dimension $\Delta$ and spin $s$,
\begin{equation}
    D|\Delta \rangle = - i \Delta | \Delta \rangle \quad \text{and} \quad M_{\mu \nu}|\Delta, s \rangle^a = \left(\Sigma\right){}^{a}{}_b | \Delta, s \rangle{}^{b}. 
\end{equation}
We can organise ${\cal H}$ in this way by defining states on the sphere though insertions of local operators at its origin,\footnote{In this language, the vacuum is prepared by inserting the identity operator.}
\begin{equation}
    {\cal O}_j\left(x\right) \: \rightarrow \: |{\cal O}_j \rangle \equiv \lim_{x \rightarrow 0} {\cal O}_j\left(x\right) |0\rangle. \label{stdef}
\end{equation}
Since we diagonalise the dilatation generator acting on operators at the origin \eqref{do}, such states will be eigenstates of $D$. We can determine their quantum numbers as follows. For a state created by a primary operator ${\cal O}$ of scaling dimension $\Delta$, we have
\begin{align}
    D|{\cal O} \rangle & = \left[D,{\cal O}\left(0\right)\right]|0\rangle + {\cal O}\left(0\right) D|0\rangle\\
    &  = -i\Delta |{\cal O}\rangle. 
\end{align}
Similarly, we can conclude
\begin{equation}
    M_{\mu \nu}|{\cal O} \rangle = \Sigma_{\mu \nu} |{\cal O} \rangle \quad \text{and} \quad K_{\mu}|{\cal O}\rangle = 0. 
\end{equation}
When we act on $|{\cal O}\rangle$ with the translation generator $P_{\mu}$, it raises the energy (scaling dimension) by one unit
\begin{align}
    D P_\mu |{\cal O}\rangle &= \left[D,P\right]|{\cal O}\rangle + P D|{\cal O}\rangle \\ \nonumber
    & = -i\left(\Delta+1\right)|{\cal O}\rangle.
\end{align}
This state is associated to the descendant operator $-i \partial_{\mu} {\cal O}$, since
\begin{align}
    P_{\mu}|{\cal O}\rangle & = \lim_{x \rightarrow 0}\left[P_{\mu},{\cal O}\left(x\right)\right]|0\rangle \\ \nonumber
    & \lim_{x \rightarrow 0} -i \partial_{\mu}{\cal O}\left(x\right)|0\rangle.
\end{align}
\begin{equation}
\partial_{\mu_1} ... \partial_{\mu_n}{\cal O}\left(x\right) = \left[iP_{\mu_1}.\right.,\left[iP_{\mu_2}\right., ...\left[iP_{\mu_n}, {\cal O}\left(x\right)\right] ...\left.\right]\left.\right].
\end{equation}
Owing to conformal symmetry, defining states in this manner accounts for the entire Hilbert space of the CFT. Any state created by an operator inserted at $x \ne 0$ is a superposition of states \eqref{stdef},
\begin{equation}
    {\cal O}\left(x\right)|0\rangle = e^{iPx}{\cal O}\left(0\right)e^{-iPx}|0\rangle = e^{iPx}|{\cal O}\rangle = \sum\limits_n \frac{1}{n!}x^{\mu_1} ... x^{\mu_n} |\partial_{\mu_1}...\partial_{\mu_n}{\cal O}\rangle.
\end{equation}
This construction also works backwards: Given a state $|\psi \rangle$ in radial quantisation, it is natural to decompose it into eigenstates $|{\cal O}_j \rangle $ of the dilatation operator,
\begin{equation}
    |\psi \rangle = \sum\limits_j a_j\, |{\cal O}_j \rangle, \qquad D |\psi \rangle = -i\sum\limits_j \Delta_j \, a_j \, |{\cal O}_j \rangle.
\end{equation}
From each eigenstate $|{\cal O}_j \rangle $ we can construct a local operator with dimension $\Delta_j$. If the state is annihilated by the superconformal generator $K_{\mu}$, this operator will be a primary one. This feature of CFTs is known as the \emph{state-operator correspondence}: States are in one-to-one correspondence with
local operators.

This is a remarkable (and, as we shall see, useful) property of CFTs. In a typical quantum field theory, states and local operators are very different objects. While local operators live at a point in space-time, states live over an entire spatial slice. 
\subsection{Unitarity bounds, conserved currents and free scalars}
\label{subsec:unitarity}
For a theory to be unitarity, all states must have positive norm. This requirement on every state in a conformal multiplet places bounds on the dimensions of primary operators, in a way that we explain in the following. We continue in radial quantisation, which has the convenient feature that $P^{\dagger}_{\mu} = K_{\mu}$.\footnote{This is most easily verified on the Minkowskian cylinder, where
\begin{align}\nonumber
    P_{\mu} = i e^{-i t}\left[-i{\bf n}_{\mu} \partial_t + \left(\delta_{\mu\nu}-{\bf n}_{\mu}{\bf n}_{\nu}\right)\frac{\partial}{\partial {\bf n}_\nu}\right], \quad 
    K_{\mu} = i e^{it}\left[i{\bf n}_{\mu} \partial_t + \left(\delta_{\mu\nu}-{\bf n}_{\mu}{\bf n}_{\nu}\right)\frac{\partial}{\partial {\bf n}_\nu}\right].
\end{align}}

For concreteness, consider the example of a scalar primary operator ${\cal O}$ of dimension $\Delta$. The condition that the first level descendant $ P_{\mu}|{\cal O} \rangle$ has positive norm gives
\begin{equation}\nonumber
  \left(P_{\mu}|{\cal O} \rangle\right)^{\dagger}P_{\nu}|{\cal O} \rangle = \langle {\cal O} | K_\mu  P_{\nu}|{\cal O} \rangle = \langle {\cal O} | \left[K_\mu,P_{\nu}\right]|{\cal O} \rangle = 2i \langle {\cal O} | \left( \delta_{\mu\nu} D + M_{\mu\nu}\right)|{\cal O} \rangle = 2\Delta \delta_{\mu\nu}.\label{l1}
    \end{equation}
    where we normalised $\langle {\cal O} | {\cal O} \rangle = 1$. For the above to be positive definite, we require $\Delta \ge 0$. The operator saturating this bound is the identity operator, which has $\Delta = 0$.

    Similarly, for spin-$s$ primaries ${\cal O}_{s}$ one can verify (see e.g. \cite{Simmons-Duffin:2016gjk}) that the same condition yields
\begin{equation}
  \left(P_{\nu}|{\cal O}_{s} \rangle\right)^{\dagger}P_{\mu}|{\cal O}_{s} \rangle \quad \text{is positive definite} \: \implies \: \Delta \ge s+d-2.
\end{equation}
Moreover, the above bound implies that spin-$s$ states with dimension $\Delta = s+d-2$ are conserved currents: $-i P^{\mu_1}|{\cal O}_{\mu_1 ...\mu_s} \rangle = 0\, \implies \, \partial^{\mu_1}{\cal O}_{\mu_1 ...\mu_s}\left(x\right)=0$. Conserved currents of arbitrary spin will appear often throughout this work.\footnote{In fact for CFTs in $d \ge 3$, assuming the existence of exactly one stress tensor, the presence of currents with spin $s \ge 3$ implies that the theory is free \cite{Maldacena:2011jn,Boulanger:2013zza,Alba:2013yda,Alba:2015upa,Friedan:2015xea}. The CFT we consider in this work is precisely of this type.}

For traceless symmetric tensors in generic CFTs, no further conditions arise by considering higher descendants. However, for scalar operators we can go one step further: Considering the second level descendant, it is possible to obtain
\begin{equation}
 \left(P^2|{\cal O} \rangle\right)^{\dagger}P^2|{\cal O} \rangle = 2\left(2\Delta + 2-d\right) \langle {\cal O}|  K \cdot  P |{\cal O} \rangle \ge 0\; \implies \; \Delta \ge \frac{d}{2}-1,
\end{equation}
where it was used that $\langle {\cal O}| K \cdot  P |{\cal O} \rangle \ge 0$, following from unitary at the first level \eqref{l1}. The scalar operator which saturates this bound satisfies the Klein-Gordon equation 
\begin{equation}
    P^2 |{\cal O} \rangle = 0 \; \implies \; \Box \,{\cal O} = 0.
\end{equation}
I.e. it is a free scalar with scaling dimension $\Delta = \frac{d}{2}-1$.

\section{Operator product expansion}
\label{sec::cftOPE}
An invaluable tool in conformal field theory is the \emph{operator product expansion} (OPE). This is a statement about what
happens as local operators approach each other. Although it has a place in any quantum field theory, in a CFT the OPE  
gains additional and very powerful properties.

More concretely, the OPE asserts that the product of any two local operators inserted at nearby points can be closely approximated by a string of operators at one of these points. In a CFT, the OPE can thus be expressed in the form
\begin{equation}\label{ope}
    {\cal O}_1\left(x_1\right){\cal O}_2\left(x_2\right) = \sum_{{\cal O}_k\, \text{primary}} {\cal F}_{12k}\left(x_{12},\partial_{x_2}\right){\cal O}_k\left(x_2\right), 
    \end{equation}
where ${\cal F}_{12k}\left(x_{12},\partial_{x_2}\right)$ is a power series in $\partial_{x_2}$, and $x_{ij} = x_i - x_j$. From a primary operator ${\cal O}_k$, this generates the contributions of all of its descendants.\footnote{For simplicity, we have suppressed all $SO\left(d\right)$ indices.}

The OPE is to be understood as a statement that holds inside correlation functions (within its radius of convergence),
\begin{equation}
    \langle \mathscr{X} {\cal O}_1\left(x_1\right){\cal O}_2\left(x_2\right)\rangle = \sum_{{\cal O}_k\, \text{primary}} {\cal F}_{12k}\left(x_{12},\partial_{x_2}\right)\langle \mathscr{X}  {\cal O}_k\left(x_2\right)\rangle, \label{corOPE}
\end{equation}
where $\mathscr{X}$ denotes any other operator insertions, $\mathscr{X} = {\cal O}_{i_1}\left(y_1\right) ...{\cal O}_{i_n}\left(y_n\right)$. This relation can in fact be established using radial quantisation:\\

Let $\mathscr{B}$ be a sphere that separates $\left\{x_1,x_2\right\}$ from all other insertion points $\left\{y_1,...,y_n\right\}$ in the correlation function \eqref{corOPE}. We can then quantise radially around the centre $z$ of $\mathscr{B}$, inserting a complete basis of states $ \mathds{1} = \sum\limits_n |n\rangle \langle n |$
\begin{equation}
    \langle 0| \mathscr{X} {\cal O}_1\left(x_1\right){\cal O}_2\left(x_2\right)|0\rangle = \sum\limits_n \langle 0| \mathscr{X}|n\rangle \langle n|{\cal O}_1\left(x_1\right){\cal O}_2\left(x_2\right)|0\rangle.
\end{equation}
By the state-operator correspondence, the states $|n\rangle$ correspond to operators ${\cal O}$ inserted at $z$. We can thus write
\begin{equation}
    \langle 0| \mathscr{X} {\cal O}_1\left(x_1\right){\cal O}_2\left(x_2\right)|0\rangle = \sum\limits_{\left\{{\cal O}\right\}} f_{{\cal O}}\left(x_{12}\right) \langle \mathscr{X} {\cal O}\left(z\right)\rangle, \quad f_{{\cal O}}\left(x_{12}\right) = \langle {\cal O}|{\cal O}_1\left(x_1\right){\cal O}_2\left(x_2\right)|0\rangle.
\end{equation}
To bring the above in the desired form \eqref{corOPE}, we separate the contributions from the primaries and descendants, and re-sum the descendants. In this way, we obtain
\begin{align}
    \langle 0| \mathscr{X} {\cal O}_1\left(x_1\right){\cal O}_2\left(x_2\right)|0\rangle & = \sum_{{\cal O}_k\, \text{primary}} {\tilde {\cal F}}_{12k}\left(x_{12},\partial_{z}\right)\langle \mathscr{X}  {\cal O}_k\left(z\right)\rangle \\ \nonumber
    & = \sum_{{\cal O}_k\, \text{primary}}  {\cal F}_{12k}\left(x_{12},\partial_{x_2}\right)\langle \mathscr{X}  {\cal O}_k\left(x_2\right)\rangle,
\end{align}
where in the second equality we expanded ${\cal O}_k\left(z\right) = {\cal O}_k\left(x_2\right) + \left(z - x_2\right) \cdot \partial_{x_2}{\cal O}_k\left(x_2\right) + ...\,$.

The above derivation indicates that for the OPE to hold, we require the existence of a sphere which separates $x_1$ and $x_2$ from all other insertion points $y_i$. In fact, in CFT the OPE is \emph{convergent}, with radius of convergence equal to the distance to the closest insertion point $y_i$ \cite{Polchinski:1998rq,Pappadopulo:2012jk}. This is by virtue of the state-operator correspondence, since convergence of the OPE is then guaranteed by the usual convergence of a complete set of states in quantum mechanics.

The form of the differential operators ${\cal F}\left(x,\partial\right)$ appearing in the OPE \eqref{ope} is fixed by conformal invariance. To see this, we simply apply the OPE to the three-point function\footnote{This is valid in the domain $|x_1-x_2| < |y-x_2|$. In the second line, we assumed an orthogonal basis of primaries.}
\begin{align} \label{3pt2pt}
    \langle {\cal O}_1\left(x_1\right){\cal O}_2\left(x_2\right){\cal O}_l\left(y\right) \rangle & = \sum_{{\cal O}_k\, \text{primary}} {\cal F}_{12 k}\left(x_{12}, \partial_{x_2}\right)\langle {\cal O}_k \left(x_2\right)  {\cal O}_l\left(y\right) \rangle \\ \nonumber
    & = {\cal F}_{12 k}\left(x_{12}, \partial_{x_2}\right)\langle {\cal O}_l \left(x_2\right)  {\cal O}_l\left(y\right) \rangle.
\end{align}
The ${\cal F}\left(x,\partial\right)$ are therefore determined by the two- and three-point functions of the corresponding primary operator, and the structures appearing in both of the latter are fixed by conformal symmetry. This is discussed in detail in the following sections.

With the above property of the OPE, higher-point functions are essentially determined with the knowledge of the two- and three-point functions in the theory. This is discussed in detail for four-point functions of scalar operators in \S \tcb{\ref{subsec:fptcft}}, and plays a key role in this work.

\section{Correlation functions of primary operators}
\label{sec::cfpo}

The structure imposed by conformal symmetry on correlation functions is a central tool in obtaining some of the results in this work. In this section we therefore review these constraints in detail for the cases relevant for the applications in this thesis.

The absence of a length scale means the concept of the standard S-matrix ill-defined in a CFT. This makes correlation functions of gauge invariant operators key observables in conformally invariant theories. As we shall see below, conformal symmetry places powerful constraints on the form they can take. We focus on the correlation functions of primary operators, since those involving descendants can be immediately determined from the latter via \eqref{desc}.

In a conformally invariant theory, correlation functions are constrained to be invariant under conformal transformations \eqref{finite}
\begin{align}
    \langle {\cal O}_1\left(x_1\right) ...\,{\cal O}_n\left(x_n\right) \rangle  & = \langle U {\cal O}_1\left(x_1\right) U^{-1} ...\, U {\cal O}_n\left(x_n\right) U^{-1} \rangle \\ \nonumber
     & = \Omega\left(x^\prime_1\right)^{\Delta_1}D_1\left(R\left(x^\prime_1\right)\right) ...\, \Omega\left(x^\prime_n\right)^{\Delta_n}D_n\left(R\left(x^\prime_n\right)\right) \langle {\cal O}_1\left(x^\prime_1\right) ...\,{\cal O}_n\left(x^\prime_n\right) \rangle,
\end{align}
where $D_i\left(R\right)$ is the SO$\left(d\right)$ representation of the primary operator ${\cal O}_i$, of scaling dimension $\Delta_i$. In the following, we investigate the consequences of the above constraints on the structure of two-, three- and four-point functions.

When considering correlators of operators with spin, the framework of ambient space provides a powerful setting to explore the consequences of conformal symmetry. We give a brief review of the ambient formalism in this chapter, however the full details are given in \S \tcb{\ref{Aambient}}. 

\subsection{Two-point functions}
\label{subsec::cfpo2}
As a warm up, consider the two-point function of scalar primary operators ${\cal O}_1$ and ${\cal O}_2$ with scaling dimensions $\Delta_1$ and $\Delta_2$. For rotation and translation invariance, we require
\begin{equation}
    \langle {\cal O}_1\left(x\right){\cal O}_2\left(y\right) \rangle =  h\left(|x-y|\right),
\end{equation}
for some function $h$. In a scale-invariant theory with scale-invariant boundary conditions, correlators must be invariant under the action of the dilatation operator $D$,
\begin{align}
 0  &= \langle \left[D,{\cal O}_1\left(x\right)\right]{\cal O}_2\left(y\right)\rangle + \langle  {\cal O}_1\left(x\right)\left[D,{\cal O}_2\left(y\right)\right] \rangle,
\end{align}
which implies the following differential equation for $h$
\begin{equation}
    i\left( x \cdot \partial +\Delta_1 +y \cdot \partial +\Delta_2 \right)h\left(|x-y|\right) = 0.
\end{equation}
It is straightforward to see that the general solution is
\begin{equation}
    h\left(|x-y|\right) = \frac{C}{|x-y|^{\Delta_1+\Delta_2}},
\end{equation}
for some constant $C$. This gives the form of the two-point function in a scale invariant theory.

For theories exhibiting full conformal symmetry, there is a further constraint from invariance under special conformal transformations:
\begin{align}
0 & =  \langle \left[K_\mu, {\cal O}_1\left(x\right)\right]{\cal O}_2\left(0\right) \rangle + \langle  {\cal O}_1\left(x\right)\left[K_\mu,{\cal O}_2\left(0\right)\right] \rangle \\ \nonumber
& = i \left(2 \Delta_1 x_\mu - x^2 \partial_\mu + x_\mu x \cdot \partial \right)\frac{C}{|x|^{\Delta_1+\Delta_2}}\\ \nonumber
& = i\left(\Delta_1 - \Delta_2\right)\frac{C}{|x|^{\Delta_1+\Delta_2}},
\end{align}
where in the second equality we used the fact that ${\cal O}_2$ is primary. For a non-trivial two-point function, we therefore require $\Delta_1 = \Delta_2$.

To summarise, two-point functions of scalar operators with dimensions $\Delta_1$ and $\Delta_2$ in a CFT take the form
\begin{equation}
    \langle {\cal O}_1\left(x\right){\cal O}_2\left(y\right) \rangle = C \frac{\delta_{12}}{|x-y|^{\Delta_1+ \Delta_2}},
\end{equation}
for some constant $C$, and Kronecker delta $\delta_{12}$.

\subsubsection{Technical interlude} 

When considering correlators of operators with spin, a direct generalisation of the approach used for the scalar operators above becomes increasingly involved as the spin increases. Before considering the analogous constraints on the two-point functions of such  operators, we make a brief pause to review some useful tools for dealing with spinning operators. The full details can be found in \S \tcb{\ref{Aambient}}.

First, when considering operators with spin it is convenient to encode them in polynomials of auxiliary vectors. For example, a spin-$s$ primary operator ${\cal O}_{\mu_1 ... \mu_s}$ can be packaged in the polynomial
\begin{equation}
    {\cal O}_{\mu_1 ... \mu_s} \quad \rightarrow \quad {\cal O}_{s}\left(x|z\right) = \frac{1}{s!}\,{\cal O}_{\mu_1 ... \mu_s}\left(x\right) z^{\mu_1} ... z^{\mu_s}, \qquad z^2 = 0, \label{oz}
\end{equation}
with the null condition on the auxiliary vector $z$ enforcing tracelessness. In this way, tensorial operations (e.g. trace and index contractions) are mapped into relatively simple differential operations.

Second, investigating the consequences of conformal invariance is more transparent in the ambient space formalism, in which the $SO\left(d,2\right)$ symmetry is manifest. The basic idea behind this framework is due to Dirac \cite{Dirac:1936fq}, and is that the natural habitat for the conformal group $SO\left(d,2\right)$ is an ambient $\left(d+2\right)$-dimensional ambient space $\mathbb{R}^{d,2}$. Here, the conformal group is realised as the group of linear isometries, and the constraints from conformal symmetry become as trivial as those from Lorentz symmetry. This is provided all CFT fields can be lifted to $\mathbb{R}^{d,2}$, which we quickly review below (see \S \tcb{\ref{Aambient}} for full details).

In the ambient framework, $d$-dimensional flat space-time is realised as a hypercone in  $\mathbb{R}^{d,2}$. Points $x^\mu$ are represented by null rays in $\mathbb{R}^{d,2}$, which can be parameterised in light-cone ambient co-ordinates as
\begin{equation}
    P^A\left(x\right) = \left(1,x^2,x^\mu\right), \quad P^A = \left(P^+,P^-,P^\mu\right).
\end{equation}
Analogously, the auxiliary variables $z^\mu$ in \eqref{oz}, are represented by
\begin{equation}
    Z^A\left(x\right) = z^{\mu} \frac{\partial P^A}{\partial x^{\mu}} = \left(0,2 x\cdot z, z^\mu\right).
\end{equation}
Therefore, the ambient counterpart of the generating function \eqref{oz} for a  spin-$s$ primary operator is given by
\begin{equation}
    {\cal O}_{s}\left(P|Z\right) = \frac{1}{s!}\,{\cal O}_{A_1 ... A_s}\left(P\right) Z^{A_1} ... Z^{A_s},
\end{equation}
where to make the mapping one-to-one, we require
\begin{equation}
    {\cal O}_{s}\left(\lambda P|Z\right) = \lambda^{-\Delta}  {\cal O}_{s}\left(P|Z\right) \quad \text{and} \quad \left(P \cdot \partial_Z\right){\cal O}_s\left(P|Z\right) = 0.
\end{equation}
In other words, ${\cal O}_{s}\left(P\right)$ must be  homogeneous and tangent to the hypercone $P^2=0$. As usual, $\Delta$ represents the scaling dimension of the operator.

\subsubsection{Two-point functions of operators with spin}
We turn to applying the above technology to the study of conformal correlators involving operators with spin, which was developed in \cite{Costa:2011mg} for symmetric and traceless operators. We begin in this section with two-point functions.

Generalising the selection rule for scalar two-point functions considered previously, it can be shown that two-point functions of primaries with different spin or conformal
dimension vanish. We therefore consider the two-point function of two identical spin-$s$ primary operators ${\cal O}_{\mu_1 ... \mu_s}$, of scaling dimension $\Delta$. Employing the ambient framework that was reviewed in the previous section, this takes the form 
\begin{align}
   & \langle {\cal O}_{s}\left(P_1|Z_1\right){\cal O}_{s}\left(P_2|Z_2\right) \rangle =
    H\left(P_1,P_2|Z_1,Z_2\right),
\end{align}
where $H$ is a function of $\left(P_1 \cdot P_2\right)$, $\left(P_1 \cdot Z_2\right)$ and $\left(P_2 \cdot Z_1\right)$.\footnote{Since the $P_i$ and $Z_i$ are null vectors, all other possible contractions are zero.} It must also be a $SO\left(d,2\right)$ singlet, satisfying homogeneity, tangentiality and spin constraints\footnote{For concision we often abbreviate functions $ {\tilde F}\left(P_1, ... , P_n\right) = {\tilde F}\left(\left\{ P_i\right\}\right)$, $i = 1, ..., n$.}
\begin{align} \label{homo}
    & H\left(\left\{\lambda_i P_i\right\}|\left\{Z_i\right\}\right) = \left(\lambda_1 \lambda_2\right)^{-\Delta} H\left(\left\{ P_i\right\}|\left\{Z_i\right\}\right), \quad \lambda_1, \lambda_2 > 0, \\ \label{trans}
    & P_1 \cdot \frac{\partial}{\partial Z_1}H\left(\left\{ P_i\right\}|\left\{Z_i\right\}\right) = P_2 \cdot \frac{\partial}{\partial Z_2}H\left(\left\{ P_i\right\}|\left\{Z_i\right\}\right) = 0, \\ \label{spinl}
    & H\left(\left\{ P_i\right\}|\left\{\alpha_i Z_i\right\}\right) = \left(\alpha_1 \alpha_2\right)^{s} H\left(\left\{ P_i\right\}|\left\{Z_i\right\}\right).
\end{align}
The first condition arises from the fact that the ambient representative of ${\cal O}_{\mu_1 ... \mu_s}$ is homogeneous of degree $-\Delta$. The second condition ensures that $H$ is tangent to the projective null cone in the $\left(d+2\right)$-dimensional ambient space. The third is just a fancy way of saying that it is a degree $s$ polynomial in null auxiliary vectors $Z_1$ and $Z_2$.

To satisfy the transversality condition \eqref{trans}, it is straightforward to see that the auxiliary vectors $Z_1$ and $Z_2$ can only appear through the following building block
\begin{equation}
    {\sf H}_{3} = \frac{1}{P_{12}}\left(\left(Z_1 \cdot Z_2\right) + \frac{2\left(Z_1 \cdot P_2\right)\left(Z_2 \cdot P_1\right)}{P_{12}}\right),
\end{equation}
where the overall factor of $P_{12} = -2 P_1 \cdot P_2$ has been chosen for future convenience. The homogeneity and spin-$s$ conditions \eqref{homo} and \eqref{spinl}, can then be solved up to an overall coefficient as
\begin{equation}
    \langle {\cal O}_{s}\left(P_1|Z_1\right){\cal O}_{s}\left(P_2|Z_2\right) \rangle = {\sf C}_{{\cal O}}\frac{\left({\sf H}_3\right)^s}{P^{\tau_s}_{12}},
    \end{equation}
where we have introduced the twist $\tau_s \equiv \Delta-s$ of a spin-$s$ operator with scaling dimension $\Delta$.

\subsection{Three-point functions}
\label{subsec::cfpo3}
Conformal invariance is also sufficiently powerful to fix three-point correlation functions. Like with the two-point functions of operators with spin above, constructing their form is most straightforward and systematic in the ambient formalism \cite{Costa:2011mg}. As a warm up, we begin with three-point functions involving only scalar primary operators.
\subsubsection{Scalar primary operators}
Consider the three-point function of scalar primary operators ${\cal O}_{i}$ with dimensions $\Delta_i$, $i=1,2,3$,
\begin{equation}
    \langle {\cal O}_{1}\left(P_1\right){\cal O}_{2}\left(P_2\right){\cal O}_{3}\left(P_3\right) \rangle = F\left(P_1, P_2, P_3\right).
\end{equation}
Since the $P_i$ are null vectors $P^2_i=0$, this is a function only of $P_{12}$, $P_{13}$ and $P_{23}$,
\begin{equation}
    F\left(P_1, P_2, P_3\right) \propto P_{23}^{k_1}P_{13}^{k_2}P_{12}^{k_3}, \label{sansatz}
\end{equation}
for some constants $k_i$ to be determined.

For scalar operators in the ambient formalism, there is only the homogeneity constraint
\begin{equation}
    F\left(\left\{\lambda_i P_i\right\}\right) = \lambda^{-\Delta_1}_1\lambda^{-\Delta_2}_2\lambda^{-\Delta_3}_3 F\left(\left\{P_i\right\}\right).
\end{equation}
This is enough to fix the $k_i$ in \eqref{sansatz} completely,
\begin{equation}
    k_1 = \frac{1}{2}\left(\Delta_1-\Delta_2-\Delta_3\right), \quad k_2 = \frac{1}{2}\left(\Delta_2-\Delta_1-\Delta_3\right), \quad k_3 = \frac{1}{2}\left(\Delta_3-\Delta_2-\Delta_1\right).
\end{equation}
Conformal symmetry thus requires the three-point function takes the form
\begin{align}
     \langle {\cal O}_{1}\left(x_1\right){\cal O}_{2}\left(x_2\right){\cal O}_{3}\left(x_3\right) \rangle = \frac{{\sf C}_{123}}{\left(P_{23}\right)^{\frac{\Delta_1 - \Delta_2-\Delta_3}{2}}\left(P_{13}\right)^{\frac{\Delta_2 - \Delta_1-\Delta_3}{2}}\left(P_{12}\right)^{\frac{\Delta_3 - \Delta_1-\Delta_2}{2}}},\label{000gen}
\end{align}
for some multiplicative constant ${\sf C}_{123}$.

As a final note, while the above form \eqref{000gen} is the  most general one dictated by conformal symmetry, an additional constraint comes from Bose symmetry. This requires that it is invariant under the exchange of any two operator insertions, from which it follows that ${\sf C}_{ijk}$ is symmetric in its indices $i$, $j$ and $k$.

\subsubsection{Two scalars and a spinning operator}
\label{00sgen}
We now turn things up a gear, and consider three-point functions involving primary operators with spin. The simplest case is given by those involving a single spinning operator and two scalars, as it involves just a single tensor structure.

For a three-point function involving scalar operators ${\cal O}_{1}$ and ${\cal O}_{2}$, and a spin-$s$ operator ${\cal O}_{\mu_1 ... \mu_s}$ of dimension $\Delta$,
\begin{equation}
    \langle {\cal O}_{1}\left(P_1\right){\cal O}_{2}\left(P_2\right){\cal O}_{s}\left(P_3|Z\right) \rangle = F\left(P_1,P_2,P_3|Z\right), \label{00l}
\end{equation}
we must satisfy the three constraints 
\begin{align} \label{00lhomo}
   & F\left(\left\{\lambda_i P_i\right\}|Z\right)  = \lambda_1^{-\Delta_1}\lambda_2^{-\Delta_2}\lambda_3^{-\Delta}F\left(\left\{P_i\right\}|Z\right),\\ \label{00ltrans}
   & P_3 \cdot \frac{\partial}{\partial Z}F\left(\left\{P_i\right\}|Z\right) = 0, \\ \label{00lspin}
   & F\left(\left\{P_i\right\}|\alpha Z\right)  = \alpha^s F\left(\left\{P_i\right\}|Z\right).
\end{align}
In this case, satisfying the transversality condition \eqref{00ltrans} is only possible if $Z$ appears with $P_1$ and $P_2$ through the new building block 
\begin{equation}
    {\sf Y}_{3} = \frac{\left(Z \cdot P_2\right)}{P_{23}} - \frac{\left(Z \cdot P_1\right)}{P_{13}},
\end{equation}
where the overall factor of $\left(P_{13}\right)\left(P_{23}\right)$ was chosen for later convenience. Combining this with the homogeneity \eqref{00lhomo} and spin conditions \eqref{00lspin}, we find that the most general form of the correlator \eqref{00l} as fixed by conformal symmetry is
\begin{align}
    \langle {\cal O}_{1}\left(x_1\right){\cal O}_{2}\left(x_2\right){\cal O}_{s}\left(x_3|z\right) \rangle = {\sf C}_{{\cal O}_1{\cal O}_2{\cal O}_s}\frac{{\sf Y}_3^{s}}{(P_{12}^2)^{\tfrac{\Delta_1+\Delta_2-\tau_s}{2}}(P_{23}^2)^{\tfrac{\Delta_2+\tau_s-\Delta_1}{2}}(P_{31}^2)^{\tfrac{\tau_s+\Delta_1-\Delta_2}{2}}}\,. \label{gen3pahhh}
\end{align}
\noindent
Just like for the scalar three-point function in the previous section, we can constrain the overall coefficient using Bose symmetry. In this case, since ${\sf Y}_3 \leftrightarrow - {\sf Y}_3$ under ${\cal O}_1 \leftrightarrow {\cal O}_2$, a similar argument shows that
\begin{equation}
    {\sf C}_{{\cal O}_1{\cal O}_2{\cal O}_s} = \left(-1\right)^s{\sf C}_{{\cal O}_2{\cal O}_1{\cal O}_s}.
\end{equation}
The immediate consequence of this condition is that the correlator \eqref{gen3pahhh} vanishes for odd spin $s$.

\subsubsection{The general case}

We now consider the general case of a three-point function involving primary operators ${\cal O}_{s_i}$ of integer spin $s_i$ and dimension $\Delta_i$. This is more involved, owing to the increased number of possible tensorial structures compatible with conformal symmetry.

We write
\begin{align}
    \langle {\cal O}_{s_1}\left(P_1|Z_1\right){\cal O}_{s_2}\left(P_2|Z_2\right){\cal O}_{s_3}\left(P_3|Z_3\right) \rangle = \frac{F\left(P_1,P_2,P_3|Z_1,Z_2,Z_3\right)}{\left(P_{12}\right)^{\frac{^{\tau_1+\tau_2-\tau_3}}{2}}\left(P_{13}\right)^{\frac{^{\tau_1+\tau_3-\tau_2}}{2}}\left(P_{23}\right)^{\frac{^{\tau_2+\tau_3-\tau_1}}{2}}}, \label{gen3pt}
\end{align}
where $\tau_i = \Delta_i - s_i$. Conformal symmetry requires\footnote{We chose the denominator such that $F\left(\left\{ P_i\right\}|\left\{ Z_i\right\}\right)$ does not scale with the $P_i$.}
\begin{align} \label{homospin}
    F\left(\left\{\lambda_i P_i\right\}|\left\{\alpha_i Z_i\right\}\right) & = F\left(\left\{ P_i\right\}|\left\{ Z_i\right\}\right)\prod_i\left(\alpha_i\right)^{s_i} \\
   P_j \cdot \frac{\partial}{\partial Z_j} F\left(\left\{ P_i\right\}|\left\{ Z_i\right\}\right) & = 0, \quad \text{for each}\: j = 1, 2, 3. \label{gen3pttrans}
\end{align}
Restricting to parity-invariant correlators,\footnote{In three- and four-dimensions, we can construct parity odd correlators using the $SO\left(d,2\right)$-invariant epsilon tensor.} in this case there are six possible transverse building blocks from which the three-point function can be constructed \cite{Costa:2011mg},
\begin{subequations}\label{6conf}
\begin{align}
    {\sf Y}_{1} & = \frac{\left(Z_1 \cdot P_3\right)}{P_{13}} -\frac{\left(Z_1 \cdot P_2\right)}{P_{12}} , \qquad {\sf H}_1 = \frac{1}{P_{23}}\left(\left(Z_2 \cdot Z_3\right) + \frac{2\left(Z_2 \cdot P_3\right)\left(Z_3 \cdot P_2\right)}{P_{23}}\right),   \\
    {\sf Y}_{2} & =  \frac{\left(Z_2 \cdot P_1\right)}{P_{21}} - \frac{\left(Z_2 \cdot P_3\right)}{P_{23}}, \qquad {\sf H}_2 = \frac{1}{P_{13}}\left(\left(Z_1 \cdot Z_3\right) + \frac{2\left(Z_1 \cdot P_3\right)\left(Z_3 \cdot P_1\right)}{P_{13}}\right),   \\
    {\sf Y}_{3} & = \frac{\left(Z_3 \cdot P_2\right)}{P_{32}}- \frac{\left(Z_3 \cdot P_1\right)}{P_{31}} , \qquad {\sf H}_3 = \frac{1}{P_{23}}\left(\left(Z_2 \cdot Z_3\right) + \frac{2\left(Z_2 \cdot P_3\right)\left(Z_3 \cdot P_1\right)}{P_{13}}\right). 
\end{align}
\end{subequations}
Each of the above structures independently satisfy the transversality constraint \eqref{gen3pttrans} in more than three dimensions.

Taking into account the spin and homogeneity conditions \eqref{homospin}, the most general three-point function compatible with conformal symmetry is thus
\begin{align}\label{ooo123321}
&    \langle {\cal O}_{s_1}\left(P_1|Z_1\right){\cal O}_{s_2}\left(P_2|Z_2\right){\cal O}_{s_3}\left(P_3|Z_3\right) \rangle  \\ \nonumber & \hspace*{5cm}= \sum\limits_{n_i} {\sf C}_{n_1, n_2, n_3}\frac{{\sf Y}_1^{s_1-n_2-n_3}{\sf Y}_2^{s_2-n_3-n_1}{\sf Y}_3^{s_3-n_1-n_2}{\sf H}_1^{n_1}{\sf H}_2^{n_2}{\sf H}_3^{n_3}}{(P_{12})^{\tfrac{\tau_1+\tau_2-\tau_3}{2}}(P_{23})^{\tfrac{\tau_2+\tau_3-\tau_1}{2}}(P_{13})^{\tfrac{\tau_3+\tau_1-\tau_2}{2}}},
\end{align}
with theory-dependent coefficients ${\sf C}_{n_1, n_2, n_3}$. 
\subsubsection{Back to Intrinsic}
In the preceding sections, with the efficiency of the ambient formalism we were able to write down the most general form of two- and three-point functions of primary operators of any scaling dimension and spin.
In \S \tcb{\ref{chapt::vec}} we apply these results to a particular CFT, and for this purpose it is useful to give the dictionary which translates them back into the standard $d$-dimensional language.

For the scalar products of ambient vectors that are ubiquitous 
in the ambient expressions for the correlators: $-2P_i\cdot P_j$, $Z_i\cdot P_j$ and $Z_i\cdot Z_j$, this is straightforward to write down from the explicit forms of the ambient space-time and auxiliary vectors  
\begin{equation}
    P^A(x)=(1,x^2,x^\mu), \qquad Z^A(x)=(0,2x\cdot z,z^\mu). 
\end{equation}
Recalling that we work in light-cone ambient co-ordinates, this establishes the following dictionary between ambient and intrinsic scalar products
\begin{align}
-2P_i\cdot P_j&=x_{ij}^2\,,&Z_i\cdot P_j&=-z_i\cdot x_{ij}\,,& Z_i\cdot Z_j=z_i\cdot z_j\,,
\end{align}
where by definition
\begin{align}
    P_i&\equiv P(x_i)\,,& Z_i&\equiv Z(x_i)\,.
\end{align}
Employing the above dictionary, one obtains the following intrinsic expressions of the building blocks found above:
\begin{subequations}
\label{6confInt}
\begin{align} 
{\sf Y}_1&=\frac{z_1\cdot x_{12}}{x_{12}^2}-\frac{z_1\cdot x_{13}}{x_{13}^2}\,,& {\sf H}_1&=\frac{1}{x_{23}^2}\left(z_2\cdot z_3+\frac{2 z_2\cdot x_{23}\,z_3\cdot x_{32}}{x_{23}^2}\right)\,,\\
{\sf Y}_2&=\frac{z_2\cdot x_{23}}{x_{23}^2}-\frac{z_2\cdot x_{21}}{x_{21}^2}\,,&{\sf H}_2&=\frac{1}{x_{31}^2}\left(z_3\cdot z_1+\frac{2 z_3\cdot x_{31}\,z_1\cdot x_{13}}{x_{31}^2}\right)\,,\\
{\sf Y}_3&=\frac{z_3\cdot x_{31}}{x_{31}^2}-\frac{z_3\cdot x_{32}}{x_{32}^2}\,,& {\sf H}_3&=\frac{1}{x_{12}^2}\left(z_1\cdot z_2+\frac{2 z_1\cdot x_{12}\,z_2\cdot x_{21}}{x_{12}^2}\right)\,.
\end{align}
\end{subequations}

\subsection{Four-point functions}
\label{subsec:fptcft}
Correlation functions with more than three operator insertions are no longer fixed by conformal kinematics alone. This is due to the existence of \emph{cross ratios}, which are conformal invariants that can be built from four or more distinct space-time points. In this section, and for the remainder of this chapter, we focus our attention on the degree to which four-point functions of scalar primary operators can be constrained by conformal symmetry.

Out of four distinct points, one can build two independent cross ratios,
\begin{equation}
    u = \frac{\left(x_1 - x_2\right)^2\left(x_3 - x_4\right)^2}{\left(x_1 - x_3\right)^2\left(x_2 - x_4\right)^2}, \qquad v = u\,\big|_{2 \leftrightarrow 4} =  \frac{\left(x_1 - x_4\right)^2\left(x_2 - x_3\right)^2}{\left(x_1 - x_3\right)^2\left(x_2 - x_4\right)^2}.
\end{equation}
Four-point functions may therefore depend non-trivially on $u$ and $v$, in a way that cannot be fixed by conformal symmetry. For a four-point function of scalar operators ${\cal O}_{i}$, reverting briefly to the ambient formalism we have
\begin{equation}
     \langle {\cal O}_{1}\left(P_1\right){\cal O}_{2}\left(P_2\right){\cal O}_{3}\left(P_3\right){\cal O}_{4}\left(P_4\right) \rangle = \frac{g_{1234}\left(u,v\right)}{P^{k_1}_{12}P^{k_2}_{13}P^{k_3}_{14}P^{k_4}_{23}P^{k_5}_{24}P^{k_6}_{34}},
\end{equation}
with the $k_i$ constrained by the homogeneity condition
\begin{align}
   &  \langle {\cal O}_{1}\left(\lambda_1 P_1\right){\cal O}_{2}\left(\lambda_2 P_2\right){\cal O}_{3}\left(\lambda_3 P_3\right){\cal O}_{4}\left(\lambda_4 P_4\right) \rangle = \\ \nonumber & \hspace*{5cm}\lambda^{-\Delta_1}_1\lambda^{-\Delta_2}_2\lambda^{-\Delta_3}_3\lambda^{-\Delta_4}_4 \langle {\cal O}_{1}\left(P_1\right){\cal O}_{2}\left(P_2\right){\cal O}_{3}\left(P_3\right){\cal O}_{4}\left(P_4\right).
\end{align}
This implies the following relations amongst the $k_i$,
\begin{align}
   & k_1 = \Delta_1 + \Delta_2 - \Delta_3 - \Delta_4 + k_6, \quad  k_2 = \Delta_1 - \Delta_2 + \Delta_3 - \Delta_4 + k_5 \\ \nonumber
  &  k_3 = 4\Delta_4 - k_5 - k_6, \quad k_4 = -\Delta_1 + \Delta_2 + \Delta_3 + \Delta_4 - k_5 - k_6.
\end{align}
For instance we can take the following convenient parametrisation
\begin{align}
   &  \langle {\cal O}_{1}\left(P_1\right){\cal O}_{2}\left(P_2\right){\cal O}_{3}\left(P_3\right){\cal O}_{4}\left( P_4\right) \rangle = \left(\frac{P_{24}}{P_{14}}\right)^{\frac{\Delta_1-\Delta_2}{2}}\left(\frac{P_{14}}{P_{13}}\right)^{\frac{\Delta_3-\Delta_4}{2}}\frac{g_{1234}{\left(u,v\right)}}{P^{\frac{\Delta_1+\Delta_2}{2}}_{12}P^{\frac{\Delta_3+\Delta_4}{2}}_{34}}.
   \end{align}
Like for the three-point functions, the function $g_{1234}\left(u,v\right)$ is not completely unconstrained as it must be consistent with Bose-symmetry. I.e. invariance under permutations of the scalar operators. Since all permutations of $\left\{x_1,x_2,x_3,x_4\right\}$ are generated by $x_1 \leftrightarrow x_2$, $x_3 \leftrightarrow x_4$ and $x_1 \leftrightarrow x_3$, it is sufficient to consider the constraints coming from ${\cal O}_{1}\left(x_1\right) \leftrightarrow {\cal O}_{2}\left(x_2\right)$, ${\cal O}_{3}\left(x_3\right) \leftrightarrow {\cal O}_{4}\left(x_4\right)$ and ${\cal O}_{1}\left(x_1\right) \leftrightarrow {\cal O}_{3}\left(x_3\right)$. These require\footnote{Note that under $x_1 \leftrightarrow x_2$ and $x_1 \leftrightarrow x_2$, the cross ratios transform as $u \rightarrow u/v$ and $v \rightarrow 1/v$. Under $x_1 \leftrightarrow x_3$, we have $u \leftrightarrow v$.}
\begin{align} \label{c1}
    g_{1234}\left(u,v\right) & = v^{\frac{\Delta_4 - \Delta_3}{2}}g_{2134}\left(\frac{u}{v},\frac{1}{v}\right), \\ \label{c2}
    g_{1234}\left(u,v\right) & = v^{\frac{\Delta_1 - \Delta_2}{2}}g_{1243}\left(\frac{u}{v},\frac{1}{v}\right), \\ \label{c3}
    g_{1234}\left(u,v\right) & = u^{\frac{\Delta_1 + \Delta_2}{2}}v^{-\frac{\Delta_2 + \Delta_3}{2}}g_{3214}\left(v,u\right),
\end{align}
respectively. This gives the most general form of a scalar four-point function in a conformally invariant theory.\footnote{We will see shortly that the functions $g_{ijkl}\left(u,v\right)$ are not arbitrary, but for a given theory are related in a non-trivial way to its three-point functions.}

The case relevant for applications in later chapters of this thesis is the four-point functions of \emph{identical} scalar operators. In general, the Bose-symmetry constraints \eqref{c1}, \eqref{c2} and \eqref{c3} relate \emph{different} functions $g_{ijkl}\left(u,v\right)$. However for four-point functions involving identical scalars ${\cal O}$, these constraints just apply to a single function $g\left(u,v\right)$,  
\begin{equation} \label{crossingsym}
    g\left(u,v\right) = g\left(\frac{u}{v},\frac{1}{v}\right), \qquad g\left(u,v\right) = \left(\frac{u}{v}\right)^\Delta g\left(v,u\right).
\end{equation}

\section{Conformal block decomposition}
\label{sec::cbd}
In the previous subsection, we observed that conformal kinematics fixes four-point functions up to a function of the cross ratios $u$ and $v$. As we shall demonstrate below, this function can be completely determined from a combination of conformal invariance and the OPE (c.f. \S \tcb{\ref{OPE}}). The OPE coefficients encode the dynamics of the CFT.

For simplicity and relevance for applications in this thesis, we simply consider a four-point function of identical scalar operators ${\cal O}$ with scaling dimension $\Delta$,
\begin{equation}
    \langle {\cal O}\left(x_1\right){\cal O}\left(x_2\right){\cal O}\left(x_3\right){\cal O}\left(x_4\right) \rangle = \frac{g\left(u,v\right)}{\left(x^2_{12}\right)^{\Delta}\left(x^2_{34}\right)^{\Delta}}, \label{4ptoooo}
\end{equation}
where $x^2_{ij} = \left(x_i - x_j\right)^2$. Applying the OPE twice, for example in the (12) and (34) channels, we can represent the four-point function as a double sum over the primary operators that appear in the OPE
\begin{equation}
    \langle {\cal O}\left(x_1\right){\cal O}\left(x_2\right){\cal O}\left(x_3\right){\cal O}\left(x_4\right) \rangle = \sum\limits_{\substack{{\cal O}_i,\,{\cal O}_j \\ \text{primary}}} {\cal F}_{{\cal O}{\cal O}i}\left(x_{12},\partial_{x_2}\right){\cal F}_{{\cal O}{\cal O}j}\left(x_{34},\partial_{x_4}\right) \langle {\cal O}_i\left(x_2\right){\cal O}_j\left(x_4\right)  \rangle. 
\end{equation}
Choosing an orthonormal basis of operators, this collapses to a single sum
\begin{equation}
    \langle {\cal O}\left(x_1\right){\cal O}\left(x_2\right){\cal O}\left(x_3\right){\cal O}\left(x_4\right) \rangle \quad = \sum\limits_{{\cal O}_i\,  \text{primary}} \left(c_{{\cal O}{\cal O}i}\right)^2W_{{\cal O}_i}\left(x_{1},x_2;x_{3},x_4\right), \label{cbe}
\end{equation}
where we have extracted the OPE coefficients $c_{{\cal O}{\cal O}i}$ from the differential operators ${\cal F}_{{\cal O}{\cal O}i}$. Owing to the relationship \eqref{3pt2pt} between ${\cal F}_{{\cal O}{\cal O}i}$ and the three-point function $\langle {\cal O}{\cal O}{\cal O}_i\rangle$, this OPE coefficient is given by
\begin{align}
    c^2_{{\cal O}{\cal O}i} = {\sf C}^2_{{\cal O}{\cal O}{\cal O}_i}/{\sf C}_{{\cal O}_i{\cal O}_i}, \label{ope32}
\end{align}
where ${\sf C}_{{\cal O}{\cal O}{\cal O}_i}$
is the overall three-point function coefficient and ${\sf C}_{{\cal O}_i{\cal O}_i}$ the overall coefficient of the two-point function $\langle {\cal O}_i{\cal O}_i \rangle$. The conformal partial wave $W_{{\cal O}_i}$ re-sums the contribution of the primary ${\cal O}_i$ and all of its descendants to the correlator, and thus represents the contribution from the entire conformal multiplet.

It is often instructive to think of a conformal partial wave $W_{{\cal O}_i}$ as the insertion of a projector onto the conformal multiplet of ${\cal O}_i$,
\begin{align} \nonumber
W_{{\cal O}_i}\left(x_{1},x_2;x_{3},x_4\right) & = \langle {\cal O}\left(x_1\right){\cal O}\left(x_2\right)|{\cal O}_i|{\cal O}\left(x_3\right){\cal O}\left(x_4\right) \rangle, \\ \label{cftcbproj} 
    |{\cal O}_i| & =  \sum\limits_n | P_{\mu_n}...P_{\mu_1} {\cal O}_i \rangle \frac{1}{{\cal N}_{{\cal O}_i}} \langle P_{\mu_n}...P_{\mu_1} {\cal O}_i |,
\end{align}
for some normalisation ${\cal N}_{{\cal O}_i}$. The identity is the sum of the projection operators over all primary operators 
\begin{equation}
    \mathds{1} = \sum\limits_{{\cal O}_i\; \text{primary}} |{\cal O}_i|. \label{po}
\end{equation}
The projector $|{\cal O}_i|$ commutes with all conformal generators, and thus the partial wave $W_{{\cal O}_i}$ has the same transformation properties as the four point function \eqref{4ptoooo}. It may then be written as
\begin{equation} 
    W_{{\cal O}_i}\left(x_{1},x_2;x_{3},x_4\right) = \frac{G_{{\cal O}_i}\left(u,v\right)}{\left(x^2_{12}\right)^{\Delta}\left(x^2_{34}\right)^{\Delta}}.
\end{equation}
The function $G_{{\cal O}_i}\left(u,v\right)$ of cross ratios is known as a \emph{conformal block}, and encodes the contribution of a primary ${\cal O}_i$ and its descendants to the function $g\left(u,v\right)$ in the correlator \eqref{4ptoooo}, 
\begin{equation}
    g\left(u,v\right) = \sum\limits_{{\cal O}_i\,  \text{primary}} \left(c_{{\cal O}{\cal O}i}\right)^2\,G_{{\cal O}_i}\left(u,v\right).
\end{equation}
The space-time dependence of $g\left(u,v\right)$ is completely encoded in the conformal blocks $G_{{\cal O}_i}\left(u,v\right)$, which are universal in the sense that they do not depend on the CFT under consideration but only on the conformal representation of ${\cal O}_i$. I.e. its spin and scaling dimension. The a priori arbitrary function $g\left(u,v\right)$ is thus completely fixed by conformal invariance and the OPE.

While conformal blocks are central objects in CFT,\footnote{For example, they are basic ingredients in the conformal bootstrap program \cite{Ferrara:1973yt,1974JETP...39...10P,Rattazzi:2008pe}.} their explicit forms have been difficult to pin down in generality. Explicit formulas are known only
for simple cases, like for external scalar operators in even spacetime dimensions \cite{Dolan:2000ut,Dolan:2003hv}.  In other cases one has to resort to more indirect methods, like recursion relations \cite{Dolan:2000ut,Dolan:2011dv,Penedones:2015aga} or efficient series expansions \cite{Hogervorst:2013sma,Hogervorst:2013kva}. There is also a general method to increase the spin of the external
operators using differential operators \cite{Iliesiu:2015qra,Costa:2011dw,Echeverri:2015rwa}.

Another indirect approach is given by so-called integral representations \cite{Ferrara:1972xe,Ferrara:1972ay,Ferrara:1972uq,Ferrara:1973vz,Dolan:2011dv,SimmonsDuffin:2012uy,Costa:2014rya,Rejon-Barrera:2015bpa}, which we introduce in the following section. This is most useful for our purposes, for it allows us to define conformal blocks as integral products of three-point structures. In this way it is more straightforward to identify conformal blocks arising from computations in theories which are not, a priori, CFTs. Such as gravity theories in anti-de Sitter space.

\subsection{Integral representation}
\label{subsec::intrep}
In this section we derive a useful representation of the projector \eqref{po}, which leads to an integral  expression for conformal blocks. This is motivated by the following observation of Dolan and Osborn \cite{Dolan:2011dv},\footnote{Later formulated in the ambient framework in  \cite{SimmonsDuffin:2012uy}.} which is underpinned by the shadow formalism of Ferrara, Gatto, Grillo, and Parisi \cite{Ferrara:1972xe,Ferrara:1972ay,Ferrara:1972uq,1972NuPhB..49...77F}.

We first define for a given primary operator $O$ with quantum numbers $\left[\Delta,s\right]$, an associated dual (or shadow) operator\footnote{ ${\hat \partial}_z$ is a differential operator which accommodates tracelessless (\S \tcb{\ref{app::ao}}), and $I_{\mu \nu}\left(x\right)$ is the inversion tensor \begin{equation}
    I_{\mu\nu}\left(x\right) = \delta_{\mu \nu} - \frac{2x_{\mu}x_{\nu}}{x^2} \: ; \qquad z_1 \cdot I\left(x\right) \cdot z_2 = z_1 \cdot z_2 - 2 \frac{z_1 \cdot x\, z_2 \cdot x }{x^2}.
\end{equation}}
\begin{equation}
    {\tilde {\cal O}}\left(x_1|z_1\right) = \kappa_{\Delta,s} \int d^dx_2\, \frac{1}{\left(x^2_{12}\right)^{d-\Delta}} \left(z_1 \cdot I\left(x_1-x_2\right) \cdot {\hat \partial}_{z_2}\right)^s {\cal O}\left(x_2|z_2\right),\label{shad}
\end{equation}
which has quantum numbers $\left[d-\Delta,s\right]$. The normalisation $\kappa_{\Delta,s}$ is chosen such that ${\tilde {\tilde {\cal O}}} = {\tilde {\cal O}}$, which requires
\begin{equation}\label{kappapoods}
    \kappa_{\Delta,s} = \left(\frac{2}{\pi}\right)^{\frac{d}{4}} \frac{\Gamma\left(d-\Delta+s\right)}{\Gamma\left(\Delta-\frac{d}{2}\right)} \frac{1}{\left(\Delta-1\right)_s}.
\end{equation}

\noindent
Dolan and Osborn observed that inserting the invariant projection operator
\begin{align}
    {\cal P}_{{\cal O}_i} = \kappa_{d-\Delta,s} \int d^dx\, {\cal O}_{i} \left(x\right)|0\rangle  \langle 0 |{\tilde {\cal O}}_{i} \left(x\right), \label{posh}
\end{align}
in a four-point function yields the contributions from the $\left[\Delta_i,s\right]$ and $\left[d-\Delta_i,s\right]$ conformal multiplets,
\begin{align} \nonumber
\langle {\cal O}\left(x_1\right){\cal O}\left(x_2\right){\cal P}_{{\cal O}_i}{\cal O}\left(x_3\right){\cal O}\left(x_4\right) \rangle = {\sf c}^2_{{\cal O}{\cal O}{\cal O}_i} W_{{\cal O}_i}\left(x_{1},x_2;x_{3},x_4\right) + {\sf c}^2_{{\cal O}{\cal O}{\tilde {\cal O}}_i} W_{{\tilde {\cal O}}_i}\left(x_{1},x_2;x_{3},x_4\right).\end{align}
This entails an integral representation of their total contribution,
\begin{align}
    & {\sf c}^2_{{\cal O}{\cal O}{\cal O}_i} W_{{\cal O}_i}\left(x_{1},x_2;x_{3},x_4\right) + {\sf c}^2_{{\cal O}{\cal O}{\tilde {\cal O}}_i} W_{{\tilde {\cal O}}_i}\left(x_{1},x_2;x_{3},x_4\right) \\ \nonumber
    & \hspace*{5cm} = \kappa_{d-\Delta,s} \int d^dx\, \langle {\cal O}\left(x_1\right){\cal O}\left(x_2\right){\cal O}_i\left(x\right) \rangle \langle {\tilde {\cal O}}_i\left(x\right){\cal O}\left(x_3\right){\cal O}\left(x_4\right) \rangle,
\end{align}
as a product of two three-point functions. The factor of $\kappa_{d-\Delta,s}$ in the above ensures that the projector \eqref{posh} acts trivially when inserted into a correlator involving ${\cal O}_i$.

An integral expression for the non-shadow contribution can be obtained by integrating over complex scaling dimensions with poles at dimensions $\Delta_i$ and $d-\Delta_i$,
\begin{align}
    \int^{\infty}_{-\infty}\frac{d\nu}{\nu^2 + \left(\Delta_i - \frac{d}{2}\right)^2} 
   \left(W_{{\cal O}_{\frac{d}{2}+i\nu}}\left(x_{1},x_2;x_{3},x_4\right) + W_{{\cal O}_{\frac{d}{2}-i\nu}}\left(x_{1},x_2;x_{3},x_4\right)\right).
\end{align}
This projects out the shadow contribution, since the $\nu$-contour may only be closed around the poles with $\frac{d}{2}\pm i\nu = \Delta_i$,\footnote{The partial wave $W_{{\cal O}_{\frac{d}{2}\pm i\nu}}\left(x_{1},x_2;x_{3},x_4\right)$ decays fast enough (exponentially) only for $\text{Im}\left(\nu\right) \rightarrow \mp \infty$. Applying the residue theorem, we must therefore close the contour in the lower half plane for $W_{{\cal O}_{\frac{d}{2} + i\nu}}$ thus picking up a contribution from the pole at $i\nu = \delta_i - \frac{d}{2}$, while for $W_{{\cal O}_{\frac{d}{2} - i\nu}}$ we close in the upper-half plane around the pole at $i\nu = \frac{d}{2}-\delta_i$.} 
\begin{align} \nonumber
  &  \int^{\infty}_{-\infty}\frac{d\nu}{\nu^2 + \left(\Delta_i - \frac{d}{2}\right)^2} 
   \left(W_{{\cal O}_{\frac{d}{2}+i\nu}}\left(x_{1},x_2;x_{3},x_4\right) + W_{{\cal O}_{\frac{d}{2}-i\nu}}\left(x_{1},x_2;x_{3},x_4\right)\right) \\ \nonumber
   & \hspace*{9.5cm}= \frac{2\pi}{\left(\Delta_i-\frac{d}{2}\right)} W_{{\cal O}_{i}}\left(x_{1},x_2;x_{3},x_4\right).
\end{align}
We then arrive to the following integral representation of a single conformal block
\begin{align}\label{factcbcont}
 &   W_{{\cal O}_{i}}\left(x_{1},x_2;x_{3},x_4\right)  =  \frac{(\Delta_i - \tfrac{d}{2})}{2\pi} \int^{\infty}_{-\infty}d\nu\,\frac{1}{\nu^2 + \left(\Delta_i - \frac{d}{2}\right)^2} \frac{\kappa_{\frac{d}{2}-i\nu,s}}{\left(2\pi\right)^{\frac{d}{4}}} \frac{\Gamma\left(\frac{\frac{d}{2}-i\nu +s}{2}\right)^2}{\Gamma\left(\frac{\frac{d}{2}+i\nu +s}{2}\right)^2} \\ \nonumber
 & \hspace*{3.8cm} \times \int d^dx\, \langle \langle {\cal O}\left(x_1\right){\cal O}\left(x_2\right){\cal O}_{\frac{d}{2}+i\nu}\left(x\right) \rangle \rangle \langle \langle {\tilde {\cal O}}_{\frac{d}{2}-i\nu}\left(x\right){\cal O}\left(x_3\right){\cal O}\left(x_4\right) \rangle \rangle,
\end{align}
where we introduced the notation 
\begin{align}
  & \langle {\cal O}\left(x_1\right){\cal O}\left(x_2\right){\cal O}_{\frac{d}{2}\pm i\nu}\left(x\right) \rangle = {\sf C}_{{\cal O}{\cal O}{\cal O}_{\frac{d}{2} \pm i\nu}}\langle \langle {\cal O}\left(x_1\right){\cal O}\left(x_2\right){\cal O}_{\frac{d}{2}  \pm i\nu}\left(x\right) \rangle \rangle,
  \end{align}
 I.e. $\langle \langle \bullet \rangle \rangle $ denotes the removal the overall coefficient from the three-point function.\footnote{In establishing \eqref{factcbcont}, we used that
 \begin{equation}
     {\sf C}_{{\cal O}{\cal O}{\cal O}_{\frac{d}{2} - i\nu}} = \frac{1}{\left(2\pi\right)^{\frac{d}{4}}} \frac{\Gamma\left(\frac{\frac{d}{2}-i\nu +s}{2}\right)^2}{\Gamma\left(\frac{\frac{d}{2}+i\nu +s}{2}\right)^2} {\sf C}_{{\cal O}{\cal O}{\cal O}_{\frac{d}{2} + i\nu}}.
 \end{equation}}

Similarly we can express the conformal block expansion of a give four-point function in the following form\footnote{Which has origins in the early literature, for example \cite{PhysRevD.13.887,1977LNP,Mack:2009mi}.}
\begin{equation}\label{contintform}
    \langle {\cal O}\left(x_1\right){\cal O}\left(x_2\right){\cal O}\left(x_3\right){\cal O}\left(x_4\right) \rangle = \sum\limits_s \int^{\infty}_{-\infty}d\nu \,g_{s}\left(\nu\right) W_{{\cal O}_{\frac{d}{2}+i\nu,s}}\left(x_{1},x_2;x_{3},x_4\right),  
\end{equation}
which we refer to as the \emph{contour integral representation}. For each spin-$s$ primary operator in the ${\cal O}{\cal O}$ OPE of scaling dimension $\Delta_i$, the meromorphic function contains a pole at $i\nu = \Delta_i - \frac{d}{2}$. The residue of $g_s\left(\nu\right)$ at this pole gives the square of the corresponding OPE coefficient, such that the usual conformal block decomposition \eqref{cbe} is recovered upon application to the residue theorem.

This way of representing the conformal block expansion and conformal blocks will prove instrumental in later sections of this thesis, in which we apply the conformal block expansion in the context of the AdS/CFT duality.

  \chapter{Correlators in the free $O\left(N\right)$ model}
\label{chapt::vec}

Having discussed the general structure and properties of two-, three- and four-point correlation functions in CFTs, in this chapter we move on to consider them in the context of the simplest examples of such theories: free conformal field theories. In this case, exact results for the correlators can be obtained by simple application of Wick's theorem.

Our consideration of these simple CFTs is motivated by the conjectured equivalence between higher-spin gauge theories on AdS and free vector model CFTs. In particular, we compute two-, three- and four-point functions of single-trace operators in the free scalar $O\left(N\right)$ vector model, which are dual to single-particle states in the higher-spin theory. Moreover, we apply the conformal partial wave expansion techniques introduced in the previous chapter to the four-point function of the scalar single-trace operator. This is in the view to compare with the corresponding processes in the dual higher-spin theory on AdS.

\section{The free scalar $O\left(N\right)$ vector model}
Some of the simplest examples of CFTs are given by free theories without a mass scale. These theories are \emph{solvable} in the sense that the path integral is Gaussian, which can therefore be straightforwardly be evaluated to determine all correlators in the theory.

In this chapter we focus on the example of the free  scalar $O\left(N\right)$ vector model, which is conjectured to be dual to the type A minimal bosonic higher-spin theory on AdS (\S \tcb{\ref{sec::ce}}). This is a theory of an $N$ component real scalar field $\phi^a$, transforming in the fundamental representation of $O\left(N\right)$. The action is simply
\begin{equation}
    S[\phi] = \sum\limits^{N}_{a=1} \int d^dx\, \frac{1}{2}\partial_{i} \phi^a \partial^{i} \phi^a,
\end{equation}
with equation of motion $\partial^2 \phi^a = 0$. As we saw in \S \tcb{\ref{subsec:unitarity}}, free massless scalar fields are primary of scaling dimension $\Delta_\phi = \frac{d}{2}-1$.

Correlation functions of operators in the theory can be defined by the path integral
\begin{equation}
    \langle {\cal O}_1\left(x_1\right) ... {\cal O}_n\left(x_n\right) \rangle = \int d[\phi]\, {\cal O}_1\left(x_1\right) ... {\cal O}_n\left(x_n\right) e^{-S[\phi]},\label{pi}
\end{equation}
where the measure is normalised such that $\langle \mathds{1} \rangle = 1 $. In particular, by evaluating the Gaussian integral the two-point function of the fundamental scalar is given by 
\begin{equation}
    \langle \phi^a\left(x_1\right)\phi^b\left(x_2\right) \rangle = \frac{1}{\left(d-2\right)\text{Vol}\left(S^d\right)} \frac{\delta^{a b}}{|x_{12}|^{d-2}}, \quad \text{Vol}\left(S^d\right) = \frac{2\pi^{\frac{d}{2}}}{\Gamma\left(\frac{d}{2}\right)}.\label{fund2pt}
\end{equation}
It is conventional to re-define $\phi \rightarrow \sqrt{\left(d-2\right)\text{Vol}\left(S^d\right)}$, to obtain a canonical unit normalisation
\begin{equation}
    \langle \phi^a\left(x_1\right)\phi^b\left(x_2\right) \rangle = \frac{\delta^{a b}}{|x_{12}|^{d-2}}.
\end{equation}

\noindent
Since the theory is free, we may determine all correlation functions of operators in the theory by expressing them in terms of the two-point function \eqref{fund2pt} through Wick's theorem.

As explained in \S \tcb{\ref{adscft::hsicft}}, from the correlation functions of operators in the singlet sector, we can study the possible interactions in the dual minimal bosonic higher-spin theory on AdS. For this purpose, in the following sections we determine the explicit forms of the appropriate 2-, 3- and 4-pt functions. We work in Euclidean signature throughout.

\subsection{Singlet sector}

By definition, operators in the singlet sector are invariants under the $O\left(N\right)$ symmetry group. Operators in this sector have no free $O\left(N\right)$ indices, and are thus composite in the elementary scalar $\phi^a$. We distinguish between \emph{single-trace} operators, consisting of a single contraction of the ${\cal O}\left(N\right)$ indices, and \emph{multi-trace} operators, consisting of two or more contractions. As discussed in \S \tcb{\ref{subsec::fom}}, these are dual to single- and multi- particle states in anti-de Sitter space, respectively. 

\subsubsection{Single-trace operators}

The spectrum of single-trace operators in the vector model are bi-linears in the elementary scalar $\phi^a$. There is just a single scalar primary operator  
\begin{equation}
    {\cal O} = \frac{1}{\sqrt{N}}: \phi^a \phi^a:\,, \qquad \Delta = 2\Delta_{\phi} = d-2, \label{sts}
\end{equation}
whose scaling dimension is simply twice the scaling dimension of the elementary scalar.\footnote{Recall that in a free theory, scaling dimensions are additive.} The rest of the single-trace spectrum comprises of an infinite tower of even spin conserved currents of the schematic form
\begin{align} \label{cc}
    {\cal J}_{i_1 ... i_s} & \:\: \sim \:\: \frac{1}{\sqrt{N}} : \phi^a \partial_{i_1} ... \partial_{i_s} \phi^a :\;+\;...\,, \quad s\, \in \, 2 \mathbb{N}.
\end{align}
The $\ldots$ denote further symmetric singlet bi-linear structures, which ensure that the operator is primary and whose form we determine explicitly below. Recall (\S \tcb{\ref{subsec:unitarity}}) that spin-$s$ conserved operators are primary with scaling dimension $\Delta_{J_s} = s+d-2 = s + \Delta$, which can be verified in the above by counting derivatives.

To compute correlation functions of the conserved currents \eqref{cc}, it is convenient to employ the index-free notation 
\begin{equation}
    {\cal J}_s\left(x|z\right) = {\cal J}_{i_1 ... i_s}\left(x\right)z^{i_1} ... z^{i_s}, \qquad z^2 = 0. \label{jz}
\end{equation}
In this way, they can be packaged in the compact form
\begin{equation}
   {\cal J}_s\left(x|z\right) = \frac{1}{\sqrt{N}}\; f^{(s)}(z\cdot\partial_{y_1},z\cdot\partial_{y_2}):\phi^a(y_1)\phi^a(y_2):\big|_{y_1,y_2 \rightarrow x},\label{jf}
\end{equation}
where the function $f^{(s)}\left(x,y\right)$ is given in terms of a Gegenbauer polynomial \cite{Craigie:1983fb},
\begin{equation}
    f^{(s)}\left(x,y\right) = \left(x+y\right)^s\,C^{\left(\Delta_{\phi}-\frac{1}{2}\right)}_s\left(\frac{x-y}{x+y}\right)\,.\label{ccf}
\end{equation}
Recall that $\Delta_{\phi}$ is the scaling dimension of $\phi^a$.

This way of representing the currents can be derived by demanding that the expression \eqref{jf} is annihilated by the conformal boost generator $K_i$, which gives rise to the differential equation,\footnote{This equation can be generalised to deal with scalar single-trace operators built out of constituents with different dimensions $\Delta_1/2$ and $\Delta_2/2$. The corresponding primary is in this case a Jacobi polynomial $f(x,y)=(x+y)^sP_s^{(\Delta_2/2-1,\Delta_1/2-1)}(\tfrac{x-y}{x+y})$.} 
\begin{equation}
\left[(\Delta_{\phi}+x\,\pl_x)\pl_x +(\Delta_{\phi}+y\,\pl_y)\pl_y\right]f^{(s)}(x,y)=0,
\end{equation}
whose solution is expressed in terms of the Gegenbauer polynomials above.

\subsubsection{Double-trace operators} 

There are also operators in the singlet sector composed of higher traces, which are dual to multi-particle states in the gravity theory on AdS. At the level of four-point functions studied in this work, we need only consider double-trace operators. 
In particular, in studying the conformal block expansion of the four-point function $\langle {\cal O}{\cal O}{\cal O}{\cal O} \rangle$, we will encounter contributions from double-trace operators of the schematic form
\begin{align}
    \left[{\cal O}{\cal O}\right]_{n,s} \: \: \sim \: \: \frac{1}{N} :\phi^a \phi^a: \, \lrpar_{i_1} \dots \lrpar_{i_s} \left(\lrpar \cdot \lrpar \,\right)^n :\phi^b \phi^b:\,,\label{dto}
\end{align}
where $s$ is the spin, and $\Delta_{n,s} = 2\Delta+2n+s$ the scaling dimension. The primary condition \eqref{pri} fixes their precise form, which we determine for all $s$ and $n$ in \S \tcb{\ref{appendix::doubletrace}}.

\section{Singlet sector correlation functions}
\label{sec::veccor}
We now consider correlation functions of operators in the singlet sector introduced above. Since the theory is free, we simply apply Wick's theorem and express them in terms of two-point functions of the fundamental scalar \eqref{fund2pt}.

Our computations are drastically simplified by using the Schwinger-parametrised form of the fundamental scalar two-point function
\begin{equation}
    \langle \phi^a\left(x_1\right)\phi^b\left(x_2\right) \rangle = \frac{\delta^{ab}}{\Gamma\left(\frac{\Delta}{2}\right)}\int^{\infty}_0 \frac{dt}{t}t^{\frac{\Delta}{2}} e^{-t\, x^2_{12}},\label{schwing}
\end{equation}
which allows for a seamless application of Wick's theorem.

Conformal symmetry fixes the form of 2- and 3-pt functions up a set of theory dependent coefficients (\S \tcb{\ref{sec::cfpo}}). The goal of the following sections is to determine them for the free scalar ${\cal O}\left(N\right)$ vector model in $d$-dimensions.

\subsection{2-point functions}
\label{sec::vec2pt}
As we reviewed in \S \tcb{\ref{subsec::cfpo2}}, conformal invariance determines two-point functions up a single overall coefficient, with those of spin-$s$ conserved currents \eqref{jf} taking the form
\begin{equation}
    \langle {\cal J}_{s}\left(x_1|z_1\right) {\cal J}_{s}\left(x_2|z_2\right)  \rangle = {\sf C}_{{\cal J}_s} \frac{({\sf H}_3)^s}{\left(x^2_{12}\right)^{\Delta}}\,. \label{js2pt}
\end{equation}
The purpose of this section is to determine the coefficient ${\sf C}_{{\cal J}_s}$, which essentially depends on the choice of normalisation of the currents \eqref{jf}.

Employing the generating function \eqref{ccf} for the conserved current, applying Wick's theorem we have 
\begin{align}
 &\langle {\cal J}_s\left(x_1|z_1\right){\cal J}_s\left(x_2|z_2\right) \rangle = \frac{1}{N}\; f^{(s)}(z_1\cdot\partial_{y_1},z_1\cdot\partial_{y_2}) f^{(s)}(z_2\cdot\partial_{{\bar y}_1},z_2\cdot\partial_{{\bar y}_2}) \\ \nonumber
& \hspace*{0.1cm}  \times \left[ \langle \phi^{a}(y_1)\phi^{a}(y_2)\rangle \langle \phi^{b}({\bar y}_1)\phi^{b}({\bar y}_2)\rangle  \,+\, \phi^{a}(y_2) \leftrightarrow  \phi^{b}({\bar y}_1) \,+\, \phi^{a}(y_2) \leftrightarrow  \phi^{b}({\bar y}_2) \right]\big|_{y_1,y_2 \rightarrow x_1; {\bar y}_1,{\bar y}_2 \rightarrow x_2 }\,.
\end{align}
To extract the overall two-point coefficient, due to  conformal invariance it is sufficient to restrict attention to terms with zero contractions between the null auxiliary vectors. We thus set to zero $z_1\cdot z_2$, and match with the corresponding term in \eqref{js2pt}. This is straightforward using the Schwinger-parameterised form of the fundamental scalar two-point function \eqref{schwing}, through which we obtain
\begin{align}
    {\sf C}_{{\cal J}_s} & = \nonumber \left[\frac{1+(-1)^s}{2}\right]\frac{2^{s+1}}{\Gamma(\frac{\Delta}{2})^2}\int^{\infty}_0 \frac{dt_1}{t_1} \frac{dt_2}{t_2}\,t_1^{\frac{\Delta}{2}}t_2^{\frac{\Delta}{2}}\,(t_1+t_2)^{2s} C_s^{\left(\Delta_{\phi}-\frac{1}{2}\right)}(\tfrac{t_1 - t_2}{t_1+t_2})C_s^{\left(\Delta_{\phi}-\frac{1}{2}\right)}(\tfrac{t_1 - t_2}{t_1+t_2})e^{-(t_1 +t_2)} \\ \nonumber
    & \overset{\text{ch. of var. \eqref{cov11}}} = \left[\frac{1+(-1)^s}{2}\right] \frac{2^{s+2-\Delta}}{\Gamma(\frac{\Delta}{2})^2} \int^{\infty}_0 \frac{dq}{q} q^{2s+\Delta}e^{-q} \int_{-1}^1 dp\,(1-p^2)^{\frac{\Delta}{2}-1} C_s^{\left(\Delta_{\phi}-\frac{1}{2}\right)}\left(p\right)C_s^{\left(\Delta_{\phi}-\frac{1}{2}\right)}\left(p\right)\\ \nonumber
    & \overset{\text{int. rep. of } \Gamma} = \left[\frac{1+(-1)^s}{2}\right] \frac{2^{s+2-\Delta}}{\Gamma(\frac{\Delta}{2})^2} \Gamma\left(\Delta+2s\right) \int_{-1}^1 dp\,(1-p^2)^{\frac{\Delta}{2}-1} C_s^{\left(\Delta_{\phi}-\frac{1}{2}\right)}\left(p\right)C_s^{\left(\Delta_{\phi}-\frac{1}{2}\right)}\left(p\right) \\
    & \overset{\text{Gegenbauer ortho.}} = \left[\tfrac{1+\left(-1\right)^s}{2}\right]2^{s+1}\frac{(\Delta-1)_s (\Delta-1)_{2s}}{\Gamma\left(s+1\right)}
\end{align}
where in the second equality we used the change of variables
\begin{equation}
    q = t_1 + t_2, \qquad p = \frac{t_1-t_2}{t_1+t_2},\label{cov11}
\end{equation}
in the third the integral representation of the Gamma function and the orthogonality relation for Gegenbauer polynomials in the fourth.

In the same way, we can determine the coefficient of the two-point functions of the double-trace operators \eqref{dto}
\begin{align}\label{dt2ptss}
 \langle \left[{\cal O}{\cal O}\right]_{n,s}\left(x_1|z_1\right)\left[{\cal O}{\cal O}\right]_{n,s}\left(x_2|z_2\right) \rangle = {\sf C}_{\left[{\cal O}{\cal O}\right]_{n,s}} \frac{\left({\sf H}_3\right)^s}{\left(x^2_{12}\right)^{2\Delta+2n}}.
\end{align}
Using the explicit form \eqref{ansatz} of the operators $\left[{\cal O}{\cal O}\right]_{n,s}$, together with the Schwinger parameterisation \eqref{schwing}, we find
\begin{align}
& {\sf C}_{\left[{\cal O}{\cal O}\right]_{n,s}} = \left(1+\left(-1\right)^n\frac{4}{N} \frac{\Gamma\left(s\right)}{2^{s}\Gamma\left(\frac{s}{2}\right)} \frac{\left(\frac{d}{2}-1\right)_{n+\tfrac{s}{2}}}{\left(\frac{d-1}{2}\right)_{\tfrac{s}{2}}\left(d-2\right)_{n+\tfrac{s}{2}}}\right) \\ \nonumber
& \hspace*{1.5cm} \times \frac{(-1)^{n} 4^{n+s} \Gamma (n+1) (-d-2 n+4)_n (d-2)_{n+s}^2 \left(\frac{d}{2}+s\right)_n \left(\frac{3 d}{2}+n+s-4\right)_{\frac{d}{2}+n+s-1}}{\Gamma (s+1) \left(\frac{d}{2}-1\right)_n^2 \left(\frac{3 d}{2}+2 n+s-4\right)_{\frac{d}{2}-1}},
\end{align}
where we inserted $\Delta = d-2$.

\subsection{3-point functions of single-trace operators}
\label{subsec::3ptcc}
\subsubsection{Warm-up: s-0-0}
As a warm up, we consider the simplest case of two scalar single trace operators ${\cal O}$ and a spin-$s$ conserved current. In \S \tcb{\ref{00sgen}}, we saw that conformal symmetry fixes its form up to an overall factor
\begin{align}
    \langle {\cal J}_s\left(x_1|z_1\right){\cal O}\left(x_2\right){\cal O}\left(x_3\right)\rangle = {\sf C}_{{\cal J}_s {\cal O}{\cal O}} \frac{1}
{\left(x^2_{12}\right)^{\frac{\Delta}{2}}\left(x^2_{13}\right)^{\frac{\Delta}{2}}\left(x^2_{23}\right)^{\frac{\Delta}{2}}}{\sf Y}_1^s,\label{vec3pt}
\end{align}
which we determine in the following. 

Applying the same tools we used for the two-point functions in the previous section, we have
\begin{align}\label{joo}
 &   \langle {\cal J}_s\left(x_1|z_1\right){\cal O}\left(x_2\right){\cal O}\left(x_3\right)\rangle \\ \nonumber
    & \hspace*{0.1cm} = \frac{4}{(\sqrt{N})^3}\, f\left(z_1 \cdot \partial_{y_1},z_1 \cdot \partial_{y_2}\right)\left[ \langle \phi^a\left(y_1\right)\phi^b\left(x_2\right) \rangle \langle \phi^a\left(y_2\right)\phi^c\left(x_3\right) \rangle\langle \phi^b\left(x_2\right)\phi^c\left(x_3\right) \rangle + \: y_1 \leftrightarrow y_2\right]\big|_{y_1,y_2 \rightarrow x_1} \\ \nonumber
    & \hspace*{.3cm} = \frac{1}{\sqrt{N}} \frac{8  }{\Gamma\left(\frac{\Delta}{2}\right)^3}\left[\frac{1+(-1)^s}{2}\right]\left(z_1 \cdot \partial_{y_1}+z_1 \cdot \partial_{y_2}\right)^sC^{\left(\Delta_{\phi}-\frac{d}{2}\right)}_s\left(\frac{z_1 \cdot \partial_{y_1}-z_1 \cdot \partial_{y_2}}{z_1 \cdot \partial_{y_1}+z_1 \cdot \partial_{y_2}}\right) \\ \nonumber
    & \hspace*{3.5cm} \times \int^{\infty}_0 \left(\prod\limits^3_{i=1} \frac{dt_i}{t_i}t^{\frac{\Delta}{2}}_i\right)\, e^{-t_1\,\left(y_1-x_2\right)^2-t_2\,\left(y_2-x_3\right)^2-t_3\,\left(x_2-x_3\right)^2}\big|_{y_1,y_2 \rightarrow x_1; {\bar y}_1,{\bar y}_2 \rightarrow x_2 }.
\end{align}
To extract ${\sf C}_{{\cal J}_s {\cal O}{\cal O}}$, as with the two-point function in the previous section, by conformal symmetry we need only identify the coefficients of  $\left(x_{12} \cdot z_1\right)^s$ in \eqref{joo} and \eqref{vec3pt}. This leads to
\begin{align} \nonumber
    {\sf C}_{{\cal J}_s {\cal O}{\cal O}} & = \frac{1}{\sqrt{N}} \frac{2^{s+2}}{\Gamma\left(\frac{\Delta}{2}\right)^3} \left[\frac{1+(-1)^s}{2}\right] \int^{\infty}_0 \left(\prod\limits^3_{i=1} \frac{dt_i}{t_i}t^{\frac{\Delta}{2}}_i\right) t^s_1\, e^{-t_1-t_2-t_3}  \\
    & = \frac{8}{\sqrt{N}} \left[\frac{1+(-1)^s}{2}\right] \frac{2^s \left(\frac{\Delta}{2}\right)_s  (\Delta-1)_s}{\Gamma (s+1)}.
\end{align}
Likewise, we can also determine the analogous three-point function for a spin-$s$ double-trace operator for general $n$,
\begin{align} \label{dt00s}
 \langle \left[{\cal O}{\cal O}\right]_{n,s}\left(x_1|z_1\right){\cal O}\left(x_2\right){\cal O}\left(x_3\right) \rangle = {\sf C}_{\left[{\cal O}{\cal O}\right]_{n,s} {\cal O}{\cal O}} \frac{\left(x^2_{23}\right)^n}
{\left(x^2_{12}\right)^{\Delta+n}\left(x^2_{13}\right)^{\Delta+n}}{\sf Y}_1^s,
\end{align}
with
\begin{align}
 {\sf C}_{\left[{\cal O}{\cal O}\right]_{n,s} {\cal O}{\cal O}} = (-1)^{n+s} 2^{n+s} \frac{\left(d-2\right)^2_{n+s}}{\Gamma\left(s+1\right)} \left(1+\left(-1\right)^n\frac{4}{N} \frac{\Gamma\left(s\right)}{2^{s}\Gamma\left(\frac{s}{2}\right)} \frac{\left(\frac{d}{2}-1\right)_{n+\tfrac{s}{2}}}{\left(\frac{d-1}{2}\right)_{\tfrac{s}{2}}\left(d-2\right)_{n+\tfrac{s}{2}}}\right).
\end{align}

\subsubsection{General case: s1-s2-s3}
With the approach clear, we now tackle the more involved general case of a three-point function involving currents \eqref{jf} for a generic triplet of spin $\left\{s_1, s_2, s_3\right\}$. As we saw in \S \tcb{\ref{subsec::cfpo3}}, the most general three-point function takes the form 
\begin{multline}\label{corr}
\langle {\cal J}_{s_1}(x_1|z_1){\cal J}_{s_2}(x_2|z_2){\cal J}_{s_3}(x_3|z_3)\rangle\\=\sum_{n_i}{\sf C}^{n_1,n_2,n_3}_{s_1,s_2,s_3}\frac{{\sf Y}_1^{s_1-n_2-n_3}{\sf Y}_2^{s_2-n_3-n_1}{\sf Y}_3^{s_3-n_1-n_2}{\sf H}_1^{n_1}{\sf H}_2^{n_2}{\sf H}_3^{n_3}}{(x_{12}^2)^{\tfrac{\tau_1+\tau_2-\tau_3}{2}}(x_{23}^2)^{\tfrac{\tau_2+\tau_3-\tau_1}{2}}(x_{31}^2)^{\tfrac{\tau_3+\tau_1-\tau_2}{2}}}\,,
\end{multline}
built from the six basic conformal structures \eqref{6confInt}. Each individual term is independently invariant under conformal transformations, and thus conformal symmetry does not determine the coefficients ${\sf C}^{n_1,n_2,n_3}_{s_1,s_2,s_3}$.

The requirement that the currents are conserved relates the coefficients ${\sf C}^{n_1,n_2,n_3}_{s_1,s_2,s_3}$ amongst each other, reducing the number of independent forms. Conservation constraints on the structure of three-point functions involving conserved vector operators and the stress-energy tensor where first studied by Osborn and Petkou in \cite{Osborn:1993cr}. This was extended to a generic triplet of spins $\left\{s_1, s_2, s_3\right\}$ in \cite{Costa:2011mg,Stanev:2012nq,Zhiboedov:2012bm}. In \cite{Zhiboedov:2012bm}, the general structure required by conservation in $d>3$ was found to be
\begin{align} \nonumber
  &  \langle {\cal J}_{s_1}(x_1|z_1){\cal J}_{s_2}(x_2|z_2){\cal J}_{s_3}(x_3|z_3)\rangle  = \sum\limits^{\frac{1+\text{min}\left(s_1, s_2, s_3\right)}{2}}_{l=0}{\sf c}_{l}\, {}_2F_{1}\left(\frac{1}{2}-l,-l,3-\frac{d}{2}-2l,-\frac{1}{2}\frac{{\sf \Lambda}}{{\sf H}^2_1{\sf H}^2_2{\sf H}^2_3}\right)  \\ \label{gencc3pt}
   & \hspace*{4cm} \times \frac{e^{{\sf Y}_1 + {\sf Y}_2 + {\sf Y}_3} {}_0 F_{1}(\Delta,-\frac{1}{2}{\sf H}_{1}){}_0 F_{1}(\Delta,-\frac{1}{2}{\sf H}_{2}){}_0 F_{1}(\Delta,-\frac{1}{2}{\sf H}_{3})}{(x_{12}^2)^{\tfrac{\tau_1+\tau_2-\tau_3}{2}}(x_{23}^2)^{\tfrac{\tau_2+\tau_3-\tau_1}{2}}(x_{31}^2)^{\tfrac{\tau_3+\tau_1-\tau_2}{2}}} {\sf \Lambda}^{2l},
\end{align}
with
\begin{align}
    {\sf \Lambda} = {\sf Y}_1{\sf Y}_2{\sf Y}_3 + \frac{1}{2}\left[{\sf Y}_1{\sf H}_1+{\sf Y}_2{\sf H}_2 + {\sf Y}_3{\sf H}_3\right],
\end{align}
and where $l$ takes both integer and half integer values. The coefficients $c_{l}$ are left unfixed by current conservation, and depend on the theory. In particular, for a generic triplet of spins there are $ 1+\text{min}\left(s_1, s_2, s_3\right)$ independent structures, which was first counted in \cite{Osborn:1993cr,Metsaev:2005ar} (see also \cite{Manvelyan:2010jr,Sagnotti:2010at,Fotopoulos:2010ay}). In the following we determine the coefficients $c_{l}$ for the $d$-dimensional free scalar $O\left(N\right)$ vector model.\footnote{These results extend to 
general dimensions the results \cite{Giombi:2011rz,Colombo:2012jx,Didenko:2012tv} for the same correlators in three-dimensions. See also \cite{Erdmenger:1996yc} for an earlier result for the stress-tensor three-point function in a free scalar theory in general dimensions.} As we shall see, our results confirm a prediction made in \cite{Zhiboedov:2012bm}, that the free scalar correlator corresponds to the $l = 0$ structure in the above.

As with the two-point functions in \S \tcb{\ref{sec::vec2pt}}, we can focus on terms with no contractions among the auxiliary vectors. In the calculations we may then set $z_i \cdot z_j =0$, giving
\begin{align}
&\langle {\cal J}_{s_1}(x_1|z_1){\cal J}_{s_2}(x_2|z_2){\cal J}_{s_3}(x_3|z_3)\rangle  \\ \nonumber
    &  = \frac{8N}{\Gamma\left(\frac{\Delta}{2}\right)^3} \int^{\infty}_0 \left(\prod\limits^3_{i=1} \frac{dt_i}{t_i}
    t^{\frac{\Delta}{2}}_i\right) f^{(s_1)}\left(-2t_3 z_1 \cdot x_{12},-2t_2 z_1 \cdot x_{13}\right)f^{(s_2)}\left(-2t_3 z_2 \cdot x_{21},-2t_1 z_2 \cdot x_{23}\right) \\ \nonumber
    & \hspace*{6.5cm} \times f^{(s_3)}\left(-2t_1 z_3 \cdot x_{32},-2t_2 z_3 \cdot x_{31}\right) e^{-t_1x^2_{23}-t_2 x^2_{31}-t_3 x^2_{12}}\\ \nonumber
    & = \sum_{n_i=0}^{s_i} {\sf D}^{n_1,n_2,n_3}_{s_1,s_2,s_3} \frac{(z_1\cdot x_{12})^{s_1-n_1}(z_2\cdot x_{21})^{s_2-n_2}}{(x_{12}^2)^{\tfrac{\Delta}{2}+s_1+s_2-n_1-n_2}}\frac{(z_1\cdot x_{13})^{n_1}(z_3\cdot x_{31})^{n_3}}{(x_{31}^2)^{\tfrac{\Delta}{2}+n_1+n_3}}\frac{(z_2\cdot x_{23})^{n_2}(z_3\cdot x_{32})^{s_3-n_3}}{(x_{23}^2)^{\tfrac{\Delta}{2}+s_3+n_2-n_3}},
\end{align}
where
{\footnotesize
\begin{align}
    & {\sf D}^{n_1,n_2,n_3}_{s_1,s_2,s_3} = \frac{N (-1)^{n_1 +n_2 +n_3 }2^{3+s_1+s_2+s_3}}{s_1!s_2!s_3!}\binom{s_1}{n_1}\binom{s_2}{n_2}\binom{s_3}{n_3} (\Delta-1)_{s_1}(\Delta-1)_{s_2}(\Delta-1)_{s_3}\\
    &\times \frac{ \Gamma \left(s_1 +\frac{\Delta }{2}\right) \Gamma \left(s_2 +\frac{\Delta }{2}\right) \Gamma \left(s_3 +\frac{\Delta }{2}\right)}{\Gamma \left(n_1 +\frac{\Delta }{2}\right) \Gamma \left(n_2 +\frac{\Delta }{2}\right)\Gamma \left(n_3 +\frac{\Delta }{2}\right) }\frac{\Gamma \left(n_1 +n_3 +\frac{\Delta }{2}\right) \Gamma \left(n_2 -n_3 +s_3 +\frac{\Delta }{2}\right) \Gamma \left(-n_1 -n_2 +s_1 +s_2 +\frac{\Delta }{2}\right)}{\Gamma \left(-n_1 +s_1 +\frac{\Delta }{2}\right) \Gamma \left(-n_2 +s_2 +\frac{\Delta }{2}\right) \Gamma \left(-n_3 +s_3 +\frac{\Delta }{2}\right)}\,.\nonumber
\end{align}}
\noindent
By matching with the corresponding expansion of the general form for the correlator \eqref{corr}, we can write down the following recursion relation for the coefficients ${\sf C}^{n_1,n_2,n_3}_{s_1,s_2,s_3}$ 
\begin{align}
{\sf C}_{s_1,s_2,s_3}^{n_1,n_2,n_3}&=2^{-n_1-n_2-n_3}\Big[{\sf D}_{s_1,s_2,s_3}^{n_2,s_2-n_3,s_3-n_1}\\
&\hspace*{1cm}-\sum_{k_1+k_2+k_3=1}^{n_1+n_2+n_3}(-1)^{k_1 +k_2 +k_3 } \binom{k_1 +k_2 -n_1 -n_2 +s_3 }{k_1 } \binom{k_1 +k_3 -n_1 -n_3 +s_2 }{k_3 }\nonumber\\
& \hspace{70pt}\times\binom{k_2 +k_3 -n_2 -n_3 +s_1 }{k_2 } 2^{-k_1 -k_2 -k_3 +n_1 +n_2 +n_3 } {\sf C}_{s_1,s_2,s_3}^{(n_1 -k_1 ,n_2 -k_2 ,n_3 -k_3 )}\Big]\,,\nonumber
\end{align}
where the summation assumes $k_i\leq n_i$. Solving for the explicit form of the coefficients, we find:
\begin{multline}
{\sf C}_{s_1,s_2,s_3}^{n_1,n_2,n_3}=-\frac{N(-1)^{n_1 +n_2 +n_3 } 2^{s_1 +s_2 +s_3-(n_1 +n_2 +n_3 ) +3}}{n_1 ! n_2 ! n_3 !(s_1-n_2-n_3)!(s_2-n_3-n_1)!(s_3-n_1-n_2)!}\\\times\frac{\Gamma \left(s_1 +\frac{\Delta }{2}\right)  \Gamma \left(s_2 +\frac{\Delta }{2}\right)  \Gamma \left(s_3 +\frac{\Delta }{2}\right)}{\Gamma \left(n_1 +\frac{\Delta }{2}\right) \Gamma \left(n_2 +\frac{\Delta }{2}\right) \Gamma \left(n_3 +\frac{\Delta }{2}\right)}\,(\Delta-1)_{s_1}(\Delta-1)_{s_2}(\Delta-1)_{s_3}.
\end{multline}
A nice observation is that the $n_i$ dependence can be re-summed in terms of a Bessel function, giving the following compact form for the correlation function
\begin{multline} \label{resvec123}
\langle {\cal J}_{s_1}(x_1|z_1){\cal J}_{s_2}(x_2|z_2){\cal J}_{s_3}(x_3|z_3)\rangle\\= N \left(\prod_{i=1}^3{\sf c}_{s_i}\,q_i^{\frac{1}{2}-\frac{\Delta }{4}}\Gamma(\tfrac{\Delta}2)\,
J_{\frac{\Delta -2}{2}}\left(\sqrt{q_i}\right) \right)
\,\frac{{\sf Y}_1^{s_1}{\sf Y}_2^{s_2}{\sf Y}_3^{s_3}}{(x_{12}^2)^{\Delta/2}(x_{23}^2)^{\Delta/2}(x_{31}^2)^{\Delta/2}}\,,
\end{multline}
where we introduced
\begin{align}
q_1&=2{\sf H}_1\pl_{{\sf Y}_2}\pl_{{\sf Y}_3}\,,&q_2&=2{\sf H}_2\pl_{{\sf Y}_3}\pl_{{\sf Y}_1}\,,&q_3&=2{\sf H}_3\pl_{{\sf Y}_1}\pl_{{\sf Y}_2}.
\end{align}
Normalising the two-point functions \eqref{js2pt} canonically,\footnote{I.e. by redefining ${\cal J}_{s_i} \rightarrow \frac{1}{\sqrt{{\sf C}_{{\cal J}_{s_i}}}} {\cal J}_{s_i}$.} the ${\sf c}_{s_i}$ are given by 
\begin{align}
 {\sf c}_{s_i}^2&= \frac{\sqrt{\pi}\,2^{-\Delta-s_i+3}\,\Gamma(s_i+\tfrac{\Delta}{2})\Gamma(s_i+\Delta-1)}{N\,s_i!\,\Gamma(s_i+\tfrac{\Delta-1}{2})\Gamma(\tfrac{\Delta}2)^2}\,.\label{g}
\end{align}
 A nice check of this result is that it is consistent with that established for the $s$-$0$-$0$ correlator in the previous section, which was already available in the literature \cite{Diaz:2006nm}. Furthermore, using the identity
 \begin{equation}
  \Gamma\left(\alpha+1\right) x^{-\alpha} J_{\alpha}\left(2 x\right) = 2^{-\alpha} {}_{0} F_{1}\left(\alpha+1;-\frac{x^2}{4}\right),
 \end{equation}
 our result confirms the conjecture made in \cite{Zhiboedov:2012bm}, that the free scalar correlator corresponds to the $l=0$ term in \eqref{gencc3pt}.

With the results \eqref{js2pt} and \eqref{resvec123} for the two- and three-point functions of higher-spin conserved currents, in \S \tcb{\ref{chapt::hr}} we are able to fix the cubic action of the dual type A minimal bosonic higher-spin theory on AdS$_{d+1}$, using the conjectured holographic higher-spin / vector model duality. Furthermore, together with the double-trace operator 3- and 2-point functions \eqref{dt00s} and \eqref{dt2ptss}, as discussed in \S \tcb{\ref{sec::cbd}} these expressions also determine the conformal block expansion of the scalar single-trace operator four-point function, to which we now turn our attention. The latter is a key result for later applications (\S \tcb{\ref{chapt::hr}}), as it will allow us to determine the on-shell quartic self interaction of the scalar in the dual higher-spin theory.

\subsection{Scalar 4-point function}
\label{subsec::4ptscalar}
\begin{figure}[h]
  \centering
  \includegraphics[scale=0.5]{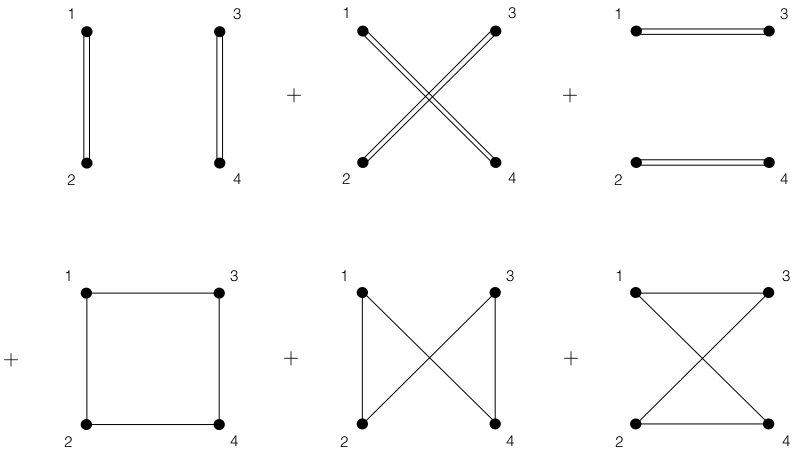}
  \caption{Contributions to the full four-point function \eqref{vec4pt2} of the scalar single-trace operator ${\cal O}$. The first line constitutes the three disconnected terms, while the second
line comprises the connected part of the correlator.} \label{fig::wick}
\end{figure}

For the remainder of this chapter we consider the four-point function of the scalar single-trace operator \cite{Dolan:2000ut},
\begin{align}
 \langle \mathcal{O}\left(x_1\right) \mathcal{O}\left(x_2\right) \mathcal{O}\left(x_3\right) \mathcal{O}\left(x_4\right) \rangle  & = \frac{g_{{\cal O}}\left(u,v\right)}{\left(x^2_{12} x^{2}_{34}\right)^{\Delta}},\label{vec4pt2}
\end{align}
which has disconnected and connected parts,
\begin{equation}
    g_{{\cal O}}\left(u,v\right) =  g_{\text{disc.}}\left(u,v\right)+g_{\text{conn.}}\left(u,v\right),
\end{equation}
\begin{align}
g_{\text{disc.}}\left(u,v\right) =  1 +  u^{\Delta} + \left(\frac{u}{v}\right)^{\Delta}, \quad g_{\text{conn.}}\left(u,v\right) = \frac{4}{N} \left( u^{\frac{\Delta}{2}}+\left(\frac{u}{v}\right)^{\frac{\Delta}{2}}+u^{\frac{\Delta}{2}}\left(\frac{u}{v}\right)^{\frac{\Delta}{2}}\right), \label{cr}
\end{align}
where we used canonical normalisation for the $\langle {\cal O} {\cal O} \rangle$ two-point function. As usual, we can obtain the above expressions by simply Wick contracting. The individual contributions can be understood pictorially in figure \ref{fig::wick}.\\

\subsubsection{OPE and conformal block expansion}
For later application, we require the conformal block expansion of the four-point function \eqref{vec4pt2}. As explained in \S \tcb{\ref{subsec:fptcft}}, although functions \eqref{cr} of the cross ratios are invariant under conformal transformations, those in four-point functions they are completely determined by the ${\cal O}{\cal O}$ OPE. Schematically, the latter takes the form\footnote{That there are no contributions from higher-trace operators is clear from the triviality of their three-point functions with two insertions of ${\cal O}$. This can be seen immediately by applying Wick's theorem.}
\begin{equation}
    {\cal O}{\cal O} \; \sim \;  \mathds{1} \;+\; \sum\limits_{s} {\sf c}_{{\cal O}{\cal O}{\cal J}_s}\,{\cal J}_s  \;+\; \sum\limits_{s,n} {\sf c}_{{\cal O}{\cal O}\left[{\cal O}{\cal O}\right]_{s,n}}\,\left[{\cal O}{\cal O}\right]_{s,n} + \text{descendents},\label{OPE}
\end{equation}
with OPE coefficients ${\sf c}_{{\cal O}{\cal O}{\cal J}_s}$ and ${\sf c}_{{\cal O}{\cal O}\left[{\cal O}{\cal O}\right]_{s,n}}$ of the conserved currents and double-trace operators, respectively. Using the results for the two- and three-point functions established in the previous sections, together with their relationship \eqref{ope32} to the corresponding OPE coefficients, we find
\begin{align} 
    {\sf c}^2_{{\cal O}{\cal O}{\cal J}_s} & = \left[\frac{\left(-1\right)^s+1}{2}\right] \frac{1}{N} \frac{2^{s+3}\left(\frac{\Delta}{2}\right)_s}{s!\left(\Delta+s-1\right)_s}, \label{cbccst} \\  
   {\sf c}^2_{{\cal O}{\cal O}\left[{\cal O}{\cal O}\right]_{s,n}} & = \frac{\left[\frac{\left(-1\right)^s+1}{2}\right]  2^{s+1}\left(\tfrac{\Delta}{2}\right)^{2}_n \left(\Delta\right)^2_{s+n}}{s! n! \left(s+\tfrac{\Delta}{2}+1\right)_n \left(\Delta-1 + n \right)_n \left(2\Delta + 2n +s-1\right)_{s} \left(\tfrac{3\Delta}{2}-1 + n +s\right)_n}\nonumber \\  
&\hspace*{5cm} \times \left(1+\left(-1\right)^n\frac{4}{N} \frac{\Gamma\left(s\right)}{2^{s}\Gamma\left(\frac{s}{2}\right)} \frac{\left(\frac{\Delta}{2}\right)_{n+\tfrac{s}{2}}}{\left(\frac{\Delta+1}{2}\right)_{\tfrac{s}{2}}\left(\Delta\right)_{n+\tfrac{s}{2}}}\right).
\end{align}
With the above results, we can write down the conformal block decomposition of the four-point function \eqref{vec4pt2} (recall the discussion \S \tcb{\ref{sec::cbd}} on the conformal block decomposition of four-point correlators). For example, in the (12)(34) channel
\begin{align} \label{cbe1234}
 &   \langle \mathcal{O}\left(x_1\right) \mathcal{O}\left(x_2\right) \mathcal{O}\left(x_3\right) \mathcal{O}\left(x_4\right) \rangle \\ \nonumber
    & \hspace*{2.25cm} = \frac{1}{\left(x^2_{12} x^{2}_{34}\right)^{\Delta}}\left(1+\sum_s {\sf c}^2_{{\cal O}{\cal O}{\cal J}_s}\,G_{\Delta+s,s}\left(u,v\right)+\sum_{s,\, n} {\sf c}^2_{{\cal O}{\cal O}\left[{\cal O}{\cal O}\right]_{s,n}}\,G_{2\Delta+2n+s,s}\left(u,v\right) \right), 
\end{align}
where the first term on the RHS is the contribution from the identity operator, the second from the conformal multiplet of each conserved currents \eqref{jf} and the third from the muliplets of each of the double-trace operators \eqref{dto}. 

To compare with the corresponding  calculations of Witten diagrams in the dual higher-spin theory in \S \tcb{\ref{chapt::witten}}, it is useful to express the expansion \eqref{cbe1234} of the correlator in a crossing-symmetric form. To this end, it is instrumental to understand the microscopic interpretation of each term in the connected part of the correlator \eqref{vec4pt2}. For this is it sufficient to recall that in, say, the limit $u \rightarrow 0$ operators of twist $\tau$ enter as $u^{\frac{\tau}{2}}$. We thus have
\begin{align}
 \frac{4}{N}\left(u^{\frac{\Delta}{2}}+ \left(\frac{u}{v}\right)^{\frac{\Delta}{2}}\right)  = \sum_s {\sf c}^2_{{\cal O}{\cal O}{\cal J}_s}\,G_{\Delta+s,s}\left(u,v\right)\label{micst},
\end{align}
and
\begin{align}
       \frac{4}{N} u^{\frac{\Delta}{2}}\left(\frac{u}{v}\right)^{\frac{\Delta}{2}} = \sum_{s,\, n} {\sf c}^2_{{\cal O}{\cal O}\left[{\cal O}{\cal O}\right]_{s,n}}\,G_{2\Delta+2n+s,s}\left(u,v\right).\label{micdt}
\end{align}
With the above understanding, we can express the four-point correlator in a manifestly crossing-symmetric form by writing,
\begin{multline}\label{manicross}
 \langle \mathcal{O}\left(x_1\right) \mathcal{O}\left(x_2\right) \mathcal{O}\left(x_3\right) \mathcal{O}\left(x_4\right) \rangle_{\text{conn.}} \\  = {\cal G}\left(x_1,x_2;x_3,x_4\right)\;+\;{\cal G}\left(x_1,x_3;x_2,x_4\right)\;+\;{\cal G}\left(x_1,x_4;x_3,x_2\right)
\end{multline}
with 
{\footnotesize \begin{multline}
 {\cal G}\left(x_1,x_2;x_3,x_4\right) = \frac{1}{\left(x^2_{12} x^{2}_{34}\right)^{\Delta}}\left(a \cdot \sum_s {\sf c}^2_{{\cal O}{\cal O}{\cal J}_s}\,G_{\Delta+s,s}\left(u,v\right) + b \cdot \sum_{s,\, n} {\sf c}^2_{{\cal O}{\cal O}\left[{\cal O}{\cal O}\right]_{s,n}}\,G_{2\Delta+2n+s,s}\left(u,v\right) \right),
\end{multline}}
where we are at liberty to choose any $a$ and $b$ with $2a+b=1$ so that, via the identifications \eqref{micst} and \eqref{micdt}, the full connected correlator \eqref{vec4pt2} is recovered. This simply parametrises the freedom to add to terms which vanish under symmetrisation. In other words ${\cal G}$ with different solutions for $a$, $b$ and $c$ are related by the addition of ${\tilde {\cal G}}$ with the property that 
\begin{equation}
    {\tilde {\cal G}}\left(x_1,x_2;x_3,x_4\right)\;+\;{\tilde {\cal G}}\left(x_1,x_3;x_2,x_4\right)\;+\;{\tilde {\cal G}}\left(x_1,x_4;x_3,x_2\right) = 0
\end{equation}

\subsubsection{Contour Integral form}

In \S \tcb{\ref{subsec::quarticextr}} we employ the conformal block expansion of the scalar singlet four-point function \eqref{cbe1234} in 3$d$ to extract the quartic self-interaction of the bulk scalar in the type A minimal bosonic higher-spin theory on AdS$_4$. This is achieved by matching with the conformal block expansions of the dual Witten diagrams in AdS. To do so, it will be convenient to employ the contour integral form \eqref{contintform}
\begin{align}\label{cif}
    &\langle \mathcal{O}\left(x_1\right) \mathcal{O}\left(x_2\right) \mathcal{O}\left(x_3\right) \mathcal{O}\left(x_4\right) \rangle =   \int^{\infty}_{-\infty}d\nu\; \sum\nolimits_s g_s\left(\nu\right) W_{\tfrac{d}{2}+i\nu,s}\left(x_1,x_2;x_3,x_4\right) ,
\end{align}
which we determine in the following for $d=3$.

As explained in \S \tcb{\ref{subsec::intrep}}, the poles of the meromorphic function $g_s\left(\nu\right)$ encode the contributions of each spin-$s$ conformal multiplet. In the case of the scalar singlet four-point function \eqref{cbe1234}, it thus takes the form
\begin{align}
    g_{s}\left(\nu\right) & = g_{{\cal J}_s}\left(\nu\right) + g_{\left[{\cal O}{\cal O}\right]_s}\left(\nu\right), 
\end{align}
with $g_{{\cal J}_s}\left(\nu\right)$ generating the contribution from the spin-$s$ conserved current multiplets, and $g_{\left[{\cal O}{\cal O}\right]_s}\left(\nu\right)$ that of each of the spin-$s$ double trace operators \eqref{dto}.

More precisely, when closing the $\nu$-contour in the lower half plane as prescribed in \S \tcb{\ref{subsec::intrep}}, $g_{{\cal J}_s}\left(\nu\right)$ contains a single simple pole corresponding to the dimension of a spin-$s$ conserved current. I.e. at $\frac{d}{2}+i\nu = \Delta+s$
\begin{align}\label{cntst}
    g_{{\cal J}_s}\left(\nu\right) & = \frac{c_{{\cal O}{\cal O}{\cal J}_s}\left(\nu\right)}{\nu^2+\left(\Delta+s-\frac{d}{2}\right)^2},\\
   c_{{\cal O}{\cal O}{\cal J}_s}\left(\nu\right) & = \left[\frac{\left(-1\right)^s+1}{2}\right] \frac{ 2^{-2 i \nu +s+5} \Gamma \left(i \nu +\frac{1}{2}\right) \Gamma \left( \frac{2 s+2 i \nu +3}{4}\right)^2}{\pi ^{3/2} N  (2 i \nu +2 s+1)\Gamma (i \nu ) \Gamma \left( \frac{2 s+2 i \nu +1}{4}\right)^2}.
\end{align}
The function $c_{{\cal O}{\cal O}{\cal J}_s}\left(\nu\right)$ is chosen so that
\begin{align}
    c_{{\cal O}{\cal O}{\cal J}_s}\left(\nu\right)\Big|_{\frac{d}{2}+i \nu = \Delta+s} = \frac{i}{2\pi}  \left(2\Delta+2s-d\right){\sf c}^2_{{\cal O}{\cal O}{\cal J}_s},
\end{align}
i.e. such that the single-trace contribution in \eqref{cbccst} is recovered upon application of Cauchy's residue theorem.

Similarly, for the spin-$s$ double-trace operators, $g_{\left[{\cal O}{\cal O}\right]_s}\left(\nu\right)$ has a string of simple poles at double-trace dimensions: $\frac{d}{2}+i\nu = 2\Delta +2n+s$, $n = 0, 1, 2, ..., \infty$
\begin{align}\label{directp}
    g_{\left[{\cal O}{\cal O}\right]_s}\left(\nu\right) = c_{{\cal O}{\cal O}\left[{\cal O}{\cal O}\right]_s}\left(\nu\right)\Gamma \left(\frac{2\Delta+s-\frac{d}{2}-i \nu}{2} \right),
\end{align}
which are accounted for by the poles of the Gamma function factor, and $c_{{\cal O}{\cal O}\left[{\cal O}{\cal O}\right]_s}\left(\nu\right)$ generates the corresponding OPE coefficients at each pole:
\begin{multline} 
 c_{{\cal O}{\cal O}\left[{\cal O}{\cal O}\right]_s}\left(\nu\right) = \left[\frac{\left(-1\right)^s+1}{2}\right] \left( \frac{\Gamma \left(i \nu +\frac{1}{2}\right) 2^{-2 i \nu +3 s+2} \Gamma \left(s+\frac{3}{2}\right) \Gamma \left(\frac{2 s+2 i \nu +3}{4}\right)^2}{\pi  \Gamma (i \nu ) (2 i \nu +2 s+1) \Gamma (s+1) \Gamma \left(s-i \nu +\frac{1}{2}\right) \Gamma \left(s+i \nu +\frac{1}{2}\right)}\right.
 \\  +\frac{1}{N}\left. \frac{i^s \Gamma \left(i \nu +\frac{1}{2}\right) 2^{-2 i \nu +3 s+\frac{3}{2}} \Gamma \left(\frac{s+1}{2}\right) \Gamma \left(s+\frac{3}{2}\right) \Gamma \left(\frac{2 s+2 i \nu +3}{4}\right)^2}{\pi  s! (s/2)! \Gamma \left(\frac{3}{2}-\frac{i \nu }{2}\right) \Gamma \left(\frac{3}{2}+\frac{i \nu }{2}\right) \Gamma (i \nu ) (2 i \nu +2 s+1) \Gamma \left(s-i \nu +\frac{1}{2}\right) \Gamma \left(s+i \nu +\frac{1}{2}\right)}\right).
\end{multline}
with 
\begin{align}
    c_{{\cal O}{\cal O}\left[{\cal O}{\cal O}\right]_s}\left(\nu\right)\Big|_{\frac{d}{2}+i \nu = 2\Delta+2n+s} = \frac{i}{2\pi} n!\, (-1)^n\,{\sf c}^2_{{\cal O}{\cal O}\left[{\cal O}{\cal O}\right]_{s,n}}.
\end{align}
  
  \chapter{Witten diagrams in higher-spin theory}
\label{chapt::witten}

With this chapter we turn to the bulk side of the story. In the preceding chapter we computed correlation functions of single-trace operators in the singlet sector of the free scalar $O\left(N\right)$ vector model. In the context of holography, correlation functions of single-trace operators correspond to so-called Witten diagrams in the weak coupling regime of the dual gravity theory on AdS. In the present context, the dual theory is one of higher-spin gauge fields on AdS. With this setting in mind, in this chapter we introduce methods for computing Witten diagrams in theories containing fields of arbitrary rank. These methods are underpinned by the ambient space framework, which is reviewed in detail in \S \tcb{\ref{Aambient}}.

In particular, we demonstrate how to compute three-point Witten diagrams involving external fields of arbitrary spin and mass (and also a parity even scalar), as well as methods to establish conformal partial wave expansions of four-point exchange and contact diagrams. This is all in the view to bring the Witten diagrams in the type A minimal bosonic higher-spin theory into the appropriate form, to match their counter-parts in the free scalar $O\left(N\right)$ vector model. All computations are at tree-level.\\

\newpage

\noindent
Before we proceed, let us first settle the notation. We denote points on AdS$_{d+1}$ by $x^{\mu}$, and often work in Poincar\'e co-ordinates $x^{\mu} = \left(z,y^i\right)$,
\begin{align}
ds^2 = \frac{R^2}{z^2}\left(dz^2+dy_i dy^i\right),
\end{align}
with $y^i$ representing points on the boundary of AdS, $i = 1, 2, ...\,, d$. We set the AdS radius $R$ is set to one for all computations in this chapter. When working in ambient space, we use $X$ to represent points on the AdS manifold and $P$ for those on the conformal boundary.

\section{Witten diagrams}
Back in \S \tcb{\ref{sec::gen}}, we saw that the AdS/CFT duality can be phrased as the equivalence between the generating functional of correlators in a CFT and the full AdS partition function of its putative dual gravity theory, which in Euclidean signature reads
\begin{equation}
    \langle \exp\left(\int d^dy\, {\bar \varphi\left(y\right){\cal O}\left(y\right)}\right)\rangle_{\text{CFT}} = \int_{\varphi|_{\partial \text{AdS}} =  {\bar \varphi}} {\cal D} \varphi \exp\left(-\frac{1}{G}S_{\text{AdS}}\left[\varphi\right]\right).
\end{equation}
The bulk field $\varphi$, which is dual to the single-trace CFT operator ${\cal O}$ of scaling dimension $\Delta$ and spin-$s$, is subject to the boundary condition \eqref{sb}
\begin{equation}
   \lim_{z \rightarrow 0} \varphi\left(z,y\right)z^{\Delta-d-s} \: = \: {\bar \varphi}\left(y\right),
\end{equation}
in order to define the path integral. In particular, this implies that correlators of single-trace operators ${\cal O}_i$ can be computed by functionally differentiating the gravity partition function with respect to ${\bar \varphi}_i$
\begin{equation}
    \langle {\cal O}_1\left(y_1\right) ... {\cal O}_n\left(y_n\right) \rangle = \frac{\delta}{\delta {\bar \varphi}_1\left(y_1\right)} ... \frac{\delta}{\delta {\bar \varphi}_n\left(y_n\right)}\int_{\varphi_i|_{\partial \text{AdS}} =  {\bar \varphi}_i} {\cal D} \varphi \exp\left(-\frac{1}{G}S_{\text{AdS}}\left[\varphi_i\right]\right)\Big|_{{\bar \varphi}_i=0}. \label{corgen}
\end{equation}
When the gravity theory in AdS is weakly coupled,\footnote{In the sense of $\hbar \rightarrow 0$, not necessarily small curvature/non-stringy.} we can perform a semi-classical expansion of the correlators generated by the gravity partition functional, 
\begin{align} \label{loopexp}
& \langle {\cal O}_1\left(y_1\right) ... {\cal O}_n\left(y_n\right) \rangle \\ \nonumber
& \hspace*{3cm}= \frac{\delta}{\delta {\bar \varphi}_1\left(y_1\right)} ... \frac{\delta}{\delta {\bar \varphi}_n\left(y_n\right)} \exp\left(-\frac{1}{G}S_{\text{AdS}}\left[\varphi_i|_{\partial \text{AdS}} =  {\bar \varphi}_i\right]\right)\Big|_{{\bar \varphi}_i=0} \quad  + \quad \text{loops},
\end{align}
in which quantum loop corrections are subdominant. This corresponds to a large $N_{\text{dof.}}$ expansion in the dual CFT, and the corresponding Feynman diagrams in AdS space are known as \emph{Witten diagrams}.

As explained in \S \tcb{\ref{adscft::hsicft}}, by studying the tree-level contribution to the relation \eqref{loopexp} for non-linear higher-spin theories on AdS, we can extract the nature of their on-shell interactions by employing their conjectured equivalence with free conformal field theories. To do so, we therefore need to develop effective techniques for the systematic evaluation of tree-level Witten diagrams in theories containing fields of arbitrary rank. We first warm up by considering the simplest index-free example of a scalar field in AdS, the amplitudes of which we later show can be related to those with external higher-spin fields.

\subsection{Warm-up example: Scalar field in AdS}
\label{scalarex}
Consider the example of a scalar field $\phi$ in AdS$_{d+1}$, with classical action
\begin{equation}
    S = \frac{1}{G} \int_{\text{AdS}} \frac{1}{2} \nabla_{\mu}\phi\nabla^{\mu}\phi +\frac{1}{2!} m^2 \phi^2 + \frac{1}{3!} g \phi^3 +  \frac{1}{4!} \lambda \phi^4 + ...\,,\label{scalarads}
\end{equation}
for some overall coupling $G$. According to the field-operator map \S \tcb{\ref{subsec::fom}}, the bulk scalar $\phi$ will be dual to some single-trace scalar operator ${\cal O}$ with scaling dimension $\Delta$, which is related to the (mass)$^2$ of $\phi$ 
\begin{equation}
    \left(m R\right)^2 = \Delta\left(\Delta-d\right).
\end{equation}
In the weak coupling regime $G << 1$ we can make the saddle point approximation 
\begin{align}
 \exp\left(-F_{\text{CFT}}\left[{\bar \phi}\right]\right)
\quad  \approx \quad \exp\left(-S\left[ {\bar \phi}\right]\right),\label{spart}
\end{align}
where we neglect loop corrections, which are suppressed in $G$. Here, $F_{\text{CFT}}\left[{\bar \phi}\right]$ is the generating function of connected correlators of ${\cal O}$ with source ${\bar \phi}$, and $S\left[{\bar \phi}\right]$ is the on-shell action \eqref{scalarads} subject to the boundary condition
\begin{equation}
    \lim_{z \rightarrow 0} \phi\left(z,y\right)z^{\Delta-d} \: = \: {\bar \phi}\left(y\right) \label{bcscalar}.
\end{equation}
In the following, we demonstrate how to to evaluate correlation functions of ${\cal O}$ at leading order in $N_{\text{dof}}$ using the holographic equality \eqref{spart}.

To do so, we must evaluate the on-shell action $S\left[ {\bar \phi}\right]$. The first step is to solve the non-linear classical equation of motion,
\begin{equation}
    \frac{\delta S}{\delta \phi} = \left(-\Box + m^2\right)\phi + \frac{g}{2!} \phi^2 + \frac{\lambda}{3!} \phi^3 + ... = 0, \label{nlsc}
\end{equation}
subject to the boundary condition \eqref{bcscalar}. To express the solution in terms of the boundary value ${\bar \phi}$, we solve perturbatively in ${\bar \phi}$ by using integral kernels.\footnote{This is justified either for small source ${\bar \phi}$ or small couplings $\left\{g, \lambda, ...\right\}$.} We write 
\begin{equation}
    \phi\left(x\right) = \phi_0\left(x\right) + \phi_1\left(x\right)+ \phi_2\left(x\right) + ...,
\end{equation}
where $\phi_n$ is the solution at order $n+1$ in ${\bar \phi}$. To begin, the solution of the linear equation
\begin{equation}
    \left(-\Box + m^2\right)\phi_0  = 0,
\end{equation}
can be constructed from the boundary data via the corresponding \emph{bulk-to-boundary propagator}. This is an integral kernel
\begin{align}
    \phi_0\left(z,y\right) & = \int_{\partial \text{AdS}} d^dy^{\prime}\, K_{\Delta}\left(z,y;y^{\prime}\right){\bar\phi}\left(y^{\prime}\right), 
\end{align}
where
\begin{equation}
     \left(-\Box + m^2\right) K_{\Delta}\left(z,y;y^{\prime}\right) = 0, \qquad \lim_{z \rightarrow 0}\left(z^{\Delta-d}K_{\Delta}\left(z,y;y^{\prime}\right)\right) = \frac{1}{2\Delta-d} \delta^d\left(y-y^\prime\right).\label{scalarbubo}
     \end{equation}
Increasing the order in ${\bar \phi}$, we must solve \begin{align}
    \left(-\Box + m^2\right)\phi_0 & = 0, \\ \nonumber
    \left(-\Box + m^2\right)\phi_1 + \frac{g}{2!} \phi^2_0 &= 0, \\\nonumber
    \left(-\Box + m^2\right)\phi_2 + g \phi_0 \phi_1 + \frac{\lambda}{3!} \phi^3_0 &= 0, \\ \nonumber
    & \hspace*{0.25cm} \vdots\,.
\end{align}
Through use of the \emph{bulk-to-bulk propagator}
\begin{equation}
    \left(-\Box + m^2\right)\Pi_{\Delta}\left(x;x^\prime\right) = \frac{1}{\sqrt{|g|}}\delta^{d+1}\left(x-x^\prime\right), \label{scbubu}
\end{equation}
the solution can be determined order-by-order in terms of the linear solution $\phi_0$ and thus the boundary value ${\bar \phi}$, 
\allowdisplaybreaks
\begin{align} 
    \phi_0\left(x\right) & = \int_{\partial \text{AdS}} d^dy^{\prime}\, K_{\Delta}\left(z,y;y^{\prime}\right){\bar\phi}\left(y^{\prime}\right), \\ \nonumber
   \phi_1\left(x\right) & = - \frac{g}{2!} \int_{\text{AdS}}d^{d+1}x^{\prime} \Pi_{\Delta}\left(x;x^\prime\right) \phi^2_0\left(x^\prime\right),\\ \nonumber
   \phi_2\left(x\right) & = - \frac{\lambda}{3!} \int_{\text{AdS}}d^{d+1}x^{\prime} \Pi_{\Delta}\left(x;x^\prime\right) \phi^3_0\left(x^\prime\right) - g \int_{\text{AdS}}d^{d+1}x^{\prime} \Pi_{\Delta}\left(x;x^\prime\right) \phi_0\left(x^\prime\right)\phi_1\left(x^\prime\right),\\ \nonumber
   & \hspace*{0.25cm} \vdots\,.
\end{align}
The value of the action for this classical solution is given by the diagrammatic expansion
\begin{equation}
\includegraphics[scale=0.46]{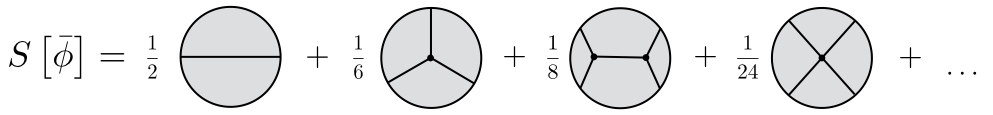}\label{expsc}\end{equation}
Correlation functions of ${\cal O}$ can be computed at leading order in $N_{\text{dof}}$ by taking functional derivatives of \eqref{expsc}. In the following subsections we give explicit examples for the two- and three-point functions, for which we only require the bulk-to-boundary propagator \eqref{scalarbubo}. We therefore first discuss the latter in more detail.

\subsubsection{Propagators}
To compute two- and three-point Witten diagrams at tree-level, as can be seen from the diagrammatic expansion \eqref{expsc} we require the form of the boundary-to-bulk propagator \eqref{scalarbubo}.\footnote{We postpone a derivation until \S \tcb{\ref{subsec::bbps}}, where it is considered together with the higher-spin propagators.} Its most familiar expression is given in the Poincar\'e patch
\begin{equation}
    K_{\Delta}\left(z,y;y^\prime\right) = C_{\Delta,0}\left(\frac{z}{z^2+\left(y-y^\prime\right)^2}\right)^{\Delta},
\end{equation}
where the overall coefficient is fixed by the near-boundary behaviour \eqref{scalarbubo}
\begin{align}
C_{\Delta,0} = \frac{\Gamma\left(\Delta\right)}{2\pi^{d/2}\Gamma\left(\Delta+1-\tfrac{d}{2}\right)}.
\end{align}
However, we consider the evaluation of Witten diagrams in ambient space \S \tcb{\ref{Aambient}}, in which the propagator takes the form
\begin{align}\label{5amsp}
K_{\Delta}\left(X;P\right) = \frac{C_{\Delta,0}}{\left(-2 X \cdot P\right)^{\Delta}}.
\end{align}
 As we shall see, the ambient framework proves effective in extending the results for the scalar to those with higher-spin external legs. In particular, the ambient expression for the propagator admits a Schwinger-Parameterised form 
\begin{align}
K_{\Delta}\left(X;P\right) = \frac{C_{\Delta,0}}{\Gamma\left(\Delta\right)} \int^\infty_0 \frac{dt}{t} t^\Delta e^{2t P \cdot X},
\end{align}
which turns out to dramatically simplify the evaluation of contact Witten diagrams.

\subsubsection{Two-point Witten diagram}
\begin{figure}[h]
  \centering
  \includegraphics[scale=0.65]{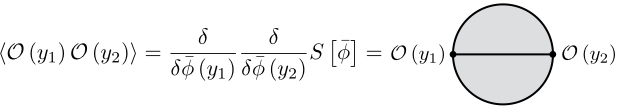}
  \caption{Bulk interpretation of the $\langle {\cal O}{\cal O} \rangle $ CFT two-point function, in the large $N_{\text{dof}}$ limit. The propagator departs from one boundary insertion point to the other.} \label{fig::holo2pt}
\end{figure}

We begin with the two-point Witten diagrams. Unlike  their higher-point counterparts, their computation via the above prescription requires careful treatment of IR divergences which arise from integrating over the infinite volume of AdS space \cite{Freedman:1998tz}. However these issues can be circumvented by noting that, as in flat space, at tree-level they can simply be given by the corresponding propagator \eqref{scbubu}/\eqref{scalarbubo}
\begin{equation}
   \langle {\cal O}\left(y_1\right)  {\cal O}\left(y_2\right) \rangle = K_{\Delta}\left(y_1;y_2\right) = \Pi_{\Delta}\left(y_1;y_2\right),
\end{equation}
which can be seen pictorially in figure \ref{fig::holo2pt}.

\subsubsection{Three-point Witten diagram}

\begin{figure}[ht]
  \centering
  \includegraphics[scale=0.45]{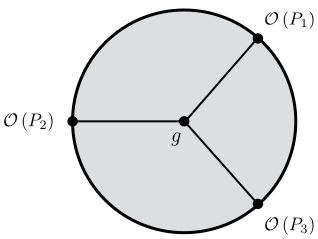}
  \caption{Tree-level three-point Witten diagram generated by the bulk vertex $g\phi^3$. This gives the holographic computation of the dual scalar operator three-point function at leading order in $1/N_{\text{dof}}$.} \label{tl3pt}
\end{figure}
\noindent
For Witten diagrams of three or more external points, there are no issues of IR divergences.\footnote{This can be verified straightforwardly by power counting.}  We thus proceed using the prescription for computing Witten diagrams as described at the beginning of this subsection. The three-point correlator of ${\cal O}$ at leading order in $1/N_{\text{dof}}$ can thus be computed holographically via
\begin{align}\nonumber
    \langle {\cal O}\left(y_1\right){\cal O}\left(y_2\right){\cal O}\left(y_3\right) \rangle 
& = \frac{\delta}{\delta {\bar \phi}\left(y_1\right)}\frac{\delta}{\delta {\bar \phi}\left(y_2\right)}\frac{\delta}{\delta {\bar \phi}\left(y_3\right)}S\left[{\bar \phi}\right]\\ 
& = g \int_{\text{AdS}} d^{d+1}x\, K_{\Delta}\left(x;y_1\right)K_{\Delta}\left(x;y_2\right)K_{\Delta}\left(x;y_3\right), \label{sc3ptint}
\end{align}
and is depicted in figure \ref{tl3pt}.
An effective way to evaluate the bulk integral \eqref{sc3ptint}, is to employ the Schwinger-parameterised form of the propagator in the ambient formalism. This gives 
\begin{align}\label{000}
\langle {\cal O}\left(P_1\right){\cal O}\left(P_2\right){\cal O}\left(P_3\right) \rangle & = g \int_{\text{AdS}} dX\, K_{\Delta}\left(X;P_1\right)K_{\Delta}\left(X;P_2\right)K_{\Delta}\left(X;P_3\right) \\ \nonumber
   & = g \left(\frac{C_{\Delta,0}}{\Gamma\left(\Delta\right)}\right)^3 \int^{\infty}_0 \prod\limits^{3}_{i=1}\left(\frac{dt_i}{t_i} t^{\Delta}\right) \int_{\text{AdS}} dX e^{2\left(t_1 P_1+t_2 P_2+t_3 P_3\right) \cdot X}.
    \end{align}
The integral over AdS is straightforward to evaluate using \eqref{appschwinwit}, and yields
\begin{align}
 &   \langle {\cal O}\left(P_1\right){\cal O}\left(P_2\right){\cal O}\left(P_3\right) \rangle \\ \nonumber
 & \hspace*{3cm}=  g\,\pi^{\frac{d}{2}}\Gamma\left(\frac{3\Delta - d}{2}\right) \left(\frac{C_{\Delta,0}}{\Gamma\left(\Delta\right)}\right)^3 \int^{\infty}_0 \prod\limits^{3}_{i=1}\left(\frac{dt_i}{t_i} t^{\Delta}_i\right) e^{\left(-t_1t_2 P_{12} - t_1t_3 P_{13}-t_2t_3 P_{23}\right)}.
 \end{align}
The remaining integrals can dealt with through the change of variables
\begin{equation}
    t_1 = \sqrt{\frac{m_2m_3}{m_1}}, \quad t_2 = \sqrt{\frac{m_1m_3}{m_2}}, \quad t_3 = \sqrt{\frac{m_1m_2}{m_3}}, \label{tov}
\end{equation}
which gives the final form for the three-point amplitude
\begin{align} \nonumber
 \langle {\cal O}\left(P_1\right){\cal O}\left(P_2\right){\cal O}\left(P_3\right) \rangle & = g\; \frac{1}{2}\pi^{\frac{d}{2}}\Gamma\left(\frac{3\Delta- d}{2}\right) \left(\frac{C_{\Delta,0}}{\Gamma\left(\Delta\right)}\right)^3 \int^{\infty}_0 \prod\limits^{3}_{i=1}\left(\frac{dm_i}{m_i} m_i^{\frac{\Delta}{2}}\right) \exp\left(-m_i P_{jk}\right) \\ 
 & = g\; \frac{1}{2}\pi^{\frac{d}{2}}\Gamma\left(\frac{3\Delta- d}{2}\right) \left(C_{\Delta,0}\frac{\Gamma\left(\frac{\Delta}{2}\right)}{\Gamma\left(\Delta\right)}\right)^3 \frac{1}{P^{\Delta/2}_{13}P^{\Delta/2}_{23}P^{\Delta/2}_{12}}.
 \end{align}
This result is straightforward to generalise to different interacting scalars $\phi_i$ of of mass $m^2_i R^2 = \Delta_i\left(\Delta_i-d\right)$, connected via the cubic vertex
\begin{equation}
    {\hat {\cal V}}_{0,0,0} = g\; \phi_1 \phi_2 \phi_3, \label{vbasic123poo}
\end{equation}
In precisely the same way as above, the corresponding amplitude at tree-level can be determined to be
\begin{align} \nonumber
   & \langle {\cal O}_1\left(P_1\right){\cal O}_2\left(P_2\right){\cal O}_3\left(P_3\right) \rangle \\ \nonumber 
  &= g\,\pi^{\frac{d}{2}}\Gamma\left(\frac{- d + \sum\nolimits^3_{i=1} \Delta_i}{2}\right) \frac{C_{\Delta_1,0}C_{\Delta_2,0}C_{\Delta_3,0}}{\Gamma\left(\Delta_1\right)\Gamma\left(\Delta_2\right)\Gamma\left(\Delta_3\right)} \int^{\infty}_0 \prod\limits^{3}_{i=1}\left(\frac{dt_i}{t_i} t^{\Delta_i}_i\right) e^{\left(-t_1t_2 P_{12} - t_1t_3 P_{13}-t_2t_3 P_{23}\right)} \\ 
   & = g\, {\sf C}\left(\Delta_1,\Delta_2,\Delta_3; 0\right)  \frac{1}{P^{\frac{\Delta_1+\Delta_3-\Delta_2}{2}}_{13}P^{\frac{\Delta_2+\Delta_3-\Delta_1}{2}}_{23}P^{\frac{\Delta_1+\Delta_2-\Delta_3}{2}}_{12}}, \label{d123}
\end{align}
where in the second equality we use the same change of variables \eqref{tov}, and we introduced
\begin{align}
   & {\sf C}\left(\Delta_1,\Delta_2,\Delta_3; 0\right) \\ \nonumber
   & \hspace*{0.5cm} = \;\frac{1}{2}\pi^{\frac{d}{2}}\Gamma\left(\frac{- d + \sum\nolimits^3_{i=1} \Delta_i}{2}\right) C_{\Delta_1,0}C_{\Delta_2,0}C_{\Delta_3,0} \frac{\Gamma\left(\frac{\Delta_1+\Delta_2-\Delta_3}{2}\right)\Gamma\left(\frac{\Delta_1+\Delta_3-\Delta_2}{2}\right)\Gamma\left(\frac{\Delta_2+\Delta_3-\Delta_1}{2}\right)}{\Gamma\left(\Delta_1\right)\Gamma\left(\Delta_2\right)\Gamma\left(\Delta_3\right)}.
\end{align}
Note that \eqref{d123} coincides with the structure \eqref{000gen} for the three-point correlator for the dual scalar operators, as required for the holographic duality to hold. In a sense this matching of space-time dependence (unlike matching the multiplicative constants) is a bit trivial at the three-point level, since everything is fixed by $SO\left(d,2\right)$ kinematics.

\section{Propagators of arbitrary spin}

With a firm understanding of tree-level three-point Witten diagrams for external scalars, for the remainder of this chapter we extend these results to four-point Witten diagrams in theories involving fields with arbitrary integer spin, and also to three-point Witten diagrams with spinning external legs. In this section we begin with the corresponding boundary-to-bulk and bulk-to-bulk propagators.

\subsection{Bulk-to-bulk propagators}
\label{subsec::bbprop}

In this subsection we derive the explicit form of the bulk-to-bulk propagators for spin-$s$ fields propagating on an AdS background.\footnote{See also \cite{Fronsdal:1978vb,Leonhardt:2003qu,Leonhardt:2003sn,Manvelyan:2005fp,Manvelyan:2008ks} for earlier works on higher-spin bulk-to-bulk propagators. For results on lower spin propagators, see: \cite{Burgess:1984ti,BURGES1986285,Allen:1985wd,Allen:1986tt,Turyn:1988af,D'Hoker:1999jc,D'Hoker:1998mz,D'Hoker:1999pj}.} In particular we employ the ambient space formalism, which is reviewed in detail in \S \tcb{\ref{Aambient}}. Let us note that this discussion only applies to gauge fields, for otherwise the only available quadratic actions require non-local terms or auxiliary fields \cite{Singh:1974qz,Singh:1974rc,Bouatta:2004kk,Francia:2007ee,Francia:2010ap}.\footnote{See however \cite{Costa:2014kfa} for the traceless and transverse parts of spinning bulk-to-bulk propagators for any mass in AdS.}

The linearised dynamics of an off shell
spin-$s$ gauge field $\varphi_s$ in AdS coupled to a conserved source ${J}_s$ is governed by the action
\begin{align}  \label{fronsdal}
    S\left[\varphi_s\right] & = \frac{s!}{2}  \int_{\text{AdS}_{d+1}}  \varphi_{s}\left(x;\partial_u\right) {\cal G}_s\left(x;u\right) -  g \, \varphi_{s}\left(x;\partial_u\right) J_{s}\left(x;u\right),
\end{align}
where ${\cal G}_s$ is the generalisation of the linearised Einstein tensor to spin-$s$ gauge fields
\begin{align}
   {\cal G}_s\left(x;u\right) & = \left(1-\frac{1}{4} \,u^2\, \partial_u \cdot \partial_u \right) \mathcal{F}_{s}\left(x; u, \nabla, \partial_u \right) \varphi_s\left(x, u\right), 
\end{align}
with Fronsdal operator \cite{Metsaev:1999ui, Mikhailov:2002bp}
\begin{align} \label{Fronsdaltensor}
{\cal F}_{s}(x,u,\nabla,\partial_u)
& =
\Box- m^2_s-u^2(\partial_u\cdot \partial_u)
-\;(u\cdot \nabla)\left((\nabla\cdot\partial_u)-\frac{1}{2}(u\cdot \nabla)
(\partial_u\cdot \partial_u)
\right), \\ \nonumber
m^2_s R^2 & = \left(s+d-2\right)\left(s-2\right) - s,
\end{align}
In order to satisfy the Bianchi identity
\begin{equation}
   \left( \partial_u \cdot \nabla \right)\mathcal{G}_{s}\left(x, u\right) = 0,
\end{equation}
we require the field $\varphi_s$ to be double-traceless, $\left(\partial_u \cdot \partial_u\right)\, \varphi_s\left(x,u\right) = 0$. The above action is invariant under the linearised spin-$s$ gauge transformations
\begin{equation}
\label{fronsdalgaugetr}
\delta\varphi_{s}(x,u)=(u\cdot\nabla)\varepsilon_{s-1}(x,u),
\end{equation}
where $\varepsilon_{s-1}(x,u)$ is a generating function for a rank-$\left(s\, \text{-}\, 1\right)$ symmetric and traceless
gauge parameter 
\begin{equation} \label{gaugepara}
\varepsilon_{s-1}(x,u)\equiv\frac{1}{(s-1)!} \varepsilon_{\mu_1\mu_2\dots \mu_{s-1}}u^{\mu_1}u^{\mu_2}\dots u^{\mu_{s-1}},
\qquad
(\partial_u\cdot \partial_u)\varepsilon_{s-1}(x,u)
= 0.
\end{equation}
The field is given in response to the source by
\begin{equation}
    \varphi_s\left(x,u\right) = g \int_{\text{AdS}} \Pi_{s}\left(x,u;x^\prime, \partial_{u^\prime}\right){J}_s\left(x^\prime;u^\prime\right), \label{bubu}
\end{equation}
where the bulk-to-bulk propagator satisfies the equation\footnote{The notation $\left\{\left\{\bullet\right\}\right\}$ signifies a double-traceless projection: 
\begin{align}
    (\partial_u\cdot \partial_u)^2\left\{\left\{f(u,x)\right\}\right\}=0, \quad \text{and} \quad \left\{\left\{f(u,x)\right\}\right\}= f(u,x)  \qquad \text{iff} \qquad (\partial_u\cdot \partial_u)^2 f(u,x)=0.
\end{align}}
\begin{align}
\notag
\left(1-
\frac{1}4{u_1^2} {\partial_{u_1}\cdot \partial_{u_1}}\right)&{\cal F}_{s}(x_1,u_1,\nabla_1,\partial_{u_1})\Pi_{s}(x_1,u_1;x_2,u_2) =\\
\label{propdef}
& \qquad -\left\{\left\{(u_1\cdot u_2)^s\right\}\right\}\delta(x_1,x_2)
+(u_2\cdot\nabla_2)\Lambda_{s,s-1}(x_1,u_1;x_2,u_2).
\end{align}
The term $\Lambda_{s,s-1}$ is pure gauge, whose effect is immaterial when integrating against a conserved current \eqref{bubu}.\footnote{In more detail, it is a bi-tensor that is traceless in tangent indices at $x_2$ and
double-traceless in tangent indices at $x_1$:\begin{align}
&(u_2\cdot \partial_{u_2})  \Lambda_{s,s-1}=s-1,\quad (u_1\cdot \partial_{u_1})  \Lambda_{s,s-1}=s,\quad (\partial_{u_2}\cdot\partial_{u_2}) \Lambda_{s,s-1}=(\partial_{u_1}\cdot\partial_{u_1})^2 \Lambda_{s,s-1}=0.
\end{align}} We thus have the following freedom in the definition of the propagator
\begin{equation}
\Pi_s(x_1,u_1;x_2,u_2)\: \sim \:  \Pi_s(x_1,u_1;x_2,u_2)+(u_2\cdot\nabla_2)\;{\cal E}_{2,s,s-1}(x_1,u_1;x_2,u_2), \label{cfreed}
\end{equation}
where ${\cal E}_2$ is defined by $\Lambda$.

However, fixing $\Lambda$ does not specify the propagator uniquely. Due to gauge
invariance \eqref{fronsdalgaugetr}, the left hand side of \eqref{propdef} is not sensitive
to  variations of the propagator of the form
\begin{equation*}
\Pi_s(x_1,u_1;x_2,u_2)\: \sim \:  \Pi_s(x_1,u_1;x_2,u_2)+(u_1\cdot\nabla_1)\;{\cal E}_{1,s-1,s}(x_1,u_1;x_2,u_2),
\end{equation*}
where ${\cal E}_1$ has the same rank and trace properties as ${\cal E}_2$, but with
`1' and `2' interchanged. This freedom is usually fixed by imposing a gauge, and is required to make the operator on the left hand side of \eqref{propdef} non-degenerate and thus invertible.

In the following sections, we determine the explicit form of the bulk-to-bulk propagator \eqref{bubu} in three different gauges: the de Donder gauge, a traceless gauge and what we define as the ``manifest trace gauge''. As the name suggests, the latter makes the trace structure of the propagator manifest. In particular, in the view of expressing exchange diagrams as partial wave expansions, we derive the propagators in a basis of AdS harmonic functions.

\subsubsection{de Donder gauge}
It is often useful to eliminate gradients and divergences from the Fronsdal tensor \eqref{Fronsdaltensor}. This can be achieved by imposing the de Donder gauge
\begin{eqnarray}
\label{dedondg1}
(\nabla\cdot\partial_u)\varphi_{s}(x,u)-\frac{1}{2}(u\cdot \nabla)
(\partial_u\cdot \partial_u)\varphi_{s}(x,u)=0,
\end{eqnarray}
With this choice, the Fronsdal tensor reads
\allowdisplaybreaks
\begin{eqnarray}
\label{FronsdaltensorDD}
{\cal F}_{s}(x,u,\nabla,\partial_u)\varphi_{s}(x,u)=(\Box- m^2_s)\varphi_{s}(x,u)-u^2(\partial_u\cdot \partial_u)\varphi_{s}(x,u).
\end{eqnarray}
To proceed, we decompose the bulk-to-bulk propagator \eqref{bubu} into symmetric and traceless components\footnote{Any double-traceless tensor $\left(\partial_u \cdot \partial_u \right)^2\,t_s\left(x,u\right) = 0$ can be expressed in the form
\allowdisplaybreaks
\begin{align}\nonumber
   & \hspace*{0.75cm} t_s\left(x,u\right) = t^{\left\{0\right\}}_s\left(x,u\right) + u^2\, t^{\left\{1\right\}}_{s-2}\left(x,u\right), \qquad \left(\partial_u \cdot \partial_u \right)  t^{\left\{0\right\}}_s\left(x,u\right) = \left(\partial_u \cdot \partial_u \right)  t^{\left\{1\right\}}_s\left(x,u\right) = 0.
\end{align}}
\allowdisplaybreaks
\begin{align}
\notag
\Pi_{s}(x_1,&u_1;x_2,u_2)=\Pi^{\{0\}}_{s}(x_1,u_1;x_2,u_2)+u_1^2 u_2^2\;\Pi^{\{1\}}_{s-2}(x_1,u_1;x_2,u_2),\\
\label{tracesplit1} 
(\partial_{u_1}\cdot\partial_{u_1})&\Pi^{\{0\}}_{s}=(\partial_{u_2}\cdot\partial_{u_2})\Pi^{\{0\}}_{s}=0=
(\partial_{u_1}\cdot\partial_{u_1})\Pi^{\{1\}}_{s-2}=(\partial_{u_2}\cdot\partial_{u_2})\Pi^{\{1\}}_{s-2}\,.
\end{align}
The equation \eqref{propdef} for the propagator thus splits in two equations for traceless bi-tensors 
\begin{align}
\label{dedonderprop1}
(\Box_1-m_s^2)\Pi^{\{0\}}_{s}&(x_1,u_1;x_2,u_2)=-\left\{(u_1\cdot u_2)^s\right\}\delta(x_1,x_2),\\ \nonumber
(\Box_1-m_t^2)\Pi^{\{1\}}_{s-2}&(x_1,u_1;x_2,u_2)=\frac{s(s-1)}{(d+2s-3)(d+2s-5)}\left\{(u_1\cdot u_2)^{s-2}\right\}\delta(x_1,x_2),
\end{align}
where $m_t^2=s^2+(d-1)s-2$, and we used
\begin{align}
\left\{\left\{(u_1\cdot u_2)^s\right\}\right\}
\label{tracesplit2}
=\left\{(u_1\cdot u_2)^s\right\}+\frac{s(s-1)}{2(d+2s-3)}u_1^2 u_2^2\left\{(u_1\cdot u_2)^{s-2}\right\}.
\end{align}
In decomposing into symmetric and traceless components, we can employ the operator algebra on traceless symmetric tensors (which can be found in \S \tcb{\ref{app::ao}}). This reduces the problem to an algebraic one: Make an ansatz in a basis of traceless harmonic functions, and apply the latter operator algebra to determine the basis coefficients. This is most streamlined using the ambient formalism (\S \tcb{\ref{Aambient}}), in which we make the ansatz 
\begin{align}
\label{tracelesspartdeDonder}
\Pi^{\{0\}}_{s} &= \sum^{s}_{\ell=0} \int^{\infty}_{-\infty} d\nu f^{\{0\}}_{s,\ell}\left(\nu\right) \left(W_1 \cdot \nabla_1\right)^{\ell}\left(W_2 \cdot \nabla_2\right)^{\ell} \Omega_{\nu,s-\ell}\left(X_1,W_1;X_2,W_2\right),\\ \nonumber
\Pi^{\{1\}}_{s-2}& = \sum^{s-2}_{\ell=0} \int^{\infty}_{-\infty}  d\nu f^{\{1\}}_{s-2,\ell}\left(\nu\right) \left(W_1 \cdot \nabla_1\right)^{\ell}\left(W_2 \cdot \nabla_2\right)^{\ell} \Omega_{\nu,s-2-\ell}\left(X_1,W_1;X_2,W_2\right),
\end{align}
where $f^{\{0\}}_{s,\ell}$ and $f^{\{1\}}_{s-2,\ell}$ are to be fixed by
\eqref{dedonderprop1}, and $W_{1,2}$ are null auxiliary vectors which enforce tracelessness. Employing the equation of motion \eqref{eomharm} for $\Omega$, the completeness relation \eqref{acomp} and the commutator \eqref{i1}, we can conclude that
\begin{align}
f^{\{0\}}_{s,\ell}\left(\nu\right)=&\;\frac{c_{s,\ell}\left(\nu\right)}{m_s^2+\tfrac{\:d^2}{4}+\nu^2+s-\ell+l(d+2s-\ell-1)},\\ \nonumber \\ 
f^{\{1\}}_{s-2,\ell}\left(\nu\right)=&-\frac{s(s-1)}{(d+2s-3)(d+2s-5)}\frac{c_{s-2,\ell}\left(\nu\right)}{m_t^2+\tfrac{\:d^2}{4}+\nu^2+s-\ell-2+\ell(d+2s-\ell-5)},
\end{align}
where
\begin{align}\label{whatcsl}
   c_{s,\ell}\left(\nu\right) = \frac{2^\ell \left(r-\ell+1\right)_{\ell}\left(\frac{d}{2}+r-\ell-\frac{1}{2}\right)_{\ell}}{\ell! \left(d+2r-2\ell-1\right)_{\ell}\left(\frac{d}{2}+r-\ell+i\nu\right)_\ell\left(\frac{d}{2}+r-\ell-i\nu\right)_\ell},
\end{align}
comes from the completeness relation \eqref{acomp}.

\subsubsection{Traceless gauge}
In a similar way, we can also solve for the propagator by imposing a traceless gauge on the spin-$s$ field. To set the trace of the spin-$s$ Fronsdal  field  to zero, the gauge parameter $\varepsilon_s\left(x,u\right)$ needs to be chosen such that 
\begin{equation}
\left(\nabla \cdot \partial_{u} \right) \varepsilon_{s-1} \left(x,u\right)  = - \left(\partial_{u} \cdot \partial_{u}\right)\varphi_{s},
\end{equation} 
after which $\left(\nabla \cdot \partial_{u} \right) \varepsilon_{s-1}$ is fixed. In this partial gauge, the field $\varphi_{s}\left(x,u\right)$ is left with residual gauge symmetry 
\begin{equation} \label{resg}
\delta\varphi_{s}(x,u)=(u\cdot\nabla)\varepsilon_{s-1}(x,u), \quad \left(\nabla \cdot \partial_{u} \right) \varepsilon_{s-1} \left(x,u\right) = \left(\partial_u \cdot \partial_u\right)  \varepsilon_{s-1} \left(x,u\right) = 0.
\end{equation}
Subject to the condition $\left(\partial_{u} \cdot \partial_{u}\right)\varphi_{s}=0$, the Fronsdal tensor then reads 
\begin{align}  \label{Tracelesseom}
& {\cal F}_{s}(x,u,\nabla,\partial_u)\varphi_{s}(x,u) \\ \nonumber
 &  \qquad \quad=\left(\Box -m^2_s\right) \varphi_{s}\left(x,u\right)- \left(u \cdot \nabla \right) \left(\nabla \cdot \partial_{u}\right) \varphi_{s} \left(x,u\right) + \frac{1}{d+2s-3} u^2 \left(\nabla \cdot \partial_{u}\right)^2 \varphi_{s}\left(x,u\right), 
\end{align}
which is simply its traceless part.

Before using the form \eqref{Tracelesseom} for the Fronsdal tensor to solve for the propagator, we need to fix the residual gauge symmetry \eqref{resg}. This can be done by further demanding that the propagator is symmetric under $u_1 \leftrightarrow u_2$.\footnote{Note that \eqref{propdef} does not assume this symmetry.} The equation for the propagator is then
\begin{align}
&\left(\Box_1 -m^2_s\right) \Pi_{s}(x_1,u_1;x_2,u_2) - \left(u_1 \cdot \nabla_1 \right) \left(\nabla_1 \cdot \partial_{u_1}\right) \Pi_{s}(x_1,u_1;x_2,u_2) \\ 
&\qquad \qquad \qquad \qquad +\frac{1}{d+2s-3} u^2_1 \left(\nabla_1 \cdot \partial_{u_{1}}\right)^2 \Pi_{s}(x_1,u_1;x_2,u_2)= -\left\{(u_1\cdot u_2)^s\right\}\delta(x_1,x_2), \nonumber
\end{align}
with
\begin{align}
\left(\partial_{u_1}\cdot \partial_{u_1}\right) \Pi_{s}(x_1,u_1;x_2,u_2) &= \left(\partial_{u_2}\cdot \partial_{u_2}\right) \Pi_{s}(x_1,u_1;x_2,u_2) = 0, \\ \qquad \Pi_{s}(x_1,u_1;x_2,u_2) &= \Pi_{s}(x_1,u_2;x_2,u_1).
\end{align}
In the ambient formalism, using the null auxiliary vectors $W_{1,2}$ this is
\begin{align} \label{eq:AmbientTraceless}
\left(\Box_1 -m^2_s\right)& \Pi_{s}-\frac{2}{d+2s-3} \left(W_1 \cdot \nabla_1\right)(\nabla_1 \cdot {\hat D}_{W_1}) \Pi_{s}= -\left(W_1 \cdot W_2\right)^s \delta\left(X_1,X_2\right), 
\end{align}
with $\Pi_{s}\left(X_1,W_1;X_2,W_2\right) = \Pi_{s}\left(X_1,W_2;X_2,W_1\right)$.

Like for the de Donder gauge, since the propagator is traceless we can make the ansatz
\begin{equation}
\Pi_{s} = \sum^{s}_{\ell=0} \int^{\infty}_{-\infty}  d\nu f_{s,\ell}\left(\nu\right) \left(W_1 \cdot \nabla_1\right)^{\ell}\left(W_2 \cdot \nabla_2\right)^{\ell} \Omega_{\nu,s-\ell}\left(X_1,W_1;X_2,W_2\right), \label{tlsplit}
\end{equation}
with $f_{s,\ell}\left(\nu\right)$ arbitrary functions to be determined. By simply employing again a combination of the equation of motion for the $\Omega$ \eqref{eomharm}, the completeness relation \eqref{acomp}, and commutators: \eqref{i1} and \eqref{i2}, the propagator equation \eqref{eq:AmbientTraceless} determines the basis coefficients to be
\begin{equation}
f_{s,\ell}\left(\nu\right) = - c_{s,\ell}\left(\nu\right) \frac{d+2s-3}{\left(\ell-1\right)\left(2s+d-\ell -3\right)}\frac{1}{\nu^2+\left(s-2+\tfrac{d}{2}\right)^2}.
\end{equation}
where we recall that $c_{s,\ell}\left(\nu\right)$ comes from the completeness relation \eqref{acomp}.
\subsubsection{Manifest trace gauge}
In the previous sections, fixing a gauge for the massless spin-$s$ field allowed us to invert the differential operator in the defining equation \eqref{bubu} for the bulk-to-bulk propagator, and solve for its form. However, as already explained, since we ultimately integrate the propagators against a conserved current \eqref{bubu}, they are defined only up to the addition of terms proportional to gradients.\footnote{In other words, we are free to add or eliminate terms of the form $\nabla \cdot f$, because they drop out after integrating by parts and employing current conservation in \eqref{bubu}.}

This freedom can be used to simplify the form of the de Donder- \eqref{tracelesspartdeDonder} and traceless- \eqref{tlsplit} propagators: By removing terms proportional to gradients, it is possible to bring them into the form 
\begin{equation}
\label{eq:ManifestTrace}
\Pi_{s} = \sum^{[s/2]}_{k=0} \int^{\infty}_{-\infty}  d\nu \; g_{s,k}\left(\nu\right) \left(u_1^2\right)^{k} \left(u_2^2\right)^{k} \Omega_{\nu,s-2k},
\end{equation}
i.e. with no differential operators in the explicit expression. We refer to this as the ``manifest trace gauge''.

Reaching this gauge, on the other hand, is a non-trivial task owing to the non-commuting covariant derivatives $\nabla$. Relegating the details to \S \tcb{\ref{app::tssp}}, starting from either the de Donder \eqref{tracelesspartdeDonder} or traceless \eqref{tlsplit} propagators, we find
\begin{align}
\notag
g_{s,0}\left(\nu\right)&=\frac{1}{(\tfrac{d}{2}+s-2)^2+\nu^2},\\
\label{answertrgauge}
g_{s,k}\left(\nu\right)&=-\frac{\left(1/2\right)_{k-1}}{2^{2k+3}\cdot k!}\frac{(s-2k+1)_{2k}}{(\tfrac{d}{2}+s-2k)_k (\tfrac{d}{2}+s-k-3/2)_k}\\
\notag
&\hspace*{2cm}\times \;
\frac{\left({(\tfrac{d}{2}+s-2k+i\nu)}/{2}\right)_{k-1}\left({(\tfrac{d}{2}+s-2k-i\nu)}/{2}\right)_{k-1}}{\left({(\tfrac{d}{2}+s-2k+1+i\nu)}/{2}\right)_{k}\left({(\tfrac{d}{2}+s-2k+1-i\nu)}/{2}\right)_{k}}, \qquad k\ne 0.
\end{align}

\subsection{Review: Boundary-to-bulk propagators}
\label{subsec::bbps}
Boundary-to-bulk propagators are limiting cases of bulk-to-bulk propagators, where one of the two bulk points is taken to the boundary. In this section we review existing results for the bosonic boundary-to-bulk propagators for particles of arbitrary spin and mass in the ambient formalism \cite{Costa:2014kfa}.\footnote{See also \cite{Fronsdal:1978vb,Dobrev:1998md,Mikhailov:2002bp,Didenko:2012vh} for earlier formulations of the propagators, which includes the intrinsic and unfolded forms.}

Generalising the scalar case \eqref{scalarbubo} presented in \S \tcb{\ref{scalarex}}, the boundary-to-bulk propagator for a symmetric and traceless rank-$s$ tensor $\varphi_{\mu_1 ... \mu_s}$ of energy $\Delta$ and spin-$s$ satisfies the wave equation
\begin{align}
    \left(-\Box + \Delta\left(\Delta-d\right)-s\right)
    K_{\Delta,\, \mu_1 ... \mu_s}{}^{i_1 ... i_s}\left(z, y;y^\prime\right) & = 0, \\
    \lim_{z \rightarrow 0}\left(z^{\Delta-d+s}K_{\Delta,\, \mu_1 ... \mu_s}{}^{i_1 ... i_s}\left(z, y;y^\prime\right)\right) & = \frac{\delta{}^{i_1 \, \ldots}_{\left\{\right.\mu_1 \, \ldots}\delta{}^{i_s }_{\mu_s\left.\right\}}}{2\Delta - d}\delta^{d}\left(y-y^\prime\right)\label{kdd}
    \end{align}
Finding the explicit form of the propagator is straightforward in the ambient formalism. Within this framework, the simple form of the constraints imposed by the boundary and bulk $SO\left(d,2\right)$ symmetry:
\begin{equation}
    K_{\Delta,s}\left(X,\alpha_1 W; \lambda P,\alpha_2 Z + \beta P\right) = \lambda^{-\Delta} \left(\alpha_1 \alpha_2 \right)^s K_{\Delta,s}\left(X, W; P, Z \right),\label{props}
\end{equation}
imply the unique structure
\begin{align}
    K_{\Delta,s}\left(X, W; P, Z \right) & = \left(W \cdot  {\cal P} \cdot Z \right)^s \frac{C_{\Delta,s}}{\left(-2 X \cdot P\right)^{\Delta}},
    \end{align}
    with projector
    \begin{align}
     {\cal P}^A{}_B
     & = \delta^{A}_B - \frac{P^A X_B}{P \cdot X}.\label{ps}
\end{align}
The role of the above projector is to ensure transversality of the propagator at both its bulk and boundary points. The normalisation $C_{\Delta,s}$ is fixed by equation \eqref{kdd}, where the coefficient of the Dirac delta function in the latter ensures consistency with the  boundary limit of the corresponding {bulk-to-bulk propagator,
\begin{align}
  &  K_{\Delta,s}\left(z,y;y^\prime\right) = \lim_{w \rightarrow 0} \Pi_{\Delta,s}\left(z,y;w, y^\prime\right).
\end{align}
This gives,
\begin{equation}\label{poonorm}
  C_{\Delta,s} = \frac{\left(s+\Delta-1\right)\Gamma\left(\Delta\right)}{2 \pi^{d/2} \left(\Delta-1\right) \Gamma\left(\Delta+1-\tfrac{d}{2}\right)}.
\end{equation}
\section{Three-point Witten diagrams}

By taking the results for the propagators in the previous section, we can proceed to evaluate Witten diagrams with spinning external legs and exchanged fields. We begin in this section with three-point Witten diagrams, which also (as we shall see) play a significant role for higher-point amplitudes.

When considering spinning external particles, the bulk integrals encountered are more complicated due to the introduction of tensor structures. Before proceeding further, we first introduce some useful tools in order to handle the latter more effectively. 

\subsection{Tool kit}
\label{subsec::tools}
The basic idea behind the techniques introduced in this subsection, is to simplify the bulk integrals encountered when evaluating Witten diagrams with spinning external legs by removing the tensor structures from the integrand. This enables amplitudes with spinning external legs to be expressed in terms of those generated by the basic scalar cubic vertex \eqref{vbasic123poo}.

The way this idea is implemented, is to express spinning propagators and their derivatives in terms of a scalar propagator:

Notice that the spin-$s$ propagator \eqref{props} is essentially the propagator of a scalar field \eqref{5amsp} of the same dimension $\Delta$, with the projector \eqref{ps} simply propagating the indices. This relationship can be expressed as the $s$-fold application of an $X$-independent differential operator
\begin{align} \label{bubos}
    & K_{\Delta,s}\left(X, W|P, Z \right) = \frac{1}{\left(\Delta-1\right)_s}\left({\cal D}_P\left(Z|W\right)\right)^s K_{\Delta,0}\left(X;P\right), \\ 
   & {\cal D}_P\left(Z|W\right) = \left(Z \cdot W\right)\left(Z \cdot \frac{\partial}{\partial Z} - P \cdot \frac{\partial}{\partial P}\right) + \left(P \cdot W \right)\left(Z \cdot \frac{\partial}{\partial P}\right).
\end{align}
Moreover, we can express the $n$-th ambient  derivative of the bulk-to-boundary propagator in terms of a scalar propagator of dimension $\Delta+n$:
\begin{align}
&(W\cdot\pl_X)^nK_{\Delta,s}(X,P|U,Z)\\
&\hspace*{1.5cm}=\frac{C_{\Delta,s}}{s!C_{\Delta+n,0}}\sum_{i=0}^s\sum_{\omega=0}^i\binom{s}{i}\binom{i}{\omega} \frac{(-1)^s\,2^n}{\left(n-\omega+1\right)_{\omega}}\frac{1}{\Gamma(\Delta+i)}\,(U\cdot P)^i(U\cdot Z)^{s-i}\nonumber\\ 
& \hspace*{3cm}\times (Z\cdot W)^\omega(P\cdot W)^{n-\omega}\left((Z\cdot\pl_P)^{i-\omega}K_{\Delta+n,0}(X,P)\right)\,.\label{schwingprop2}
\end{align}
With \eqref{bubos} and \eqref{schwingprop2}, we can shift the tensor structures from the integrand in a Witten diagram with spinning external legs, allowing the amplitudes to be expressed in terms of a basic scalar amplitude (i.e. generated by an interaction involving only scalars and with no derivatives). As we shall demonstrate in the following, this provides a powerful approach to evaluating bulk amplitudes with external fields of non-zero spin.

\subsection{Two scalars and a spin-s}

We begin with the simplest case of three-point Witten diagrams involving two external scalars and a bosonic spin-$s$ field.

Any such amplitude can be evaluated at tree-level with the knowledge of that  generated by the basic vertex,
\begin{equation}
    {\hat {\cal V}}_{s,0,0} = \varphi_{\mu_1 ... \mu_s}\, \phi_1 \nabla^{\mu_1} ...  \nabla^{\mu_s} \phi_2, \label{basic}
\end{equation}
where the covariant derivatives just act on the scalar $\phi_2$. The reason for this is that external legs are on-shell, and there is just a single on-shell non-trivial vertex involving two scalars and a spin-$s$ field \cite{Metsaev:2005ar}.

The example that we encounter in later sections is the following: On-shell, a spin-$s$ field is traceless and transverse. Through integrating by parts and using transversality, the following types of vertices can then be identified on-shell with the basic vertex \eqref{basic}:
\begin{align}
&\nabla^{\mu_1} \varphi_{\mu_1 ... \mu_s} \approx 0 \\ \nonumber 
& \implies \quad \int_{\text{AdS}} \; \varphi_{\mu_1 ... \mu_s} \nabla^{\mu_1} ... \nabla^{\mu_r} \phi_1 \nabla^{\mu_{r+1}} ... \nabla^{\mu_s} \phi_2 \quad \approx \quad   \int_{\text{AdS}} \; {\hat {\cal V}}_{s,0,0}\left(x\right), \quad r \in \left\{1, ..., s\right\}.
\end{align}
We therefore spend the remainder of this section on evaluating tree-level three-point amplitudes generated by the basic vertex \eqref{basic}, which in the ambient formalism can be found in \cite{Costa:2014kfa,Bekaert:2015tva}. According to the usual recipe, this is given by
\begin{align}\label{intamp00s}
  &  {\hat {\cal A}}_{\Delta_1, \Delta_2,\Delta_3, s}\left(P_1,P_2,P_3|Z\right) \\ \nonumber
    & \hspace*{2cm} = \frac{1}{s!\left(\frac{d}{2}-1\right)_s} \int_{\text{AdS}}\, K_{\Delta_3,s}\left(X,{\hat D}_W;P_3,Z\right)  K_{\Delta_1,0}\left(X;P_1\right) \left(W \cdot \nabla\right)^s K_{\Delta_2,0}\left(X;P_2\right). \end{align}
As explained in \S \tcb{\ref{subsec::tools}}, with the help of \eqref{bubos} and \eqref{schwingprop2} which express any propagator or its derivatives in terms of a scalar propagator, this amplitude can be generated from that with $s=0$. In this way we can use the result already obtained in \S \tcb{\ref{scalarex}}, equation \eqref{d123}. More explicitly, we have
\begin{align} \label{damp0}
  &  {\hat {\cal A}}_{\Delta_1, \Delta_2,\Delta_3, s}\left(P_1,P_2,P_3|Z\right) \\ \nonumber
    & \hspace*{1.5cm} = \frac{1}{s!\left(\frac{d}{2}-1\right)_s} \frac{\left({\cal D}_{P_3}(Z|{\hat D}_W)\right)^s}{\left(\Delta_3-1\right)_s} \int_{\text{AdS}}\, K_{\Delta_3,0}\left(X;P_3\right)  K_{\Delta_1,0}\left(X;P_1\right) \left(W \cdot \nabla\right)^s K_{\Delta_2,0}\left(X;P_2\right) \\ \nonumber
    & \hspace*{1.5cm} = \frac{\left(1-\frac{d}{2}+\Delta_2\right)_s}{\left(\Delta_3-1\right)_s} \left(2\,{\cal D}_{P_3}(Z|P_2)\right)^s\:{\hat {\cal A}}_{\Delta_1, \Delta_2+s,\Delta_3, 0}\left(P_1,P_2,P_3\right).
\end{align}
This relationship is illustrated in figure \ref{sto0}.\\
\begin{figure}[h]
  \centering
  \includegraphics[scale=0.5]{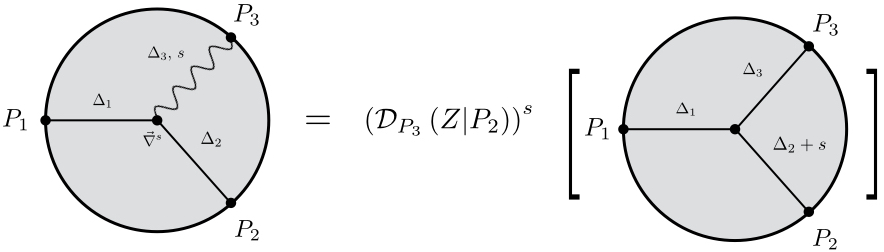}
  \caption{Through the techniques given in \S \tcb{\ref{subsec::tools}}, the three-point Witten diagram generated by the basic vertex \eqref{vbasic123poo}, with a single spin-$s$ external leg and scaling dimensions $\left(\Delta_1,\Delta_2,\Delta_3\right)$, may be expressed in terms of that with only external scalars and scaling dimensions $\left(\Delta_1,\Delta_2+s,\Delta_3\right)$. The latter we already computed in \S \tcb{\ref{scalarex}}. The notation $\vec{\nabla}$ indicates that the covariant derivative acts on the rightmost external scalar field.} \label{sto0}
\end{figure}

The amplitude \eqref{intamp00s} can then be expressed in the more familiar form \eqref{gen3pahhh} dictated by conformal symmetry by using
\begin{align}
&  \left({\cal D}_{P_3}\left(Z|P_2\right)\right)^s \;\frac{1}{P^{\frac{\Delta_1+\Delta_3-\Delta_2 - s }{2}}_{13}P^{\frac{\Delta_2+\Delta_3-\Delta_1+s}{2}}_{23}P^{\frac{\Delta_1+\Delta_2-\Delta_3+s}{2}}_{12}} \\ \nonumber 
 & \hspace*{4.5cm} = \left(\frac{\Delta_1 + \Delta_3-\Delta_2 - s}{2}\right)_s \frac{\left(\left(Z \cdot P_1\right)P_{23} - \left(Z \cdot P_2\right)P_{13}  \right)^s}{P^{\frac{\Delta_1+\Delta_3-\Delta_2 + s }{2}}_{13}P^{\frac{\Delta_2+\Delta_3-\Delta_1+s}{2}}_{23}P^{\frac{\Delta_1+\Delta_2-\Delta_3+s}{2}}_{12}}.
\end{align}
We therefore obtain
\begin{align} \label{damp}
  &  {\hat {\cal A}}_{\Delta_1, \Delta_2,\Delta_3, s}\left(P_1,P_2,P_3\right) =  {\sf C}\left(\Delta_1,\Delta_2,\Delta_3;s\right) \; \frac{\left(\left(Z \cdot P_1\right)P_{23} - \left(Z \cdot P_2\right)P_{13}  \right)^s}{P^{\frac{\Delta_1+\Delta_3-\Delta_2 + s }{2}}_{13}P^{\frac{\Delta_2+\Delta_3-\Delta_1+s}{2}}_{23}P^{\frac{\Delta_1+\Delta_2-\Delta_3+s}{2}}_{12}},
\end{align}
with
\begin{align}\label{00sW}
& {\sf C}\left(\Delta_1,\Delta_2,\Delta_3;s\right) \\ 
& \hspace*{0.25cm} = \frac{2^s \left(1-\frac{d}{2}+\Delta_2\right)_s}{\left(\Delta_3-1\right)_s} \left(\frac{\Delta_1 + \Delta_3-\Delta_2 - s}{2}\right)_s {\sf C}\left(\Delta_1,\Delta_2+s,\Delta_3;0\right) \nonumber \\ \nonumber
&  \hspace*{0.25cm} = C_{\Delta_1,0}C_{\Delta_2,0}C_{\Delta_3,s} \frac{2^s \pi^{\frac{d}{2}}\Gamma\left(\frac{\Delta_1+\Delta_2+\Delta_3-d+s}{2}\right)\Gamma\left(\frac{\Delta_1+\Delta_2-\Delta_3+s}{2}\right)\Gamma\left(\frac{\Delta_1+\Delta_3-\Delta_2+s}{2}\right)\Gamma\left(\frac{\Delta_2+\Delta_3-\Delta_1+s}{2}\right)}{2\Gamma\left(\Delta_1\right)\Gamma\left(\Delta_2\right)\Gamma\left(\Delta_3+s\right)}.
\end{align}

\subsection{General case: s1-s2-s3}
\label{subsec::witt123}
We now consider the more involved general case of tree-level three-point Witten diagrams with external fields of spin $\left(s_1,s_2,s_3\right)$, based on the original work \cite{Sleight:2016dba}. The logic is the same as for the simpler  $\left(0,0,s\right)$ amplitudes considered in the previous section: Reduce the task to the straightforward evaluation of basic three-point Witten diagrams with external scalars.

The most general cubic vertex involving symmetric fields of spin $\left(s_1,s_2,s_3\right)$ in AdS$_{d+1}$ is parameterised by six basic contractions, given by \eqref{6cont} in the ambient formalism (\S \tcb{\ref{Aambient}}) below. These are the bulk counterparts of the six conformal structures on the boundary \eqref{6conf}. Using point-splitting, the most general (on-shell) cubic vertex takes the form \cite{Joung:2011ww}\footnote{In AdS$_4$, dimensional dependent identities reduce the number of independent structures, and allow for parity violating vertices. However these are not relevant in the context of this thesis.} 
\begin{align}
    {\cal V}_{s_1,s_2,s_3}\left(\varphi_i\right) = \sum\limits^{s_i}_{n_i=0} g_{s_1,s_2,s_3}^{n_1,n_2,n_3}\, {\cal I}_{s_1,s_2,s_3}^{n_1,n_2,n_3}(\varphi_i),
\end{align}
with structures
\begin{multline}\label{is0}
    {\cal I}_{s_1,s_2,s_3}^{n_1,n_2,n_3}(\varphi_i)=\mathcal{Y}_1^{s_1-n_2-n_3}\mathcal{Y}_2^{s_2-n_3-n_1}\mathcal{Y}_3^{s_3-n_1-n_2}\\\times\mathcal{H}_1^{n_1}\mathcal{H}_2^{n_2}\mathcal{H}_3^{n_3}\,\varphi_{s_1}(X_1,U_1)\varphi_{s_2}(X_2,U_2)\varphi_{s_3}(X_3,U_3)\Big|_{X_i=X}\,,
\end{multline}
and
\begin{subequations}\label{6cont}
\begin{align}
    \mathcal{Y}_1&=\pl_{U_1}\cdot\pl_{X_2}\,,&\mathcal{Y}_2&=\pl_{U_2}\cdot\pl_{X_3}\,,&\mathcal{Y}_3&=\pl_{U_3}\cdot\pl_{X_1}\,,\\
    \mathcal{H}_1&=\pl_{U_2}\cdot\pl_{U_3}\,,&\mathcal{H}_2&=\pl_{U_3}\cdot\pl_{U_1}\,,&\mathcal{H}_3&=\pl_{U_1}\cdot\pl_{U_2}\,.
\end{align}
\end{subequations}
Note that in the above we employed the partial derivative $\partial_A$ (which acts in the ambient $\mathbb{R}^{d,2}$ space), as opposed to the  covariant derivative $\nabla$ on AdS$_{d+1}$. For obvious reasons, this is for ease of computation e.g. avoiding the non-commutativity of covariant derivatives.
However, it is explained in detail in \S \tcb{\ref{radialred}} how the above expressions can be re-expressed in terms of  $\nabla$, and this is just a change of basis for representing the vertices in the ambient framework. The above is simply more convenient of packaging the vertices.

To compute the tree-level amplitude generated by the structure \eqref{is0}, using \eqref{schwingprop2} we again generate it from the result \eqref{d123} for the basic scalar vertex. The final expression for the amplitude is quite involved, but for completeness we include it below
\begin{align}
& {\cal A}_{\left(\Delta_1,s_1\right),\left(\Delta_2,s_2\right),\left(\Delta_3,s_3\right)}^{n_1,n_2,n_3}\left(\left\{P_i\right\}|\left\{Z_i\right\}\right) \\ \nonumber
 & \hspace*{1.5cm} =  \int_{\text{AdS}} dX \, {\cal I}_{s_1,s_2,s_3}^{n_1,n_2,n_3}(K_{\Delta_i,s_i}) \\ \nonumber
 &  \hspace*{1.5cm} =  f^{n_1,n_2,n_3}_{s_1,s_2,s_3}\left(\left\{\partial_{P_i}\right\},\left\{\partial_{P_i}\right\},\left\{Z_i\right\} \right) {\cal A}_{\left(\Delta_1+s_3-k_3,0\right),\left(\Delta_2+s_1-k_1,0\right),\left(\Delta_3+s_2-k_2,0\right)}^{n_1,n_2,n_3}\left(\left\{P_i\right\}\right),
 \end{align}
where
\begin{multline}\label{fn1n2n3s1s2s3}
   f^{n_1,n_2,n_3}_{s_1,s_2,s_3}\left(\left\{\partial_{P_i}\right\},\left\{\partial_{P_i}\right\},\left\{Z_i\right\} \right) = \frac{2^{{\tilde s}_1+{\tilde s}_2+{\tilde s}_3}\left(\Delta_1+1-\tfrac{d}{2}\right)_{{\tilde s}_3}\left(\Delta_2+1-\tfrac{d}{2}\right)_{{\tilde s}_1}\left(\Delta_3+1-\tfrac{d}{2}\right)_{{\tilde s}_2}}{\left(\Delta_1-1\right)_{s_1}\left(\Delta_2-1\right)_{s_2}\left(\Delta_3-1\right)_{s_3}\left({\tilde s}_1\right)!\left({\tilde s}_2\right)!\left({\tilde s}_3\right)!}  \\ 
   \times  {\cal H}^{n_1}_1{\cal H}^{n_2}_2{\cal H}^{n_3}_3  {\bar {\cal H}}^{{\tilde s}_2}_1 {\bar {\cal H}}^{{\tilde s}_3}_2 {\bar {\cal H}}^{{\tilde s}_1}_3 {\cal D}_{P_1}^{s_1}
  {\cal D}_{P_2}^{s_2}
   {\cal D}_{P_3}^{s_3}\left({\bar U}_1 \cdot P_1 \right)^{{\tilde s}_3}\left({\bar U}_2 \cdot P_2 \right)^{{\tilde s}_2}\left({\bar U}_3 \cdot P_3 \right)^{{\tilde s}_1}
\end{multline}
where for concision we defined ${\tilde s}_i = s_i - n_{i-1}-n_{i+1}$ and $k_1 = n_2 + n_3$, $k_2 = n_1 + n_3$ and $k_3 = n_1 + n_2$, and introduced the auxiliary vector ${\bar U}_i$ which enters the contraction ${\bar {\cal H}}_i = \partial_{U_{i-1}} \cdot \partial_{{\bar U}_{i+1}}$. We illustrate this relation in figure \ref{fig::s1s2s3}.

\begin{figure}[h]
  \includegraphics[scale=0.45]{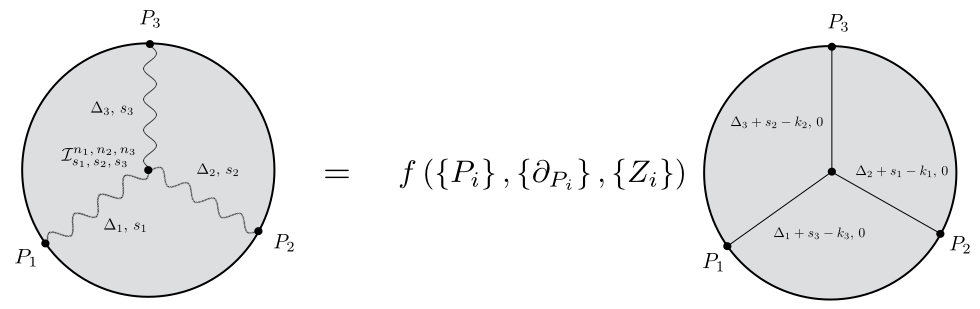} 
  \caption{Employing the tools given in \S \tcb{\ref{subsec::tools}}, we may relate the tree-level Witten diagram with external fields of spins $(s_1,s_2,s_3)$ and scaling dimensions $(\Delta_1,\Delta_2,\Delta_3)$ to that with only external scalars and dimensions $(\Delta_1+s_3-n_1-n_2,\Delta_2+s_1-n_2-n_3,\Delta_3+s_2-n_1-n_3)$. }\label{fig::s1s2s3}
\end{figure} 

Using the explicit result \eqref{d123} for the scalar Witten diagram, after evaluating all derivatives and some lengthy algebra, we can see how the amplitude generated by each individual bulk structure \eqref{is0} decomposes into conformal structures \eqref{6conf} on the boundary:\footnote{Where we defined:
\begin{equation}
\sum_{\alpha,\beta,\delta,\omega,\gamma}\equiv \sum_{\alpha_\kappa=0}^{s_\kappa-k_\kappa}\sum_{\beta_\kappa=0}^{k_{\kappa}}\sum_{\delta_{\kappa}=0}^{n_{\kappa}}\sum_{\omega_{\kappa}=0}^{\alpha_{\kappa-1}+\beta_{\kappa-1}}\sum_{\gamma_{\kappa}=0}^{\alpha_{\kappa-1}+\beta_{\kappa-1}}\,,
\end{equation}
} 
{\footnotesize \begin{multline}
    {\cal A}_{\left(\Delta_1,s_1\right),\left(\Delta_2,s_2\right),\left(\Delta_3,s_3\right)}^{n_1,n_2,n_3}\left(\left\{P_i\right\}|\left\{Z_i\right\}\right) \\ = \text{P}_3 \sum_{\alpha,\beta,\delta,\omega,\gamma}\prod_{i=1}^3\, (-1)^{s_i-n_i-\delta_i +\alpha_i +\beta_i} 2^{s_i-n_i-\gamma_i  -\delta_i -\omega_i }\frac{n_i! (\alpha_i +\beta_i)! (s_i-n_{i+1} -n_{i-1} )! }{\gamma_i!\delta_i!\alpha_i!\omega_i!(\beta_i+\delta_{i+1} -n_{i+1}   +1)!}\\
   \times\frac{\left(\alpha_i +\beta_i+\Delta_i\right)_{s_i+\delta_{i(i+1)} -\gamma_{i+1} -n_{i+1}  -\omega_{i+1} -\Delta_i}}{(\alpha_i +\beta_i -\gamma_{i+1}  -\gamma_{i-1}  -\omega_{i+1}+1)! (s_i-\alpha_i -n_{i+1} -n_{i-1} -\omega_{i-1}  +1)!(n_{i+1} +n_{i-1}-\beta_i -\delta_{i+1}  -\delta_{i-1}  +1)!}\\
     \times {\sf H}_1^{\gamma_1 +\delta_1 +\omega_1} {\sf H}_2^{\gamma_2 +\delta_2 +\omega_2 } {\sf H}_3^{\gamma_3 +\delta_3 +\omega_3 } {\sf Y}_1^{s_1-\gamma_2 -\gamma_3 -\delta_2 -\delta_3 -\omega_2 -\omega_3 } {\sf Y}_2^{s_2-\gamma_1 -\gamma_3 -\delta_1 -\delta_3 -\omega_1 -\omega_3 } {\sf Y}_3^{s_3-\gamma_1 -\gamma_2 -\delta_1 -\delta_2 -\omega_1 -\omega_2}\,,\nonumber
\end{multline}}
with the prefactor
\begin{multline}
\text{P}_3=\frac{1}{16\,\pi ^{d}}\frac1{(y_{12})^{\delta_{12}}(y_{23})^{\delta_{23}}(y_{31})^{\delta_{31}}}\,\Gamma \left(\sum_{\alpha}(\tfrac{\tau_\alpha}{2} +s_\alpha - n_\alpha ) -\tfrac{d}2\right)\prod_{i=1}^3\frac{ \Gamma ( \Delta_i-1)  ( \Delta_i+s_i -1) }{\, \Gamma \left(\Delta_i+1-\frac{d}{2}\right) }\,,\nonumber
\end{multline}
and
\begin{align}
    \delta_{ij}&=\frac12(\tau_i+\tau_j-\tau_k)\,,& \tau_i&=\Delta_i-s_i\,.
\end{align}
Although this result is quite involved, it will play an instrumental role in \S \tcb{\ref{subsec::cubicrecon}} to fix the cubic action of higher-spin gauge fields in AdS, by matching with the dual three-point correlators in the free scalar $O\left(N\right)$ vector model. As a double-check, with the above result we recover the same coefficient as for the $(s,0,0)$ case \eqref{00sW} when any two of the $s_i$ are set to zero.

\section{Four-point Witten diagrams}
\label{sec::4ptwittpoo}

In this section we consider four-point Witten diagrams with identical parity even external scalar fields. At tree level, there are two possible processes: A contact interaction or the exchange of a field between two pairs of the scalars. These two possibilities are illustrated in figure \ref{fig:4pt}.\\

\begin{figure}[h]
\centering
\begin{subfigure}{0.31\textwidth}
\includegraphics[width=\linewidth]{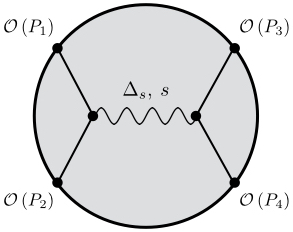}
\caption{Exchange of a spin-$s$ field of dimension $\Delta_s$.} \label{fig:4ptexch}
\end{subfigure}
\hspace*{3cm} 
\begin{subfigure}{0.31\textwidth}
\includegraphics[scale=0.45]{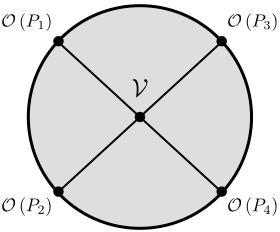}
\caption{Contact diagram generated by a quartic self  interaction ${\cal V}$ of the bulk scalar.} \label{fig:4ptcont}
\end{subfigure}
\caption{Possible tree-level bulk processes with four identical external scalars.} \label{fig:4pt}
\end{figure}

\noindent
In the present context of higher-spin gauge theories on AdS, the exchanged fields are bosonic gauge fields of integer spin, in addition to the bulk scalar itself. The former have dimension $\Delta_s = s+d-2$, while the parity even scalar of the type A theory has dimension $\Delta = d-2$.\footnote{See \cite{Francia:2007qt,Francia:2008hd} for previous work on exchange diagrams in higher-spin gauge theories on anti de Sitter backgrounds.}

Ultimately, our goal is to identify the above four-point Witten diagrams in the type A minimal bosonic higher-spin theory with the dual scalar operator four-point function in the free scalar $O\left(N\right)$ vector model. This is in order to extract the on-shell quartic self-interaction of the bulk scalar from its corresponding contact diagram (figure \ref{fig:4ptcont}). To compare the Witten diagrams with their dual CFT correlator, we put them both on an equal footing by decomposing them into conformal blocks. For the  correlator of the scalar operator ${\cal O}$ in the free scalar $O\left(N\right)$ model, this was carried out in \S \tcb{\ref{subsec::4ptscalar}}. In the following we introduce a method to decompose four-point Witten diagrams into their constituent conformal blocks, which draws upon the results for the three-point Witten diagrams obtained in the previous sections.

\subsection{Conformal partial wave expansion}

The key idea behind our method to determine the conformal block expansion of four-point Witten diagrams, is to decompose the them into partial waves of the isometry group $SO\left(d+1,1\right)$. I.e. in terms of AdS harmonic functions labelled by energy and spin,
\begin{equation}\label{graphi::pwe}
    \includegraphics[scale=0.45]{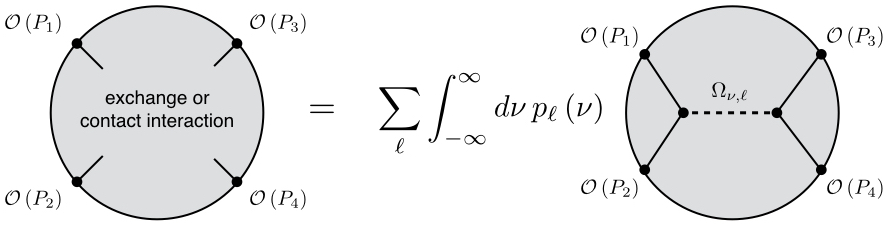}.
\end{equation}
Since $SO\left(d+1,1\right)$ is also the $d$-dimensional Euclidean conformal group, this way of decomposing Witten diagrams is equivalent to a conformal block expansion on the boundary. We make this identification more concrete in the following, which draws upon a particular feature of harmonic functions, known as their \emph{split representation}.\footnote{See also an alternative approach \cite{Hijano:2015zsa}, centered on the bulk objects ``geodesic Witten diagrams'' interpreted as the holographic counterparts to conformal blocks. This was extended to the case with a single external spinning field in: \cite{Nishida:2016vds}. See also \cite{Chen:2017yia} for some work towards linking these two approaches.} See e.g. \cite{Penedones:2007ns,Costa:2014kfa} for previous related investigations.

\subsubsection{Split representation}
As projectors onto angular momentum and energy components, harmonic functions admit the factorisation \cite{PhysRevD.10.589,Fronsdal:1978vb,Leonhardt:2003qu,Leonhardt:2003sn,Costa:2014kfa}
\begin{equation} \label{splitharm}
    \Omega_{\nu,\ell}\left(x_1, w_1;x_2, w_2\right) = \frac{\nu^2}{\pi \ell!\left(\frac{d}{2}-1\right)_\ell} \int_{\partial\text{AdS}}d^dy\, K_{\frac{d}{2}+i\nu,\ell}(x_1,w_1;y,{\hat \partial}_w) K_{\frac{d}{2}-i\nu,\ell}\left(y,w;x_2,w_2\right),\end{equation}
which is depicted in figure \ref{fig::harmonic}.

\begin{figure}[h]
  \centering
  \includegraphics[scale=0.5]{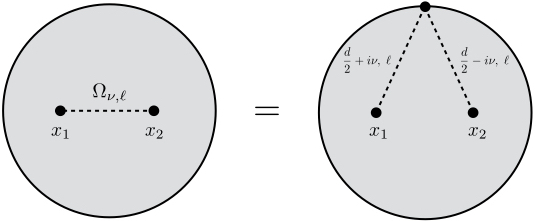}
  \caption{Factorisation of an AdS harmonic function $\Omega_{\nu,\ell}$ into a product of two spin-$\ell$ boundary-to-bulk propagators of dimension $\frac{d}{2} \pm i\nu$, integrated over their common boundary point.} \label{fig::harmonic}
\end{figure}

\noindent
Drawing comparisons with the conformal block expansion in CFT, the split representation \eqref{splitharm} makes manifest that harmonic functions $\Omega_{\nu,\ell}$ are the bulk analogue of the boundary projector \eqref{cftcbproj} onto a given conformal multiplet. Indeed, inserting this form for the harmonic functions into the partial wave expansion \eqref{graphi::pwe} of a four-point Witten diagram causes each partial wave to factorise into a product of two three-point Witten diagrams,
\begin{equation}
    \includegraphics[scale=0.5]{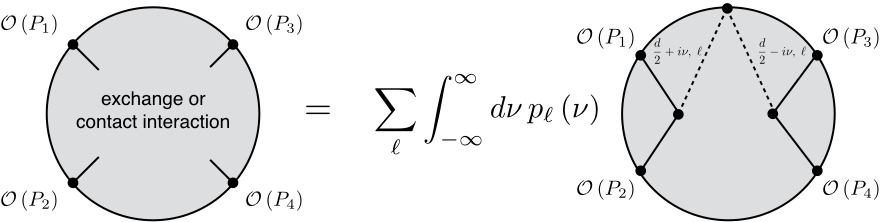}.
\end{equation}
Evaluating the two integrals over AdS in each bulk partial wave gives the unique structure of a product of two three-point correlation functions with two scalar operator insertions, as in the analogous representation of boundary conformal partial waves \eqref{factcbcont}. The bulk partial wave expansion \eqref{graphi::pwe} thus induces a conformal partial wave expansion on the boundary, in the contour integral form \eqref{contintform}.

In the following we apply this technology to derive the conformal partial wave expansions of four-point Witten diagrams with identical external scalars (figure \ref{fig:4pt}). In particular, for the exchange diagrams we focus on the exchange of bosonic gauge fields of arbitrary spin.

\subsection{Exchange diagrams}
\label{subsec::4ptexch}
We first determine the conformal partial wave expansion of the Witten diagram corresponding to the tree-level exchange of a spin-$s$ gauge field between two pairs of real scalars in the s-channel (figure \ref{fig:4ptexch} for $\Delta_s = s+d-2$).\footnote{Since the external scalars are identical, expansions for the t- and u- channel exchanges can be obtained from the s-channel result by permuting the external legs.}
As previously explained, the cubic vertex mediating this exchange is unique on-shell, and for consistency with the gauge symmetry is required to be of the Noether type
\begin{equation} \label{witten::jvert}
    {\cal V}_{s,0,0}\left(x\right) = g_s\, s!\, \varphi_s\left(x;\partial_u\right) J_s\left(x,u\right), \quad \left(\nabla \cdot \partial_u\right) J_s\left(x,u\right) \approx 0,
\end{equation}
with coupling $g_s$, and where \cite{Bekaert:2010hk}
\begin{equation}
    J_s\left(x,u\right) = \sum\limits^s_{k=0} \frac{\left(-1\right)^k}{s!\left(s-k\right)!} \left(u \cdot \nabla \right)^k \phi\left(x\right) \left(u \cdot \nabla \right)^{s-k} \phi\left(x\right) + \Lambda\, u^2\left(...\right).\label{cur1}
\end{equation}
The second term that is proportional to the cosmological constant $\Lambda$ is pure trace and vanishes in the flat-space limit, $\Lambda \rightarrow 0$.

Putting the ingredients together, the four-point amplitude takes the form
\begin{align}\label{exchfirst}
  &  {\cal A}^{\text{exch.}}_s\left(y_1,y_2;y_3,y_4\right) \\ \nonumber
    & \hspace*{1.25cm} = g^2_s\,\int_{\text{AdS}} d^{d+1}x_1 \int_{\text{AdS}} d^{d+1}x_2\, \Pi_{s}\left(x_1, \partial_{u_1};x_2, \partial_{u_2}\right) J_{s}\left(x_1,u_1;y_1;y_2\right)J_{s}\left(x_2,u_2;y_3;y_4\right),
\end{align}
where $\Pi_s$ is the spin-$s$ bulk-to-bulk propagator, and the notation $J_{s}\left(x,u;y_i;y_j\right)$ denotes the insertion of scalar boundary-to-bulk propagators anchored at the boundary points $y_i$ and $y_j$.

In \S \tcb{\ref{subsec::bbprop}}, we derived bulk-to-bulk propagators gauge fields of arbitrary integer spin in a basis of harmonic functions. This way of representing the propagators therefore turned out to be beneficial, as it readily establishes the partial wave expansion of the bulk amplitude and the subsequent factorisation of each partial wave into three-point Witten diagrams. In practice, it is most convenient to employ the propagator in the manifest trace gauge \eqref{eq:ManifestTrace}, as it eliminates the need to deal directly with derivatives. The exchange then takes the form
\begin{align}
    {\cal A}^{\text{exch.}}_s\left(y_1,y_2;y_3,y_4\right) =  \sum\limits^{\left[s/2\right]}_{k=0}\int^{\infty}_{-\infty} d\nu\, g_{s,k}\left(\nu\right)\,{\bar {\cal A}}^{\text{exch.}}_{s,s-2k}\left(y_1,y_2;y_3,y_4\right),
\end{align}
with factorised partial amplitudes
\begin{align}\label{witt::prod3pt}
    &{\bar {\cal A}}^{\text{exch.}}_{s,s-2k}\left(y_1,y_2;y_3,y_4\right) = \frac{\nu^2}{\pi \left(s-2k\right)!\left(\frac{d}{2}-1\right)_{s-2k}}\int_{\partial \text{AdS}}d^dy \\ \nonumber
    & \hspace*{5cm} g_s\, s!\,\int_{\text{AdS}}d^{d+1}x_1 K_{\frac{d}{2}+i\nu,s-2k}(x_1, \partial_{u_1};y,{\hat \partial}_z)J^{\left(k\right)}_{s}\left(x_1,u_1;y_1;y_2\right) \\ \nonumber
    & \hspace*{4.5cm} \times g_s\, s!\, \int_{\text{AdS}}d^{d+1}x_2 K_{\frac{d}{2}-i\nu,s-2k}(x_2, \partial_{u_2};y,z) J^{\left(k\right)}_{s}\left(x_2,u_2;y_3;y_4\right), \\ \nonumber
    & \hspace*{3.75cm} = \frac{\nu^2}{\pi \left(s-2k\right)!\left(\frac{d}{2}-1\right)_{s-2k}}\int_{\partial \text{AdS}}d^dy \\ \nonumber
    & \hspace*{6cm} \times {\cal A}_{\Delta, \Delta, \frac{d}{2} + i\nu,s-2k}(y_1,y_2,y|{\hat \partial}_z)\,  {\cal A}_{\Delta, \Delta, \frac{d}{2} - i\nu,s-2k}\left(y,y_3,y_4|z\right)
    \end{align}
where $J^{\left(k\right)}_{s}$ denotes the $k$-th trace of $J_s$. Each of the above partial amplitudes is a product of three-point Witten diagrams ${\cal A}_{\Delta, \Delta, \frac{d}{2}\pm i\nu,s-2k}$, generated by cubic vertices of the form
\begin{align} \label{exchvertbar}
    {\bar {\cal V}}_{\frac{d}{2}\pm i\nu,s-2k} & = g_s\, \varphi_{\frac{d}{2}\pm i\nu,\,s-2k}\left(x,\partial_u\right) J^{\left(k\right)}_s\left(x,u\right),
    \end{align}
where $\varphi_{\frac{d}{2}\pm i\nu,\,s-2k}$ is a spin-($s\,$--$\,2k$) field of dual scaling dimension $\frac{d}{2} \pm i\nu$.

The next step is to evaluate the amplitudes  ${\cal A}_{\Delta, \Delta, \frac{d}{2}\pm i\nu,s-2k}$ in \eqref{witt::prod3pt}. As previously noted, cubic vertices involving two scalars and a (massive or massless) spinning field are unique on-shell. This has the implication that the vertex \eqref{exchvertbar} above can be expressed in terms of the basic vertex \eqref{basic}, modulo terms which do not contribute to the tree-level three-point amplitudes. Relegating the details to \S \tcb{\ref{app::hsccads}}, this relation is
    \begin{align}\label{sec::wiit::ta}
   {\bar {\cal V}}_{\frac{d}{2}\pm i\nu,s-2k} & \quad \approx \quad g_s\,2^{s-2k}\,\tau_{s,k}\left(\nu\right) \varphi_{\frac{d}{2}\pm i\nu,s-2k}\left(x,\partial_u\right) \varphi_0\left(x\right) \left(u \cdot \nabla\right)^{s-2k}\varphi_0\left(x\right),
    \end{align}
with 
\begin{multline}
    \tau_{s,k}\left(\nu\right)=\sum_{m=0}^k 
\frac{2^{2k}k!}{m!(k-m)!}
(-1)^{k-m}\left(\tfrac{d}{2}-\tfrac{3}{2}\right)_{k-m}\left(\tfrac{5}{2}-\tfrac{d}{2}\right)_{k-m}\\ \times \left(\frac{\tfrac{d}{2}+s-2m+1+i\nu}{2}\right)_{m}\left(\frac{\tfrac{d}{2}+s-2m+1-i\nu}{2}\right)_{m}.
\end{multline}
In this way we may we the result \eqref{damp} for the three-point amplitude of the basic vertex \eqref{basic}, which gives\footnote{We divide by the bulk-to-boundary propagator normalisation $C_{\Delta,0}$, equation \eqref{poonorm}, to normalise the two-point functions canonically:
\begin{equation}
    \langle {\cal O}\left(y_1\right) {\cal O}\left(y_2\right)  \rangle = \frac{1}{\left(y^2_{12}\right)^{\Delta}}.
\end{equation}}
\begin{align}
   & {\cal A}_{\Delta, \Delta, \frac{d}{2}\pm i\nu,s-2k}\left(y_1,y_2,y|z\right) \\ \nonumber
    & \hspace*{2cm} = \frac{g_s\, 2^{s-2k}\, \tau_{s,k}\left(\nu\right)}{C_{\Delta,0}}{\hat {\cal A}}_{\Delta, \Delta, \frac{d}{2}\pm i\nu,s-2k}\left(y_1,y_2,y|z\right) \nonumber \\ \nonumber
    & \hspace*{2cm} = \frac{g_s\, 2^{s-2k}\, \tau_{s,k}\left(\nu\right)}{C_{\Delta,0}} {\sf  C}(\Delta,\Delta,\tfrac{d}{2}\pm i \nu;s-2k) \langle \langle {\cal O}\left(y_1\right) {\cal O}\left(y_2\right) {\cal O}_{\frac{d}{2}\pm i\nu,s-2k}\left(y|z\right) \rangle  \rangle ,
\end{align}
where the $\langle \langle \bullet \rangle \rangle$ notation was introduced in \S \tcb{\ref{subsec::intrep}}.\footnote{
Note that 
\begin{align}
{\sf C}(\Delta ,\Delta ,\tfrac{d}{2}-i \nu ,s-2k) = \frac{i {\sf C}\left(\Delta ,\Delta ,\frac{d}{2}+i \nu ,s-2k\right)}{\nu \,C_{\frac{d}{2}+i\nu,s-2k}} \frac{ \Gamma \left(\frac{\frac{d}{2}-i\nu +s-2k}{2}  \right)^2 }{2  \Gamma \left(\frac{\frac{d}{2}+i\nu +s-2k}{2} \right)^2}\frac{\kappa_{\frac{d}{2}-i \nu ,s-2k}}{ \left(2 \pi \right)^{\frac{d}{4}}},
\end{align}
with $\kappa_{\frac{d}{2}-i \nu ,s-2k}$ defined in \eqref{kappapoods}.}

With this way of decomposing the exchange amplitude, we may readily apply the techniques introduced in \S \tcb{\ref{subsec::intrep}} to establish the conformal partial wave expansion. By simply recalling that 
{\footnotesize\begin{align}
& W_{\frac{d}{2}+i\nu,s-2k}\left(y_1,y_2; y_3,y_4\right) + W_{\frac{d}{2}-i\nu,s-2k}\left(y_1,y_2; y_3,y_4\right) \\ \nonumber
& \hspace*{0.35cm} =  \frac{\kappa_{\frac{d}{2}-i\nu,s-2k}}{\left(2\pi\right)^{\frac{d}{4}}} \frac{\Gamma\left(\frac{\frac{d}{2}-i\nu +s-2k}{2}\right)^2}{\Gamma\left(\frac{\frac{d}{2}+i\nu +s-2k}{2}\right)^2} \int d^dy\, \langle \langle {\cal O}\left(y_1\right){\cal O}\left(y_2\right){\cal O}_{\frac{d}{2}+i\nu,s-2k}\left(y\right) \rangle \rangle \langle \langle {\tilde {\cal O}}_{\frac{d}{2}-i\nu,s-2k}\left(y\right){\cal O}\left(y_3\right){\cal O}\left(y_4\right) \rangle \rangle, 
\end{align}}
\noindent
where $W_{\frac{d}{2}\pm i\nu,s-2k}$ denote conformal partial waves for $\left[\frac{d}{2}\pm i\nu,s-2k\right]$ conformal multiplets, we may write down the conformal partial wave expansion for the spin-$s$ exchange diagram in the contour integral form:
\begin{align} \label{tlpwe}
\mathcal{A}^{\text{exch.}}_{s}\left(y_1,y_2; y_3,y_4\right) =   \sum^{s}_{k=0} \int^{\infty}_{-\infty} d\nu \; b_{s-2k}\left(\nu\right) W_{\frac{d}{2}+i\nu,s-2k}\left(y_1,y_2;y_3,y_4\right),  
\end{align} with
\begin{align}
b_{s-2k}(\nu)= \frac{i \nu}{\pi} g_{s,k}(\nu)\tau^2_{s,k}(\nu) \frac{\left(g_{s}\,2^{s-2k}{\sf C}(\Delta,\Delta,\tfrac{d}{2}+i\nu,s-2k)\right)^2}{C_{\frac{d}{2}+i\nu,s-2k}C^2_{\Delta,0}}.  
\label{bnu}
\end{align}
We recall (\S \tcb{\ref{subsec::intrep}}) that the poles of the function $b_{s-2k}(\nu)$ in the lower-half complex plane encode the contributions to the conformal block expansion: A pole located at $\frac{d}{2}+i\nu = \Delta$ corresponds to the contribution from a conformal multiplet with quantum numbers $\left[\Delta,s-2k\right]$. We discuss these contributions in detail for the spin-$s$ exchange in the following. 

\subsubsection{Conformal block expansion contributions}

It is illuminating to understand the contributions to the conformal block expansions of Witten diagrams, for they give a means to quantify the properties of the process that occurs in the bulk. For example, the properties of the intermediate states.

Focusing first on the spin-$s$ contribution, \eqref{bnu} for $k=0$, explicitly we have
\begin{align} \label{bsnu}
b_{s}\left(\nu\right) & = \frac{i \nu}{\pi} \frac{1}{\nu^2+\left(\Delta+s-\frac{d}{2}\right)^2}\frac{\left(g_{s}\,2^{s}{\sf C}(\Delta,\Delta,\tfrac{d}{2}+i\nu,s)\right)^2}{C_{\frac{d}{2}+i\nu,s}C^2_{\Delta,0}}.
\end{align}
This encodes two types of contribution: Single trace and double-trace of spin-$s$.
\begin{itemize}
    \item {\bf Single-trace} \\
    
    The factor $g_{s,0}\left(\nu\right) = \frac{1}{\nu^2+\left(\Delta+s-\frac{d}{2}\right)^2}$ coming from the on-shell  (traceless and transverse) part of the spin-$s$ bulk-to-bulk propagator contains a pole for Im$\left(\nu\right) < 0$ at $\frac{d}{2}+i\nu = \Delta+s = s+d-2$. This corresponds to a contribution from a conformal multiplet of a spin-$s$ single-trace conserved current ${\cal J}_s$. It is consistent with the notion that a single-trace operator in the CFT corresponds to a single-particle state in the bulk: The exchange of a single-trace conserved current in the boundary four-point function corresponds to the exchange of the dual gauge field  in AdS.

    Furthermore, recalling that $g_s 2^s {\sf C}(\Delta,\Delta,\Delta+s,s)$ is the overall coefficient of the dual $\left(0,0,s\right)$ CFT correlator as generated by the bulk cubic vertex \eqref{witten::jvert}, we may identify the following combination with the OPE coefficient ${\sf c}_{{\cal O}{\cal O}{\cal J}_s}$ at leading order in $N_{\text{dof.}}$:\footnote{The definition of the OPE coefficient is given in terms of those of three- and two-point functions \eqref{ope32}. Recall also that the bulk-to-boundary propagator coefficients $C_{\Delta,s}$ give the normalisations of the two-point functions from the bulk side.}
    \begin{equation}
        \frac{\left(g_{s}\,2^{s}{\sf C}(\Delta,\Delta,\Delta+s,s)\right)^2}{C_{\Delta+s,s}C^2_{\Delta,0}} = {\sf c}^2_{{\cal O}{\cal O}{\cal J}_s}.
    \end{equation}
   In applying Cauchy's residue theorem to \eqref{tlpwe} about the pole $\frac{d}{2}+i\nu = \Delta+s$, the spin-$s$ single-trace contribution to the spin-$s$ exchange is thus
    \begin{equation}\label{stcontr}
        \mathcal{A}^{\text{exch.}}_{s}\left(y_1,y_2; y_3,y_4\right)\Big|_{\text{spin}-s\, \text{single-trace}} = {\sf c}^2_{{\cal O}{\cal O}{\cal J}_s} W_{\Delta+s,s}\left(y_1,y_2;y_3,y_4\right).
    \end{equation}
    This precisely coincides with the form of the spin-$s$ single-trace contribution in the (12)(34) channel expansion of the dual CFT correlator.
    \item {\bf Double-trace} \\
        
    The remaining contributions come from the factor 
   \begin{align}
      &  {\sf C}(\Delta,\Delta,\tfrac{d}{2}+i\nu,s)\\ \nonumber
        & \hspace*{1cm} = \frac{ 2^{-\frac{d}{2}-i \nu -2} \Gamma \left(\frac{d}{2}+i \nu -1\right) \Gamma \left(\frac{d+2 s+2 i \nu }{4}\right) \Gamma \left(-\frac{d}{4}+\frac{s}{2}+\Delta +\frac{i \nu }{2}\right)}{\pi ^{d-\frac{1}{2}}\Gamma (i \nu +1) \Gamma \left(-\frac{d}{2}+\Delta +1\right)^2 \Gamma \left(\frac{d+2 s+2 i \nu -2}{4}\right)}\Gamma \left(\tfrac{s+2\Delta -\frac{d}{2}-\frac{i \nu }{2}}{2}\right),
    \end{align}
    in which the Gamma function $ \Gamma \left(\frac{s+2\Delta -\frac{d}{2}-\frac{i \nu }{2}}{2}\right)$ generates an infinite string of poles at $\frac{d}{2}+i\nu = 2\Delta + 2n+s$, for each $n = 0, 1, 2, 3, ...\,.$\footnote{Recall that the gamma function $\Gamma\left(x\right)$ has poles at $x = 0, -1, -2, -3, ...\,$.} These correspond to contributions from the family of spin-$s$ double-trace operators \eqref{dto}.

    Double-trace contributions originate from contact terms in the bulk-to-bulk propagator (note the Dirac delta function on the RHS of its equation of motion \eqref{propdef}), and therefore arise from the limit of coincidence between the two-points where the spin-$s$ particle is exchanged in AdS. This is consistent with the interpretation that double-trace operators are dual to two-particle states of the scalar in the bulk.
\end{itemize}

As we shall discuss below, for the exchange in AdS$_4$ we need only consider the spin-$s$ contributions above. However let us comment that the spin $s-2k$ contributions in \eqref{bnu} for $k \ne 0$ take the form
\begin{align} \nonumber
b_{s-2k}\left(\nu\right) = \left(\text{no poles in Im}\left(\nu\right) < 0\right) \times \Gamma \left(\tfrac{2\Delta+s-2k-\frac{d}{2}-i \nu}{2}\right)^2,
\end{align}
and therefore give rise to contributions from double-trace operators of scaling dimension $2\Delta+2n+s-2k$ and spin $s-2k$. In the same way as for the contributions from the spin-$s$ double-trace operators above, these also arise from limit of coincidence between the two-points in the bulk.

\begin{figure}[h]
  \centering
  \includegraphics[scale=0.45]{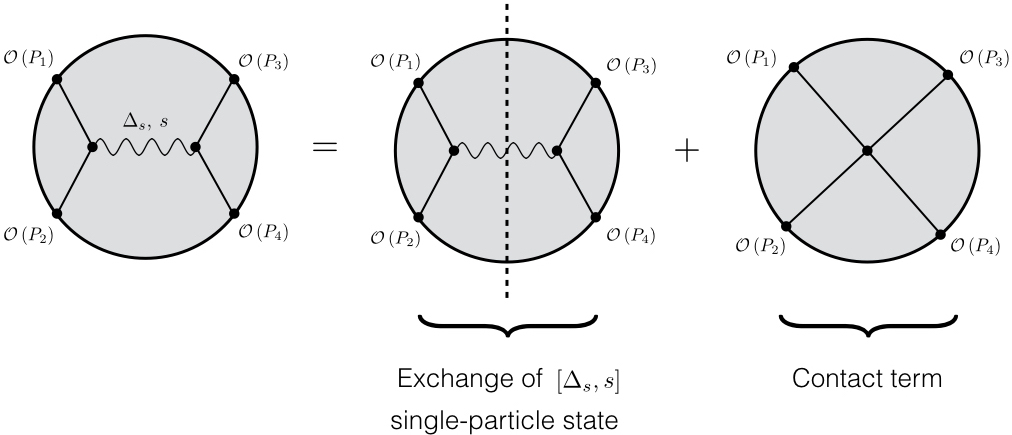}
  \caption{An exchange diagram can be divided into two types of contributions: When the points are separated (represented by a dashed line), there is a genuine exchange of a single-particle state. At coinciding points, contact terms arise from the off-shell bulk-to-bulk propagator: $\Box\,  \Pi_s\left(x_1,x_2\right) + ... = \delta\left(x_1,x_2\right)$. The latter generate double-trace contributions in the conformal block expansion, while the exchanged single-particle state generates the contribution from its dual single-trace operator.} 
\end{figure}

\subsubsection{Exchange in AdS$_4$}
 
In \S \tcb{\ref{subsec::quarticextr}} we will employ the above result for the conformal block expansion of the spin-$s$ exchange to extract the quartic self interaction of the scalar in the type A minimal bosonic higher-spin theory on AdS$_4$. That we work in AdS$_4$ is no coincidence, for in this case the bulk scalar is conformally coupled. This provides certain simplifications, which we discuss below.

A particular consequence of having a conformally coupled scalar, is that the spin-$s$ conserved currents \eqref{cur1} built from it can be taken to be traceless. Since the scalar is conformal, its energy momentum tensor can be made traceless. The currents of higher-spin sit in the same higher-spin symmetry multiplet as the energy momentum tensor, and they are also traceless as a result. Choosing a traceless conserved current to enter the cubic vertex \eqref{witten::jvert}, means that the contributions to the conformal block expansion of spin less than $s$ are vanishing: ${\bar {\cal A}}^{\text{exch.}}_{s,s-2k} \equiv 0$. This is a dramatic simplification, for then only spin-$s$ conformal blocks contribute to the spin-$s$ exchange; there is no sum over $k$ in \eqref{tlpwe}.  Moreover, these contributions are encoded in a single meromorphic function $b_{s}\left(\nu\right)$, equation \eqref{bsnu}.

\subsection{Contact diagrams}
\label{subsec::cd}
In this section we determine the conformal partial wave expansion of a general four-point contact Witten diagram with identical external scalars.
For similar reasons as described in the previous section, we focus our attention on contact amplitudes in AdS$_4$. As will be explained, since the scalar is conformal in AdS$_4$ a particularly useful basis for their quartic interaction may be chosen.

A generic quartic self-interaction of a scalar $\varphi_0$ in AdS space has the form
\begin{align}
   \Box^{n_1}\left( \nabla ...\nabla \varphi_0\right)\, \Box^{n_2} \left( \nabla ...\nabla \varphi_0\right)\, \Box^{n_3} \left( \nabla ...\nabla \varphi_0\right)\, \Box^{n_4} \left( \nabla ...\nabla \varphi_0\right), \quad n_i \in \mathbb{N},  
\end{align}
i.e. each $\varphi_0$ is dressed with covariant derivatives $\nabla$, which are contracted amongst each other. Not all such vertices are independent: First of all, since the covariant derivatives do not commute there is a choice of ordering. There is also freedom to distribute derivatives amongst the four scalars by integrating by parts. Going  on-shell, as relevant for four-point contact Witten diagrams, vertices can also be related using the equation of motion for $\varphi_0$.

An example of a complete basis of on-shell quartic self-interactions can be constructed from the spin-$s$ conserved currents \eqref{cur1}, which are bi-linear in the bulk scalar 
\begin{equation}
    J_{\mu_1 ...\mu_s}\left(x\right) = \varphi_0\left(x\right) \nabla_{\left(\mu_1\right.}...\nabla_{\left.\mu_s\right)} \varphi_0\left(x\right) + ...\,.
\end{equation}
In \S \tcb{\ref{appendix::quarticbasis}} it is shown that an example of such a basis is given by the collection of structures
\begin{equation}\label{contwbasistf}
    {\hat {\cal V}}_{s,m} = J_s\left(x\right) \cdot \Box^m(J_s\left(x\right)), \quad s=2k, \quad k \ge m \ge 0, \quad k,\, m \in \mathbb{N},
\end{equation}
and accounts for all independent quartic self-interactions of the scalar in AdS$_{d+1}$. 
As will become clear, this choice of basis allows us to re-cycle some of the steps taken in computing the partial wave expansion of the exchange diagrams in the previous section. While on the one hand this reduces the amount of extra work, it also facilitates a comparison between the features of exchanges and contact diagrams.

 A generic quartic self-interaction in AdS$_{d+1}$ can therefore be expressed as the linear combination
\begin{equation}
    {\cal V}_{0,0,0,0} = \sum\limits_{s,m} a_{s,m} {\hat {\cal V}}_{s,m}, \qquad a_{s,m} \in \mathbb{R},
\end{equation}
for some coefficients $a_{s,m}$. To determine the conformal partial wave expansion of a generic four-point contact diagram, it is therefore sufficient to determine the conformal partial wave expansions of the contact amplitudes generated by the basis vertices \eqref{contwbasistf}.

For the remainder of this section, we therefore consider the tree-level four-point contact diagram associated to the vertex ${\hat {\cal V}}_{s,m}$. Furthermore, for simplicity we restrict to the AdS$_4$ case, where we may instead use the traceless improvement ${\tilde J}_s$ in place of $J_s$. That this still provides a complete basis is shown in \S \tcb{\ref{appendix::quarticbasis}}.

The first step in determining the conformal partial wave expansion of a given contact diagram, is to decompose the corresponding interaction into into AdS harmonic functions. This can be achieved by simply inserting the completeness relation \eqref{acomp} for harmonic functions between the two conserved currents in the vertex \eqref{contwbasistf}, through which we obtain 
\begin{align} \label{harmovert}
   {\hat {\cal V}}_{s,m}\left(x_1\right) & = ({\hat \partial}_{w_1} \cdot {\hat \partial}_{w_2})^s{\tilde J}_s(x_1,w_1)\, \Box^m({\tilde J}_s\left(x_1,w_2\right)) \\ \nonumber
    &  =  \quad \sum\limits^s_{\ell=0} \int^{\infty}_{-\infty} d\nu\, c_{s,\ell}\left(\nu\right)\int_{\text{AdS}}d^{4}x_2\, ({\hat \partial}_{w_1} \cdot \nabla)^\ell ({\hat \partial}_{w_2} \cdot \nabla)^\ell \Omega_{\nu,s-\ell}(x_1,{\hat \partial}_{w_1};x_2,{\hat \partial}_{w_2}) \\ \nonumber
    & \hspace*{9cm} \times  {\tilde J}_s(x_1,w_1)\Box^m({\tilde J}_s\left(x_2,w_2\right)) \\ \nonumber
    &   \approx \quad  \int^{\infty}_{-\infty} d\nu \left(-1\right)^m\left(\nu^2+s+\tfrac{9}{4}\right)^m \int_{\text{AdS}}d^{4}x_2\;  \Omega_{\nu,s}(x_1,{\hat \partial}_{w_1};x_2,{\hat \partial}_{w_2}) \\ \nonumber 
    &  \hspace*{9cm} \times {\tilde J}_s(x_1,w_1){\tilde J}_s\left(x_2,w_2\right),
\end{align}
where in the second equality we used conservation of the spin-$s$ current, and the equation of motion for the harmonic function $\Omega_{\nu,s}$. The function $c_{s,\ell}\left(\nu\right)$ originates from the completeness relation \eqref{acomp} for the harmonic functions. Owing to the tracelessness of the currents, only the term $\ell=s$ contributes.

With the above form for the vertex, we can determine the partial wave expansion for the corresponding bulk amplitude in the same way as for the exchange diagrams in the previous section. For a given configuration of external legs, we have
\begin{align} 
&\mathcal{A}^{\text{cont.}}_{m,s}\left(y_1,y_2; y_3,y_4\right) = \int^{\infty}_{-\infty} d\nu \left(-1\right)^m \left(\nu^2 + s + \tfrac{9}{4}\right)^m \frac{\nu^2}{\pi} \int_{\partial \text{AdS}} d^{3}y \; \frac{1}{s!\left(\tfrac{1}{2}\right)_{s}} \\ \nonumber
& \hspace*{5cm} \times \int_{\text{AdS}} \sqrt{\left|g\right|}\; d^{4}x_1\; {\tilde J}_{s}\left(x_1, \partial_{u_1}; y_1, y_2\right)  K_{\tfrac{3}{2} + i\nu, s}(x_1,u_1; y, \hat{\partial}_z) \\ \nonumber
& \hspace*{5cm} \times  \int_{\text{AdS}} \sqrt{\left|g\right|}\; d^{4}x_2\;  {\tilde J}_{s}\left(x_2, \partial_{u_2}; y_3, y_4\right)  K_{\tfrac{3}{2} - i\nu, s}\left(x_2,u_2; y, z \right).
\end{align}
The utility of the basis \eqref{contwbasistf} becomes manifest when one notices that the product of three-point Witten diagrams in the above is precisely the same as that appearing in the corresponding result for the spin-$s$ exchange \eqref{witt::prod3pt}. This allows us to immediately write down the conformal partial wave expansion 
\begin{align} \label{contactpwe}
& \mathcal{A}^{\text{cont.}}_{m,s}\left(y_1,y_2; y_3,y_4\right) \\ \nonumber
& \hspace*{2.5cm} =  \int^{\infty}_{-\infty} d\nu \left(-1\right)^m\left(\nu^2 + s + \tfrac{d^2}{4}\right)^m  \frac{i \nu}{\pi} \frac{\left(2^{s}{\sf C}(\Delta,\Delta,\tfrac{d}{2}+i\nu,s)\right)^2}{C_{\frac{d}{2}+i\nu,s}C^2_{\Delta,0}}\,
W_{\tfrac{d}{2}+i \nu,s}\left(u,v\right),
\end{align}
where we must keep in mind that $d=3$.

In this case, since the only poles with Im$\left(\nu\right) < 0$ come from the factor ${\sf C}(\Delta,\Delta,\tfrac{d}{2}+i\nu,s)$, the only contributions to the conformal block expansion come from spin-$s$ double-trace operators. The absence of any single-trace contribution is consistent, since there is no exchange of a single-particle state in a local contact interaction.

With the above result \eqref{contactpwe} for a given basis vertex, the conformal block expansion of a contact diagram generated by a generic quartic scalar self interaction follows by expressing it in the basis \eqref{contwbasistf}.

Let us note that while in this section we restricted for simplicity to the AdS$_4$ case, the same procedure applies in general dimensions. The difference is that generally it is not possible to use traceless currents in the basis \eqref{contwbasistf}. In general dimensions, the use of \emph{traceful} currents introduces contributions for all $\ell \leq s$ in \eqref{harmovert}. This gives rise  to contributions from double-trace operators for all spins $\ell \leq s$ in the subsequent conformal block expansion \eqref{contactpwe}. As a final comment, in applying the above approach it is not necessary to use a basis of quartic vertices built from conserved currents. It applies to any quartic vertex, and the choice \eqref{contwbasistf} was made for ease of comparison with the exchange diagrams.
  
 \chapter{Holographic reconstruction}
\label{chapt::hr}
This chapter is a culmination of the results established in the previous, in which we fix the complete cubic action and the quartic scalar self-interaction in the type A minimal bosonic higher-spin theory on AdS. We do so by employing its conjectured duality with the free scalar $O\left(N\right)$ vector model, by matching three- and four-point functions of single-trace operators in the latter (computed in \S \tcb{\ref{sec::veccor}}) with their dual Witten diagrams in the higher-spin theory (computed in \S \tcb{\ref{chapt::witten}}). In extracting the quartic scalar interaction, we employ the conformal partial wave expansion in CFT \S\tcb{\ref{sec::cbd}} and \S \tcb{\ref{subsec::4ptscalar}}, and for four-point Witten diagrams \S \tcb{\ref{sec::4ptwittpoo}}.

We begin by reviewing the standard non-holographic Noether approach to constructing Lagrangian interactions in higher-spin theories, exhibiting the existing results upon which we build in this work.

\section{Interactions in higher-spin theory}

One of the main tools for constructing interactions in higher-spin theories is the Noether method. Any theory whose non-linear form is determined by a gauge principle can be constructed by a Noether procedure, which was first understood in the context of gravity in \cite{Papapetrou:1948jw,Gupta:1952zz,Kraichnan:1955zz,0143-0807-24-3-702,Deser:1969wk} and  proved highly successful in the construction of supergravity theories \cite{Freedman:1976xh,Deser:1976eh,Stelle:1978ye,Ferrara:1978em,Cremmer:1978km}. This is a perturbative approach whose aim is to give a systematic prescription for constructing interacting theories, order-by-order in the fields, as deformations of free theories.

The starting point is a given spectrum of free fields, which we denote collectively by $\varphi$, on a fixed background. These are described by a quadratic action $S^{\left(2\right)}\left[\varphi\right]$, which is invariant under the linear gauge transformations $\delta^{\left(0\right)}_{\xi}\varphi$. One then postulates the existence of a full non-linear action with corresponding non-linear gauge transformations, and expands them both in a weak field expansion around the background in powers of the fields
\begin{align}
    S[\varphi] &= S^{\left(2\right)}\left[\varphi\right] + \,S^{\left(3\right)}\left[\varphi\right] + \,S^{\left(4\right)}\left[\varphi\right] + ...\, ,\\
    \delta_{\xi}\varphi &= \delta^{\left(0\right)}\varphi + \,\delta^{\left(1\right)}\varphi + \,\delta^{\left(2\right)}\varphi + ...\,.
\end{align}
The superscript $\left(n\right)$ means that the corresponding term is power $n$ in the fields. The requirement $\delta_{\xi} S[\varphi]=0$ that the action is gauge-invariant provides a set of recursive equations 
\begin{subequations}\label{noether}
\begin{align} 
    \delta^{\left(0\right)}S^{\left(2\right)}\left[\varphi\right]&=0,\\
    \delta^{\left(0\right)}S^{\left(3\right)}\left[\varphi\right]+\delta^{\left(1\right)}S^{\left(2\right)}\left[\varphi\right]&=0, \label{cubicequ} \\
    \delta^{\left(0\right)} S^{\left(4\right)}\left[\varphi\right]+\delta^{\left(1\right)}S^{\left(3\right)}\left[\varphi\right]+\delta^{\left(2\right)}S^{\left(2\right)}\left[\varphi\right]&=0,\\ \label{quarticcondnoe}
    ...\,,
\end{align}
\end{subequations}
whose solution would iteratively fix the full non-linear action.

In this way, all possible cubic vertices involving symmetric gauge fields of arbitrary integer spin on an AdS$_{d+1}$ background that are consistent with the first order constraint \eqref{cubicequ} have been determined \cite{Fradkin:1986qy,Fradkin:1987ks,Alkalaev:2002rq,Bekaert:2010hk, Vasilev:2011xf,Joung:2011ww,Joung:2012rv,Joung:2012fv,Taronna:2012gb,Joung:2012hz,Boulanger:2012dx,Joung:2013doa,Joung:2013nma}.

The approach to establishing consistent cubic vertices is similar to that of constraining correlation functions involving conserved currents of arbitrary spin (\S \tcb{\ref{subsec::3ptcc}}): Working in the ambient framework \cite{Joung:2011ww,Joung:2012rv,Joung:2012fv,Taronna:2012gb,Joung:2012hz,Joung:2013doa,Joung:2013nma}, for a given triplet of spins $\left\{s_1, s_2, s_3\right\}$, one considers the most general cubic vertex 
\begin{align}
    {\cal V}_{s_1,s_2,s_3}\left(\varphi_i\right) = \sum\limits^{s_i}_{n_i=0} g_{s_1,s_2,s_3}^{n_1,n_2,n_3}\, {\cal I}_{s_1,s_2,s_3}^{n_1,n_2,n_3}(\varphi_i),\label{ansatzcubic}
\end{align}
where
\begin{multline}\label{is}
    {\cal I}_{s_1,s_2,s_3}^{n_1,n_2,n_3}(\varphi_i)=\mathcal{Y}_1^{s_1-n_2-n_3}\mathcal{Y}_2^{s_2-n_3-n_1}\mathcal{Y}_3^{s_3-n_1-n_2}\\\times\mathcal{H}_1^{n_1}\mathcal{H}_2^{n_2}\mathcal{H}_3^{n_3}\,\varphi_{s_1}(X_1,U_1)\varphi_{s_2}(X_2,U_2)\varphi_{s_3}(X_3,U_3)\Big|_{X_i=X}\,,
\end{multline}
built from the six basic contractions \eqref{6cont},
\begin{subequations}\label{6conthr}
\begin{align}
    \mathcal{Y}_1&=\pl_{U_1}\cdot\pl_{X_2}\,,&\mathcal{Y}_2&=\pl_{U_2}\cdot\pl_{X_3}\,,&\mathcal{Y}_3&=\pl_{U_3}\cdot\pl_{X_1}\,,\\
    \mathcal{H}_1&=\pl_{U_2}\cdot\pl_{U_3}\,,&\mathcal{H}_2&=\pl_{U_3}\cdot\pl_{U_1}\,,&\mathcal{H}_3&=\pl_{U_1}\cdot\pl_{U_2}\,.
\end{align}
\end{subequations}
This general cubic interaction \eqref{ansatzcubic} is the analogue of \eqref{ooo123321}, for a correlation function involving operators of spin $\left\{s_1, s_2, s_3\right\}$ in CFT. To constrain the coefficients $g_{s_1,s_2,s_3}^{n_1,n_2,n_3}$, one first considers the first non-trivial constraint \eqref{cubicequ}, which leads to the requirement
\begin{align}
\delta^{\left(0\right)}S^{\left(3\right)} \approx 0, \label{s3noether}
\end{align}
since the rightmost term in \eqref{cubicequ} is proportional to the free equations of motion. By inserting the ansatz \eqref{ansatzcubic} into the above, it is possible to identify all independent cubic interactions consistent with the linear gauge symmetries $\delta^{\left(0\right)} \varphi_{s_i}$ .\footnote{\label{foo::1}Each such structure is a linear combination of the ${\cal I}_{s_1,s_2,s_3}^{n_1,n_2,n_3}$. The number of independent structures is equal to: $1+\text{min}\left(s_1,s_2,s_3\right)$, which (as required by the duality) coincides with the number demanded by current conservation on the CFT three-point correlator $\langle {\cal J}_{s_1} {\cal J}_{s_2} {\cal J}_{s_3} \rangle$, with $\partial \cdot {\cal J}_{s_i} \approx 0$.}

For example, in the case of spin-$1$ gauge fields there are two independent interaction terms of the schematic form
\begin{align}\label{beee}
S^{(3)} = {\sf b}^{1}_{1,1,1} \int A\, A\, F + {\sf b}^{0}_{1,1,1} \int F\, F\, F.
\end{align}
The first is the one-derivative Yang-Mills vertex, while the second is the three-derivative Born Infeld vertex. For spin-$2$ gauge fields, there are three independent interactions
\begin{align}
S^{(3)} = {\sf b}^{2}_{2,2,2} \int h\, \nabla h\, \nabla h \: + \: {\sf b}^{1}_{2,2,2} \int R\, R \: + \: {\sf b}^{0}_{2,2,2} \int R\, R\, R,
\end{align}
where the first is the two-derivative gravitational minimal coupling, while the other two come from the expansions of (Riemann)$^2$ and (Riemann)$^3$. The latter involve four and six derivatives, respectively. More generally, solutions to \eqref{s3noether} are vertices involving fields of spin $\left(s_1,s_2,s_3\right)$, with different number of derivatives associated to the coupling constants ${\sf b}^{n}_{s_1,s_2,s_3}$, where $0 \leq n \leq \text{min}\left(s_1,s_2,s_3\right)$.

Let us note that the coupling constants ${\sf b}^{n}_{s_1,s_2,s_3}$ are not fixed at this level: equation \eqref{cubicequ} is linear in the deformation, and so any linear combination of solutions is also a solution. This freedom left over in the cubic action can in principle be fixed by considering higher-order consistency conditions. This has led to results \cite{Fradkin:1986qy,Vasilev:2011xf,Boulanger:2013zza,Kessel:2015kna} for the relative coefficients between a certain class of cubic vertices, known as non-abelian vertices.\footnote{As their name suggests, such vertices deform the linearised gauge transformations and give rise to the non-zero structure constants of the higher-spin algebra. The relative coefficients between such vertices can be constrained by demanding that the linearised theory carries a representation of the global symmetry algebra. This is known as \emph{admissibility} and is a second order constraint.}
However, due to the increasing complexity of higher order conditions in the Noether procedure, neither the complete cubic action nor higher-order interactions have been determined in a metric-like Lagrangian form without auxiliary fields.\footnote{See however \cite{Vasiliev:1989yr} for results for higher-order vertices in the so-called unfolded formulation, which employs an infinite set of auxiliary fields.}

In this chapter we push the above results further by taking an alternative route, using the conjectured holographic duality between higher-spin theories in AdS$_{d+1}$ and free CFTs in $d$-dimensions. As we saw in \S \tcb{\ref{sec::ce}}, the generating function of connected correlators in a free CFT can be identified with the on-shell classical action of the dual higher-spin theory in AdS,
\begin{align}\label{gencor2}
\langle {\cal O}_1\left(y_1\right) ... {\cal O}_n\left(y_n\right) \rangle_{\text{conn.}} = \frac{\delta}{\delta {\bar \varphi}_1\left(y_1\right)} ... \frac{\delta}{\delta {\bar \varphi}_n\left(y_n\right)} S_{\text{HS AdS}}\left[\varphi_i \big|_{\partial AdS} = {\bar \varphi_i}\right]\Big|_{{\bar \varphi}=0}.
\end{align}
With the result for the correlators of single-trace operators in the free CFT,  one should then in principle be able to determine the on-shell interactions of their dual higher-spin fields in AdS by requiring that \eqref{gencor2} holds.

Here, we work in the context of the conjectured duality between the type A minimal bosonic higher-spin theory on AdS and the free scalar $O\left(N\right)$ vector model. In \S \tcb{\ref{subsec::3ptcc}}, we determined the three-point functions of all single-trace operators in the latter. With these results, combined with those for tree-level 3-point Witten diagrams with spinning external fields in \S \tcb{\ref{subsec::witt123}}, by employing the holographic requirement \eqref{gencor2} we are able to extract the explicit form of \emph{all} coefficients in the cubic vertex \eqref{ansatzcubic} in the higher-spin theory on AdS$_{d+1}$. In more detail, we solve the equation shown diagrammatically below for $g^{n_1, n_2, n_3}_{s_1, s_2, s_3}$, for any triplet of even integer spins $\left\{s_1,s_2,s_3\right\}$
\begin{align}\nonumber
 \includegraphics[scale=0.4]{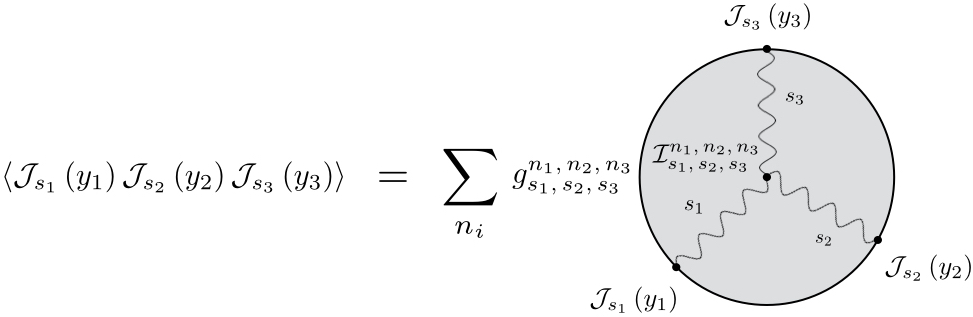}.
\end{align}

\noindent
With the action fixed up to cubic order, we take the holographic reconstruction to quartic order.\footnote{In general, the tree-level Witten diagram computation of $n$-point CFT correlators involves bulk interactions of orders $n$, $n-1$, ..., 3. The holographic reconstruction therefore provides an alternative (and more efficient) iterative procedure to the Noether procedure in constructing interactions.} The simplest example to begin with is given by solving for the quartic self interaction of the scalar, as it minimises the number of tensor structures that one needs to confront. The scalar quartic self-interaction contributes to the holographic computation of the scalar single-trace operator 4-pt function \S \tcb{\ref{subsec::4ptscalar}}, together with the bulk exchanges mediated by the $0$-$0$-$s$ cubic vertices \eqref{ansatzcubic}. The latter we determine at lower order in the  holographic reconstruction procedure. The equation we solve for the quartic interaction is shown in figure \ref{fig::4ptequ}.
\begin{figure}[h]
  \centering
  \includegraphics[scale=0.425]{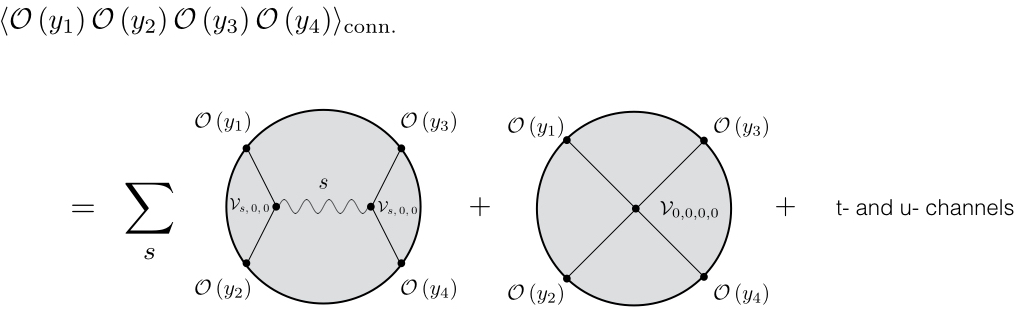}
  \caption{Assuming validity of the HS / free vector model duality, the four-point function of the boundary scalar single trace operator ${\cal O}$ can be computed from the sum of all four-point exchanges of the spin-$s$ gauge fields between two pairs of scalars and the four-point contact diagram generated by the quartic scalar self-interaction ${\cal V}_{0,0,0,0}$.} \label{fig::4ptequ}
\end{figure}

To extract the scalar quartic self-interaction, we proceed in the same way as at cubic order by making the most general ansatz
\begin{align}
    {\cal V}_{0,0,0,0} = \sum_{r,\,m} a_{r,m} \left( \varphi_0\, \nabla_{\mu_1} ...\nabla_{\mu_r}\, \varphi_0 + ... \right) \Box^m\left( \varphi_0\, \nabla^{\mu_1} ...\nabla^{\mu_r}\, \varphi_0 + ...  \right) 
\end{align}
where a completion of the $\ldots$ is given for example by the basis of quartic vertices \eqref{contwbasistf}. For technical reasons that were explained in the previous section, we restrict to extracting the quartic vertex in AdS$_4$. To solve the holographic equation shown in figure \ref{fig::4ptequ}
for the coefficients $a_{r,m}$ in the derivative expansion, it is effective to decompose the bulk and boundary amplitudes into conformal partial waves. We therefore employ the results \S \tcb{\ref{subsec::4ptscalar}} for the expansion of the scalar single-trace four-point function, and those established in \S \tcb{\ref{subsec::4ptexch}} and \S \tcb{\ref{subsec::cd}} for tree-level four-point Witten diagrams with identical parity even external scalars.

\section{Holographic reconstruction}

Before we proceed, let us note that in order for the holographic relation \eqref{gencor2} which we use to extract the interactions to be meaningful, it must be ensured that the two-point functions have the same normalisation from both the bulk and boundary perspectives. In all that follows, we choose the unit normalisation
\begin{align} \label{jjnorm}
\langle {\cal J}_{s}\left(y_1|z_1\right){\cal J}_{s}\left(y_2|z_2\right) \rangle = \frac{\left({\sf H}_3\right)^s}{\left(y^2_{12}\right)^{\Delta}}.
\end{align}

\subsection{Cubic couplings}
\label{subsec::cubicrecon}

\subsubsection{Simple illustrative example: s-0-0 coupling}

As customary, we begin the simplest example \cite{Bekaert:2015tva}: Fixing holographically the bulk $0$-$0$-$s$ interactions, which are of the Noether type 
\begin{align}\label{vnt}
{\cal V}_{s,0,0} \: = \: g_{s,\,0,\,0}\, \varphi_s\left(x,\partial_u\right) \cdot J_s\left(x,u\right), \quad \nabla \cdot \partial_u \, J_s\left(x,u\right) \approx 0,
\end{align}
where $J_s$ is a conserved current that is bi-linear in the bulk scalar $\varphi_0$ (e.g. \eqref{cur1}), and $g_{s,\,0,\,0}$ is the coupling we would like to determine. As we already saw, for each $s$ the form of these vertices is unique on-shell, and the structure is on-shell equivalent to the basic structure \eqref{basic} 
\begin{align}\label{ex00s}
 {\cal V}_{s,0,0} \: & = \: g_{s,\,0,\,0}\, \varphi_s\left(x,\partial_u\right) \cdot J_s\left(x,u\right) \: \approx \:  g_{s,\,0,\,0}\, 2^s\, {\hat {\cal V}}_{s,0,0},
 \end{align}
 where
 \begin{align}
 {\hat {\cal V}}_{s,0,0} & = \varphi_s\left(x,\partial_u\right)  \varphi_0\left(x\right) \left(u \cdot \nabla\right)^s \varphi_0\left(x\right).\label{holorecbasic}
\end{align}
The vertex \eqref{ex00s} in the type A minimal bosonic theory is the unique bulk structure responsible for the holographic computation of the $0$-$0$-$s$ correlator \eqref{vec3pt} in the free scalar $O\left(N\right)$ vector model. Using AdS/CFT dictionary \eqref{gencor2}, we have 
\begin{align}\label{00sequ}
    \langle {\cal O}\left(y_1\right) {\cal O}\left(y_2\right) {\cal J}_s\left(y_3|z\right) \rangle & = \frac{\delta}{\delta {\bar \varphi_0}\left(y_1\right)}\frac{\delta}{\delta {\bar \varphi_0}\left(y_2\right)}\frac{\delta}{\delta {\bar \varphi}_s\left(y_3\right)} S_{\text{HS AdS}}\left[\varphi_i \big|_{\partial \text{AdS}} = {\bar \varphi}_{i}\right] \\ \nonumber
    & = g_{s,\,0,\,0}\, 2^s {\hat {\cal A}}_{\Delta,\Delta,\Delta+s,s}\left(y_1,y_2,y_3|z\right),
\end{align}
where ${\hat {\cal A}}_{\Delta,\Delta,\Delta+s,s}$ is the tree-level 3-pt amplitude \eqref{damp} generated by the basic vertex \eqref{holorecbasic}. Together with the result \eqref{vec3pt} for the dual vector model correlator on the LHS, normalising the two-point functions consistently \eqref{jjnorm}, we can extract $g_{s,\,0,\,0}$ from \eqref{00sequ}: 
\begin{align} \label{couplgend}
g_{s, 0, 0} =  \frac{2^{\tfrac{3d-s-1}{2}}\pi^{\tfrac{d-3}{4}}\Gamma\left(\frac{d-1}{2}\right)\sqrt{\Gamma\left(s+\tfrac{d}{2}-\tfrac{1}{2}\right)}}{\sqrt{N}\sqrt{s!}\,\Gamma\left(d+s-3\right)}. 
\end{align}
In particular, for $d=3$ we have 
\begin{align} \label{coupl3d}
g_{s, 0, 0} =  \frac{2^{4-\frac{s}{2}}}{\sqrt{N}\,\Gamma\left(s\right)}.
\end{align}
This is consistent with the known vanishing of the cubic scalar self coupling ($s=0$) in AdS$_4$ \cite{Sezgin:2003pt}, which itself provided an early check of the holographic duality by comparing with earlier CFT results in \cite{Petkou:1994ad}. In \cite{Skvortsov:2015pea} this result was generalised to the type B theory.

In \S \tcb{\ref{subsec::quarticextr}} we shall employ the result \eqref{coupl3d} to complete the computation of the tree-level spin-$s$ exchange diagram \eqref{exchfirst}, mediated by the $\left(0,0,s\right)$ cubic vertices \eqref{vnt} in the type A minimal bosonic higher-spin theory on AdS$_4$.

\subsubsection{Full cubic action: s1-s2-s3 couplings}

Let us now consider the problem in full generality \cite{Sleight:2016dba}. Our goal is to determine holographically the complete set of cubic couplings in the action of the type A minimal bosonic higher-spin theory on AdS$_{d+1}$. As discussed in the preceding section, this takes the form
\begin{align}\nonumber
S_{\text{HS AdS}}\left[\varphi\right] = \sum\limits_{s_i} \int_{\text{AdS}_{d+1}}\, \frac{1}{2}\varphi_{s_i} \left(\Box-m^2_{s_i} + ...\right) \varphi_{s_i} + \sum\limits_{s_i,\,n_i}g^{n_1,n_2,n_3}_{s_1,s_2,s_3} \int_{\text{AdS}_{d+1}}\, {\cal I}_{s_1,s_2,s_3}^{n_1,n_2,n_3}(\varphi_{s_i}) + ...\,,
\end{align}
for some coefficients $g^{n_1,n_2,n_3}_{s_1,s_2,s_3}$ to be determined. We also display the kinetic terms, to emphasise that the couplings we determine correspond to the canonical normalisation of the former. This is ensured by the normalisation of the boundary-to-bulk propagators.

In this general case, we have to solve the holographic equation in the figure below
\begin{align}\label{holoeq123}
 \includegraphics[scale=0.385]{3ptcouplings.jpeg},
\end{align}
for the coefficients $g^{n_1,n_2,n_3}_{s_1,s_2,s_3}$. All the ingredients are at our disposal: In \S \tcb{\ref{subsec::witt123}} we computed the tree-level 3-pt amplitudes generated by each vertex structure \eqref{is}, and in \S \tcb{\ref{subsec::3ptcc}} we computed all three-point correlation functions of the single-trace conserved currents in the free scalar $O\left(N\right)$ vector model. For convenience, we recall here that the latter are given by \eqref{resvec123} 
\begin{multline} \label{resvec1231}
\langle {\cal J}_{s_1}(y_1|z_1){\cal J}_{s_2}(y_2|z_2){\cal J}_{s_3}(y_3|z_3)\rangle\\= N \left(\prod_{i=1}^3{\sf c}_{s_i}\,q_i^{\frac{1}{2}-\frac{\Delta }{4}}\Gamma(\tfrac{\Delta}2)\,
J_{\frac{\Delta -2}{2}}\left(\sqrt{q_i}\right) \right)
\,\frac{{\sf Y}_1^{s_1}{\sf Y}_2^{s_2}{\sf Y}_3^{s_3}}{(y_{12}^2)^{\Delta/2}(y_{23}^2)^{\Delta/2}(y_{31}^2)^{\Delta/2}}\,,
\end{multline}
with 
\begin{align}
q_1&=2{\sf H}_1\pl_{{\sf Y}_2}\pl_{{\sf Y}_3}\,,&q_2&=2{\sf H}_2\pl_{{\sf Y}_3}\pl_{{\sf Y}_1}\,,&q_3&=2{\sf H}_3\pl_{{\sf Y}_1}\pl_{{\sf Y}_2}.
\end{align}
The ${\sf c}_{s_i}$ are given by
\begin{align}
 {\sf c}_{s_i}^2&= \frac{\sqrt{\pi}\,2^{-\Delta-s_i+3}\,\Gamma(s_i+\tfrac{\Delta}{2})\Gamma(s_i+\Delta-1)}{N\,s_i!\,\Gamma(s_i+\tfrac{\Delta-1}{2})\Gamma(\tfrac{\Delta}2)^2}\,,\label{ghrecon}
\end{align}
for canonical normalisation \eqref{jjnorm} of the two-point functions.

To solve equation \eqref{holoeq123} for the $g^{n_1,n_2,n_3}_{s_1,s_2,s_3}$, the crucial observation is that the 3-pt amplitude generated by the simplest vertex structure \eqref{is} with $n_i = 0$ and $\tau_i=d-2$ can be re-summed. The resulting expression precisely produces the free scalar conformal structure \eqref{resvec1231} above, up to an overall coefficient
\begin{multline}\label{resbulk}
  \int_{\text{AdS}_{d+1}}\,{\cal I}_{s_1,s_2,s_3}^{0,0,0}(K_{s_i+d-2,s_i})=\,\frac{\pi ^{\frac{3}{2}-d}\,(-1)^{s_1+s_2+s_3}\,2^{-3 d-s_1 -s_2 -s_1 +8}\,\Gamma (d+s_1 +s_2 +s_1 -3)}{(y_{12}^2)^{d/2-1}(y_{23}^2)^{d/2-1}(y_{31}^2)^{d/2-1}}\,\\\times\frac{ \Gamma (d-3+s_1) \Gamma (d-3+s_2) \Gamma (d-3+s_1) }{\Gamma \left(\frac{d-3}{2}+s_1 \right) \Gamma \left(\frac{d-3}{2}+s_2 \right) \Gamma \left(\frac{d-3}{2}+s_1 \right)}\prod_{i=1}^3\left(q_i^{1-\frac{d}{4}}
J_{\frac{d}{2}-2}\left(\sqrt{q_i}\right) \right)
\,{\sf Y}_1^{s_1}{\sf Y}_2^{s_2}{\sf Y}_3^{s_3}\,.
\end{multline}
Since the mapping between consistent cubic vertices in the bulk and conserved 3-pt structures in the dual CFT is one-to-one (cf. footnote \ref{foo::1} of this chapter), the above vertex is the only one which can possibly generate the CFT result \eqref{resvec1231}.\footnote{Other vertex structures \eqref{is} may generate the free scalar conserved structure \eqref{resvec1231} on the boundary, but since the mapping is one-to-one there will also be contributions from the other conserved structures in \eqref{gencc3pt} for $l \ne 0$, which cannot be removed by any linear combination of bulk cubic vertices. The vertex with $n_i = 0$ is thus singled out when demanding agreement with the result \eqref{resvec1231} in the free scalar $O\left(N\right)$ model.} This immediately gives $g^{n_1,n_2,n_3}_{s_1,s_2,s_3} = 0$ for all $n_i$ except when $n_1=n_2=n_3 = 0$.

The final step is then to determine $g^{0,0,0}_{s_1,s_2,s_3}$. Normalising the two-point functions in both the bulk and boundary computations consistently \eqref{jjnorm}, and combining the CFT \eqref{resvec1231} and bulk \eqref{resbulk} results, we obtain the following coupling constants
\begin{align}
   g_{s_1,s_2,s_3} :=  g^{0,0,0}_{s_1,s_2,s_3} = \frac{1}{\sqrt{N}}\frac{\pi ^{\frac{d-3}{4}}2^{\tfrac{3 d-1+s_1+s_2+s_3}{2}}}{ \Gamma (d+s_1+s_2+s_3-3)}\prod_{i=1}^3\sqrt{\frac{\Gamma(s_i+\tfrac{d-1}{2})}{\Gamma\left(s_i+1\right)}}\,.
\end{align}

The simplest form for the above coupling constants manifests itself in AdS$_4$, where the spin-dependence coincides with the one obtained in \cite{Metsaev:1991mt,Metsaev:1991nb} from a flat space quartic analysis
\begin{align}
   g_{s_1,s_2,s_3}= \frac{2^{\tfrac{s_1+s_2+s_3}{2}+4}}{\sqrt{N}\,\Gamma(s_1+s_2+s_3)}\,.
\end{align}
This is to be expected, since this is moreover the coefficient of the highest derivative part of the $s_1$-$s_2$-$s_3$ vertex in AdS$_4$, which is the part of a vertex which survives when taking the flat limit of AdS. For discussions on this see e.g.  \cite{Skvortsov:2015pea}. This is a non-trivial consistency check of our results, and also a check of the higher-spin / vector model duality itself: Upon taking the flat limit $\Lambda \rightarrow 0$, we recover the known results in flat space. 

\subsubsection{Discussion / Back to intrinsic}

As mentioned earlier in \S \tcb{\ref{subsec::witt123}}, for ease of computation we established the cubic couplings in the ambient formalism. In this framework,  the complete bulk cubic coupling for given $\left(s_1,s_2,s_3\right)$ in ambient space reads
\begin{equation}
    \mathcal{V}_{s_1,s_2,s_3}=\sum_{s_1,s_2,s_3} g_{s_1,s_2,s_3}\,{\cal I}_{s_1,s_2,s_3}^{0,0,0}\,,\label{v123hrecon}
\end{equation}
where explicitly
\begin{align}\label{is0result}
   & {\cal I}_{s_1,s_2,s_3}^{0,0,0}(\varphi_i)\\ \nonumber
    & \hspace*{1cm} =\left(\pl_{U_1}\cdot\pl_{X_2}\right)^{s_1}\left(\pl_{U_2}\cdot\pl_{X_3}\right)^{s_2}\left(\pl_{U_3}\cdot\pl_{X_1}\right)^{s_3}\,\varphi_{s_1}(X_1,U_1)\varphi_{s_2}(X_2,U_2)\varphi_{s_3}(X_3,U_3)\Big|_{X_i=X}\,.
\end{align}
In particular, we employed the partial derivative $\partial_A$ which acts in the flat $\mathbb{R}^{d,2}$ ambient space, as opposed to the covariant derivative $\nabla$ that is intrinsic to the AdS$_{d+1}$ manifold. In \S \tcb{\ref{radialred}} we explain how to obtain the intrinsic expressions for the interactions \eqref{v123hrecon}, i.e.  in terms of $\nabla$.

One should note that although the interaction \eqref{is0result} contains $s_1+s_2+s_3$ \emph{ambient} partial derivatives, when re-expressing it in terms intrinsic to AdS one picks up lower derivative terms. In particular, the quasi-minimal interactions with the lowest number $s_1+s_2 -s_3$ of derivatives ($s_1 \geq s_2 \geq s_3$). In the spin-$2$ case for example, one recovers the Einstein-Hilbert terms supplemented by those with higher-derivatives.

\subsection{Quartic couplings}
\label{subsec::quarticextr}

In the previous section, we fixed holographically the complete cubic action for the type A minimal bosonic higher-spin theory on AdS$_{d+1}$. With this result, we may now extend this reconstruction of bulk interactions to the quartic order. The simplest case is given by extracting the scalar self-interaction \cite{Bekaert:2015tva}, whose four-point contact diagram contributes to the holographic computation of the scalar single-trace operator four-point function \eqref{vec4pt2}, with the total contributing Witten diagrams shown in figure \ref{fig::total4pt}.\\
\begin{figure}[h]
  \centering
  \includegraphics[scale=0.45]{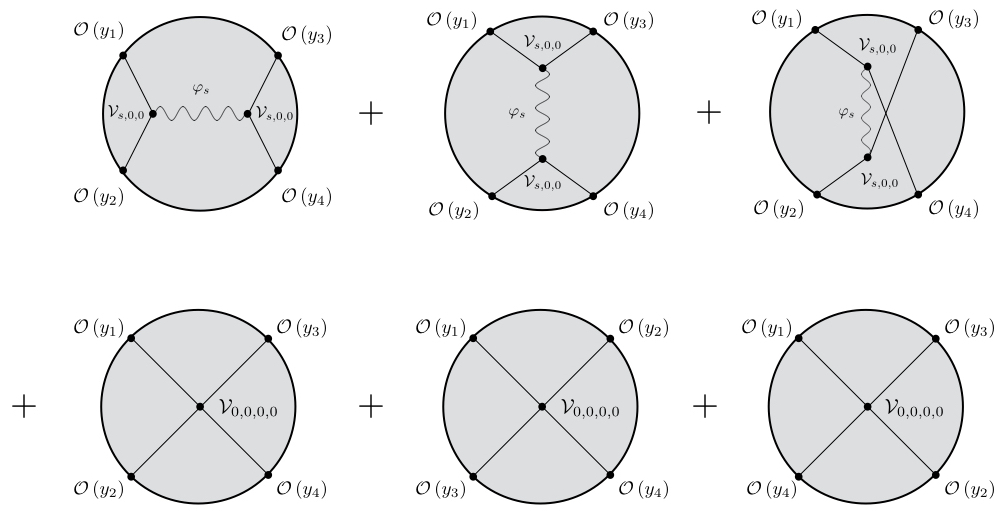}
  \caption{Total contributing four-point Witten diagrams to the holographic computation of $\langle {\cal O}{\cal O}{\cal O}{\cal O} \rangle_{\text{conn.}}$. The four-point exchange diagrams (top line), were computed in \S \tcb{\ref{subsec::4ptexch}} as a conformal partial wave expansion. The analogous result for tree-level contact diagrams for a general scalar quartic self-interaction (second line) was computed in \S \tcb{\ref{subsec::cd}}. All permutations of the external legs are taken into account.} \label{fig::total4pt}
\end{figure}

\noindent
In \S \tcb{\ref{subsec::cd}} we saw that any bulk quartic scalar self-interaction can be expressed in the form
\begin{align}\label{vv}
    {\cal V}_{0,0,0,0} & = \sum\limits_{r,m} a_{r,m} \,{\hat {\cal V}}_{r,m} \\ \nonumber
    & = \sum\limits_{r,m} a_{r,m}\left(\varphi_0 \nabla_{\mu_1} ...\nabla_{\mu_r}\varphi_0 + ... \right) \Box^m \left(\varphi_0 \nabla_{\mu_1} ...\nabla_{\mu_r}\varphi_0 + ...\right),
\end{align}
built from the complete set of vertices ${\hat {\cal V}}_{r,m}$ defined in \eqref{contwbasistf}. In this section we determine the coefficients $a_{r,m}$ for the higher-spin theory on AdS$_4$. As discussed in \S \tcb{\ref{sec::4ptwittpoo}}, working in AdS$_4$ allowed for certain simplifications in computing the relevant Witten diagrams, which will also manifest themselves in extracting the quartic vertex in this section. The coefficients are fixed by requiring that its four-point contact diagram correctly reproduces the dual vector model correlator \eqref{vec4pt2} when supplemented with the exchange diagrams \eqref{tlpwe} (figure \ref{fig::total4pt}). In other words, they must satisfy the equation,
\begin{align} \label{qv1}
& \langle {\cal O}(y_1) {\cal O}(y_2) {\cal O}(y_3) {\cal O}(y_4) \rangle_{\text{conn.}} \\ \nonumber
& \hspace*{2.5cm} = \sum\limits_s {\cal A}^{\text{exch.}}_s(y_1,y_2;y_3,y_4)\;+\; \sum\limits_{s,\,m} a_{s,m}\;{\hat {\cal A}}^{\text{cont.}}_{s,m}(y_1,y_2;y_3,y_4)\;+\;y_2\;\leftrightarrow\;y_{3},\:y_4,
\end{align}
which we solve in the following. The role of $y_2\;\leftrightarrow\;y_{3},\:y_4$ is to include all permutations of the external legs -- i.e. all diagrams in figure \ref{fig::total4pt}.

First, notice that the Witten diagram computation of the dual CFT correlator is automatically crossing symmetric (as can be seen from the RHS of \eqref{qv1}). Using the manifestly crossing symmetric form \eqref{manicross} for the scalar single-trace four-point function, we may thus reduce the above problem to solving the equation 
\begin{equation}
    {\cal G}\left(y_1,y_2;y_3,y_4\right) = \sum\limits_s {\cal A}^{\text{exch.}}_s(y_1,y_2;y_3,y_4)\;+\; \sum\limits_{s,\,m} a_{s,m}\;{\hat {\cal A}}^{\text{cont.}}_{s,m}(y_1,y_2;y_3,y_4), \label{qv2}
\end{equation}
where we recall that 
 \begin{equation}
 \label{13oooo}
  \langle {\cal O}(y_1){\cal O}(y_2){\cal O}(y_3){\cal O}(y_4)\rangle_{\text{conn.}} = {\cal G}(y_1,y_2;y_3,y_4)+y_2\;\leftrightarrow\;y_{3},\:y_4.
   \end{equation}
If a vertex that solves \eqref{qv2} is found, then the definition of ${\cal G}$ above ensures that equation \eqref{qv1} is automatically satisfied. Since the non-trivial quartic vertex is completely determined by \eqref{qv1}, any solution of \eqref{qv2} would give the only solution of \eqref{qv1}. The two problems are thus equivalent.

To proceed, we solve for the coefficients $a_{s,m}$ by matching the contributions in the conformal block expansion of both sides of equation \eqref{qv2}. Solving equation \eqref{qv2} is therefore much simpler, as it eliminates the issue of comparing conformal block expansions in different channels. This is essentially the key technical difficulty behind the conformal bootstrap.  Another crucial simplification is provided by the conformally coupled scalar in  AdS$_4$: As we saw in \S \tcb{\ref{sec::4ptwittpoo}}, in this case each bulk amplitude for fixed $s$ in equation \eqref{qv2} only generates spin-$s$ contributions. We may thus solve the equation spin-by-spin: 
\begin{align}\label{sbys}
  &  {\cal A}^{\text{exch.}}_s(y_1,y_2;y_3,y_4)\;+\; \sum\limits_{m} a_{s,m}\;{\hat {\cal A}}^{\text{cont.}}_{s,m}(y_1,y_2;y_3,y_4) \\ \nonumber
  & \hspace*{1.5cm} =  a \cdot  {\sf c}^2_{{\cal O}{\cal O}{\cal J}_s}\,W_{\Delta+s,s}(y_1,y_2;y_3,y_4) + c \cdot \sum_{n} {\sf c}^2_{{\cal O}{\cal O}\left[{\cal O}{\cal O}\right]_{s,n}}\,W_{2\Delta+2n+s,s}(y_1,y_2;y_3,y_4),
\end{align}
where we used the conformal block expansion for ${\cal G}$ derived in \S \tcb{\ref{subsec::4ptscalar}}, which we repeat for convenience below:
 \begin{align}
    & {\cal G}(y_1,y_2;y_3,y_4)  \\ \nonumber
    & \hspace*{1.25cm} =  \quad a \cdot \sum_s {\sf c}^2_{{\cal O}{\cal O}{\cal J}_s}\,W_{\Delta+s,s}(y_1,y_2;y_3,y_4) + c \cdot \sum_{s,\, n} {\sf c}^2_{{\cal O}{\cal O}\left[{\cal O}{\cal O}\right]_{s,n}}\,W_{2\Delta+2n+s,s}(y_1,y_2;y_3,y_4),
\end{align}
with $a$ and $c$ constrained by $2a+c = 1$. As previously explained, this freedom in the choice of $a$ and $c$ represents the ambiguity in the definition \eqref{13oooo} of ${\cal G}$, in which we are at liberty to add terms ${\cal G}_{\text{trivial}}$ with the property
\begin{align}
    {\cal G}_{\text{trivial}}(y_1,y_2;y_3,y_4)+{\cal G}_{\text{trivial}}(y_1,y_3;y_2,y_4)+ {\cal G}_{\text{trivial}}(y_1,y_4;y_3,y_2) = 0.
\end{align}
In the bulk, this translates to the freedom of adding trivial vertices to the scalar self interaction \eqref{vv}, which give a vanishing total contact amplitude 
\begin{equation}
 {\cal A}^{\text{cont.}}_{\text{trivial}}(y_1,y_2;y_3,y_4)+{\cal A}^{\text{cont.}}_{\text{trivial}}(y_1,y_3;y_2,y_4)+ {\cal A}^{\text{cont.}}_{\text{trivial}}(y_1,y_4;y_3,y_2) = 0.
\end{equation}
In other words, the different solutions for the quartic vertex which arise from the ambiguity in the definition of ${\cal G}$ differ by trivial vertices which vanish on-shell.

A particularly illuminating choice is $a=-c=1$:
 \begin{align}
   & {\cal G}(y_1,y_2;y_3,y_4)  \\ \nonumber
    & \hspace*{1.25cm} =  \quad   \sum_s {\sf c}^2_{{\cal O}{\cal O}{\cal J}_s}\,W_{\Delta+s,s}(y_1,y_2;y_3,y_4) -  \sum_{s,\, n} {\sf c}^2_{{\cal O}{\cal O}\left[{\cal O}{\cal O}\right]_{s,n}}\,W_{2\Delta+2n+s,s}(y_1,y_2;y_3,y_4).
\end{align}
Recalling (equation \eqref{stcontr}) that the spin-$s$ exchange generates precisely the spin-$s$ single-trace contribution in the (12)(34) channel expansion of the dual CFT correlator, the above choice of $a$ and $c$ exhibits the explicit cancellation of the single-trace contributions in equation \eqref{sbys}. This leaves only double-trace contributions, to be matched by the contact amplitude of the quartic vertex.

Solving for the coefficients is simplified by using the contour-integral form of the conformal block expansion \eqref{cif}, as it eliminates the sum over $n$.\footnote{This is because the contributions from double-trace operators of the same spin are packaged in a single gamma function.} In this way, the coefficients $a_{s,\,m}$ are given implicitly by 
\begin{align}\label{answervertex}
&\sum\limits_m a_{s,m}\left(-1\right)^m\left(\nu^2+s+\tfrac{9}{4}\right)^m = \frac{2^{8-s}}{N} \frac{1}{\nu ^2+(s-\tfrac{1}{2})^2}\left[\frac{\pi}{\Gamma \left(\frac{2 s-2 i \nu +1}{4}\right)^2 \Gamma \left(\frac{2 s+2 i \nu +1}{4} \right)^2} - \frac{1}{ \Gamma\left(s\right)^2}\right] \\ \nonumber
 &\hspace*{1.5cm} - \frac{1}{N} \frac{ \left(-1\right)^{\tfrac{s}{2}} \pi ^{\frac{3}{2}} 2^{s+5} \Gamma \left(s+\frac{3}{2}\right)\Gamma \left(\frac{s}{2}+\frac{1}{2}\right)}{\sqrt{2}\Gamma \left(\frac{s}{2}+1\right)\Gamma\left(s+1\right)\Gamma \left(\frac{3}{4}-\frac{i \nu }{2}\right) \Gamma \left(\frac{3}{4}+\frac{i \nu }{2}\right) \Gamma \left(s+\tfrac{1}{2}+ i \nu \right) \Gamma \left(s+\tfrac{1}{2}- i \nu \right)},
\end{align}
which we obtain by inserting the contour integral forms \eqref{cif}, \eqref{tlpwe} and \eqref{contactpwe} into equation \eqref{sbys} for $a = -c = 1$. Through setting $z = -\left(\nu^2+s+\tfrac{9}{4}\right)$, one can establish a generating function
\begin{align}
a_s(z)\equiv\sum\limits_m a_{s,m}\;z^m,
\end{align}
for the coefficients for each $s$:
\begin{align} \nonumber
 a_{s}\left(z\right) \:  = \quad &  \frac{\Lambda^{s}}{ N} \frac{\left(\pi e^{\gamma\left(2s+1\right)}\Gamma\left(s\right)^2\left(\left(2s+1\right)^2+\left(\Lambda z+s+\tfrac{9}{4}\right)^2\right)^2 P\left(s+\tfrac{1}{2};z\right)^2   - 4^2\right)}{2^{s-4}\Gamma\left(s\right)^2\left(\left(\Lambda z+s+\tfrac{9}{4}\right)^2+\left(s-\tfrac{1}{2}\right)^2\right)}
\\  \nonumber
&-\frac{ i^s \pi ^{2} }{2\sqrt{2}N} \frac{ \Gamma\left(s+\frac{3}{2}\right)}{\Gamma\left(\frac{s}{2}+1\right)^2} e^{\gamma\left(2s+\frac{5}{2}\right)}P\left(s+\tfrac{1}{2};z\right)P\left(\tfrac{3}{4}; z\right)\\ \nonumber & \hspace*{3.5cm}\times \left(\left(2s+1\right)^2+\left(2\Lambda z+2s+\tfrac{9}{2}\right)^2\right) \left(9+\left(2\Lambda z+2s+\tfrac{9}{2}\right)^2\right),\label{boredgf}
\end{align}
where we introduced
\begin{align}
P\left(a|z\right) = \prod^{\infty}_{k=0} \left[\left(1+\frac{a}{k}\right)^2+\left(\frac{4z+4s+9}{4k}\right)^2\right]e^{-2a/k},
\end{align}
and $\gamma$ is the Euler-Mascheroni constant. In particular, we employed the Weierstrass infinite product representation of the Gamma function and reinserted the dependence on the cosmological constant.

 This determines the quartic vertex \eqref{vv} of the parity even scalar in the minimal bosonic higher-spin theory on AdS$_4$, in the form
\begin{equation}
{\cal V}_{0,0,0,0}= \sum\limits_{s\in 2\mathbb{N}} {\tilde J}_{s}\left(x, \partial_u\right) a_s(\Box) {\tilde J}_{s}\left(x, u\right),  \label{quartverts}
\end{equation}
where the ${\tilde J}_{s}$ is the traceless improvement of the spin-$s$ conserved currents \eqref{cur1}, bi-linear in the bulk scalar $\varphi_0$.

We conclude this chapter with some comments:

\begin{itemize}
    \item To solve for the vertex \eqref{quartverts}, we relaxed the constraint $m \le k$ in the basis \eqref{contwbasistf}. I.e. we used a redundant set of vertices, which should be re-expressed purely in terms of those for $m\le k$ using the freedom of integration by parts and the free equations of motion. This would help in completing the locality discussion of \S \tcb{\ref{epi::local}}.

\item Since we solve for the coefficients $a_{s,m}$ at the level of the contour integral integrand in \eqref{answervertex}, na\"ively it may seem that the vertex is not defined uniquely: One is free to add functions of $\nu$ that are entire within the contour. In doing so, this would change the form  of $a_{s,m}$ and thus the solution for the quartic vertex. However, the addition of such terms does not change the four-point amplitude and therefore the different vertices obtained in this manner would differ only by trivial terms which vanish on the free mass shell.

\item Let us also comment on the presence of anomalous scaling dimensions for the double-trace operators in the partial wave decompositions of Witten diagrams: These manifest themselves by the presence of \emph{double-poles} in (\ref{bnu}) and (\ref{contactpwe}) for $\frac{d}{2}+i\nu = 2\Delta +2n+s$, $n = 0, 1, 2, ...\,$. Owing to the double-poles, evaluating the $\nu$-integral using Cauchy's theorem produces not only the ``non-anomalous'' double-trace conformal blocks $G_{\Delta_{n,s},s}$, but also their derivatives with respect to the dimension 
  $(\partial/\partial \Delta_{n,s})G_{\Delta_{n,s},s}$. These two terms together originate from the conformal block $G_{\Delta_{n,s}+\gamma_{n,s},s}$, with an anomalous scaling dimension $\gamma_{n,s}$ for the double-trace operator. Indeed, this can be seen by expanding as a Taylor series in $\gamma_{n,s}$,
 \begin{align}
 G_{\Delta_{n,s}+\gamma_{n,s},s} = G_{\Delta_{n,s},s} + \gamma_{n,s}\frac{\partial}{\partial \Delta_{n,s}}G_{\Delta_{n,s},s} + \mathcal{O}\left(1/N^2\right).
 \end{align}
The presence of anomalous dimensions for a given Witten diagram is natural, since the former are dual to the binding energies of two-particle states in the bulk. At the same time however, the boundary theory is free and the double-trace operators \eqref{dto} do not receive any anomalous dimensions. This fact manifests itself in the contour integral representation \eqref{cif} for the dual CFT four-point function, which contains only single poles at $\frac{d}{2}+i\nu = 2\Delta +2n+s$, $n = 0, 1, 2, ...\,$. The holographic duality between higher spin theories and free CFTs therefore requires a delicate cancellation of the anomalous conformal blocks generated by each individual Witten diagram. That the quartic vertex that we found solves (\ref{qv1}), implies that
this cancellation indeed takes place.

In this respect, the contour integral representation for the conformal block decomposition
turned out to be a powerful tool: First of all, it allowed us to solve for the quartic interaction by
performing simple algebraic manipulations with the functions of $\nu$. Moreover, this representation allowed us to treat conformal
blocks with anomalous dimensions on the same footing as non-anomalous ones, simply
by controlling the degree of the pole in $\nu$ at the associated point.

\item As a final point: One motivation for studying higher-spin interactions using holography, was owing due to the technical difficulties encountered when restricted to standard methods of approach. Indeed, using the Noether procedure it has not yet been possible to fix all cubic couplings, or to determine quartic or higher-order vertices in a standard Lagrangian form. In comparison, the holographic approach we took seems to be more efficient. Furthermore, solving for the quartic scalar self-interaction \eqref{epiquartic} holographically even skips steps which are necessary in the Noether procedure. The first Noether-consistency condition in which the quartic action enters is \eqref{quarticcondnoe}
    \begin{equation}
     \delta^{(2)}S^{(2)} + \delta^{(1)}S^{(3)} + \delta^{(0)}S^{(4)} = 0.
    \end{equation}
    However, since the scalar is invariant under $\delta^{(0)}$ (it is not a gauge field itself), to extract the quartic scalar self-interaction \eqref{epiquartic} using the Noether method one needs to (at  least) consider the quintic-order condition
    \begin{align}
      \delta^{(3)}S^{(2)} + \delta^{(2)}S^{(3)} + \delta^{(1)}S^{(4)} + \delta^{(0)}S^{(5)} = 0. \label{epiq}
\end{align}
Furthermore, in principle one may require the knowledge of other quartic interactions to solve \eqref{epiq}  for the scalar quartic self-interaction.
\end{itemize}
 
 \chapter{Summary and discussion}
\label{chapt::summ}
\section{Summary of results}

In this thesis, employing the conjectured duality between higher-spin gauge theories on anti-de Sitter space and free conformal field theories, we studied  higher-spin interactions on an AdS background.

In the context of the duality between the type A minimal bosonic higher-spin theory on AdS and the free scalar $O\left(N\right)$ vector model, we determined the complete on-shell cubic action (on AdS$_{d+1}$) and the quartic self interaction of the scalar (on AdS$_4$). These interactions were previously unknown in a metric-like Lagrangian form, without auxiliary fields. For this we drew on the equivalence between correlation functions of single-trace operators in CFT and Witten diagrams in the dual theory on AdS in the large $N$ limit. The approach is therefore particularly appealing, as correlation functions are straightforward to compute in free CFTs.

We repeat the explicit results for the interactions in the higher-spin action (expanded about AdS) below:
\begin{align}\nonumber
    S\left[\varphi\right] & = \sum\limits_{s} \int_{\text{AdS}}  \frac{1}{2} \varphi_{s} \left(\Box - m^2_{s} + ... \right) \varphi_{s} + \sum\limits_{s_i}\int_{\text{AdS}}  {\cal V}_{s_1,s_2,s_3}\left(\varphi\right) + \int_{\text{AdS}}  {\cal V}_{0,0,0,0}\left(\varphi_0\right) + ...\,.
\end{align}
The cubic interaction for each triplet of spins takes the schematic form
\begin{align} \label{epivert12345}
  &{\cal V}_{s_1,s_2,s_3}\left(\varphi\right)\\
  & \hspace*{1cm}= g_{s_1, s_2, s_3} \left[ \nabla^{\mu_1} ... \nabla^{\mu_{s_3}} \varphi_{\nu_1 ... \nu_{s_1}} \nabla^{\nu_1} ... \nabla^{\nu_{s_1}} \varphi_{\rho_1 ... \rho_{s_2}} \nabla^{\rho_1} ... \nabla^{\rho_{s_2}} \varphi_{\mu_1 ... \mu_{s_1}} + {\cal O}(\sqrt{|\Lambda|}) \right],\nonumber
\end{align}
where the terms ${\cal O}(\sqrt{|\Lambda|})$ and higher in the bracket are descending in the number of derivatives, and are fixed uniquely. The overall coupling is given by 
\begin{align}\label{g0123}
    g_{s_1, s_2, s_3} = \frac{1}{\sqrt{N}}\frac{\pi ^{\frac{d-3}{4}}2^{\tfrac{3 d-1}{2}}\sqrt{-2\Lambda}^{s_1+s_2+s_3}}{ \Gamma (d+s_1+s_2+s_3-3)}\sqrt{\frac{\Gamma(s_1+\tfrac{d-1}{2})}{\Gamma\left(s_1+1\right)}\frac{\Gamma(s_2+\tfrac{d-1}{2})}{\Gamma\left(s_2+1\right)}\frac{\Gamma(s_3+\tfrac{d-1}{2})}{\Gamma\left(s_3+1\right)}}.
\end{align}
In \cite{Sleight:2016dba} it was verified that the rigid structure constants implied by the holographically reconstructed cubic action \eqref{epivert12345} coincide with the
the known expressions \cite{Vasiliev:2003ev,Joung:2014qya} for higher-spin algebra structure constants, which are unique
in general dimensions \cite{Boulanger:2013zza}. This extends to general dimensions (and to the metric-like formalism) the tree-level
three-point function test \cite{Giombi:2009wh} of Giombi and Yin in AdS$_4$. In particular, the test confirms
that the holographically reconstructed cubic couplings solve the Noether procedure at the
quartic order.

The quartic self-interaction of the scalar takes the schematic form
\begin{align} \label{epiquartic}
    {\cal V}_{0,0,0,0}\left(\varphi_0\right) = \sum\limits_{r,\,m}\, a_{r, m} \left( \varphi_0 \nabla_{\mu_1} ... \nabla_{\mu_r} \varphi_0 + ... \right) \Box^m \left( \varphi_0 \nabla^{\mu_1} ... \nabla^{\mu_r} \varphi_0 + ... \right),
\end{align}
where the $\dots$ denote a finite number of terms with no more than $r$ derivatives. In \S \tcb{\ref{subsec::quarticextr}} we gave a generating function \eqref{boredgf} for the coefficients $a_{r,m}$.

Establishing the above results involved a number of non-trivial intermediate steps. We list those which may be of interest for further applications below. Afterwards (\S \tcb{\ref{epi::local}}), in the context of the above results we discuss the issue of locality in higher-spin theories.

\subsubsection{Three-point Witten diagrams with external fields of arbitrary spin}

In order to evaluate tree-level three-point Witten diagrams involving external fields of arbitrary integer spin, we developed an approach which allowed them to be  re-expressed in terms of a differential operator acting on three-point diagrams involving only external scalars. This was an instrumental result, in particular because the latter scalar amplitudes are well-known and straightforward to compute. This method is summarised in the figure below.

In this work we focused on symmetric fields, while for broader applications it may be useful to extend this result to more general representations of the isometry group. For instance, those of the anti-symmetric or mixed-symmetry type, which can be found for example in string theory.

\begin{figure}[h]
  \centering
  \includegraphics[scale=0.45]{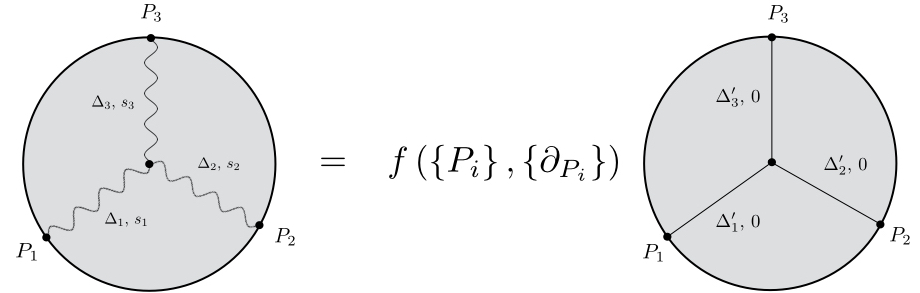}
  \caption{Tree-level three-point Witten diagrams with external fields of arbitrary integer spin and scaling dimensions, can be expressed in terms of tree-level three-point diagrams involving only external scalars. This is achieved through the application of a differential operator $f\left(\left\{P_i\right\},\left\{\partial_{P_i}\right\}\right)$, which encodes the tensor structure of the original amplitude.} \label{fig::epi3pt}
\end{figure}

\subsubsection{Conformal partial wave expansion of four-point Witten diagrams}

In order to extract the quartic self-interaction of the scalar \eqref{epiquartic}, we developed tools to determine  conformal block expansions of tree-level four-point Witten diagrams. See e.g. \cite{Penedones:2007ns,Costa:2014kfa} for previous related investigations. This was in order to match with the corresponding decomposition of the dual CFT correlator. The idea was centred upon the decomposition of four-point Witten diagrams into partial waves, 
\begin{equation}
  \includegraphics[scale=0.5]{factor.jpeg},
\end{equation}
whose factorised form could be identified with the integral representation \eqref{factcbcont} of conformal blocks,
\begin{align}
 &   W_{{\cal O}_{\Delta_i,\ell}}\left(y_{1},y_2;y_{3},y_4\right)  =  \frac{(\Delta_i - \tfrac{d}{2})}{2\pi} \int^{\infty}_{-\infty}d\nu\,\frac{1}{\nu^2 + \left(\Delta_i - \frac{d}{2}\right)^2} \frac{\kappa_{\frac{d}{2}-i\nu,\ell}}{\left(2\pi\right)^{\frac{d}{4}}} \frac{\Gamma\left(\frac{\frac{d}{2}-i\nu +\ell}{2}\right)^2}{\Gamma\left(\frac{\frac{d}{2}+i\nu +\ell}{2}\right)^2} \\ \nonumber
 & \hspace*{3.5cm} \times \int d^dy\, \langle \langle {\cal O}\left(y_1\right){\cal O}\left(y_2\right){\cal O}_{\frac{d}{2}+i\nu,\ell}\left(y\right) \rangle \rangle \langle \langle {\tilde {\cal O}}_{\frac{d}{2}-i\nu,\ell}\left(y\right){\cal O}\left(y_3\right){\cal O}\left(y_4\right) \rangle \rangle.
\end{align}

\noindent
In particular, for exchange diagrams we expressed the bulk-to-bulk propagators for gauge fields of arbitrary spin in a basis of AdS harmonic functions, shown below. This form for the propagator is also applicable in extending this approach to conformal block expansions of diagrams with loops. This would allow, for example, further tests of the higher-spin AdS/CFT duality at loop-level. 

\begin{equation}
  \includegraphics[scale=0.5]{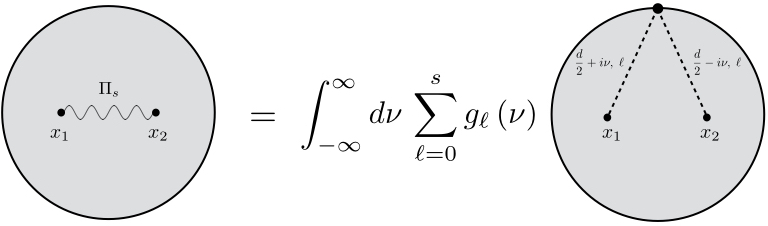},
\end{equation}

\noindent
While in this work we focused on four-point Witten diagrams involving external scalars, the above method is applicable to more general scenarios. For example, to determine the conformal block expansions of four-point Witten diagrams with spinning external legs -- see e.g. \cite{Sleight:2017fpc}. The latter result would be crucial to extract quartic vertices involving fields with non-zero spin. Moreover, it would utilise the result for three-point Witten diagrams mentioned above.\\

\section{Discussion: Locality}
\label{epi::local}
The most distinguishable feature of our quartic vertex \eqref{epiquartic} is that it is unbounded in its number of derivatives. 
In a sense, this is to be expected: In string theory there is an infinite number of higher-derivative $\alpha^\prime$ corrections, which play a key role in the soft behaviour of high energy scattering amplitudes. However, the implications of non-localities for higher-spin gauge theories are still to be fully understood. In this section, we discuss how the AdS/CFT correspondence provides a framework in which these issues may be elucidated. But let us begin by revisiting the concept of locality in the more familiar framework of QFT.

In QFT, locality is a crucial requirement for causality. 
Since the space-time manifold is non-dynamical in non-gravitational theories, causality in this context is the requirement that the theory has localisable observables which commute at space-like separations, 
\begin{equation}
  \left[{\cal O}\left(x\right),{\cal O}\left(y\right)\right] = 0, \quad \left(x-y\right)^2 > 0,
\end{equation}
since space-like separated measurements should not interfere. This is a direct consequence of locality, since observables are functions of local fields which, as a consequence of Lorentz invariance and locality of interactions, when considered as quantum operators must (anti-)commute at space-like separations. However in gravitational theories there are no local gauge (diffeomorphism) invariant observables, which leaves open the possibility for fields to interact in a non-local way. As we shall discuss below, in such theories these issues can be translated instead into analyticity properties of the S-matrix.

A well known example of a consistent theory that is not local in the standard QFT sense is string theory. Indeed, the covariant string field theory action governing the dynamics of the infinite set of fields contains space-time derivatives of all orders. For example, one finds interactions of the form
\begin{align} 
{\cal V}_3 = \left[e^{\frac{1}{2}\alpha^\prime \partial^2} \phi \left(x\right)\right]^3,
\end{align}
where we can write
\begin{align}
  e^{\frac{1}{2}\alpha^\prime\partial^2} \phi \left(x\right) = \frac{1}{2 \pi \alpha^\prime} \int dy \, e^{\frac{1}{2 \alpha^\prime}\left(x-y\right)^2} \phi\left(y\right).
\end{align}
Na\"ively, this seems to be problematic: The field $\phi$ at the point $x$ not only couples to itself, but also to its values arbitrarily far in the future,  past, and even at space-like separations. Most troubling is the presence of an infinite number of time derivatives. The issues that generally accompany theories containing such interactions are well-known, and include \cite{Ostrogradsky,Eliezer:1988rr,Eliezer:1989cr,ANDP:ANDP19915030806,Simon:1990ic,Schmidt:1994iz,Nakamura:1995qz,Bennett:1997wj,Woodard:2000bt,Bering:2000hc,Cheng:2001du}: Possible violations of unitarity and causality, problems in setting up an initial value problem, and difficulties quantising the theory and/or finding a stable Hamiltonian. However, for string theory there is evidence which indicates that the higher-derivative structure is not problematic. Indeed, when passing to the light-cone gauge it is possible to make field redefinitions which render the theory completely local in light-cone time \cite{PhysRevD.10.1110,Kaku:415848,Erler:2004hv}.\footnote{See also \cite{Taronna:2016ats} for a recent covariant formulation of locality in String Theory.}

\subsubsection{Locality in higher-spin theories, and Mellin amplitudes}

Our understanding of these issues in higher-spin gauge theories is comparably limited. Recently, classes of admissible field redefinitions (i.e. those which do not alter physical observables) have been proposed in \cite{Vasiliev:2015wma,Skvortsov:2015lja,Taronna:2016ats,Vasiliev:2016xui,Taronna:2016xrm}, however their implications are yet to be fully realised. The point is whether the field re-definitions required to reduce the number of time derivatives are physical (admissible).

More generally, questions about locality can be translated into certain analytical properties of the S-matrix \cite{1966asm..book.....E}. It is well-known that local interactions give rise to amplitudes which are polynomial in Mandelstam variables. In turn, amplitudes for exchange
diagrams, which are distinct from local contact interactions, contain simple poles in the Mandelstam variables.
So one criterion for consistency in higher-spin theories may be to demand that amplitudes of contact interactions be free from poles, for such non-localities cannot be tamed by admissible field redefinitions. In fact, already in higher-spin theory on flat space it has been observed that gauge invariance requires the quartic contact amplitudes to generate poles in the Mandelstam variables whenever one has external higher-spin particles \cite{Taronna:2010qq}. I.e. they are of the form
\begin{align}
  \varphi^2 \frac{1}{\Box-m^2} \varphi^2,
\end{align}
and the corresponding class of field-redefinitions is unphysical, as they would allow the removal of non-trivial cubic couplings -- i.e. map to a free theory.

On the other hand it is well known that higher-spin theories in flat space are plagued by unresolved existence issues, so one might hope for a better picture when turning to higher-spin theories on AdS. While the standard notion of an S-matrix in AdS is ill defined,\footnote{AdS space-time has a time-like conformal
boundary, and thus does not admit in and out states; particles
in AdS live in a box and interact forever.} we can create and annihilate particles in AdS by changing the boundary conditions at the time-like boundary. Via the AdS/CFT correspondence, the transition amplitudes between such states are identified with the correlation functions of the dual CFT, which suggests an interpretation of CFT correlators as AdS scattering amplitudes \cite{Polchinski:1999ry,Susskind:1998vk,Giddings:1999qu,Giddings:1999jq}. In fact, it has been put forward that the Mellin representation of a CFT correlator of Mack \cite{Mack:2009mi,Mack:2009gy} may provide a definition for an S-matrix of the dual gravity theory in AdS \cite{Mack:2009mi,Mack:2009gy,Penedones:2010ue,Fitzpatrick:2011ia,Paulos:2011ie}:

Consider the example of a Euclidean four-point CFT correlator with identical scalar fields,
\begin{equation} \label{epi4pt}
  \langle {\cal O}\left(y_1\right) {\cal O}\left(y_2\right){\cal O}\left(y_3\right){\cal O}\left(y_4\right) \rangle = \frac{g\left(u,v\right)}{\left(y^2_{12} y^2_{34}\right)^{\Delta}}.
\end{equation}
The Mellin representation of \eqref{epi4pt} is given by inverting the Mellin transform
\begin{align} \label{epi4ptmellin}
  \langle {\cal O}\left(y_1\right) {\cal O}\left(y_2\right){\cal O}\left(y_3\right){\cal O}\left(y_4\right) \rangle = \frac{{\cal N}}{\left(2\pi i\right)^2}\int d\delta_{ij}\, M\left(\delta_{ij}\right) \prod^{4}_{i < j} \left( y^{2}_{ij}\right)^{-\delta_{ij}},
\end{align}
with some normalisation ${\cal N}$, and where the integration contour runs parallel to the imaginary axis with Re$\left(\delta_{ij}\right) > 0$. By conformal symmetry, the Mellin variables $\delta_{ij}$ are constrained by
\begin{align}
 \sum_{j} \delta_{ij} = \Delta, \qquad \:   \delta_{ii} = 0, \label{auxmom}
\end{align}
The object ${\cal M}\left(\delta_{ij}\right)$ defined by
\begin{align}
 M\left(\delta_{ij}\right) = {\cal M}\left(\delta_{ij}\right) \prod^{4}_{i < j} \Gamma\left(\delta_{ij}\right), \label{mellinrep}
\end{align}
is known as the \emph{Mellin amplitude}, and has been proposed to be understood as the AdS scattering amplitude \cite{Penedones:2010ue}. To help motivate this identification, let us first solve the conditions \eqref{auxmom} in a suggestive way by introducing auxiliary ``momenta'' $k_i$, satisfying 
\begin{align}
 \sum^4_{i=4} k_i = 0, \qquad \quad k^2_i = -\Delta.
\end{align}
One can then define analogues of Mandelstam variables as
\begin{align}
 s  & = -\left(k_1+k_2\right)^2 = 2\Delta - 2 \delta_{12} \\
 t  & = -\left(k_1+k_3\right)^2 = 2\Delta - 2 \delta_{13} \\
 u  & = -\left(k_1+k_4\right)^2 = 2\Delta - 2 \delta_{14},
\end{align}
 satisfying the relation $s+t+u = 4\Delta$. Let us now consider the structure of the Mellin amplitude in terms of these variables. As observed by Mack, in order to reproduce the contribution of a conformal multiplet with quantum numbers $\left[\Delta_k,J\right]$ in the conformal partial wave expansion of \eqref{epi4pt} in, say, the (12)(34) channel, the Mellin representation \eqref{mellinrep}
\begin{align}\label{mellinmandel}
 M\left(s,t,u\right) = {\cal M}\left(s,t,u\right)  \Gamma\left(\Delta - \tfrac{s}{2}\right)^2\Gamma\left(\Delta - \tfrac{t}{2}\right)^2\Gamma\left(\Delta - \tfrac{u}{2}\right)^2,
\end{align}
must contain simple poles at
\begin{align}\label{epi::poles}
 s = \Delta_k - J + 2m, \qquad m = 0, 1, 2, 3, ... \,.
\end{align}
The contributions from double-trace operators are thus taken care of by the Gamma function factors in \eqref{mellinmandel}, leaving the Mellin amplitude ${\cal M}\left(s,t,u\right)$ to take care of the single-trace contributions. Under the AdS/CFT correspondence, single-trace CFT operators are identified with single-particle bulk states \S \tcb{\ref{subsec::fom}}, and therefore the Mellin amplitude of a four-point exchange Witten diagram contains the corresponding simple poles in the Mandelstam variables, analogous to flat space S-matrix amplitudes for exchanges.\footnote{This was verified explicitly in: \cite{Penedones:2010ue} for the scalar and graviton exchange.} Furthermore, local bulk contact interactions give rise to polynomial Mellin amplitudes \cite{Penedones:2010ue}, in perfect analogy with flat space scattering amplitudes. More substance to this identification of the Mellin amplitude as an S-matrix for AdS was given by the observation that the standard flat space scattering amplitude is recovered in the appropriate flat limit \cite{Penedones:2010ue,Fitzpatrick:2011hu} (see also: \cite{Fitzpatrick:2011dm}).

With the above analogy in place, let us explore its implications in the context of higher-spin holography. According to the above prescription, we may interpret the Mellin amplitude of the connected four-point correlator \eqref{vec4pt2} in the free scalar $O\left(N\right)$ vector model
\begin{align}
  \langle {\cal O}\left(y_1\right) {\cal O}\left(y_2\right){\cal O}\left(y_3\right){\cal O}\left(y_4\right) \rangle_{\text{conn.}}  = \frac{4}{N} \frac{1}{\left(y^2_{12} y^2_{34}\right)^{\Delta}}\left[ u^{\frac{\Delta}{2}}+\left(\frac{u}{v}\right)^{\frac{\Delta}{2}}+u^{\frac{\Delta}{2}}\left(\frac{u}{v}\right)^{\frac{\Delta}{2}}\right], \label{epi4ptvec}
\end{align}
as a scattering amplitude with external scalar legs in the type A minimal bosonic higher-spin theory on AdS. While the Mellin amplitudes of correlators in strongly coupled CFTs are well understood \cite{Costa:2012cb}, there are some subtleties in applying the definition \eqref{epi4ptmellin} to free (or weakly coupled) CFTs. To see this more clearly, consider the latter in terms of the ``Mandelstam variables'', 
\begin{equation}
  \langle {\cal O}\left(y_1\right) {\cal O}\left(y_2\right){\cal O}\left(y_3\right){\cal O}\left(y_4\right) \rangle = \frac{{\cal N}}{\left(y^2_{12} y^2_{34}\right)^{\Delta}}\int \frac{ds}{2\pi i} \int \frac{dt}{2\pi i}\,u^{\tfrac{s}{2}} v^{\Delta-\tfrac{s+t}{2}} M\left(s,t\right). 
\end{equation}
The main issue is that a power function (which is characteristic of free CFT correlators as a result of Wick contractions)
\begin{align}
  f\left(w\right) = w^\alpha, \qquad \alpha \in \mathbb{R},
\end{align}
does not have a well defined Mellin transform \cite{Flajolet19953}
\begin{align}
  M\left(z\right) = \int^{\infty}_0 dw\, w^z f\left(w\right) \frac{dw}{w}.
\end{align}
One possibility is to understand the Mellin transform in this case as a distribution \cite{mellin}
\begin{align}
  M\left(z\right) = \delta\left(z+\alpha\right).
\end{align}
With this interpretation, the Mellin transform of the  vector model correlator \eqref{epi4ptvec} is given by \cite{Taronna:2016ats,Bekaert:2016ezc}
\begin{align}\label{distmellin}
  M\left(s,t,u\right) = \frac{4}{N} \delta\left(\tfrac{s}{2}-\tfrac{\Delta}{2}\right)\delta\left(\tfrac{t}{2}-\tfrac{\Delta}{2}\right)\delta\left(\tfrac{u}{2}-\Delta\right) + \text{cycl.}
\end{align}
Owing to its distributional nature, there may be subtleties in applying the definition \eqref{mellinmandel} to obtain the corresponding Mellin amplitude (and hence scattering amplitude in AdS higher-spin theory). The problem is that each value in the support of the LHS in \eqref{mellinmandel} corresponds to a pole of one of the Gamma functions on the RHS. Formally, treating the Mellin amplitude as a distributional product one would conclude
that it must \emph{vanish} identically.

Assuming the validity of the above observation, let us briefly explore its implications in the context of the results obtained in this thesis. Recall (\S \tcb{\ref{subsec::4ptexch}}) that the single-trace contributions from the bulk exchanges\footnote{To avoid confusion with the Mandelstam variable ``$s$'', in this section we use the label $J$ to denote spin.}
\begin{align} \label{epiexhsum}
  {\cal A}^{\text{exch.}}_{\text{total}}\left(y_1, y_2;y_3,y_4\right) & = \sum\limits_J {\cal A}^{\text{exch. }}_{J}\left(y_1, y_2;y_3,y_4\right) \\ \nonumber
  & = \sum\limits_J {\sf c}^2_{{\cal O}{\cal O}{\cal J}_J} W_{\Delta+J,J}\left(y_1, y_2;y_3,y_4\right) + \text{double-trace},
\end{align}
precisely account for all single-trace contributions in the conformal block expansion of the dual CFT correlator \eqref{cbe1234},
\begin{align}
  {\cal G}\left(y_1, y_2;y_3,y_4\right) = \sum\limits_J {\sf c}^2_{{\cal O}{\cal O}{\cal J}_J} W_{\Delta+J,J}\left(y_1, y_2;y_3,y_4\right) + \text{double-trace}.
\end{align}
The Witten diagram generated by the quartic vertex \eqref{epiquartic} therefore should not generate single-trace contributions. Recalling that single-trace contributions are the only ones which can generate simple poles in the corresponding Mellin amplitude,\footnote{For clarity, the Mellin transform of the sum over exchanges \eqref{epiexhsum} takes the form,
\begin{align}
  M^{\text{exch.}}_{\text{total}}\left(s,t\right) & = \Gamma\left(\Delta - \tfrac{s}{2}\right)^2\Gamma\left(\Delta - \tfrac{t}{2}\right)^2\Gamma\left(\tfrac{s+t}{2}-\Delta\right)^2 \sum^\infty_{J=0}{\cal M}^{\text{exch.}}_{J}\left(s,t\right) ,\label{mellinexchsum}\\
{\cal M}^{\text{exch.}}_{J}\left(s,t\right)  & = \sum^{\infty}_{m=0} g_{J}\frac{{\cal Q}_{J,m}\left(t\right)}{s-\Delta-2m} + \text{Pol}_{J}\left(s,t\right)
\end{align}
where ${\cal Q}_{J,m}\left(t\right)$ is a kinematical polynomial \cite{Mack:2009mi}, ${\cal Q}_{J,m}\left(t\right) = t^J + {\cal O}\left(t^{J-1}\right)$, and $\text{Pol}_{J}\left(s,t\right)$ is a possible degree $J-1$ polynomial in $s$ and $t$.} one is led to conclude that the scattering amplitude generated by the quartic vertex \eqref{epiquartic} does not give rise to simple poles in the ``Mandelstam variables''.

While it seems that there are no simple poles in the four-point contact amplitude in AdS, we encounter the following issue. Should the re-summation \eqref{mellinexchsum} of the Mellin transforms of the bulk exchanges be an analytic function, with simple poles corresponding to the respective intermediate states \eqref{epi::poles}, the above discussion may be in contradiction with the form \eqref{distmellin} of the dual CFT correlator's Mellin transform. In this case, the absence of these poles in the full amplitude \eqref{distmellin} would then imply that the bulk contact amplitude contains simple poles to compensate.\footnote{This is parallel to results in flat space \cite{Taronna:2011kt}.} On the other hand, the appearance of Dirac delta functions in the Mellin transform \eqref{distmellin} indicates possible issues of convergence for the conformal partial wave expansion of the scalar four-point function when summing over the spin. Therefore, one subtlety may lie in subtracting, term-by-term, a divergent sum over bulk exchanges from a divergent series \cite{Taronna:2016ats,Bekaert:2016ezc}. In this delicate scenario, to study the Mellin amplitude generated by the quartic vertex \eqref{epiquartic}, the most rigorous way to proceed may be to instead re-sum the bulk exchanges (sum over spins) and \emph{then} subtract them from the re-summed CFT amplitude \eqref{distmellin}. 
Further clarity may also be afforded by expressing the quartic vertex \eqref{epiquartic} in a non-redundant basis.

As a final comment, let us note that potential divergent behaviour of higher-spin exchanges discussed above has been observed in the case of conformal higher-spin gauge theories in flat space \cite{Joung:2015eny,Beccaria:2016syk}. There, the regularised sum over the spin-$J$ exchanges gives
\begin{align}
  {\cal M}_{\text{total exch.}}\left(s,t,u\right) & = \frac{1}{\left(-s\right)^{\frac{d-4}{2}}}\, \sum^{\infty}_{J=0}(J+\tfrac{d-3}{2})C^{\left(\frac{d-3}{2}\right)}_J\left(\frac{t-u}{t+u}\right) \\
  & \sim \quad \frac{1}{\left(-s\right)^{\frac{d-4}{2}}}\, \frac{\left(-1\right)^{d-4}}{\left(d-4\right)!}\delta^{\left[d-4\right]} \left(-\frac{2u}{t+u}\right), 
\end{align}
where $\delta^{\left[d-4\right]}\left(z\right)$ denotes the $\left(d-4\right)$-th derivative of the Dirac delta function, and we recall that $C^{\left(\frac{d-3}{2}\right)}_J\left(z\right)$ is a Gegenbauer polynomial.

\begin{appendix}

\chapter{Ambient formalism}
\label{Aambient}
\section{Ambient Space}

\begin{figure}[htb]
\centering
  \includegraphics[scale=0.5]{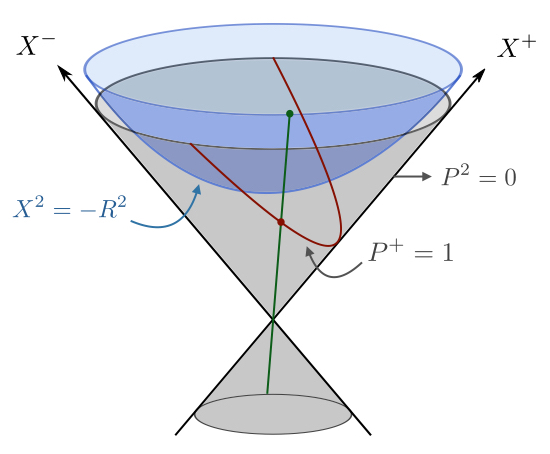}
  \caption{Euclidean AdS and its boundary in ambient space. This figure displays the AdS surface $X^2 = - R^2 = - 1$ and the identification of a (green) boundary point with a (green) light ray of the light cone $P^2 = 0$, which intersects the Poincar\'e section \eqref{Asect} on a (red) point.}\label{ambcone}
\end{figure}

\noindent
The basic idea behind the ambient formalism is to regard AdS space as a one-sheeted hyperboloid $H_{d+1}$ with curvature radius $R$
\begin{equation}
   H_{d+1}: \quad X^2 = -R^2, \qquad X^{0} > 0,
\end{equation}
embedded into an ambient $\left(d+2\right)$-dimensional flat space-time, parameterised by Cartesian co-ordinates $X^{A}$ with $A = 0, 1, ..., d+1$.

To be more precise, denoting the intrinsic coordinates on $H_{d+1}$ by $x^{\mu}$, we are considering the isometric smooth embedding
\begin{equation}
i\; : \quad H_{d+1} \: \longhookrightarrow \: \mathbb{R}^{d+2} : \quad x^{\mu} \longmapsto X^{A}\left(x^{\mu}\right).
\end{equation}
The ambient space itself is endowed with a flat metric $\eta_{AB}$, which defines the quadratic form $X^2 = -R^2$. The choice of signature dictates whether one describes Euclidean or Lorentzian AdS:
\begin{align}
 \text{Lorentzian signature:} \quad   &\eta_{AB} = \left(- + + \, ... \, + \right) \quad \rightarrow  \quad \text{Euclidean AdS} \label{Aeuclid}\\
  \text{Conformal signature:} \quad   &\eta_{AB} = \left(-  + \, ... \, + -\right) \quad \rightarrow  \quad \text{Lorentzian AdS}, \label{Alorentz}
\end{align}
where in both cases we adopt the ``mostly plus'' convention.

Towards the boundary of AdS, the hyperboloid $H_{d+1}$ asymptotes to the light cone $X^2 = 0$. The boundary is thus identified with the ambient projective cone of light rays, which can be identified by setting
\begin{equation}
    P^A \equiv \epsilon X^A,
\end{equation}
in the limit $\epsilon \rightarrow 0$. Since $X^2$ is fixed, these ambient co-ordinates satisfy
\begin{equation}
P^2=0, \qquad P \: \sim \: \lambda P, \qquad \lambda \neq 0, \label{lightcone}
\end{equation}
where the equivalence relation expresses the fact that we are dealing with rays. This quotienting of the light cone allows us to describe the $d$-dimensional boundary. We can identify this projective null cone with the $d$-dimensional space by ``gauge-fixing'' this rescaling. We give an example of this below.

A particular  virtue of the ambient formalism is that the mutual $SO\left(d,2\right)$ (or $SO\left(d+1,1\right)$ in the Euclidean case) symmetry of AdS and its conformal boundary is made manifest, by which we mean that the symmetry transformations act linearly. Simply put, the ambient formalism makes the kinematics in AdS or its conformal boundary as simple as Lorentz-invariant kinematics. For co-ordinates, we have
\begin{equation}
    X^{\prime \,A} = e^{i \omega ^{CD} J_{CD}} X^A, \qquad iJ_{CD} = X_C \frac{\partial}{\partial X_D} -X_D \frac{\partial}{\partial X_C},\label{a7}
\end{equation}
with $SO\left(d,2\right)$ generators $J_{CD}$, which satisfy the commutation relations \eqref{sod2alg}.

\subsubsection*{Example: Euclidean AdS in Poincar\'e co-ordinates}
A simple example is given by Euclidean AdS in Poincar\'e co-ordinates $x^{\mu} = \left(z,\,y^i\right)$, $i = 1, ..., d$. The points on AdS are parameterised by
\begin{align}
    X^0 &= \frac{z^2+y^2+R^2}{2z} \\
    X^{d+1} &=\frac{R^2-z^2-y^2}{2z} \\
    X^i & = \frac{R}{z} y^i
\end{align}
Pulling the ambient metric $\eta_{AB}$ with signature \eqref{Aeuclid} back onto the AdS manifold one recovers
\begin{equation}
    ds^2 = \left(\frac{\partial X^A}{\partial x^\mu}\frac{\partial X^B}{\partial x^\nu}\eta_{AB}\right) dx^\mu dx^\nu = \frac{R^2}{z^2}\left(dz^2+\delta_{ij}dy^idy^j\right).
\end{equation}

\noindent
With the gauge choice $P^+ = P^0 + P^{d+1} = 1$, the boundary can be parameterised by
\begin{align}
    P^0 = \frac{1}{2}\left(1+y^2\right), \quad  P^{d+1} = \frac{1}{2}\left(1-y^2\right), \quad P^i = y^i.  \label{Asect}
\end{align}
This corresponds to a Poincar\'e section of the light cone, and is illustrated in figure \ref{ambcone} for the  Euclidean conformal group $SO\left(d+1,1\right)$. Conformal transformations relate this section to others with $dP^+ =0$.

\section{Fields in the ambient formalism}

To describe fields in the bulk or on the boundary of AdS in this formalism, they need to be identified with the appropriate representatives in the ambient space. Uplifting to the higher-dimensional ambient space introduces extra degrees of freedom, and thus the representatives of bulk and boundary fields must be constrained to keep the number of degrees of freedom constant. In the following subsections we detail the ambient description of scalar and tensorial bulk and boundary fields. For definiteness, we work with the Euclidean AdS isometry group / conformal group $SO\left(d+1,1\right)$.

\subsection{Bulk fields}

A smooth rank-$r$ covariant tensor
field $t_{\mu_1 ...\mu_r}\left(x\right)$ on AdS$_{d+1}$ is represented in ambient space $\mathbb{R}^{d+1,1}$ by a $SO\left(d+1,1\right)$-tensor $T_{A_1...A_r}\left(X\right)$, whose pullback onto the AdS manifold satisfies
\begin{align}
i^{*}\: : \: T_{A_1...A_r}\left(X\right) \: \longmapsto \: t_{\mu_1...\mu_r}\left(x\right) = \frac{\partial X^{A_1}\left(x\right)}{\partial x^{\mu_1}}\, ...\, 
\frac{\partial X^{A_r}\left(x\right)}{\partial x^{\mu_r}} T_{A_1...A_r}\left(X\left(x\right)\right). \label{pullback}
\end{align}
They not represented uniquely: The pullback is surjective, so every such tensor on the AdS manifold has an ambient representative, but it is not injective. Indeed, since for $H_{d+1}$
\begin{equation}
X^2 = -R^2 \: \implies \: \frac{\partial X}{\partial x^{\mu}} \cdot X \: \bigg|_{H_{d+1}}\,= 0, 
\end{equation}
the kernel of the pullback \eqref{pullback} contains ``pure gauge'' tensors with components normal to the AdS manifold, which have no influence in the theory defined 
on AdS.

To obtain a unique representation of AdS tensors in ambient space, the extra components need to be eliminated \cite{Fronsdal:1978vb}. This is achieved by demanding:
\begin{itemize}
    \item {\bf Tangentiality} to surfaces of constant radial co-ordinate $\rho = \sqrt{-X^2}$,
    \begin{align}
    X^{A_i} T_{A_1...A_i ... A_r}\bigg|_{\rho = \text{const.}} = 0.
    \end{align}
    Explicitly, we can apply the projection operator
\begin{equation}
\mathcal{P}^{B}_{A} = \delta^{B}_{A} - \frac{X_{A} X^{B}}{X^2}, \label{oproj}
\end{equation}
which acts on ambient tensors as
\begin{align}
\left(\mathcal{P} T\right)_{A_1...A_r} \; := \; \mathcal{P}^{B_1}_{A_1}...\mathcal{P}^{B_r}_{A_r}T_{B_1 ...B_r}, \qquad X^{A_i} \left(\mathcal{P} T\right)_{A_1...A_i...A_r}=0.
\end{align}
\item {\bf Homogeneity} in $X^A$. I.e. prescribed extension along the extra radial direction $\rho$ 
\begin{align}
T_{A_1...A_r}\left(\lambda X\right) = \lambda^{-\mu}\, T_{A_1...A_r}\left(X\right),
\end{align}
where $\mu$ is fixed by the representation of $SO\left(d+1,1\right)$.
\end{itemize}
\subsubsection*{Example I: AdS metric}
A straightforward application of the above is to the intrinsic AdS metric
\begin{align}
g_{\mu \nu} = \frac{\partial X^{A}}{\partial x^{\mu}} \frac{\partial X^{B}}{\partial x^{\nu}} \eta_{AB}.
\end{align}
The induced AdS metric is given by
\begin{align}\nonumber
G_{AB} = \mathcal{P}^{C}_{A}\mathcal{P}^{D}_{B} \eta_{CD} & = \eta_{AB} + \frac{X_{A} X_{B}}{R^2}, \label{ambmetric}
\end{align}
confirming its role as a projector.
\subsubsection*{Example II: Covariant derivative}
Following the above prescription, in ambient representative of the covariant derivative is
\begin{equation}
    \nabla_{A} = {\cal P}^{B}_{A}\partial_{B},\label{ambcovd}
\end{equation}
and its action is given by
\begin{equation}
\nabla = {\cal P} \circ \partial \circ {\cal P}.
\end{equation}
This ensures that the result is tangent to the AdS manifold, and crucially that the partial derivative is acting on an object which already represents an AdS tensor.

For example,
\begin{equation}
\nabla_{B} T_{A_1...A_r} = \mathcal{P}_{B}{}^{C}\mathcal{P}_{A_1}{}^{C_1}...\mathcal{P}_{A_r}{}^{C_r} \frac{\partial}{\partial X^{C}} \left(\mathcal{P}T\right)_{C_1...C_r}\left(X\right).
\end{equation}

\subsubsection{Example III: Symmetric and traceless spin-s fields}

An important example for this work is the ambient description of symmetric and traceless spin-$s$ fields in AdS$_{d+1}$. Such fields satisfy the Fierz system (generalised to AdS)
\begin{align}
\left(-\nabla^{\mu} \nabla_{\mu} + m^2_s\right)\varphi_{\mu_1 ... \mu_s}\left(x\right) = 0, \qquad \nabla^{\mu_1} \varphi_{\mu_1 ...\mu_s}\left(x\right) = 0, \qquad \varphi_{\mu_1 ...\mu_{s-1}}{}^{\mu_{s-1}}\left(x\right) = 0,
\end{align}
where $\nabla_{\mu}$ is the intrinsic covariant derivative on AdS and $m^2_{s}R^2 = \Delta\left(\Delta-d\right)-s$, with $\left[\Delta,s\right]$ labelling representation of $SO\left(d,2\right)$: $\Delta$ is the lowest energy eigenvalue, and $s$ the spin.

The ambient description $\varphi_{A_1 ... A_s}\left(X\right)$ of the above intrinsic system, is naturally given by
\begin{align}
\left(-\nabla^{A} \nabla_{A} + m^2_s\right)\varphi_{A_1 ... A_s}\left(X\right) = 0, \qquad \nabla^{A_1} \varphi_{A_1 ... A_s}\left(X\right) = 0, \qquad \varphi_{A_1 ... A_{s-1}}{}^{A_{s-1}}\left(X\right) = 0,
\end{align}
where $\nabla^A$ is the ambient representative of the covariant derivative. These are supplemented by the tangentiality and homogeneity conditions
\begin{align}
X^{A_1} \varphi_{A_1 ...A_s}\left(X\right) = 0, \qquad \varphi_{A_1 ...A_s}\left(\lambda X\right) = \lambda^{-\Delta} \varphi_{A_1 ...A_s}\left( X\right),\label{homot}
\end{align}
with the requirement that $\varphi_{A_1 ... A_s}$ carries the same representation of $SO\left(d,2\right)$ fixing the degree of homogeneity to be $\Delta$ or $d-\Delta$.

A nice observation is that the above is equivalent to the flat ambient Fierz system
\begin{align}
\partial^2 \varphi_{A_1 ...A_s}\left(X\right) = 0, \qquad \partial^{A_1} \varphi_{A_1 ...A_s}\left(X\right) = 0,\qquad  \eta^{A_1A_2} \varphi_{A_1 ...A_s}\left(X\right) = 0.
\end{align}
when supplemented with the homogeneity and tangentiality conditions \eqref{homot}, and employing the definition \eqref{ambcovd} of the ambient representative of the covariant derivative.

\subsection{Boundary fields}
\label{Aboundaryfields}
In this thesis we consider only symmetric and traceless primary fields in CFT, though the ambient description can be extended to other representations of $SO\left(d\right)$.

A spin-$r$ primary field $f_{i_1 ...i_r}\left(y\right)$ of dimension
$\Delta$ is represented in the ambient formalism by a $SO\left(d+1,1\right)$-tensor $F_{A_1...A_r}\left(P\right)$ which lives on the light cone $P^{2} = 0$. This ambient tensor is also symmetric and traceless, 
as well as homogeneous of degree $-\Delta$
\begin{equation}
F_{A_1...A_r}\left(\lambda P \right) = \lambda^{-\Delta} F_{A_1...A_r}\left(P\right), \qquad \lambda > 0. \label{AF}
\end{equation}
To be tangent to the light cone, $F_{A_1...A_r}\left(P\right)$ must satisfy
\begin{equation}
P^{A_1} F_{A_1...A_r}=0. \label{Atrans}
\end{equation}
We can define $f_{i_1 ...i_r}\left(y\right)$ is related to $F_{A_1...A_r}\left(P\right)$ by a projection onto the Euclidean section \eqref{Asect}\footnote{Note that this relation preserves tracelessness: To compute the trace of $f_{i_1 ...i_r}\left(y\right)$, we need the contraction
\begin{equation}
    \delta^{ij}\frac{\partial P^A}{\partial y^i} \frac{\partial P^B}{\partial y^j} = \eta^{AB} + P^A Q^B + P^B Q^A, \quad \text{where} \quad Q^A = (1,0,-1).
\end{equation}
This gives vanishing trace of $f_{i_1 ...i_r}$ owing to the tracelessness and transversality \eqref{Atrans} of $F_{A_1...A_r}$.}
\begin{equation}
    f_{i_1 ...i_r}\left(y\right) = \frac{\partial P^{A_1}\left(y\right)}{\partial y^{i_1}} ... \frac{\partial P^{A_r}\left(y\right)}{\partial y^{i_r}} F_{A_1...A_r}\left(P\left(y\right)\right), \label{Apullback}
\end{equation}
whose kernel contains arbitrary tensors proportional to $P^A$.

To check that the number of degrees of freedom is unchanged, note that the transversality condition \eqref{Atrans} eliminates one of the two extra components per ambient index. Since we are on the light cone $P^2=0$, unlike for the bulk case considered in the previous section, the transversality condition \eqref{Atrans} is not sufficient to fix $F_{A_1...A_r}$ uniquely: Since the kernal of \eqref{Apullback} contains tensors proportional to $P^A$, the tensor $F_{A_1...A_r}$ is defined up to arbitrary tensors proportional to $P^A$
\begin{equation}
F_{A_1...A_r} \; \sim \; F_{A_1...A_r} + P_{\left(A_1\right.}S_{A_2...A_r\left.\right)}, \quad P^{A_2}S_{A_2...A_r}=0.\end{equation}
This ``gauge invariance'' eliminates the residual extra degrees of freedom.

\subsection{Operations with ambient tensors}
\label{app::ao}
Tensor operations such as contractions can be greatly simplified by encoding tensors in generating functions. In $d$-dimensional intrinsic space, we can encode symmetric tensors in polynomials
\begin{equation}
t_{\mu_1...\mu_r}\left(x\right) \: \longrightarrow \: t\left(x,u\right) = \frac{1}{r!} t_{\mu_1...\mu_r}\left(x\right) u^{\mu_1}...u^{\mu_r}, \label{Asymgen}
\end{equation}
where we have introduced the constant $d$-dimensional auxiliary vector $u^{\mu}$. Furthermore, \emph{traceless} tensors can be encoded as
\begin{equation}
   t_{\mu_1...\mu_r}\left(x\right) \: \longrightarrow \: t\left(x,w\right) = \frac{1}{r!} t_{\mu_1...\mu_r}\left(x\right) w^{\mu_1}...w^{\mu_r}, \label{Atlessgen}
\end{equation}
where now the auxiliary vector is {\it null}, $w^2 =0$.

In terms of generating functions, tensor operations are then translated into an operator calculus, which simplifies manipulations significantly. For example, the contraction between two symmetric rank-$r$ tensors is implemented via
\begin{equation}
    t_{\mu_1 ... \mu_r}\left(x\right)s^{\mu_1 ... \mu_r}\left(x\right) = r!\, t\left(x,\partial_u\right)s\left(x,u\right)\big|_{u=0} = r!\, s\left(x,\partial_u\right)t\left(x,u\right)\big|_{u=0}.\label{Aintsymcont}
\end{equation}
The covariant derivative also gets modified when acting on functions of the intrinsic auxiliary variables,
\begin{equation}
    \nabla_\mu \rightarrow \nabla_\mu+\omega_\mu^{ab}\, u_a\,\tfrac{\partial}{\partial u^b}\,,
\end{equation}
with $\omega_\mu^{ab}$ the spin-connection. With this, the following further operations can be represented: 
\begin{equation}
\text{divergence:} \quad \nabla \cdot \partial_u, \hspace*{1cm} \text{symmetrised gradient:} \quad u \cdot \nabla, \hspace*{1cm} \text{trace:} \quad \partial_u \cdot \partial_u, \label{divgradtr}
\end{equation}
The symmetric metric $g_{\mu \nu}$ is denoted simply by $u^2$, and thus terms proportional to $u^2$ are pure trace.

For symmetric and {\it traceless} tensors, one instead uses the Thomas derivative \cite{Thomas01051926}, which we denote by $\hat \partial_w$.\footnote{This is often referred to as the the Todorov differential \cite{Dobrev:1975ru} in the CFT literature.} This takes into account $w^2=0$,
\begin{equation}
    t_{\mu_1 ... \mu_r}\left(x\right)s^{\mu_1 ... \mu_r}\left(x\right) = \frac{r!}{\left(\tfrac{d}{2}-1\right)_r} t\left(x,\partial_w\right)s\left(x,w\right)\big|_{w=0} = \frac{r!}{\left(\tfrac{d}{2}-1\right)_r} s\left(x,\partial_w\right)t\left(x,w\right)\big|_{w=0},\label{Astc}
\end{equation}
where
\begin{equation}
\hat{\partial}_{w^\mu} = \left(\tfrac{d}{2} - 1+ w \cdot \frac{\partial}{\partial w}\right)\frac{\partial}{\partial w^\mu} - \frac{1}{2} w_\mu \frac{\partial^2}{\partial w \cdot \partial w}. \label{Athomas}
\end{equation}
\noindent
This formalism can be extended to ambient representatives of bulk and boundary tensor fields discussed in the previous section, which we outline in the following.

\subsubsection*{Ambient AdS tensors}
The ambient counterpart of the intrinsic generating function \eqref{Asymgen} for symmetric rank-$r$ tensors is,
 \begin{equation}
 T_{A_1...A_r}\left(X\right) \: \longrightarrow \: T\left(X,U\right) = \frac{1}{r!} T_{A_1...A_r}\left(X\right) U^{A_1}...U^{A_r}, \qquad X \cdot U = 0,
 \end{equation} 
where $T_{A_1 ...A_r}$ is the ambient representative of $t_{\mu_1...\mu_r}$. In this case, the auxiliary vector $U^A$ is constrained to ensure that we are working modulo components which drop out after projection onto the tangent space of $H_{d+1}$.

Contraction between symmetric tensors slightly modified in contrast to the intrinsic case. The generalisation of \eqref{Aintsymcont} being
\begin{equation}
T_{A_1...A_r}\left(X\right) S^{A_1...A_r}\left(X\right) = r!\, T\left(X,D_{U} \right) S\left(X,U\right) \Big|_{U=0} = r!\, S\left(X,D_{U} \right) T\left(X,U\right)\Big|_{U=0},
\end{equation}
where the derivative $D_U$ is given by,
\begin{equation}
    D_U^A = {\cal P}^B_A \frac{\partial}{\partial U^B} = \frac{\partial}{\partial U^A} - \frac{X_A}{X^2} \left(X \cdot \frac{\partial}{\partial U} \right),
\end{equation}
and accommodates for the condition $X \cdot U = 0$.

The ambient representative of the covariant derivative in the generating formalism takes the form
\begin{align}
\nabla^A = \partial^A_{X} - \frac{1}{X^2}\left(X^A X \cdot \partial_X + U^A X \cdot \partial_U - U \cdot X \partial^A_{U}\right),
\end{align}
which satisfies the useful operator algebra
\begin{align}\label{app::amsscomm}
    [X\cdot\pl_U,\nabla_A]&=0\,,& [\pl_U\cdot\pl_U,\nabla_A]&=0\,,&  [\nabla_A,X^2]&=0\,,&
    X\cdot\nabla&=0\,,
\end{align}
together with
\begin{align}
    [D_{U}^A,\nabla^B]=\tfrac{X^A}{X^2}\,D_U^B\,.
\end{align}
For symmetric and traceless tensors, the ambient counterpart of the intrinsic generating function \eqref{Atlessgen} is
 \begin{equation}
 T_{A_1...A_r}\left(X\right) \: \longrightarrow \: T\left(X,W\right) = \frac{1}{r!} T_{A_1...A_r}\left(X\right) W^{A_1}...W^{A_r}, \quad X \cdot W = W^2=0. 
 \end{equation}
 Like for the symmetric ambient tensors above, the Thomas derivative \eqref{Athomas} that implements contractions between symmetric traceless tensors is slightly modified, to account for the transversality condition $X \cdot W = 0$. We denote it by ${\hat D}_W$, and it is given explicitly by
 \begin{align}
     {\hat D}_{W^A} & = \left(\tfrac{d+1}{2}-1 + W \cdot \frac{\partial}{\partial W}\right){\cal P}^{B}_{A} \frac{\partial}{\partial W^B}-\frac{1}{2}W_A\,  \left({\cal P} \circ {\cal P}\right)^{CD}  \frac{\partial}{\partial W^C} \frac{\partial}{\partial W^D}\\ \nonumber
 & =  \frac{d-1}{2} \left( \frac{\partial}{\partial W^{A}} - \frac{X_{A}}{X^2} \left(X \cdot \frac{\partial}{\partial W} \right)\right) + \left(W \cdot \frac{\partial}{\partial W} \right)\frac{\partial}{\partial W^{A}} \\ \nonumber 
 &\hspace{0.45cm}- \frac{X_{A}}{X^2}  \left(W \cdot \frac{\partial}{\partial W}\right)  \left(X \cdot \frac{\partial}{\partial W}\right) - \frac{1}{2} W_{A} \left(\frac{\partial^{2}}{\partial W \cdot \partial W} - \frac{1}{X^2} \left(X \cdot \frac{\partial}{\partial W}\right) \left(X \cdot \frac{\partial}{\partial W}\right)\right). 
 \end{align}
The ambient generalisation of \eqref{Astc} is then
\begin{align}
T_{A_1...A_r}\left(X\right) S^{ A_1...A_r}\left(X\right) & = \frac{r!}{\left(\tfrac{d}{2}-\frac{1}{2}\right)_r} T\left(X, {\hat D}_W \right) S\left(X,W\right)\Big|_{W=0} \\ \nonumber 
& = \frac{r!}{\left(\tfrac{d}{2}-\frac{1}{2}\right)_r} S\left(X, {\hat D}_W \right) T\left(X,W\right)\Big|_{W=0}.
\end{align}
When working modulo traces, we have the operations
\begin{equation}
  \text{divergence:} \quad \nabla \cdot {\hat D}_W, \hspace*{0.75cm} \text{symmetrised gradient:} \quad W \cdot \nabla,\hspace*{0.75cm} \text{Laplacian:} \quad \nabla^2, \label{tldivgradtr}  
\end{equation}
and $W \cdot \partial_W$, which returns the spin of a tensor. These satisfy the operator algebra
 \begin{align} \label{divcim}
& \left[\nabla \cdot {\hat D}_W,W \cdot \nabla \right] \\ \nonumber
 & \hspace*{2.2cm} = \left(\tfrac{d}{2}-\tfrac{1}{2}+W \cdot \partial_W \right)\nabla^{2}-\left(\left(W \cdot \partial_W\right)^{2}+3\left(\tfrac{d}{2}-\tfrac{1}{2}\right)W \cdot \partial_W+
\left(\tfrac{d}{2}-\tfrac{1}{2}\right)^{2}\right)W \cdot \partial_W,  \\ \label{nablacom}
&\left[\nabla^2,W \cdot \nabla \right] = -2\left(\tfrac{d}{2}-1+W \cdot \partial_W\right) W \cdot \nabla.
\end{align}
The following commutators will come in use,
\begin{equation} \label{i1}
\left[\nabla^2,\left(W \cdot \nabla\right)^{n} \right] = -n\left(d-1+2 W \cdot \partial_W-n\right) \left(W \cdot \nabla\right)^{n},
\end{equation}
and 
{\footnotesize
\begin{align} \label{i2}
&\left[\nabla \cdot {\hat D}_W,\left(W \cdot \nabla\right)^{n} \right] \\ \nonumber
& \hspace*{1.5cm} =  \frac{n}{2} \left(W \cdot \nabla\right)^{n-1} \left(d+n+ 2 W \cdot \partial_W-2\right) \left(1-n-\left(n+W \cdot \partial_W-1\right)\left(d+n+W \cdot \partial_W-2\right)+\nabla^{2}\right). \nonumber
\end{align}}
\subsubsection*{Ambient boundary tensors}

The ambient representative \eqref{AF} of a symmetric traceless spin-$r$ primary field $f_{i_1 ... i_r}$ can be encoded in the generating polynomial
\begin{equation}
F_{A_1 ...A_r}\left(P\right) \; \longrightarrow \; F\left(P,Z\right)=\frac{1}{r!}  F_{A_1 ...A_r}\left(P\right) Z^{A_1}...Z^{A_r}, \quad Z^2=0.
\end{equation}
Tangentiality to the light cone, 
\begin{equation}
    \left(P \cdot \frac{\partial}{\partial Z}\right)F\left(P,Z\right) = 0,
\end{equation}
 can be enforced by requiring $F\left(P,Z+\alpha P\right) = F\left(P,Z\right)$ for any $\alpha$.  The ``gauge freedom'' is represented by the orthogonality condition $Z \cdot P=0$.

Contractions between ambient representatives of primary fields can be implemented via the Thomas derivative \eqref{Athomas}, acting in the $\left(d+2\right)$-dimensional ambient space
\begin{equation}
{\hat \partial}^{A}_{Z} = \left(\tfrac{d}{2}-1+Z\cdot \frac{\partial}{\partial Z}\right)  \frac{\partial}{\partial Z_{A}} - \frac{1}{2} Z^{A} \frac{\partial^2}{\partial Z \cdot \partial Z}.
 \end{equation}

\chapter{Appendix of integrals}

\section{Bulk integrals}
\label{app::bis}

In using the Schwinger parameterised form \eqref{schkapp} of boundary-to-bulk propagators in the ambient formalism, a key integral often encountered is of the form
\begin{align}\label{appschwinwit}
& \int^{+\infty}_0 \prod^n_{i=1}\left(\frac{dt_i}{t_i} t^{\Delta_i}\right)\int_{\text{AdS}}dX \exp\left(2\sum\limits^n_{i=1}t_i\,P_i \cdot X \right) \\ \nonumber
& \hspace*{4cm} = \pi^{d/2}\Gamma\left(-\frac{d}{2}+\frac{1}{2}\sum\limits^n_{i=1}\Delta_i\right) \int^{+\infty}_0 \prod^n_{i=1}\left(\frac{dt_i}{t_i} t^{\Delta_i}\right) \exp\left(-\sum\limits_{i < k} t_i t_k P_{ik}\right).
\end{align}
In this appendix we explain its application to tree-level contact diagrams, and give a proof. We also discuss its utility in establishing Mellin amplitudes of Witten diagrams.

\subsubsection{Application: Tree-level contact diagrams}

A straightforward application of the above integral is in evaluating tree-level Witten diagrams. Recall that the Schwinger parameterised form of a scalar boundary-to-bulk propagator of dimension $\Delta$ is
\begin{align}
K_{\Delta,0}\left(X,P\right) = \frac{C_{\Delta,0}}{\left(-2 X\cdot P\right)^\Delta} = \frac{C_{\Delta,0}}{\Gamma\left(\Delta\right)} \int^\infty_0 \frac{dt}{t}t^{\Delta} \exp\left(2 t P \cdot X\right).\label{schkapp}
\end{align}
Considering for example the tree-level $n$-point contact diagram 
\begin{align}
{\cal A}^{\text{cont}}\left(P_1, ..., P_n\right) & = \int_{\text{AdS}} dX  K_{\Delta_1,0}\left(X,P_1\right) ...\,  K_{\Delta_n,0}\left(X,P_n\right) 
\\ \nonumber
&  = \left(\prod^n_{i=1} C_{\Delta_i,0}\right) \int^{+\infty}_0 \prod^n_{i=1}\left(\frac{dt_i}{t_i} t^{\Delta_i}\right)\int_{\text{AdS}}dX \exp\left(2\sum\limits^n_{i=1}t_i\,P_i \cdot X \right), 
\end{align}
we see that the integral \eqref{appschwinwit} naturally arises and can be employed to evaluate the amplitude. This is particularly straightforward for three-point diagrams, for which no integration is required. This will be shown in the subsequent, after proving the formula \eqref{appschwinwit}.

\subsubsection{Proof of \eqref{appschwinwit}:}

To prove \eqref{appschwinwit}, we first evaluate the AdS integral. Defining $T = \sum\limits^n_{i=1}t_i\,P_i$, by Lorentz invariance we can make the simple choice $T = |T|\left(1,1,0\right)$, where we recall that we parameterise AdS$_{d+1}$ in the $\left(d+2\right)$-dimensional ambient space by
\begin{equation}
X = \left(X^+,X^-,X^\mu\right) = \frac{1}{z}\left(1,z^2 + y^2,y^i\right).
\end{equation}
In this way we obtain
\begin{align}
\int_{\text{AdS}}dX \exp\left(2\sum\limits^n_{i=1}t_i\,P_i \cdot X \right) & = \int^{+\infty}_0 \frac{dz}{z} z^{-d} \int d^{d}x\, e^{-\left(1+z^2+y^2\right)|T|/z} \\
& = \pi^{d/2} \int^{+\infty}_0 \frac{dz}{z} z^{-d/2}\, e^{-z+T^2/z},
\end{align}
where in the second line we evaluated the Gaussian integral over $x$. Returning to the RHS of the full expression \eqref{appschwinwit}, and rescaling $t_i \rightarrow t_i/\sqrt{z}$, 
\begin{align}
& \pi^{d/2} \int^{+\infty}_0 \prod^n_{i=1}\left(\frac{dt_i}{t_i} t^{\Delta_i}\right) \int^{+\infty}_0 \frac{dz}{z} z^{-d/2}\, e^{-z+T^2/z} \\ \nonumber
& \hspace*{5cm} = \pi^{d/2} \int^{+\infty}_0 \prod^n_{i=1}\left(\frac{dt_i}{t_i} t^{\Delta_i}\right) e^{T^2}\int^{+\infty}_0 \frac{dz}{z}\, z^{-\frac{d}{2}+\frac{1}{2}\sum\limits^n_{i=1}\Delta_i}\, e^{-z} \\ \nonumber
& \hspace*{5cm} = \pi^{d/2} \Gamma\left(-\frac{d}{2}+\frac{1}{2}\sum\limits^n_{i=1}\Delta_i\right) \int^{+\infty}_0 \prod^n_{i=1}\left(\frac{dt_i}{t_i} t^{\Delta_i}\right)e^{T^2},
\end{align}
where in the second equality we simply employ the integral representation of the Gamma function. Noting that $T^2 = - \sum_{i < k} P_{ik}$, the above establishes \eqref{appschwinwit}.

\subsubsection{The Symanzik star formula}

To establish the Mellin amplitude of Witten diagrams, it is illuminating to employ the Symanzik star formula \cite{Symanzik:1972wj}
\begin{align}
\int^{+\infty}_0 \prod^n_{i=1}\left(\frac{dt_i}{t_i} t^{\Delta_i}\right) \exp\left(-\sum\limits_{i < k} t_i t_k P_{ik}\right) = \frac{\pi^{d/2}/2}{\left(2\pi i\right)^{\frac{1}{2}n\left(n-3\right)}} \int d\delta_{ij} \prod_{1 \leq i \leq j \leq n} \Gamma\left(\delta_{ij}\right) \left(P_{ij}\right)^{-\delta_{ij}},\label{}
\end{align}
which allows to massage Witten diagrams in the form of a Mellin transform \cite{Mack:2009mi,Penedones:2010ue}. The integration contour runs parallel to the imaginary axis with Re$\,\delta_{ij} > 0$, with the integration variables constrained by
\begin{align}
\sum\limits^n_{j\ne i} \delta_{ij} = \Delta_i,
\end{align}
so that the integrand is conformally covariant with scaling dimension $\Delta_i$ at the point $P_i$. This gives $n\left(n-3\right)/2$ integration variables.

In particular, this implies that the integration variables are completely fixed for three-point integrals, allowing the straightforward evaluation of three-point Witten diagrams.

\chapter{Harmonic functions in AdS}

In this appendix we give an overview of the harmonic decomposition of tensorial functions of geodesic distance in AdS$_{d+1}$.

Consider bi-tensors $t_{\mu_1 ... \mu_r , \nu_1 ... \nu_r}\left(x_1,x_2\right)$, which depend only on the geodesic distance between $x_1,\;x_2 \in \text{AdS}_{d+1}$, and are square integrable
\begin{align}
\int_{\text{AdS}_{d+1}} \left|t\left(x_1,x_2\right)\right|^2 \: < \: \infty.
\end{align}
This includes bulk-to-bulk propagators $\Pi_{\Delta,r}\left(x_1,x_2\right)$.

Such tensors can be decomposed in a basis of regular eigenfunctions of the AdS Laplacian,
\begin{align}\label{eomharm}
    \left(\Box_{\text{AdS}}+\left(\frac{d}{2}+i\nu\right)\left(\frac{d}{2}-i\nu\right)+\ell\right)\Omega_{\nu,\ell}\left(x_1,x_2\right) & \equiv 0, \quad \nu \in \mathbb{R}, \quad \ell \in \mathbb{N} \\
    \nabla \cdot \Omega_{\nu,\ell} & = 0,
\end{align}
which are symmetric and traceless spin-$\ell$ irreducible representations of $SO\left(d+1,1\right)$, with $\nu$ labelling the energy.

Crucially, they satisfy orthogonality and (traceless) completeness relations \cite{Costa:2014kfa}
\begin{align}
& \hspace*{1cm}  \int_{\text{AdS}} d^{d+1}x^\prime\; \Omega_{{\bar \nu},\ell}\left(x_1;x^\prime\right) \cdot \Omega_{ \nu,\ell}\left(x^\prime;x_2\right) = \frac{1}{2}\left[\delta\left(\nu + {\bar \nu}\right)+\delta\left(\nu - {\bar \nu}\right)\right]\Omega_{\nu,\ell}\left(x_1;x_2\right),\\
 &  \sum\limits^{s}_{\ell=0} \int^{\infty}_{-\infty} d\nu \,c_{r,\ell}\left(\nu\right)\left(w_1 \cdot \nabla_1 \right)^{\ell}\left(w_2 \cdot \nabla_2 \right)^{\ell} \Omega_{\nu,r-\ell}\left(x_1,w_1;x_2,w_2\right) = \left(w_1 \cdot w_2\right)^r \delta^{d+1}\left(x_1,x_2\right),\label{acomp}
\end{align}
where auxiliary variables $w_i^2=0$ enforce tracelessness, and 
\begin{equation}
    c_{r,\ell}\left(\nu\right) = \frac{2^\ell \left(r-\ell+1\right)_{\ell}\left(\frac{d}{2}+r-\ell-\frac{1}{2}\right)_{\ell}}{\ell! \left(d+2r-2\ell-1\right)_{\ell}\left(\frac{d}{2}+r-\ell+i\nu\right)_\ell\left(\frac{d}{2}+r-\ell-i\nu\right)_\ell}.
\end{equation}
The set of tensors 
\begin{equation}
    \left\{\left(w_1 \cdot \nabla_1 \right)^{\ell}\left(w_2 \cdot \nabla_2 \right)^{\ell} \Omega_{\nu,r-\ell}\left(x_1,w_1;x_2,w_2\right) | \:w^2_i=0,\; \nu \in \mathbb{R}, \: \ell = 0, 1, .., r \right\}, \label{hbasis}
\end{equation}
thus form a basis for symmetric traceless bi-tensors in AdS,
\begin{align}
    t\left(x_1,w_1;x_2,w_2\right) = \sum\limits^r_{\ell=0} \int^{\infty}_{-\infty} f_{r,\ell}\left(\nu\right)\, \left(w_1 \cdot \nabla_1 \right)^{\ell}\left(w_2 \cdot \nabla_2 \right)^{\ell} \Omega_{\nu,r-\ell}\left(x_1,w_1;x_2,w_2\right),
\end{align}
for some $f_{r,\ell}\left(\nu\right) = f_{r,\ell}\left(-\nu\right)$.

As corollary, any symmetric bi-tensor can be expressed in the basis \eqref{hbasis} supplemented with products of the metric; any symmetric bi-tensor $t\left(x_1,u_1;x_2,u_2\right)$ admits the trace decomposition of the schematic form,\footnote{I.e. omitting for concision the relative coefficients between the terms.}
\begin{align}
  & t\left(x_1,u_1;x_2,u_2\right) \\ \nonumber
    & \hspace*{0.5cm} \sim \: t^{\left\{0\right\}}\left(x_1,u_1;x_2,u_2\right) + u^2_1 u^2_2 \; t^{\left\{1\right\}}\left(x_1,u_1;x_2,u_2\right) + ... + \left(u^2_1 u^2_2\right)^{\left[\frac{r}{2}\right]}\; t^{\left\{\left[\frac{r}{2}\right]\right\}}\left(x_1,u_1;x_2,u_2\right),
\end{align}
where here $t^{\left\{n\right\}}$ denotes the traceless part of the $n$-th trace of $t$. Each of the $t^{\left\{n\right\}}$ can then be decomposed in the basis \eqref{hbasis}.

\chapter{Double trace operators}

\label{appendix::doubletrace}

In this appendix we construct primary operators bi-linear in two scalar primary operators
${\cal O}_1$ and ${\cal O}_2$
of scaling dimensions $\Delta_1$ and $\Delta_2$. These have the schematic form
\begin{equation}
\label{schem1}
[{\cal O}_1{\cal O}_2]_{n,s} = {\cal O}_1\partial_{\mu(s)}\Box^n {\cal O}_2 +\dots,
\end{equation}
where the ellipsis denote terms which ensure that $[{\cal O}_1{\cal O}_2]_{n,s}$ is primary. The double trace operator $[{\cal O}_1{\cal O}_2]_{n,s}$ has spin $s$, and in a free theory (or otherwise at leading order in the $1/N$ expansion) has scaling dimension $\Delta_{n,s}=\Delta_1+\Delta_2+2n+s$.

Our goal is to specify implicit terms in (\ref{schem1}). To this end, we make the 
most general ansatz\footnote{There are three independent summuations in the above sum, since $s=s_1+s_2$ and $n=b_1+b_2+b_{12}$.}
\begin{equation}
\label{ansatz}
[{\cal O}_1{\cal O}_2]_{n,s} = \sum_{s_1,b_1,b_2} a_{n,s}(s_1,s_2;b_1,b_2,b_{12})
\partial_{\mu(s_1)}\Box^{b_1}\partial^{\nu(b_{12})}{\cal O}_1 \partial_{\mu(s_2)}\Box^{b_2}\partial_{\nu(b_{12})}{\cal O}_2-\text{traces},
\end{equation}
and specify $a_{n,s}$ from the requirement \eqref{pri} that $[{\cal O}_1{\cal O}_2]_{n,s}$
is primary. In other words, we solve for the coefficients by imposing the condition 
\begin{equation}
    K_{\mu}[{\cal O}_1{\cal O}_2]_{n,s} = 0.
\end{equation}

\subsubsection{The action of the special conformal generator}
For ease of computation, we contract the free indices with the auxiliary traceless symmetric tensors $z^{\mu}$ to ensure that
$[{\cal O}_1{\cal O}_2]_{n,s} $ is traceless and symmetric. A typical term in the sum (\ref{ansatz}) then takes the form
\begin{equation}
\label{oneterm}
T_{n,s}(s_1,s_2;b_1,b_2,b_{12})=z^{\mu(s_1)\nu(s_2)}(\eta^{\mu\nu})^{b_1}(\eta^{\mu\nu})^{b_{12}}(\eta^{\nu\nu})^{b_2}P_{\mu(s_1+2b_1+b_{12})}{\cal O}_1 P_{\nu(s_2+2b_2+b_{12})}{\cal O}_2,
\end{equation}
where we employ the notation $u^{\mu(s)} = u^{\mu_1} ... u^{\mu_s}$.

Using the conformal algebra \eqref{confalg}, one can show that
\begin{equation}
\label{specialconformal}
K_\nu P_{\mu(n)}{\cal O}_1=2n (\Delta_1+n-1)\eta_{\mu\nu}P_{\mu(n-1)}{\cal O}_1-n(n-1)\eta_{\mu\mu}P_\nu P_{\mu(n-3)}{\cal O}_1,
\end{equation}
which use to determine the action of $K_{\mu}$ on each term in the ansatz \eqref{ansatz}. We find
\begin{align}
\notag
K_\nu T_{n,s}(s_1,s_2;b_1,b_2,b_{12})
&=2[\Delta_1+s_1+2b_1+b_{12}-1]s_1 C_z(s_1-1,s_2;b_1,b_2,b_{12})\\
\notag
&+2b_1[2\Delta_1+2b_1-d] C_1(s_1+1,s_2;b_1-1,b_2,b_{12})\\
\notag
&-b_{12}(b_{12}-1)C_{1}(s_1+1,s_2;b_1,b_2+1,b_{12}-2)\\
\notag
&-2s_1 b_{12} C_1(s_1,s_2+1;b_1,b_2,b_{12}-1)\\
\notag
&+2[\Delta_1+s_1+2b_1+b_{12}-1]b_{12}C_2(s_1,s_2+1;b_1,b_2,b_{12}-1)\\
\notag
&+2[\Delta_2+s_2+2b_2+b_{12}-1]s_2 C_z(s_1,s_2-1;b_1,b_2,b_{12})\\
\notag
&+2b_2[2\Delta_2+2b_2-d] C_2(s_1,s_2+1;b_1,b_2-1,b_{12})\\
\notag
&-b_{12}(b_{12}-1)C_{2}(s_1,s_2+1;b_1+1,b_2,b_{12}-2)\\
\notag
&-2s_2 b_{12} C_2(s_1+1,s_2;b_1,b_2,b_{12}-1)\\
\label{Kaction}
&+2[\Delta_2+s_2+2b_2+b_{12}-1]b_{12}C_1(s_1+1,s_2;b_1,b_2,b_{12}-1),
\end{align}
where 
\begin{align}
\notag
&C_z(s_1-1,s_2;b_1,b_2,b_{12})\\
\notag
&\hspace*{1.5cm}=V^{\nu\mu(s_1-1)\sigma(s_2)}(\eta^{\mu\mu})^{b_1}(\eta^{\mu\sigma})^{b_2}(\eta^{\sigma\sigma})^{b_2}P_{\mu(s_1+2b_1+b_{12}-1)}{\cal O}_1 P_{\sigma(s_2+2b_2+b_{12})}{\cal O}_2,\\
\notag
&C_{1}(s_1+1,s_2;b_1-1,b_2,b_{12})\\
\notag
&\hspace*{1.5cm}=z^{\mu(s_1)\sigma(s_2)}(\eta^{\mu\mu})^{b_1-1}(\eta^{\mu \sigma})^{b_{12}}(\eta^{\sigma\sigma})^{b_2}P_\nu P_{\mu(s_1+2b_1+b_{12}-2)}{\cal O}_1P_{\sigma(s_2+2b_2+b_{12})}{\cal O}_2,\\
\label{structures}
&C_2(s_1,s_2+1;b_1,b_2,b_{12}-1)\notag\\
&\hspace*{1.5cm}=z^{\mu(s_1)\sigma(s_2)}(\eta^{\mu\mu})^{b_1}(\eta^{\mu\sigma})^{b_{12}-1}(\eta^{\sigma\sigma})^{b_2}P_{\mu(s_1+2b_1+b_{12}-1)}{\cal O}_1P_{\nu}P_{\sigma(s_2+2b_2+b_{12}-1)}{\cal O}_2.
\end{align}

\subsubsection{Imposing the primary condition}

Having understood how the special conformal generator $K_\nu$ acts on each term of the 
ansatz (\ref{ansatz}), we act with $K_\nu$ on the entire expression. Using (\ref{Kaction}), the condition 
\begin{equation}
    K_\nu [{\cal O}_1{\cal O}_2]_{n,s} = 0,
\end{equation}
can be expressed in terms of the three independent structures $C_z$, $C_1$ and $C_2$ defined in (\ref{structures}). By setting the prefactors of each of these structures to zero we obtain three conditions on $a_{n,s}$.

The equation corresponding to $C_z$ reads
\begin{align}
\notag
&a_{n,s}(s_1,s_2;b_1,b_2,b_{12}) 2[\Delta_1+s_1+2b_1+b_{12}-1]s_1\\
\label{cv}
&\qquad+
 a_{n,s}(s_1-1,s_2+1;b_1,b_2,b_{12})2[\Delta_2+s_2+2b_2+b_{12}](s_2+1)=0.
\end{align}
It can be used to fix the dependence of $a_{n,s}$ on $s_1$ for fixed $b_1$ and $b_2$
\begin{equation}
\label{sdep}
a_{n,s}(s_1,s_2;b_1,b_2,b_{12})=(-1)^{s_2} \frac{s!}{s_2! s_1!}\frac{(\Delta_1+2b_1+b_{12}+s_1)_{s_2}}{(\Delta_2+2b_2+b_{12})_{s_2}}a_{n,s}(s,0;b_1,b_2,b_{12}).
\end{equation}
A second equation, which sets the prefactor of $C_1$ to zero is
\begin{align}
\notag
&a_{n,s}(s_1,s_2;b_1,b_2,b_{12})2b_1 (2\Delta_1+2b_1-d)\\
\notag
&\quad-a_{n,s}(s_1,s_2;b_1-1,b_2-1,b_{12}+2)(b_{12}+2)(b_{12}+1)\\
\notag
&\quad\quad-a_{n,s}(s_1+1,s_2-1;b_1-1,b_2,b_{12}+1)2(s_1+1)(b_{12}+1)\\
\label{2}
&\quad\quad\quad+a_{n,s}(s_1,s_2;b_1-1,b_2,b_{12}+1)2(\Delta_2+s_2+2b_2+b_{12})(b_{12}+1)=0.
\end{align}
The third term contains $a_{n,s}(s_1+1,s_2-1;b_1-1,b_2,b_{12}+1)$, which can be 
expressed in terms of $a_{n,s}(s_1,s_2;b_1-1,b_2,b_{12}+1)$ by (\ref{cv}). This results in
\begin{align}
\notag
&a_{n,s}(s_1,s_2;b_1,b_2,b_{12})2b_1 (2\Delta_1+2b_1-d)\\
\notag
&\quad-a_{n,s}(s_1,s_2;b_1-1,b_2-1,b_{12}+2)(b_{12}+2)(b_{12}+1)\\
\notag
&\quad\quad+a_{n,s}(s_1,s_2;b_1-1,b_2,b_{12}+1)2(b_{12}+1)(\Delta_2+s_2+2b_2+b_{12})
\\
\label{rec1}
&\qquad\qquad\qquad\qquad\qquad\qquad\qquad\qquad\qquad\times \frac{\Delta_1+s+2b_1+b_{12}-1}{\Delta_1+s_1+2b_1+b_{12}-1}=0,
\end{align}
which relates $a_{n,s}$'s for the same values of arguments $s_1$ and $s_2$ and different
values of $b_1$, $b_2$ and $b_{12}$.

In a similar way, the equation corresponding to $C_2$ reads 
\begin{align}
\notag
&a_{n,s}(s_1,s_2;b_1,b_2,b_{12})2b_2 (2\Delta_2+2b_2-d)\\
\notag
&\quad-a_{n,s}(s_1,s_2;b_1-1,b_2-1,b_{12}+2)(b_{12}+2)(b_{12}+1)\\
\notag
&\quad\quad+a_{n,s}(s_1,s_2;b_1,b_2-1,b_{12}+1)2(b_{12}+1)(\Delta_1+s_1+2b_1+b_{12})
\\
\label{rec2}
&\qquad\qquad\qquad\qquad\qquad\qquad\qquad\qquad\qquad\times \frac{\Delta_2+s+2b_2+b_{12}-1}{\Delta_2+s_2+2b_2+b_{12}-1}=0.
\end{align}
Together with \eqref{rec1}, this defines  $b$-dependence of $a_{n,s}$ for fixed $s_1$ and $s_2$.

\subsubsection{Dependence on contracted derivatives}
In this section we solve the recurrence equations (\ref{rec1}), (\ref{rec2}) subject to the boundary condition
\begin{equation}
\label{boundcond}
a_{n,s}(s_1,s_2;b_1,b_2,b_{12})=0, \quad \text{for $b_1<0$ or $b_2<0$ or $b_1+b_2>n$}.
\end{equation}
More precisely, we express the unknown coefficients in terms of 
$a_{n,s}(s_1,s_2;0,0,n)$. For brevity, in the following we will keep only the arguments
$b_1$ and $b_2$ of $a_{n,s}$ explicit.

First, we consider (\ref{rec1}) for $b_2=0$. In this case the recursion relation simplifies, since the boundary condition causes the second term to drop out. We can then express $a(b_1,0)$ with arbitrary $b_1$ in terms of $a(0,0)$
\begin{equation}
\label{b2eq0}
a(b_1,0)= (-1)^{b_1}\frac{n!}{b_1! b_2!}\frac{(\Delta_2+s_2+n-b_1)_{b_1}(\Delta_1+s+n)_{b_1}}{2^{b_1}(\Delta_1+1-h)_{b_1} (\Delta_1+s_1+n)_{b_1}}a(0,0).
\end{equation}
To establish the $b_2$-dependence, we use (\ref{rec2}): First, for any $b_1$ we express $a(b_1,1)$ in terms of $a(b_1,0)$ and $a(b_1-1,0)$, which are known. We then solve for $a(b_1,2)$ in terms of $a(b_1,1)$ and $a(b_1-1,1)$, which were determined in the previous step. This process can be continued to express $a(b_1,b_2)$
for any $b_2$ in terms of $a(0,0)$. We find 
\begin{align}
\notag
& a(b_1,b_2)=\left(-\frac{1}{2}\right)^{b_1+b_2}\frac{n!}{b_1!b_2!b_{12}!}\\
\notag
&\quad\times
\frac{(\Delta_1+s+n)_{b_1}(\Delta_2+s+n-b_1)_{b_2}}{(\Delta_1+1-h)_{b_1}(\Delta_2+1-h)_{b_2}(\Delta_1+s_1+n)_{b_1-b_2}(\Delta_2+s_2+n)_{b_2-b_1}}\\
\label{result}
&\quad\quad\quad \quad\times \sum_{k=0}^{b_2}\frac{b_2!}{k!(b_2-k)!}\frac{(b_1-k+1)_k(\Delta_1+b_1-h-k+1)_k}{(\Delta_2+s+n-b_1)_k(\Delta_1+s+n+b_1-k)_k}a(0,0).
\end{align}
It can be checked that (\ref{result}) satisfies (\ref{rec1}) and (\ref{rec2}).

Combining the $s$- and $b$-dependences, we obtain the final result
\begin{align}
\notag
&a_{n,s}(s_1,s_2;b_1,b_2,b_{12})=\frac{(-1)^{s_2+b_1+b_2}}{2^{b_1+b_2}}\frac{s!}{s_1!s_2!}\frac{(\Delta_1+s_1+2b_1+b_{12})_{s_2}}{(\Delta_2+2b_2+b_{12})_{s_2}}\\
\notag
&\quad\times \frac{n!}{b_1!b_2!b_{12}!}
\frac{(\Delta_1+s+n)_{b_1}(\Delta_2+s+n-b_1)_{b_2}}{(\Delta_1+1-h)_{b_1}(\Delta_2+1-h)_{b_2}(\Delta_1+s_1+n)_{b_1-b_2}(\Delta_2+s_2+n)_{b_2-b_1}}\\
\label{completedep}
&\qquad\quad \times  \sum_{k=0}^{b_2}\frac{b_2!}{k!(b_2-k)!}\frac{(b_1-k+1)_k(\Delta_1+b_1-h-k+1)_k}{(\Delta_2+s+n-b_1)_k(\Delta_1+s+n+b_1-k)_k}
a_{n,s}(s,0;0,0,n)\,,
\end{align}
where $a_{n,s}(s,0;0,0,n)$ is an arbitrary factor.

\chapter{Trace structure of spinning propagators}
\label{app::tssp}

With the freedom to remove gradients within the propagators \eqref{tracelesspartdeDonder} and \eqref{tlsplit} already derived, it should be possible to bring them into the form
\begin{equation}
\label{eq:ManifestTraceapp}
\Pi_{s} = \sum^{[s/2]}_{k=0} \int^{\infty}_{-\infty}  d\nu \; g_{s,k}\left(\nu\right) \left(u_1^2\right)^{k} \left(u_2^2\right)^{k} \Omega_{\nu,s-2k},
\end{equation}
by making gauge transformations. This is the goal of this section. We refer to this as the ``manifest trace gauge'', 
because here the propagator is presented as a sum of terms, each being essentially a product of certain number of background metrics and a
harmonic function $\Omega$, which is traceless and transverse.

To reach the form \eqref{eq:ManifestTraceapp}, naively one might expect that, for example in \eqref{tlsplit}, one can gauge away all the terms
except the one for which $\ell=0$. However, closer inspection reveals that this is not the case: According to our conventions,
 contractions with $W$ implicitly make a projection onto the traceless part,
 so a generic term in \eqref{tlsplit} is of the form
  \begin{equation}
  \label{tomanifesttraces}
\left\{ \left(u_1 \cdot \nabla_1\right)^{\ell}\left(u_2 \cdot \nabla_2\right)^{\ell}
 \Omega_{\nu,s-\ell}(u_1,x_1,u_2,x_2)\right\}.
\end{equation}
For a general rank-$s$ tensor $T_s$, the explicit action of the operator that implements the traceless projection $\left\{\bullet\right\}$, is
 \begin{eqnarray} \label{tlprojector}
 \left\{T_{s}(x,u)\right\} = \sum_{j=0}^{[s/2]}\frac{(-1)^j}{4^j j! (\tfrac{d}{2}+s-3/2)_j}(u^2)^j (\partial_u\cdot \partial_u)^j 
 T_{s}(x,u),
 \end{eqnarray}
 which makes various contractions of its argument $T_s$. It is then clear that for non-zero $\ell$ there are terms in  \eqref{tomanifesttraces} which cannot be gauged away. For example, terms in which all $\nabla$'s are contracted because then no gradients will be present. In fact, the story is even more complicated because for other terms in the projection the commutators of derivatives that appear yield extra lower derivative
 terms, some of which are also not pure gauge. To summarise, eliminating pure gradient terms is non-trivial, and our aim in the following
 is to compute \eqref{tomanifesttraces} explicitly modulo such gradient terms. We do this by studying the details of the contractions under \eqref{tlprojector}.

It is straightforward to see that when  ${\ell}$ is odd, \eqref{tomanifesttraces} can be 
gauged away since it is a gradient. Let us then consider examples for when $\ell$ is even:
\begin{itemize}
 \item \underline{$\ell = 2$}
 
  Using \eqref{tlprojector}, the explicit form of the traceless projection for $\ell=2$ is
 \begin{align} \label{l2}
&\left\{ \left(u_1 \cdot \nabla_1\right)^{2}\left(u_2 \cdot \nabla_2\right)^{2}
 \Omega_{\nu,s-2}(u_1,u_2)\right\}&\\\nonumber
 &\quad =\left(1-\frac{u_1^2 (\partial_{u_1}\cdot \partial_{u_1})}{2(d+2s-3)}\right)
 \left(1-\frac{u_2^2 (\partial_{u_2}\cdot \partial_{u_2})}{2(d+2s-3)}\right)&
 \left(u_1 \cdot \nabla_1\right)^{2}\left(u_2 \cdot \nabla_2\right)^{2}
 \Omega_{\nu,s-2}(u_1,u_2). \nonumber
 \end{align} 
 Dropping gradient terms, it is straightforward to compute that 
\begin{equation}
\label{trace1time}
(\partial_{u_1}\cdot \partial_{u_1})(u_1\cdot \nabla_1)^2\Omega_{\nu,n}(u_1,u_2)\sim
2\left(\Box_1-n(d+n-1)\right)\Omega_{\nu,n}(u_1,u_2),
\end{equation}
where we used the fact that $\Omega$ traceless and divergence-less.\footnote{The dropping of gradient terms is denoted by $``\sim"$. Note that in appendix \ref{apptraceofcur1} we also use this notation to instead indicate that equalities hold modulo gradients \emph{and} traces.} Therefore the only terms that are not pure gradient in \eqref{l2} come from the product of the second terms in each of the brackets:
 \begin{align}
 \notag
 \frac{u_1^2 (\partial_{u_1}\cdot \partial_{u_1})}{2(d+2s-3)}(u_1\cdot \nabla_1)^2
& \frac{u_2^2 (\partial_{u_2}\cdot \partial_{u_2})}{2(d+2s-3)}(u_2\cdot \nabla_2)^2
 \Omega_{\nu,s-2}(u_1,u_2)\\
 \label{ell2almostans}
\; \sim \; & u_1^2 u_2^2\left(\frac{\Box_1-(s-2)(d+s-3)}{d+2s-3}\right)^2 \Omega_{\nu,s-2}(u_1,u_2),
\end{align}
Finally, employing the equation of motion \eqref{eomharm} for $\Omega$ we find
\begin{equation}
\label{ell2ans}
\left\{ \left(u_1 \cdot \nabla_1\right)^{2}\left(u_2 \cdot \nabla_2\right)^{2}
 \Omega_{\nu,s-2}(u_1,u_2)\right\} \: \sim \:
u_1^2 u_2^2 \left(\frac{\nu^2+(\tfrac{d}{2}+s-2)^2}{d+2s-3}\right)^2 \Omega_{\nu,s-2}(u_1,u_2).
\end{equation}
\item \underline{$\ell = 4$}
Explicitly, traceless projection in this case is
\begin{align}
\notag
\left\{ \left(u_1 \cdot \nabla_1\right)^{4}\left(u_2 \cdot \nabla_2\right)^{4}
 \Omega_{\nu,s-4}(u_1,u_2)\right\}& \\
 \label{justanotherformula}
 =\Big(1-\frac{u_1^2 (\partial_{u_1}\cdot \partial_{u_1})}{2(d+2s-3)}&
 +\frac{u_1^4 (\partial_{u_1}\cdot \partial_{u_1})^2}{8(d+2s-3)(d+2s-5)}\Big)
 \Big(1\leftrightarrow 2\Big)\\
 \notag
 &\qquad \quad \times \;\left(u_1 \cdot \nabla_1\right)^{4}\left(u_2 \cdot \nabla_2\right)^{4}
 \Omega_{\nu,s-4}(u_1,u_2).
 \end{align}
In order to evaluate the traces in the above, a tedious but straightforward computation shows that
\begin{align}
\notag
&(\partial_{u_1}\cdot \partial_{u_1})(u_1\cdot \nabla_1)^4\Omega_{\nu,n}(u_1,u_2)
= 2\Big( 6 (u_1\cdot \nabla_1)^2\Box_1\\
 \label{justanotherformula1}
& \hspace*{0.25cm}-2(3n^2+3dn+4d+5n+2)
(u_1\cdot \nabla_1)^2+4u^2_1\Box_1-4n(d+n-1)u^2_1\Big)\Omega_{\nu,n}(u_1,u_2),
\end{align}
and
\begin{align}
\notag
& (\partial_{u_1}\cdot \partial_{u_1})^2(u_1\cdot \nabla_1)^4\Omega_{\nu,n}(u_1,u_2) \quad 
\sim \quad 4\Big(6\Box^2_1-4(d+3dn-n-3n^2)\Box_1\\
 \label{justanotherformula2}
&\hspace*{6cm}+2n(d+n-1)(2d+n+3dn+n^2)\Big)\Omega_{\nu,n}(u_1,u_2).
\end{align}
Modulo gradient terms, each bracket in 
 \eqref{justanotherformula} therefore produces a second order polynomial in $\Box$. 
 Using the equations of motion \eqref{eomharm}, this eventually gives
 \begin{align}
\notag
&\left\{ \left(u_1 \cdot \nabla_1\right)^{4}\left(u_2 \cdot \nabla_2\right)^{4}
 \Omega_{\nu,s-4}(u_1,u_2)\right\}\qquad \quad &\\
  \label{ell4ans}
& \qquad \sim
\left(u_1^2\right)^2 \left(u_2^2\right)^23^2 \left(\frac{\nu^2+(\tfrac{d}{2}+s-2)^2}{d+2s-3}\right)^2 &
\left(\frac{\nu^2+(\tfrac{d}{2}+s-4)^2}{d+2s-5}\right)^2 \Omega_{\nu,s-4}(u_1,u_2).
\end{align}
\item \underline{$\ell = 2k$}

After studying explicitly cases the $\ell=2k$ with  $k=1,2$ above, we conjecture that
\begin{align}  \label{tomanifesttraces1}
& \left\{ \left(u_1 \cdot \nabla_1\right)^{2k}\left(u_2 \cdot \nabla_2\right)^{2k}
 \Omega_{\nu,s-2k}(u_1,x_1,u_2,x_2)\right\}\\ \notag
 &\hspace*{1.5cm}=\left(N\left(k,s\right)\right)^2(u_1^2)^k(u_2^2)^k \left(4^k \left(\frac{\tfrac{d}{2}+s-2k+i\nu}{2}\right)_k\left(\frac{\tfrac{d}{2}+s-2k-i\nu}{2}\right)_k\right)^2\\
 \notag 
 & \hspace*{2cm} + (u_1\cdot \nabla_1) (\dots ) + (u_2\cdot \nabla_2) (\dots ),
\end{align}
where 
\begin{equation}
\label{tomanifesttraces2}
N(k,s)=\frac{(2k)!}{4^k k! (\tfrac{d}{2}+s-k-1/2)_k}
\end{equation}
and $\left(a\right)_{r} = \Gamma\left(a+r\right)/\Gamma\left(a\right)$ is the rising Pochhammer symbol.
\end{itemize}
 Let us note that in spite of the fact that equation \eqref{tomanifesttraces1} is a conjecture, the pre-factor
$N(k,s)$ is determined exactly. We explain how in the following. The combinatorial factor $N(k,s)$ can be derived by studying
only the contractions that produce the maximal power of $\Box$. The issue of non-commutativity
of covariant derivatives for this computation is irrelevant. Indeed, let us consider an analogous
computation in the flat space. Then in \eqref{trace1time} one has only a $\Box$-term
so that instead of \eqref{ell2almostans} we find
\begin{equation}
\label{ell2ansflat}
\left\{ \left(u_1 \cdot \nabla_1\right)^{2}\left(u_2 \cdot \nabla_2\right)^{2}
 \Omega_{\nu,s-2}(u_1,u_2)\right\}\Big|_{flat}\sim
u_1^2 u_2^2 \left(\frac{\Box}{d+2s-3}\right)^2 \Omega_{\nu,s-2}(u_1,u_2).
\end{equation}
Analogously, for the $\ell=4$ flat case one finds
 \begin{align}
\notag
\left\{ \left(u_1 \cdot \nabla_1\right)^{4}\left(u_2 \cdot \nabla_2\right)^{4}
 \Omega_{\nu,s-4}(u_1,u_2)\right\}\Big|_{flat}& \\
  \label{ell4ansflat}
 \sim
\left(u_1^2\right)^2 \left(u_2^2\right)^23^2& \left(\frac{\Box^2}{(d+2s-3)(d+2s-5)}\right)^2 \Omega_{\nu,s-4}(u_1,u_2).
\end{align}
This computation can be easily generalised to the case of any $\ell=2k$, and the result is
 \begin{align}
\left\{ \left(u_1 \cdot \nabla_1\right)^{2k}\left(u_2 \cdot \nabla_2\right)^{2k}
 \Omega_{\nu,s-2k}(u_1,u_2)\right\}\Big|_{flat}
  \label{ellanyansflat}
 \sim
\left(u_1^2\right)^k \left(u_2^2\right)^kN(k,s) \;\Box^{2k} \Omega_{\nu,s-2k}(u_1,u_2),
\end{align}
with $N(k,s)$ given in \eqref{tomanifesttraces2}. In fact, this justifies its explicit form.

On the other hand, in AdS the $\Box^{2k}$-term will receive lower derivative corrections,
which originate from the non-commutativity of the covariant derivatives.
By generalising the $k=1,2$ cases, what we conjecture is that these lower derivative terms are
such that after evaluating $\Box$ on $\Omega_{\nu, s-2k}$ one finds
\begin{equation*}
\left((\nu^2+(\tfrac{d}{2}+s-2)^2)(\nu^2+(\tfrac{d}{2}+s-4)^2) \dots (\nu^2+(\tfrac{d}{2}+s-2k)^2)\right)^2.
\end{equation*}
Combining this with the pre-factor found previously, one obtains \eqref{tomanifesttraces1}.

\chapter{Higher-spin conserved currents in AdS}
\label{app::hsccads}

\section{Single trace of the currents}
\label{apptraceofcur}
In this appendix, we show how a single trace of the spin-$s$ current
\begin{equation}
    J_s\left(x,u\right) = \sum\limits^s_{k=0} \frac{\left(-1\right)^k}{s!\left(s-k\right)!} \left(u \cdot \nabla \right)^k \varphi_0\left(x\right) \left(u \cdot \nabla \right)^{s-k} \varphi_0\left(x\right) + \Lambda\, u^2\left(...\right),\label{cur1app}
\end{equation}
can be expressed in terms of currents of the same form, but with lower spin. For simplicity we work in ambient space, with the necessary ingredients reviewed below.

\subsubsection*{Scalar fields in AdS}

Since the currents \eqref{cur1app} are bi-linear in the bulk scalar, we first recall the representation of scalar fields in AdS and their ambient formulation. The lowest weight unitary irreducible scalar representations of $so(d,2)$ 
\begin{equation}
\label{massdimension}
\Box\varphi_0(x)-m^2\varphi_0(x)=0, \qquad m^2\equiv \Delta (\Delta-d)
\end{equation}
can be realised as
the evaluation $\varphi_0(x)$ on AdS${}_{d+1}$ of
ambient homogeneous harmonic functions 
$\Phi_0(X)$
\begin{equation}
\label{masslessamb}
(\partial_X\cdot \partial_X) \Phi_0(X) =0, \qquad      \Phi_0(X)=\left(X^2\right)^{-\frac{\Delta}{2}}\varphi_0(x).
\end{equation} 
For later use, let us denote 
\begin{equation}
\Phi_0^\dagger(X)= \left(X^2\right)^{-\frac{\Delta_-}{2}}\varphi_0(x), \label{dagger}
\end{equation}
where $\Delta_-=d-\Delta$, and we assume $\Delta\ge\Delta_-$. Throughout, $\Delta$ will often be referred to as $\Delta_+$.
\subsubsection*{The currents}
Towards an ambient representation of the currents \eqref{cur1app} on AdS, it is instructive to consider analogous conserved currents  in the flat ambient space. I.e. bi-linear currents conserved with respect to the flat ambient derivative.

It is straightforward to verify that for any 
ambient massive scalar fields $\Phi_1(X)$ and $\Phi_2(X)$ of the same mass $M$
\begin{equation}
(\partial_X\cdot \partial_X-M^2) \Phi_1(X)=0=(\partial_X\cdot \partial_X-M^2) \Phi_2(X),
\end{equation}
the currents, given by the generating function \cite{Bekaert:2009ud,Bekaert:2010hk}
\begin{equation}
\label{ambcur1}
I(X,U)=\Phi_1(X+U)\Phi_2(X-U),
\end{equation}
are conserved with respect to flat ambient space derivative
\begin{equation}
(\partial_X\cdot \partial_U)I(X,U)= 0.
\end{equation}
Explicitly, the rank-$s$ current generating function is given by
\begin{equation}
\label{rankscur1}
I_s(X,U)=\sum_{k=0}^s \frac{(-1)^k}{k!(s-k)!}
(U\cdot \partial_X)^{s-k}\Phi_1(X)(U\cdot \partial_X)^k\Phi_2(X),
\end{equation}
which is obtained by extracting from \eqref{ambcur1} the $\mathcal{O}\left(U^s\right)$ coefficient. I.e.
\begin{align}  
I\left(X,U\right) &= \sum_{s} \frac{1}{s!}I_{A_1 ... A_s} U^{A_1} ... U^{A_s}, \qquad I_s\left(X,U\right)  = \frac{1}{s!}I_{A_1 ... A_s} U^{A_1} ... U^{A_s}.
\end{align}
However, a conserved current in the flat ambient space does not necessarily define a conserved current on AdS. In other words, 
the pull-back of \eqref{ambcur1} onto the AdS manifold is not in general conserved with respect to the covariant derivative on AdS. To do so, it must satisfy some 
additional constraints, which we specify in the following.

\noindent
We require \eqref{ambcur1} to be conserved with respect to the ambient representative \eqref{ambcovd} of the covariant derivative, $\nabla = {\cal P} \circ \partial \circ {\cal P}$. First, the projection of the ambient current onto AdS is of the form
\begin{equation}
{\cal P} \; I(X,U) = I(X,U)+ (X\cdot U) L(X,U)
\end{equation}
for some $L(X,U)$.
Further, the commutation relation 
\begin{equation*}
[\left(\partial_X\cdot\partial_U),X\cdot U\right]=X\cdot\partial_X+U\cdot\partial_U+d+2
\end{equation*}
implies that 
\begin{align}
\label{conservpr}
(\partial_X\cdot \partial_U){\cal P} \; I(X,&U)
= (X\cdot U)\left(\partial_X\cdot\partial_U\right)
L(X,U)+
\left(X\cdot\partial_X+U\cdot\partial_U+d+2\right) L(X,U).
\end{align}
The first term in \eqref{conservpr} is transversal to $H_{d+1}$, so it drops out upon application of the second projection in the covariant derivative \eqref{ambcovd}. We then demand that $I\left(X,U\right)$ is
conserved in AdS,
\begin{equation}
 \left( \nabla \cdot \partial_{U} \right) I\left(X,U\right) = 0,
\end{equation}
which yields the condition:
\begin{equation*}
\left(X\cdot\partial_X+U\cdot\partial_U+d+2\right) L(X,U)=0.
\end{equation*}
This can be satisfied by imposing the following homogeneity condition on the current
\begin{equation}
\label{homcur}
\left(X\cdot\partial_X+U\cdot\partial_U+d\right) I(X,U)=0.
\end{equation}
The demonstrate that the current \eqref{ambcur1}
is conserved in AdS if it obeys \eqref{homcur}. In particular for $\Phi_1=\Phi$, and $\Phi_2=\Phi^{\dagger}$ as introduced in \eqref{dagger}, the current
\begin{equation}
\label{ambcur2}
J(X,U)=\Phi(X+U)\Phi^\dagger(X-U)
\end{equation}
is conserved on AdS. Note that the more familiar form \eqref{cur1app},  intrinsic to AdS, can be obtained from the above by expressing the ambient partial derivatives in terms of AdS covariant derivatives and the metric.

\subsubsection{Trace of the currents}

We are now ready to proceed in ambient space to express the trace of the current \eqref{cur1app} in terms of those of lower ranks. The formula we want to establish generalises to AdS the expression
\begin{equation}
\label{BBVDcons}
(\partial_u\cdot \partial_u)\;I(x,u)=\left(-\Box+ 4M^2\right){I}(x,u),
\end{equation}
for the trace in flat space.

On AdS, the trace is given by 
\begin{align}
\notag
(\partial_u\cdot\partial_u)J(X,U)&\equiv\left(\partial_U\cdot\partial_U+(X\cdot \partial_U)^2\right)J(X,U)\\
\label{trcur1} 
&=\;
\left(-\partial_X\cdot\partial_X+(X\cdot \partial_U)^2\right)J(X,U),
\end{align}
where on the first line we expressed the intrinsic AdS trace in terms of ambient quantities, and in the second equality we used that \begin{equation}
(\partial_X\cdot\partial_X+\partial_U\cdot\partial_U)J(X,U)= 0,
\end{equation}
taking into account that the ambient scalar $\Phi_0$ is massless \eqref{masslessamb}.

The next step is to rewrite the RHS of \eqref{trcur1},
\begin{align}
\notag
(\partial_u\cdot\partial_u)J(X,U) =
\left(-\partial_X\cdot\partial_X+(X\cdot \partial_U)^2\right)J(X,U),
\end{align}
in terms of the ambient representatives of AdS covariant 
operators: The Laplacian $\Box$, rank of the current $(U\cdot\partial_U)$, multiplication by the 
metric $U^2$ and the covariant gradient
$(U\cdot\nabla)$:

\begin{enumerate}
    \item We first focus on the term $\left(X\cdot \partial_U\right)^2J\left(X,U\right)$. We may eliminate $(X\cdot \partial_U)$ in favour of $(U\cdot \partial_X)$, using that
\begin{equation*}
(X\cdot \partial_U)+(U\cdot \partial_X)=X^+\partial_+-X^-\partial_-.
\end{equation*}
This gives\footnote{Let us note that the identity above can be used only when $(X\cdot \partial_U)$
acts directly on the current.}
\begin{equation}
\label{trcur2}
(X\cdot \partial_U)J= (\Delta_+-\Delta_--(U\cdot \partial_X))J.
\end{equation}
Applying this twice, we find 
\begin{align*}
(X\cdot \partial_U)^2J&=(X\cdot \partial_U) (\Delta_+-\Delta_--(U\cdot \partial_X))J\\
&=(\Delta_+-\Delta_--(U\cdot \partial_X))(X\cdot \partial_U) J-[(X\cdot \partial_U),(U\cdot \partial_X)]
J\\
&=(\Delta_+-\Delta_--(U\cdot \partial_X))^2J-(X\cdot \partial_X-U\cdot \partial_U)J.
\end{align*}
The $(X\cdot\partial_X)$ can be expressed in terms of $(U\cdot \partial_U)$ via \eqref{homcur},
\begin{equation}
(X\cdot\partial_X)J=-((U\cdot\partial_U)+d)J.
\end{equation}
As a final step, we express $(U\cdot \partial_X)$ in terms of covariant gradients $(U\cdot \nabla)$, 
using
\begin{equation*}
(U\cdot \nabla)= U^2 (X\cdot\partial_U)+(U\cdot \partial_X).
\end{equation*}
Together with \eqref{trcur2}, this yields
\begin{equation}
(U\cdot \partial_X)J=\frac{1}{1-U^2}\left((U\cdot\nabla)-U^2(\Delta_+-\Delta_-)\right)J,
\end{equation}
where the fraction should be understood as a power series
\begin{equation*}
\frac{1}{1-U^2}=1+U^2+U^4+\dots.
\end{equation*}
We finally arrive to, 
\begin{align}
(U\cdot \partial_X)^2J=\left(\frac{1}{1-U^2}\left((U\cdot\nabla)-U^2(\Delta_+-\Delta_-)\right)\right)^2J
-\frac{U^2}{1-U^2}(X\partial_X-U\partial_U)J.
\end{align}
\item To express $\left(\partial_X \cdot \partial_X\right) J\left(X,U\right)$ in terms of AdS covariant quantities, we need only employ
\begin{align}
\label{laplacian}
\Box=\partial_X\cdot\partial_X+(X\cdot \partial_X)\left((X\cdot \partial_X)+d-1\right)
-\;(U\cdot\partial_U) +2(U\cdot\partial_X)(X\cdot\partial_U)+U^2(X\cdot\partial_U)^2,
\end{align}
in combination with the above results. 
\end{enumerate}    

The generalisation of \eqref{BBVDcons} to AdS is thus
\begin{align}
\label{traceofXEcurrent}
\notag
\left(\partial_u\cdot\partial_u\right)J
=
\Big(-\Box+(u\cdot\partial_u+1)(&u\cdot\partial_u+d)\\
&+\;\frac{1}{1-u^2}\left((\Delta_+-\Delta_-)^2-(u\cdot\nabla)^2\right) \Big)J.
\end{align}
Recalling that 
\begin{align}
J\left(X,U\right) = \sum^\infty_{s=0} J_{s}\left(X,U\right),
\end{align}
i.e. it is the generating function of conserved currents, equation \eqref{traceofXEcurrent} serves as a starting point to express the trace of a current of given spin in terms of other currents.
\section{Multiple traces of the currents}
\label{apptraceofcur1}

In this appendix we compute $\tau_{s,k}\left(\nu\right)$ in \eqref{sec::wiit::ta}. 
To this end, we take repeated traces of \eqref{traceofXEcurrent}.
This allows us to express a $k$-fold trace of rank-$s$ current $J_{s}$ in terms of lower degree traces of lower-spin currents 
\begin{eqnarray}
\label{schematically}
(\partial_u\cdot\partial_u)^{k}J_{s}\quad \rightarrow \quad
(\partial_u\cdot\partial_u)^{k-1}J_{s-2}, \quad(\partial_u\cdot\partial_u)^{k-2}J_{s-4},
\quad\dots,\quad
J_{s-2k}
\end{eqnarray}
and terms of the form
\begin{equation}
\label{tracesandgrad}
(u\cdot\nabla)(\dots) \quad \text{or} \quad  u^2  \cdot (\dots).
\end{equation}
The relation \eqref{schematically} can then be iteratively applied to eliminate all the traces, thereby expressing $(\partial_u\cdot\partial_u)^{k}J_{s}$
in terms of $J_{s-2k}$ and terms of the form \eqref{tracesandgrad}.
Our main goal is to compute the four-point spin-$s$ exchange \S \tcb{\ref{subsec::4ptexch}}, so it is
enough to know $(\partial_u\cdot\partial_u)^{k}J_{s}$ modulo terms which vanish upon contraction against the traceless and divergence-free $\Omega_{s-2k}$.
These are precisely the terms in \eqref{tracesandgrad}. We will often drop such terms where they are unimportant. Equalities that hold modulo
these terms will be denoted by ``$\sim$''.

We shall employ the following useful identities
\begin{align}
\notag
(\partial_u\cdot \partial_u)^i\left(u^2\right)^k J_l
=\sum_{n=0}^{i}&4^{i-n}\frac{i!}{n!(i-n)!}(k-i+n+1)_{i-n}\\
&\times\;(\tfrac{d+1}{2}+k+l-i)_{i-n}
\label{appb0}
\left(u^2\right)^{k-i+n}\left(\partial_u\cdot\partial_u\right)^nJ_l,
\end{align}
and
\begin{align}
\notag
(\partial_u\cdot\partial_u)^m(u\cdot\nabla)^2J_k=
2m\cdot\frac{k-2m+2}{k-2m+3}&\;(\partial_u\cdot \nabla)( u\cdot\nabla)(\partial_u\cdot\partial_u)^{m-1}J_k
\\
\label{traceofgrad}
+\;
2m\;\Box\; (\partial_u&\cdot\partial_u)^{m-1}J_k
+
(u\cdot \nabla)^2(\partial_u\cdot\partial_u)^mJ_k.
\end{align}
Here and throughout we use that the current is conserved.
To commute divergence and gradient in \eqref{traceofgrad}, we employ
\begin{equation}
\label{comdivandgrad}
\frac{1}{n+1}(\partial_u\cdot \nabla)(u\cdot\nabla)J_n=
\frac{1}{n}(u\cdot\nabla)(\partial_u\cdot \nabla)J_n-(n+d)J_n+\frac{1}{n}
u^2(\partial_u\cdot\partial_u)J_n,
\end{equation}
which entails
\begin{equation}
\label{appb1}
(\partial_u\cdot\partial_u)^m(u\cdot\nabla)^2J_k\sim2m\left(\Box-(k-2m-2)(k-2m+d+1) \right)(\partial_u\cdot\partial_u)^{m-1}J_k.
\end{equation}
With these formulas at hand, we are prepared to compute the $i$-fold trace
$(\partial_u\cdot\partial_u)^{i}$ of both sides of \eqref{traceofXEcurrent}.
Using \eqref{appb0}, keeping only terms with $k-i+n=0$, we obtain 
\begin{align*}
&(\partial_u\cdot\partial_u)^{i+1}J_{s+2}\sim-\;\Box\; (\partial_u\cdot\partial_u)^iJ_s+
(s+1)(s+d)(\partial_u\cdot\partial_u)^iJ_s\\
& \hspace*{.1cm} +
\sum_{k=0}^i\frac{i!}{(i-k)!}4^k \left((d+1)/2+s-i-k\right)_k (\partial_u\cdot\partial_u)^{i-k}\big((\Delta_+-\Delta_-)^2J_{s-2k}-(u\cdot\nabla)^2J_{s-2k-2}\big).
\end{align*}
Which, employing \eqref{appb1}, becomes
\begin{align}
\notag
(\partial_u\cdot\partial_u)^{i+1}J_{s+2}\quad \sim \quad & - \quad \Box (\partial_u\cdot\partial_u)^iJ_s+
(s+1)(s+d)(\partial_u\cdot\partial_u)^iJ_s\\
\notag
& +\; \sum_{k=0}^i(\partial_u\cdot\partial_u)^{i-k}(\Delta_+-\Delta_-)^2J_{s-2k}
\frac{i!}{(i-k)!}4^k \left(\tfrac{d+1}{2}+s-i-k\right)_k\\
\notag
-\;&\sum_{k=0}^{i-1}\left(\Box-(s-2i)(s-2i+d-1)\right)(\partial_u\cdot\partial_u)^{i-k-1}J_{s-2k-2}\\
\label{appb2}
&\qquad\qquad\times\;
\frac{i!}{(i-k-1)!}4^{k+1} \left(\tfrac{d+1}{2}+s-i-k\right)_{k+1}.
\end{align}
The crucial observation is that applying \eqref{appb2} to the right combination of 
$(\partial_u\cdot\partial_u)^{i+1}J_{s+2}$ and $(\partial_u\cdot\partial_u)^{i}J_{s}$,
the tails of terms on the right hand side cancel. To wit,
\begin{align}\label{appb3}
& (\partial_u\cdot\partial_u)^{i+1}J_{s+2}-2i(d+2s-2i-1)(\partial_u\cdot\partial_u)^{i}J_{s} \\ \nonumber
&\hspace*{3cm}\sim \quad \left((\Delta_+-\Delta_-)^2-\Box+(s+1)(s+d)\right)(\partial_u\cdot\partial_u)^{i}J_{s} \\ \nonumber
& \hspace*{4cm} -2i\left(\Box-(s-2i)(s-2i+d-1)\right)(\partial_u\cdot\partial_u)^{i-1}J_{s-2}\\  \nonumber
& \hspace*{4cm}-2i(d+2s-2i-1)\left(\Box-(s-1)(s+d-2)\right)(\partial_u\cdot\partial_u)^{i-1}J_{s-2}.
\end{align}
To deal with the $\Box$ terms, from this point onwards we assume that both sides of \eqref{appb3} are integrated against the traceless and divergenceless harmonic
function $\Omega_{s-2k}$. Then one can integrate by parts all $\Box$'s that appear
 on the right hand side of \eqref{appb3}, thus making them act on $\Omega_{s-2k}$. We can then use the equation of motion \eqref{eomharm} for $\Omega_{s-2k}$ to eliminate all $\Box$'s. This establishes the following iterative equation
\begin{equation}
\label{appb4}
(\partial_u\cdot\partial_u)^{i+1}J_{s+2}\sim f(i,s)(\partial_u\cdot\partial_u)^{i}J_{s}+
g(i,s)(\partial_u\cdot\partial_u)^{i-1}J_{s-2},
\end{equation}
where 
\begin{align*}
&f(i,s)=(2\Delta-d-1)(2\Delta-d+1)\; +\;\left(\nu^2+(\tfrac{d}{2}+(s-2i)+1)^2\right)+8i(\tfrac{d}{2}+(s-2i)+i),\\
&g(i,s)=-4i(\tfrac{d}{2}+(s-2i)+i-1)\left(\left(\nu^2+(\tfrac{d}{2}+(s-2i)+1)^2\right)+4(i-1)(\tfrac{d}{2}+(s-2i)+i)\right).
\end{align*}
As ``boundary conditions'' we take $(\partial_u\cdot\partial_u)^{-1}J_{s-2k-2}\equiv 0$
and assume that $J_{s-2k}$ is given. Then the iterative equation \eqref{appb4}
allows to express $(\partial_u\cdot\partial_u)^{k}J_{s}$ in terms of  $J_{s-2k}$.
The result is
\begin{eqnarray*}
(\partial_u\cdot\partial_u)^{k}J_{s}\sim \tau_{s,k}\left(\nu\right)\;J_{s-2k}, 
\end{eqnarray*}
with
\begin{multline}
\tau_{s,k}\left(\nu\right)=\sum_{m=0}^k 
\frac{2^{2k}\cdot k!}{m!(k-m)!}
(-1)^{k-m}\left(1/2-d/2+\Delta\right)_{k-m}\left(1/2+d/2-\Delta\right)_{k-m}\\ \times \left(\frac{\tfrac{d}{2}+s-2m+1+i\nu}{2}\right)_{m}\left(\frac{\tfrac{d}{2}+s-2m+1-i\nu}{2}\right)_{m}.
\end{multline}

\chapter{Intrinsic expressions for vertices}

\label{radialred}

In this appendix we give the recipe of how to express vertices in ambient space in intrinsic AdS terms. In \S \tcb{\ref{subsec::cubicrecon}} we used holography to fix the cubic vertices of minimal bosnic higher-spin theory on AdS$_{d+1}$ in terms of the ambient structures    
\begin{align}
  &  {\cal I}_{s_1,s_2,s_3}^{n_1,n_2,n_3}(\varphi_{s_i})=\mathcal{Y}_1^{s_1-n_2-n_3}\mathcal{Y}_2^{s_2-n_3-n_1}\mathcal{Y}_3^{s_3-n_1-n_2}\\ \nonumber
    & \hspace*{6cm} \times\mathcal{H}_1^{n_1}\mathcal{H}_2^{n_2}\mathcal{H}_3^{n_3}\,\varphi_{s_1}(X_1,U_1)\varphi_{s_2}(X_2,U_2)\varphi_{s_3}(X_3,U_3)\Big|_{X_i=X}\,.\label{innnapp}
\end{align}
where we used point splitting, and introduced
\begin{align}
    \mathcal{Y}_1&=\pl_{U_1}\cdot\pl_{X_2}\,,&\mathcal{Y}_2&=\pl_{U_2}\cdot\pl_{X_3}\,,&\mathcal{Y}_3&=\pl_{U_3}\cdot\pl_{X_1}\,,\\
    \mathcal{H}_1&=\pl_{U_2}\cdot\pl_{U_3}\,,&\mathcal{H}_2&=\pl_{U_3}\cdot\pl_{U_1}\,,&\mathcal{H}_3&=\pl_{U_1}\cdot\pl_{U_2}\,.
\end{align}
While this representation allowed for a more straightforward manipulation of tensor structures, since they are vertices for a theory on AdS naturally one would like to express them in terms intrinsic to the AdS manifold.  In other words, in terms of contractions of the form
\begin{align}
    \widetilde{\cal Y}_1&=\pl_{U_1}\cdot\nabla_2\,,& \widetilde{\cal Y}_2&=\pl_{U_2}\cdot\nabla_3\,,&\widetilde{\cal Y}_3&=\pl_{U_3}\cdot\nabla_1\,.
\end{align}
In order to achieve this, we consider the structure
\begin{multline}
    I_{l_1,l_2,l_3,k_1,k_2,k_3}^{n_1,n_2,n_3,m_1,m_2,m_3}(\varphi_{s_i})\equiv {\cal H}_1^{n_1} {\cal H}_2^{n_2} {\cal H}_3^{n_3}\,{\cal Y}^{l_1}_1{\cal Y}^{l_2}_2{\cal Y}^{l_3}_3\,\widetilde{\cal Y}^{k_1}_1\widetilde{\cal Y}^{k_2}_2\widetilde{\cal Y}^{k_3}_3\\\times\,(X_1^2)^{-m_1}(X_2^2)^{-m_2}(X_3^2)^{-m_3}\,\varphi_{s_1}(X_1,U_1)\varphi_{s_2}(X_2,U_2)\varphi_{s_3}(X_3,U_3),
\end{multline}
where
\begin{align}
I_{s_1-n_2-n_3,s_2-n_1-n_3,s_3-n_1-n_2,0,0,0}^{n_1,n_2,n_3,0,0,0}(\varphi_{s_i}) = {\cal I}_{s_1,s_2,s_3}^{n_1,n_2,n_3}(\varphi_{s_i}). \label{startiterapp}
\end{align}
Employing the commutation relations \eqref{app::amsscomm}, one can establish the following recursion relations\footnote{Recall that $\Delta_i$ is the homogeneity degree of the field $\varphi_{s_i}$, and $\lambda$ is an auxiliary variable to be replaced at the very end of the recursion procedure as follows:
\begin{equation}\label{lambda}
    \lambda^n\equiv(-1)^n(\Delta+d)(\Delta+d-2)\ldots(\Delta+d-2n+2)\,,
\end{equation}
where $\Delta$ is the total degree of homogeneity of the given term of the vertex. The reason $\lambda$ appears is that in order to establish the recursion relations one has to integrate by parts in ambient space (see e.g. \cite{Taronna:2012gb}).}
{\allowdisplaybreaks
\begin{subequations}
\begin{align}
    I_{l_1,l_2,l_3,0,0,k_3}^{n_1,n_2,n_3,m_1,0,0}&\!= I_{l_1,l_2,l_3-1,0,0,k_3+1}^{n_1,n_2,n_3,m_1,0,0}\!-l_2(\Delta_1\!-\!k_3\!-\!n_2\!-\!2m_1)\,I_{l_1,l_2-1,l_3-1,0,0,k_3}^{n_1+1,n_2,n_3,m_1+1,0,0}\\&-\!n_3(l_3\!-\!1)\,I_{l_1,l_2,l_3-2,0,0,k_3}^{n_1+1,n_2+1,n_3-1,m_1+1,0,0}\nonumber\\\nonumber
    &-l_1(\Delta_2\!-\!l_1\!-\!n_3\!-\!l_3\!+\!2)\,I_{l_1-1,l_2,l_3-1,0,0,k_3}^{n_1,n_2+1,n_3,m_1+1,0,0}\\
    &+\!\lambda\,l_1  l_2(l_3-1)\,I_{l_1-1,l_2-1,l_3-2,0,0,k_3}^{n_1+1,n_2+1,n_3,m_1+1,0,0}\,,\nonumber\\
    I_{l_1,l_2,0,0,k_2,k_3}^{n_1,n_2,n_3,m_1,0,m_3}&\!= I_{l_1,l_2-1,0,0,k_2+1,k_3}^{n_1,n_2,n_3,m_1,0,m_3}\!\!-l_1(\Delta_3\!-\!k_2\!-\!n_1\!-\!2m_3)\,I_{l_1-1,l_2-1,l_3,0,k_2,k_3}^{n_1,n_2,n_3+1,m_1,0,m_3+1}\nonumber\\&-n_2(l_2\!-\!1)\,I_{l_1,l_2-2,l_3,0,k_2,k_3}^{n_1\!+\!1,n_2-1,n_3\!+\!1,m_1,0,m_3\!+\!1}\\
    I_{l_1,0,0,k_1,k_2,k_3}^{n_1,n_2,n_3,m_1,m_2,m_3}&\!= I_{l_1-1,0,0,k_1+1,k_2,k_3}^{n_1,n_2,n_3,m_1,m_2,m_3}-n_1(l_1\!-\!1)\,I_{l_1-2,0,0,k_1,k_2,k_3}^{n_1-1,n_2+1,n_3+1,m_1,m_2+1,m_3}\,,
\end{align}
\end{subequations}}
which can be solved in terms of the intrinsic structures 
\begin{align}
I_{0,0,0,k_1,k_2,k_3}^{n_1,n_2,n_3,0,0,0}(\varphi_{s_i}) = {\cal H}_1^{n_1} {\cal H}_2^{n_2} {\cal H}_3^{n_3}\,\widetilde{\cal Y}^{k_1}_1\widetilde{\cal Y}^{k_2}_2\widetilde{\cal Y}^{k_3}_3\,\varphi_{s_1}(X_1,U_1)\varphi_{s_2}(X_2,U_2)\varphi_{s_3}(X_3,U_3).
\end{align}
Starting with the original basis \eqref{innnapp} for the vertices, with the above recursions their intrinsic forms can therefore be established.

 Below we give some examples obtained by implementing the above recursion relations in Mathematica for massless fields ($\Delta_i=2-d-s_i$) 
{\allowdisplaybreaks
\begin{subequations}
\begin{align}
    {\cal Y}_1{\cal Y}_2&=\widetilde{\cal Y}_1\widetilde{\cal Y}_2-(d-2) {\cal H}_3\,,\\
    {\cal Y}_1{\cal Y}_2{\cal Y}_3&=\widetilde{\cal Y}_1\widetilde{\cal Y}_2\widetilde{\cal Y}_3-(d-1)({\cal H}_1\widetilde{\cal Y}_1+{\cal H}_2\widetilde{\cal Y}_2+{\cal H}_3\widetilde{\cal Y}_3)\,,\nonumber\\
    {\cal Y}^2_1{\cal Y}^2_2{\cal Y}^2_3
    &=\widetilde{\cal Y}_1^2 \widetilde{\cal Y}_2^2 \widetilde{\cal Y}_3^2-2(2 d+1) {\cal H}_1  \widetilde{\cal Y}_1 ^2 \widetilde{\cal Y}_2  \widetilde{\cal Y}_3 -2(2 d+3) {\cal H}_2  \widetilde{\cal Y}_1  \widetilde{\cal Y}_2 ^2 \widetilde{\cal Y}_3 -2(2 d+1) {\cal H}_3  \widetilde{\cal Y}_1  \widetilde{\cal Y}_2  \widetilde{\cal Y}_3 ^2\nonumber\\
    &+2 d (d+2) {\cal H}_1^2 \widetilde{\cal Y}_1^2+2 d (d+2) {\cal H}_3^2 \widetilde{\cal Y}_3^2+2 d (d+2) {\cal H}_2^2 \widetilde{\cal Y}_2^2\nonumber\\
    &+8 \left(d^2+2 d+2\right) {\cal H}_1  {\cal H}_2  \widetilde{\cal Y}_1  \widetilde{\cal Y}_2 +2\left(4 d^2+4 d-1\right) {\cal H}_1  {\cal H}_3  \widetilde{\cal Y}_1  \widetilde{\cal Y}_3 \nonumber\\
    &+4 (d+1) (2 d+1) {\cal H}_2  {\cal H}_3  \widetilde{\cal Y}_2  \widetilde{\cal Y}_3 -8 \left(d^3+2 d^2+d+1\right) {\cal H}_1  {\cal H}_2  {\cal H}_3\,,\\
    {\cal Y}^3_1{\cal Y}^3_2&{\cal Y}^3_3=\widetilde{\cal Y}_1^3 \widetilde{\cal Y}_2^3 \widetilde{\cal Y}_3^3\nonumber\\
    &-9 (d+2) {\cal H}_1 \widetilde{\cal Y}_1^3 \widetilde{\cal Y}_2^2 \widetilde{\cal Y}_3^2-9 (d+4) {\cal H}_2 \widetilde{\cal Y}_1^2 \widetilde{\cal Y}_2^3 \widetilde{\cal Y}_3^2\nonumber\\
    &-9 (d+2) {\cal H}_3 \widetilde{\cal Y}_1^2 \widetilde{\cal Y}_2^2 \widetilde{\cal Y}_3^3+18 \left(d^2+5 d+5\right) {\cal H}_1^2 \widetilde{\cal Y}_1^3 \widetilde{\cal Y}_2 \widetilde{\cal Y}_3\nonumber\\
    &+27 \left(2 d^2+12 d+21\right) {\cal H}_1 {\cal H}_2 \widetilde{\cal Y}_1^2 \widetilde{\cal Y}_2^2 \widetilde{\cal Y}_3\nonumber\\
    &+18 \left(d^2+7 d+11\right) {\cal H}_2^2 \widetilde{\cal Y}_1 \widetilde{\cal Y}_2^3 \widetilde{\cal Y}_3+18 \left(d^2+5 d+5\right) {\cal H}_3^2 \widetilde{\cal Y}_1 \widetilde{\cal Y}_2 \widetilde{\cal Y}_3^3\nonumber\\
    &+54 (d+1) (d+3) {\cal H}_1 {\cal H}_3 \widetilde{\cal Y}_1^2 \widetilde{\cal Y}_2 \widetilde{\cal Y}_3^2+27 (d+2) (2 d+7) {\cal H}_2 {\cal H}_3 \widetilde{\cal Y}_1 \widetilde{\cal Y}_2^2 \widetilde{\cal Y}_3^2\nonumber\\
    &-54 (d+3) \left(d^2+6 d+10\right) {\cal H}_1^2 {\cal H}_2 \widetilde{\cal Y}_1^2 \widetilde{\cal Y}_2-54 (d+3) \left(d^2+6 d+10\right) {\cal H}_1 {\cal H}_2^2 \widetilde{\cal Y}_1 \widetilde{\cal Y}_2^2\nonumber\\
    &-54 (2 d+7) \left(2 d^2+8 d+9\right) {\cal H}_1 {\cal H}_2 {\cal H}_3 \widetilde{\cal Y}_1 \widetilde{\cal Y}_2 \widetilde{\cal Y}_3-54 (d+4) \left(d^2+3 d+1\right) {\cal H}_1 {\cal H}_3^2 \widetilde{\cal Y}_1 \widetilde{\cal Y}_3^2\nonumber\\
    &-27 \left(2 d^3+14 d^2+28 d+11\right) {\cal H}_1^2 {\cal H}_3 \widetilde{\cal Y}_1^2 \widetilde{\cal Y}_3-27 \left(2 d^3+16 d^2+38 d+25\right) {\cal H}_2^2 {\cal H}_3 \widetilde{\cal Y}_2^2 \widetilde{\cal Y}_3\nonumber\\
    &-54 \left(d^3+8 d^2+21 d+17\right) {\cal H}_2 {\cal H}_3^2 \widetilde{\cal Y}_2 \widetilde{\cal Y}_3^2-6 (d+1) (d+3) (d+5) {\cal H}_1^3 \widetilde{\cal Y}_1^3\nonumber\\
    &-6 (d+1) (d+3) (d+5) {\cal H}_2^3 \widetilde{\cal Y}_2^3-6 (d+1) (d+3) (d+5) {\cal H}_3^3 \widetilde{\cal Y}_3^3\nonumber\\
    &+54 (d+3) \left(2 d^3+15 d^2+34 d+27\right) {\cal H}_1^2 {\cal H}_2 {\cal H}_3 \widetilde{\cal Y}_1\nonumber\\
    &+54 (d+3) \left(2 d^3+14 d^2+28 d+19\right) {\cal H}_1 {\cal H}_2^2 {\cal H}_3 \widetilde{\cal Y}_2\nonumber\\
    &+54 (d+3) \left(2 d^3+14 d^2+29 d+19\right) {\cal H}_1 {\cal H}_2 {\cal H}_3^2 \widetilde{\cal Y}_3\,.
\end{align}
\end{subequations}}

\chapter{Basis of quartic vertices}
\label{appendix::quarticbasis}
In this appendix we demonstrate completeness of the basis  \eqref{contwbasistf} of quartic scalar self-interactions.

A given vertex can be brought into different forms via integration by parts. For our purposes, we need only find a complete set of quartic vertices which are independent on-free-shell. We therefore need to take into account all possible relations between such vertices via integration by parts and using the free equations of motion. In counting the number of independent on-shell vertices, it is simpler to consider the equivalent problem of counting the associated
four-point amplitudes. This is how we proceed in the following. For simplicity, following the approach of \cite{Heemskerk:2009pn} we carry out this analysis in flat space.

As shown in \cite{Heemskerk:2009pn}, a complete non-redundant basis of flat space four-point amplitudes is given by the set monomials of the form
\begin{equation}
\label{verticesfromHPPS}
 s^kt^ku^m\quad \text{with integers}\quad  k\ge m\ge 0\,,
 \end{equation}
where $s$, $t$ and $u$ are the Mandelstam variables
\begin{equation*}
s=(p_1+p_2)^2, \quad t=(p_1+p_3)^2, \quad u=(p_2+p_3)^2\,.
\end{equation*}
For massless particles $p^2_i=0$, one has
\begin{equation}
s=2\,p_1\cdot p_2=2\,p_3\cdot p_4, \quad t=2\,p_1\cdot p_3=2\,p_2\cdot p_4, \quad u=2\,p_2\cdot p_3=2\,p_1\cdot p_4\,.
\end{equation}
Our goal is to replace the basis (\ref{verticesfromHPPS}) with an equivalent one,
that is better suited to our purposes - i.e. in terms of amplitudes generated by quartic vertices built from conserved currents. Let us first consider such vertices in position space
\begin{equation}
\label{verticesequivalent1}
\Box_{12}^m J_{\mu_1\cdots\mu_\ell}\big(\varphi_0(x_1),\varphi_0(x_2)\big)J^{\mu_1\cdots\mu_\ell}\big(\varphi_0(x_3),\varphi_0(x_4)\big),
\end{equation}
with $\Box_{12}=(\partial_{x_1}+\partial_{x_2})^2$ and
\begin{equation}
\label{bbvd}
J_{\mu_1\cdots\mu_\ell}\big(\varphi_0(x_1),\varphi_0(x_2)\big)\equiv \varphi_0(x_1)\,
\lrpar_{\mu_1}...\,\lrpar_{\mu_\ell}\,
\varphi_0(x_2)\,,
\end{equation}
are the conserved currents of \cite{Berends:1985xx}, with $\lrpar=\overleftarrow\partial_{x_1}-\overrightarrow\partial_{x_2}$.
The amplitude in momentum space associated to the vertex \eqref{verticesequivalent1} reads
\begin{align}
(p_1+p_2)^m\left[(p_1-p_2)\cdot (p_3-p_4)\right]^l=s^m (t-u)^\ell.
\end{align}
For $\ell=2k$, this amplitude contains a term $s^m t^k u^k$. Up to a Bose symmetry transformation,
this reproduces the form of the vertices 
 (\ref{verticesfromHPPS}).
We therefore conclude that vertices (\ref{verticesequivalent1})
with $\ell=2k$ and $k\ge m\ge 0$ generate the basis of quartic vertices.

For our purposes it is more convenient to use the traceless improvement of \eqref{bbvd}, given in  \cite{Anselmi:1999bb}\footnote{Note that the normalisation here is such that it is consistent with the normalisation of \eqref{bbvd}.}
\begin{align} \label{anselmi}
& {\tilde J}_{\mu_1 ...\mu_{\ell}}\big(\varphi_0(x_1),\varphi_0(x_2)\big) \\ \nonumber
& \hspace*{1cm}=  \frac{\sqrt{\pi}\;\Gamma\left(d+s-2\right)}{2^{s+d-3}\Gamma\left(\tfrac{d-1}{2}\right)\Gamma\left(s+\tfrac{d}{2}-1\right)} \sum_{k=0}^l a_k\, \partial_{\mu_1} ... \partial_{\mu_k} \varphi_0(x_1) \partial_{\mu_{k+1}} ... \partial_{\mu_{\ell}} \varphi_0(x_2)- \text{traces},
\end{align}
where 
\begin{equation*}
a_k=(-1)^k \frac{l!}{k! (l-k)!}\frac{(\frac{d}{2}-1)_l}{(\frac{d}{2}-1)_k(\frac{d}{2}-1)_{l-k}}\,.
\end{equation*}
The improvement \eqref{anselmi} can always be obtained from \eqref{bbvd} when the scalar field is conformal.

Analogously to (\ref{verticesequivalent1}), the flat space amplitude
corresponding to  a vertex
\begin{equation}
\label{verticesequivalent2}
{\cal V}_{m,\ell}\equiv\Box_{12}^m\, {\tilde J}_{\mu_1 ...\mu_{\ell}}\big(\varphi_0(x_1),\varphi_0(x_2)\big){\tilde J}^{{\mu_1 ...\mu_{\ell}}}\big(\varphi_0(x_3),\varphi_0(x_4)\big)
\end{equation}
with $\ell=2k$ generates a term $s^mt^ku^k$. Thus, vertices (\ref{verticesequivalent2})
with $\ell=2k$ and  $k\ge m\ge 0$ may be used as a basis for quartic interactions.

\end{appendix}

 \backmatter
  \bibliographystyle{unsrt}
\bibliography{refs}

\chapter*{Acknowledgements}
\addcontentsline{toc}{chapter}{\numberline{}Acknowledgements}
\markboth{Acknowledgements}{Acknowledgements}

\noindent
\emph{This work would not have been possible without a special set of boundary conditions:}

\emph{I am especially grateful to my supervisor Johanna Erdmenger, in many respects. For her continuous support, guidance and encouragement throughout these years. For introducing me to this fascinating subject and encouraging me to be independent from the very beginning.}

\emph{It is a privilege to acknowledge Dieter L\"ust, for his constant support and interest in my work. Moreover, for creating the fruitful and friendly atmosphere that permeates the String Theory group. Thank you also for refereeing this thesis.}

\emph{I am deeply indebted to Hugh Osborn, not only for valuable discussions about conformal field theory, but also for his unwavering support since my undergraduate years in Cambridge. In this regard I also owe a lot to Irena Borzym, Wayne Boucher and David Tong.}

\emph{My very special thanks go to my collaborators on the work which made this thesis: Xavier Bekaert, Mitya Ponomarev and Massimo Taronna. You guys taught me most of what I know about higher-spin theory, and I am also thankful for your patience, support and good advice.}

\emph{I have also benefited greatly from discussions and correspondence with: Martin Ammon, Francesco Azzurli, Glenn Barnich, Thomas Basile, Nicolas Boulanger, Andrea Campoleoni, Alejandra Castro, Shai Chester, Daniele Dorigoni, Dario Francia, Olga Gelfond, Gaston Giribet, Vasco Gon\c{c}alves, Thomas Grimm, Pan Kessel, Wei Li, Ruslan Metsaev, Miguel Paulos, Eric Perlmutter, Tassos Petkou, Ergin Sezgin, Jon Shock, Zhenya Skvortsov, Dmitri Sorokin, Stephan Stieberger, Bo Sundborg, Per Sundell, Marika Taylor, Misha Vasiliev and Alessandro Vichi. As well as my fellow colleagues and gauge-gravity group comrades, who have accompanied me over the years at the MPI and the LMU.}

\emph{I owe a lot to the Max-Planck-Institut f\"ur Physik, the Ludwigs Maximilians-Universit\"at M\"unchen and the IMPRS program. It was a great privilege to be able to work for my Ph.D. in such a stimulating scientific environment, in a charming city. At the MPI I am particularly grateful to the ever-welcoming Annette, Monika and Rosita, for providing continuous support which helped smoothen the navigation through all possible administrative and bureaucratic
labyrinths.}

\emph{My deepest gratitude goes to my family. Without their endless support,
I would have never made it close to this point. For continuously reminding me of what is truly important in life. This thesis is dedicated to them.}

\emph{Finally, I would like to thank all of my other teachers, colleagues and friends in Hull, Cambridge, Berlin, Munich and those scattered elsewhere around the world, who have helped me along the way.}

\emph{I also thank Eva Llabr\'es, Aninda Sinha and Rajesh Gopakumar for pointing out some typos after v1 of this thesis appeared on the arXiv.}

\end{document}